# Modeling biological networks

From single gene systems to large microbial communities

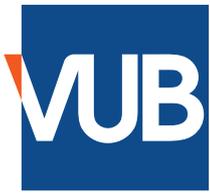
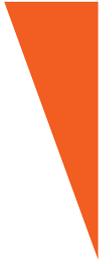

Thesis submitted in partial fulfilment of the requirements for the academic degree of Doctor of Sciences

# MODELING BIOLOGICAL NETWORKS

From single gene systems to large microbial communities

Lana Descheemaeker

October 2020

| | |
|---|---|
| Supervisor | Prof. dr. Sophie de Buyl |
| Jury members | Prof. dr. Alberto Mariotti (Vrije Universiteit Brussel) |
| | Prof. dr. ir. Eveline Peeters (Vrije Universiteit Brussel) |
| | Prof. dr. ir. Lendert Gelens (Katholieke Universiteit Leuven) |
| | Prof. dr. Aleksandra Walczak (École Normale Superieur, Paris) |
| | Dr. Jacopo Grilli (International Centre for Theoretical Physics, Trieste) |

**Sciences and Bio-Engineering Sciences**

This document was typeset with the help of KOMA-Script and LATEX using the kaobook class.

*If I have seen further, it is by standing on the shoulders of giants.*

<div align="right">Isaac Newton (1675)</div>

# Abstract

Biological systems are composed of many small components that interact in a non-trivial way. To study these systems as a whole, we can translate them into mathematical models. In this way, we can test different hypotheses and make predictions. In this research, we study biological networks at different scales: a gene autoregulatory network at the single-cell level and the gut microbiota at the population level.

Proteins are the main actors in cells, they are the building blocks, act as enzymes and antibodies. The production of proteins is mediated by transcription factors. In some cases, a protein acts as its own transcription factor, this is called autoregulation. It is known that autorepression speeds up the response and that autoactivation can lead to multiple stable equilibria. In this thesis, we study the effects of the combination of activation and repression in autoregulation, as a case study we investigate the possible dynamics of the leucine responsive protein B of the archaeon *Sulfolobus solfataricus* (Ss-LrpB), a protein that regulates itself in a unique and non-monotonic way via three binding boxes. We examine for which conditions this type of network leads to oscillations or bistability.

In the second part, much larger biological systems are considered. Ecological systems, among which the human gut microbiome, are characterized by heavy-tailed abundance profiles, this means that there are few abundant species and many rare. We study how these distributions can arise from population-based models by adding saturation effects and linear noise. Moreover, we examine different characteristics of experimental time series of microbial communities, such as the noise color and neutrality of the biodiversity, and look at the influence of the parameters on these characteristics.

With the first research topic we want to lay a foundation for the understanding of non-monotonic gene regulation and take the first steps toward synthetic biology in archaea. In the second part of the thesis, we investigate experimental time series from complex ecosystems and seek theoretical models reproducing all observed characteristics in view of building predictive models.

# Samenvatting

Biologische systemen bestaan uit vele kleinere onderdelen die op een niet-triviale manier met elkaar interageren. Om deze systemen als een geheel te bestuderen, kunnen we deze naar wiskundige modellen vertalen. Op die manier kunnen we verschillende hypotheses testen en voorspellingen maken. In dit onderzoek bestuderen we biologische netwerken van verschillende groottes: een genregulatorisch systeem op de schaal van een enkele cel tot het darmmicrobioom dat bestaat uit een 100 biljoen interagerende cellen.

Eiwitten behoren tot de belangrijkste actoren in cellen, ze vormen bouwstenen, handelen als enzymen en afweerstoffen. De productie van eiwitten wordt geregeld door transcriptiefactoren. In sommige gevallen handelt een eiwit als zijn eigen transcriptiefactor, dit heet autoregulatie. Het is bekend dat zelf-repressie de respons versnelt en dat zelf-activatie kan leiden tot meerdere stabiele evenwichtspunten. In deze thesis, bestuderen wij de effecten de combinatie van activatie en repressie in autoregulatie, als voorbeeld onderzoeken we de mogelijke dynamische toestanden van het leucine responsive protein B van het archaeon *Sulfolobus solfataricus* (Ss-LrpB), een eiwit dat zichzelf reguleert op een unieke en niet-monotone wijze via drie bindplaatsen. We bestuderen in welke gevallen dit type netwerk tot oscillaties of bistabiliteit leidt.

In het tweede luik van deze thesis worden biologische systemen van een veel grotere schaal beschouwd. Ecologische systemen, waaronder het menselijke darmmicrobioom, worden gekenmerkt door abundantie verdelingen met een dikke staart, dit wil zeggen dat een merendeel van de individuen uit slechts een een klein aantal soorten bestaat en dat vele soorten erg zeldzaam zijn. Wij bestuderen hoe deze verdelingen kunnen volgen uit veelgebruikte modellen door toevoeging van saturatie-effecten en lineaire ruis. Verder beschouwen we verschillende kenmerken van experimentele tijdsreeksen van microbiële gemeenschappen, zoals de kleur van de ruis en de neutraliteit van de biodiversiteit, en kijken we welke parameters deze kenmerken beïnvloeden.

Met het eerste onderzoek willen we een basis leggen voor het begrip van niet-monotone genregulatie en de eerste stappen nemen naar synthetische biologie in archaea. In het tweede deel van de thesis bestuderen we experimentele tijdsreeksen van complexe ecosystemen en zoeken we naar theoretische modellen die alle waargenomen eigenschappen kunnen reproduceren met het oog op het opstellen van voorspellende modellen.

# Résumé

Les systèmes biologiques sont composés de nombreux composants qui interagissent de manière non triviale. Une manière d'étudier ces systèmes dans leur ensemble est de les décrire via des modèles mathématiques. De cette manière, nous pouvons tester différentes hypothèses et faire des prédictions. Dans cette recherche, nous étudions les réseaux biologiques à différentes échelles: un réseau d'autorégulation d'un gène au niveau de la cellule unique et le microbiote intestinal au niveau de la population.

Les protéines sont les principaux acteurs dans les cellules, elles sont les éléments constitutifs, agissent comme des enzymes et des anticorps. La production de protéines est médiée par des facteurs de transcription. Dans certains cas, une protéine agit comme son propre facteur de transcription, c'est ce qu'on appelle l'autorégulation. On sait que l'autorépression accélère la réponse et que l'autoactivation peut conduire à de multiples équilibres stables. Dans cette thèse, nous étudions les effets de la combinaison de l'activation et de la répression dans l'autorégulation, comme cas d'étude, nous étudions la dynamique possible de Ss-LrpB (leucine responsive protein B) de l'archée *Sulfolobus solfataricus* (Ss-LrpB), une protéine qui régule elle-même de manière unique et non monotone via trois boîtes de liaison. Nous examinons pour quelles conditions ce type de réseau conduit à des oscillations ou à la bistabilité.

Dans la deuxième partie, des systèmes biologiques plus complexes sont considérés. Les systèmes écologiques, parmi lesquels le microbiome intestinal humain, sont caractérisés par des profils d'abondance à queue lourde, ce qui signifie qu'il y a peu d'espèces abondantes mais de nombreuses espèces rares. Nous étudions comment ces distributions peuvent provenir de modèles basés sur la population en ajoutant des effets de saturation et du bruit linéaire. De plus, nous examinons différentes caractéristiques des séries temporelles expérimentales des communautés microbiennes, telles que la couleur du bruit et la neutralité de la biodiversité, et examinons l'influence des paramètres sur ces caractéristiques.

Avec le premier sujet de recherche, notre but est de construire les bases de la compréhension de la régulation des gènes non monotoniques et faire les premiers pas vers la biologie synthétique chez les archées. Dans la deuxième partie de la thèse, nous étudions des séries temporelles expérimentales à partir d'écosystèmes complexes et proposons des modèles théoriques reproduisant toutes les caractéristiques observées en vue de construire des modèles prédictifs.

# Publication list

Some of the chapters of this thesis are published:

▶ Chapter 5

Descheemaeker, L., Peeters, E., & de Buyl, S. (2019). Non-monotonic auto-regulation in single gene circuits. *PLOS ONE*, *14*(5), e0216089. https://doi.org/10.1371/journal.pone.0216089

▶ Chapter 8

Descheemaeker, L., & de Buyl, S. (2020). Stochastic logistic models reproduce experimental time series of microbial communities. *eLife*, *9*, e55650. https://doi.org/10.7554/eLife.55650

Other publication:

▶ Descheemaeker, L., Ginis, V., Viaene, S., & Tassin, P. (2017). Optical Force Enhancement Using an Imaginary Vector Potential for Photons. *Physical Review Letters*, *119*(13), 137402. https://doi.org/10.1103/PhysRevLett.119.137402

# Acknowledgments

This thesis would not have been possible without the help of many people. Therefore, I would like to thank everyone who has offered help or supported me along the way.

I'm extremely grateful to my supervisor, Prof. Sophie de Buyl. Sophie, thank you for introducing me to the field of systems biology, for sharing all your scientific knowledge. Thank you for your neverending enthusiasm for research, for being kind and always ready to discuss.

I want to thank Alberto Mariotti, Eveline Peeters, Lendert Gelens, Aleksandra Walczak, and Jacopo Grilli for accepting the invitation to be a member of my PhD thesis committee. I want to thank Eveline Peeters for kindly introducing me to microbiology and the world of archaea. I am grateful to Aleksandra Walczak and Jacopo Grilli for the insightful scientific discussions. My sincere thanks go to Lendert Gelens for his valuable feedback both on the scientific content as its presentation.

Next, I want to show my deepest gratitude to the people who inspired me to become a PhD candidate and join the Applied Physics Research Group, the supervisors of my master's thesis: Sophie Viaene, Vincent Ginis, and Philippe Tassin. Thank you for sharing your passion for scientific research, for teaching me scientific writing and presentation skills, and for welcoming me in APHY. Thank you for encouraging me to pursue a PhD.

I have much enjoyed being a student in APHY. I want to thank Prof. Jan Danckaert, head of APHY, for giving me the opportunity to pursue a PhD. I want to thank all old and new members of APHY: Guy, Guy, Vincent, Lars, Mulham, Pedro, Gabin, Sophie, Lieve, Stefan, Krishan, Jaël, Fabian, Pieter, Ali, and Zaïd. I was able to share the joys and frustrations that come with research, teaching and academic life with you. I especially want to thank Jaël and Krishan with whom I shared an office during my 4-year-long PhD journey. I will definitely miss the games and puzzles.

For administrative support, I could always count on Merel Fabré, Sofie Van den Bussche, and Rebekah Hajir.

I would like to warmly thank my friends for the emotional support and the entertainment. Danku Maarten, Ruben, om de treinritten aangenaam en interessant te maken met politiek, sterrenkunde, go en fietsverhalen. Bedankt Lore, Febe, Lucie, Hélène, Sofie en Veerle voor de circusavonturen. Bedankt aan Hilde, Liesbet, Ann en Anja om de passie voor muziek en de cello te delen. Merci à Quentin, Mehdi, Arnaud et Nicolas pour les nombreuses soirées jeux. Dankjewel Céline om al zovele jaren een goede vriendin te zijn, bedankt voor alle ontspannende, creatieve en sportieve momenten.

Olivier, merci, merci pour tout! Merci aussi à Anne et Luc pour leur support.

I also had a lot of support from my family. I want to thank my parents and my sisters, Kari, Cloé, and Luca. Bedankt, mama en papa, om mij de vrijheid te geven om mijn eigen (studie)keuzes te maken en mij hierin te steunen.

<div style="text-align: right;">
Lana Descheemaeker<br>
Brussels, 2 October 2020
</div>

# Contents

## MICROBIAL COMMUNITIES 83

# List of Figures

# List of Tables

# Introduction | 1

> In physics we have dealt hitherto only with periodic crystals. To a humble physicist's mind, these are very interesting and complicated objects; they constitute one of the most fascinating and complex material structures by which inanimate nature puzzles his wits. Yet, compared with the aperiodic crystal, they are rather plain and dull. The difference in structure is of the same kind as that between an ordinary wallpaper in which the same pattern is repeated again and again in regular periodicity and a masterpiece of embroidery, say a Raphael tapestry, which shows no dull repetition, but an elaborate, coherent, meaningful design traced by the great master.
>
> (Erwin Schrödinger, What is life?)

LIVING CELLS ARE AMONG THE MOST COMPLEX SYSTEMS FOUND IN THE UNIVERSE. Their structure is rather peculiar. Unlike atoms in crystals, the components of living matter are not frozen in a lattice, but neither are they disordered, following random paths like molecules of gases. Furthermore, living systems often operate far from equilibrium with a constant flow of energy and information. Citing Alan Turing:

> Most of an organism, most of the time is developing from one pattern into another, rather than from homogeneity into a pattern.
>
> (Turing, 1952)

This concept of life existing on the border of stability is also known as self-organized criticality. It is found in ecological communities, neural networks, bird flocks, etc. (Mora & Bialek, 2011). One of the topics of this thesis is self-organized instability in the context of microbial communities.

## 1.1 Big questions in biology

Three of the big questions in biology, as presented by Erwin Frey (Frey, 2019), are

> - *the origin of life:* Living systems are self-replicating patterns: they evolve and copy themselves. We know how cells replicate, but no experiment could yet create a self-replicating system from inanimate matter[1]. How did life emerge from a jungle of molecules?
> - *the laws governing self-organization:* Many systems are combinations of a large number of smaller components. Ecosystems consist of many animals, plants, bacteria, etc. How do they remain stable? What are the principles governing complex systems?
> - *how to build synthetic cells:* Cells contain numerous genes, and the function of many of them is still left unexplained. How do all processes in a cell work together? Which parts can be removed? How can we add parts? Both bottom-up and top-down approaches are addressed by synthetic biologists to build a minimal viable cell. Moreover, synthetic biologists are building a toolbox of genetic circuits with specific dynamical behaviors. By combining these building blocks, they can engineer a cell to perform any function.

The subjects of this thesis are related to the two last questions. We study the self-organization of microbial communities and a genetic circuit of an archaeon to initiate the development of a toolbox for synthetic biology in archaea.

Biology studies living systems ranging from the small scale of the cell up to the large scale of communities and ecosystems. Four important challenges that are encountered when studying the big questions in biology are:

1. *Large numbers:* With the advent of high-throughput sequencing, large amounts of data have become available, such as the human genome. Homo sapiens has about 21,000 protein-coding genes. Not only the number of different proteins is large, also the copy number of the proteins is large. Because there is a linear correlation between the volume of a cell and its protein count, the number of proteins per cell ranges from $5 \cdot 10^4$ for *M. Pneumoniae*[2], to $2 \cdot 10^6$ in *E. Coli* and even $2 \cdot 10^9$ in *H. Sapiens*. In the same way as single cells are defined by large

[1]: Experiments have shown that proteins can be created in a mud pool, but they never combined to make a functional self-replicating system (Miller, 1953; Miller & Urey, 1959).

[2]: *Mycoplasma pneumoniae* are among the smallest organisms. They are only 100-200nm in width and 1-2µm in length. Stereomicroscopes are needed to examine these cells as they cannot be detected by light microscopy (Waites & Talkington, 2004).

amounts of different proteins, microbial communities such as the gut microbiome are defined by numerous microbes of around 500 different species (Guarner & Malagelada, 2003).
2. *Complex interactions:* At the level of a single cell, all genes work tightly together to allow the cell to grow and divide. At the larger scale of a multicellular organism, cells also need to communicate among each other to make the organism functional. For example, during embryogenesis, cell-cell communication is extremely important, but still in later stages is communication important to induce apoptosis in malignant or damaged cells to prevent tumor growth. Likewise, in the dynamics of communities interaction plays a large role: species appear as predator-prey couples, they compete for resources or cooperate through cross-feeding.
3. *Processes at different scales of time and space:* While the binding of molecules to DNA happens at the subsecond scale, the production of a protein takes minutes. Furthermore, cells often show a circadian rhythm[3] tracking the moment of the day or annual cycles that follow the seasons of the year. Combining all these timescales in one model is challenging. Similarly, the spatial scales of biological systems range from the size of the DNA to the size of the cell and even large ecosystems, such as forests or oceans (Figure 1.1).
4. *Stochasticity:* Some biological processes have small copy numbers of one or more of their components, think of the DNA in a cell or rare but important species in a community. In these cases, one cannot simply make approximations through averages. The inherent randomness of proteins binding to DNA can give rise to emergent properties and different dynamical behaviors. Also, mutations are dominated by stochasticity, but they are crucial for evolution.

3: Circadian means "almost daily". The circadian rhythm is an oscillation with a period of around 24 hours.

These four aspects of biology make it difficult to understand systems intuitively or to make qualitative or quantitative predictions. Different approaches to give meaning to the large amount of noisy data of interacting systems at different time and spatial scales have been developed. They belong to the field of *quantitative biology*. This interdisciplinary field combines techniques from mathematics, statistics, physics and computer science to solve biological problems. In physics, many analytical and numerical techniques have been devel-

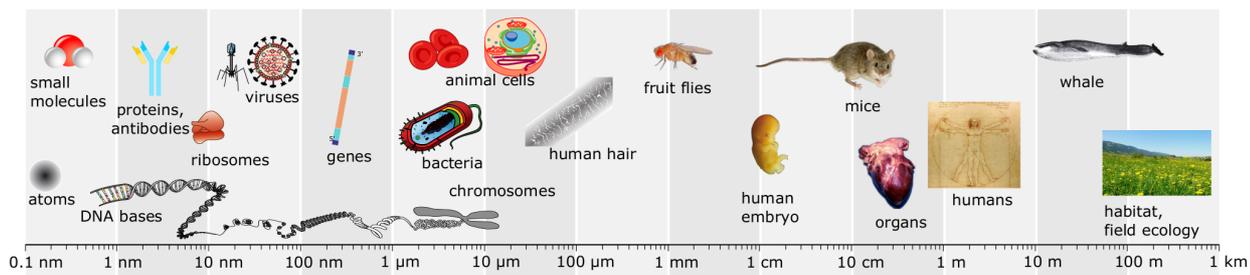

**Figure 1.1:** Biological systems at different scales

oped to make abstractions, build models, and predict qualitative behavior of complex systems. For this reason, applied physics has been used as a tool in biology for centuries. The interdisciplinary field that uses techniques of nonlinear dynamics to study a wide variety of complex biological systems is *systems biology*. Another subfield of quantitative biology is for example bioinformatics which uses techniques from data science and machine learning to find structure in the data via a top-down approach. Here follow some examples to illustrate the importance of quantitative biology:

▶ The field of quantitative biology is not young. Already at the end of the 18th century did Thomas Malthus, an English economist, propose a simple model for the growth of the human population (Malthus, 1798). This was the start of the study of population dynamics and the model, also called the exponential law, is considered as its first principle in analogy to Newton's first law for mechanics.

▶ The spread of infections can be modeled. Recently, much research has been done in the context of the Covid-19 pandemic. Dynamic and stochastic models have been proposed to understand its epidemic characteristics, predict the inflection point and ending time depending on different measures taken (Kucharski et al., 2020; Peng et al., 2020). Some models have been based on the Lotka-Volterra equations (Younes & Hasan, 2020), equations that were also studied in this thesis in the context of microbial community dynamics.

▶ Another topic that is extensively studied is cancer dynamics. A myriad of models for the origin, development, spread and treatment of this group of diseases exist (Azmi, 2012).

Above we gave some examples of complex systems that are studied by systems biology, but what exactly makes a system complex? This question is addressed in the next section.

## 1.2 Complex systems

For a long time, biologists have used the reductionist method and studied smaller and smaller parts of complex systems. However, the difficulty of biological questions does not only lie in understanding the details of all small processes, but also in creating a holistic view and understanding how all these processes work together. A famous saying goes "The whole is more than the sum of parts."[4] and this is without doubt applicable to biology. Understanding all the parts does not always mean understanding the whole (Goldstein, 1999). The difficulty arises through nonlinearities and sensitivity to initial conditions. A well-known example is weather prediction. Although different parts of the problem (temperature, air pressure, humidity, etc.) are elaborately studied, meteorologists can only predict the weather for 2 weeks at most and even then, they often get it wrong.

4: This statement is often attributed to Aristotle, but this is a misinterpretation (Goddu, 2010).

The complex systems that are studied in this thesis are a gene regulatory network and microbial communities. Because of the nonlinear feedback of the gene regulatory system, we cannot intuitively predict the dynamics of the system, and the behavior will depend on the parameters. Microbial communities exhibit emergent properties just like many ecosystems. These characteristics, such as the community's diversity, functional redundancy of different species, the stability of the community arise from the community as a whole and not from the individual species. The patterns emerging from the collective behavior of many smaller subunits are determined by processes occuring at different scales in time, space, and taxonomy. This problem of identifying the ecological origins of the structure of microbial communities is therefore also called the *problem of pattern and scale* (Levin, 1992).

## 1.3 Importance of modeling

Scientific models are representations of our understanding in an abstract and quantitative form. In this thesis, we use the language of mathematics to describe biological processes. We make assumptions about the studied systems, translate our knowledge into equations, and run simulations. Subsequently, we verify whether our predictions match reality. If this is the case, we can extend the model and test new aspects of the problem. If not, we discovered that at least one of the assumptions we made is incorrect and that the understanding of the underlying biological processes is still incomplete. Consequently, alterations to the assumptions and model can be made. This cycle of proposing and verifying models allows us to enlarge our knowledge.

Verifying a model through experiments might be very challenging, but theoretical models and simulations can explore different setups easily. Next, these in silico experiments provide a guide for experimentalists to the most promising measurements. In this theoretical framework, we also aim to obtain predictive power which is challenging in view of the nonlinearity of a vast majority of most biological systems.

The advantage of mathematical models is that in this framework, complex and nonlinear systems can be described by equations. These equations can deal with the large numbers by a mean-field approximation. Without modeling every individual protein, qualitative and quantitative predictions of the protein dynamics can be made. Although these equations can often not be solved analytically, Alan Turing postulated in 1952 that solutions can be offered by using numerical techniques and a machine (Turing, 1952). Ever since the advent of the computer, many results have been obtained through numerical simulations. Also techniques of statistical mechanics have been used to describe collective behavior in biology. Models expose the dynamics of complex systems, a task that is difficult to do intuitively.

## 1.4 Outline

Because the topic of this thesis is interdisciplinary—joining biology and techniques from nonlinear dynamics and statistical physics—there are two introductory chapters: the first one is a crash course on biology and the second one introduces nonlinear dynamics applied to biological problems. Next, the two main parts of this thesis follow: "Gene regulatory systems" and "Microbial communities".

In the first part, there are two chapters. The first chapter summarises the state-of-the-art of mathematics of single gene regulation. In the second chapter we present our work about non-monotonic single gene autoregulation which was published in PLOS ONE.

The second part of this thesis comprises four chapters. The first chapter gives an introduction to the biology of microbial communities, the second one resumes the state-of-the-art of community models. The third chapter presents our work on the characteristics of time series of microbial communities which was published in eLife, and in the fourth chapter our work about heavy-tailed abundance distributions of microbial communities is described.

In the final chapter, there is an outline of our conclusions and the outlook for future research.

After these main chapters, two appendices follow with technical details about dissimilarity and neutrality, and statistical distributions and methods used throughout the thesis. In the end, there are three appendices with detailed material and methods and supporting results for the three chapters describing our research.

We want to encourage open-source research and reproducibility. Therefore, chapters of this work were published in open-access journals (PLOS ONE and eLife). Additionally, we used the free and open-source programming language python to generate our results and made the code publicly available on github repositories:

https://github.com/lanadescheemaeker/

- Chapter 5: single_gene_systems
- Chapter 8: logistic_models
- Chapter 9: rank_abundance

# Introduction to biology | 2

There are millions of different organisms, and all are build out of cells. In this chapter, we first introduce cell biology and the central dogma of molecular biology. Next, we address the different organisms and their classification. We finish with a discussion of synthetic biology.

## 2.1 Introduction to cell biology

In the 19th century, the German physician, Rudolf Virchow, stated that "Omnis cellula e cellula", by which he meant that all cells stem from other pre-existing cells and that no free cells can be produced. Up until the day of today this statement holds, raising the question of how life could have formed from inanimate matter.

### DNA and genes

Cells reproduce through division and all information they need to live, grow, and reproduce is contained in a long polymer known as the *DNA*, short for deoxyribonucleic acid[1]. This information is contained in the molecular sequence of four different nucleotides which are defined by their nucleobases[2], adenine, cytosine, thymine and guanine, abbreviated to A, C, T and G respectively. The structure of DNA is the familiar double helix (Figure 2.1). Its structure was discovered by James Watson and Francis Crick in 1953 (Watson & Crick, 1953). Also Rosalind Franklin played a crucial role in the discovery, but only Watson and Crick were awarded the Nobel prize. The two composing strands of the DNA are bound together by hydrogen bonds. Moreover, they are complementary as the nucleobases appear in pairs: A and T, and, C and G will always be at opposite sides of the hydrogen bonds. Before cell division, all DNA is duplicated and every daughter cell receives one copy[3].

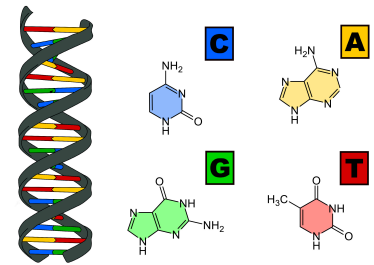

**Figure 2.1:** DNA is double stranded helix composed of four different nucleotides: adenine (A), thymine (T), cytosine (C) and guanine (G).

[1]: Only a limited number of cells such as red blood cells do not contain DNA and some viruses use RNA instead of DNA. Apart from these exceptions, all self-reproducing cellular organisms contain DNA and whether life without DNA can exist is still subject to debate (Hiyoshi et al., 2011).

[2]: Next to the nucleobase, each nucleotide contains a deoxyribose (sugar) and a phosphate group.

[3]: We here discussed the *vegetative division* or *mitosis*. In eukaryotes, there is also the *reproductive cell division* or *meiosis*, where the DNA is not duplicated before division such that each daughter cell ends up with only half of the DNA.

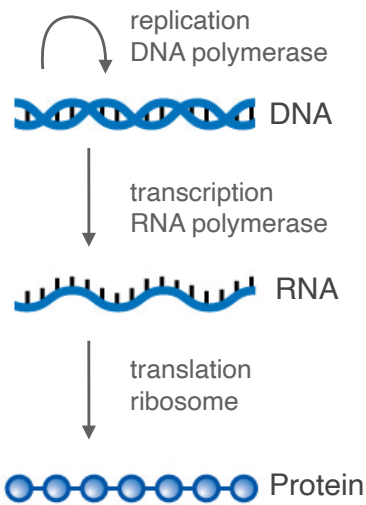

**Figure 2.2:** The central dogma of molecular biology: information flows from the DNA to proteins through transcription and translation processes.

4: In the human genome, the size of a gene ranges from $10^2$ to $2 \cdot 10^6$.

5: Changes in the DNA are known as mutations. Their origin can either be spontaneous or due to exposure to mutagens, such as chemical agents or ultraviolet light. Spontaneous mutations are rare and can be explained by the instability of nucleotides or errors during DNA replication (Lodish et al., 2000).

Although most cells will not be able to live and replicate without their DNA, DNA in itself is not at all sufficient to maintain life (Noble, 2008). The main actors in a cell are the proteins. They give structure to the cell, control DNA replication, act as enzymes and antibodies, etc. Typically, a cell has 20,000 different proteins of which 10% has copy numbers as high as 50,000, which means a cell contains at least $10^8$ proteins. DNA merely serves as the instruction book for when and how to produce proteins. How the information flows from the DNA to the proteins is described by the *central dogma of molecular biology*. The latter states that in cells, the information of the DNA is first copied into *mRNA*, short for messenger ribonucleic acid, by a protein called RNA polymerase (RNAP). This process is called transcription. Subsequently, the mRNA can be translated into proteins by the ribosomes. The flow of information thus goes from the DNA to the proteins. This mechanism is unidirectional and no RNA can be constructed from a given protein.

We said that most information a cell needs is contained in its DNA. This is in no way the same as saying that all nucleotides of the DNA are informative. In fact, more than 98% of the human genome is non-coding (Elgar & Vavouri, 2008), which means it is never transcribed. Some of this DNA has been identified to act as the 'metadata' for the coding regions. It tells the cell where the location of the information is, where the RNAP can bind for transcription. It also tells the cell when the proteins can be produced, some parts of the DNA serve as binding sites for signaling molecules to initiate or block transcription of their proteins.

The parts of the DNA that are transcribed are referred to as *genes*. However, there is no exact definition of a gene because of the many subtleties concerning transcription (Pearson, 2006). Furthermore, many genes do not code for proteins and their sizes vary by orders of magnitude[4]. The number of genes of a human body is vast (35,000-40,000 genes), but this does not necessarily reflect the complexity of the organism as the number of genes in grass is similar.

During the life time of a cell, the DNA remains mostly invariable[5], living cells, however, are highly dynamic. They need to respond to changing environments, adapt their strategy when the available nutrients are altered. Also, in multicellular organisms, the DNA is the same in all cells. However, cells can be highly different, a skin cell and an eye cell do not look alike and they also have to fulfill different functions.

The characteristics of a cell, such as its shape and function, are essentially determined by its protein composition. The information for protein synthesis is found in the DNA, but as stated by the central dogma of molecular biology, the information first needs to pass through an intermediate step, the mRNA. Regulation of protein concentration can thus happen at different stages.

## Transcription and post-transcriptional regulation

The first step in gene expression is the *transcription* process. Here a copy of a DNA sequence is made by the enzyme called RNA polymerase. This enzyme will bind to the promoter, the starting site of transcription which is upstream of the gene and where transcription machinery is assembled. Next, transcription is initiated: the strands of the DNA helix are separated and a sequence of nucleic acids is copied by finding the complementary nucleobases[6]. The resulting copy is called the mRNA and it contains the same information as the DNA. It is an instruction guide for making proteins.

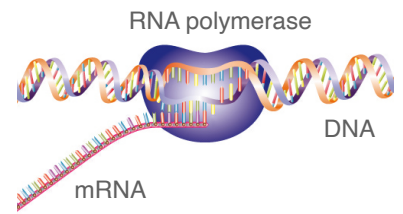

**Figure 2.3:** Transcription: an RNA polymerase copies the DNA sequence into an RNA strand.

6: There is one exception: for RNA adenine (A) gets complemented by uracil (U) which corresponds to thymine (T) in DNA.

The more mRNA there is, the more protein can be produced, as will be explained in the following section about translation. Therefore, the regulation of protein levels starts at the transcription level. The number of mRNAs is proportional to the rate at which RNAP binds to the promoter of the gene and initiates transcription. This rate is controlled by the regulatory sequence. Next to the promoter site, the regulatory sequence can also contain sites for *regulatory proteins*. We distinguish two main categories: repressors and activators. Repressor proteins bind to DNA and thereby prevent RNAPs from binding and initiating transcription. On the other hand, activator proteins can recruit RNAPs and increase the transcription of the gene.

In prokaryotes, the mRNA is immediately ready for translation after transcription. In eukaryotes, this primary transcript, called the precursor mRNA (pre-mRNA) often needs modifications to form mature mRNA. For example, many eukaryotic genes contain non-coding regions, called introns, which are removed from the primary transcript in a splicing process.

If the cell has a nucleus, the mRNA needs to be transported outside the nucleus to initiate translation by the ribosomes in the cytoplasm. If the cell has no nucleus, translation can occur simultaneously with the transcription. Coupled transcription and translation has been found in bacteria as well as archaea (French et al., 2007).

### Translation and post-translational regulation

Proteins are built out of around 20 different amino acids, but mRNA only contains four different nucleotides. The theory of combinatorics dictates that at least three nucleotides are needed to encode a minimum of 20 amino acids[7]. Experiments have shown that every amino acid is encoded by a set of three nucleobases, which is called a *codon*. Because there are more unique codons than amino acids, multiple codons result in the same amino acid[8]. This absence of a one-to-one mapping between codons and amino acids brings along multiple advantages (Itzkovitz & Alon, 2007).

The translation process takes place in the ribosomes, which in eukaryotes are situated in the cytoplasm outside the nucleus. During translation, a polypeptide chain, *i.e.* a long string of amino acids is created. In order to make a functional protein, it needs to be folded in the correct way to form a 3-dimensional structure[9]. Even after folding, some proteins need more post-translational regulation to activate them. A common post-translational mechanism is phosphorylation. Here one or more phosphoryl groups attach to the protein to (de)activate it. Another common post-translational modification is *multimerisation*: some proteins need to be combined into larger complexes. In our research, we study the system of Ss-LrpB, a homo-dimeric protein which thus consists of two identical monomers. Dimers are more stable than monomers, their degradation rate is lower. This effect is referred to as *cooperative stability* (Buchler et al., 2005).

In this section, an overview of the main components of gene regulation was presented. Of course, the field of gene expression is vast and we will not go into all the details and the innumerous exceptions. For our modeling approach of the gene regulatory system, we will not consider all biological details of the transcription and translation mechanisms. What is important, however, are the timescales of the considered

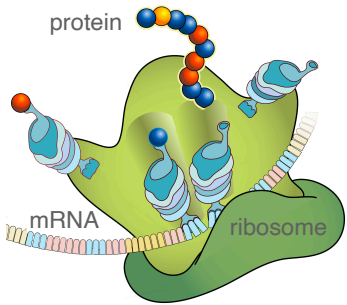

**Figure 2.4:** Translation: a ribosome copies the mRNA sequence into a protein. Every amino acid is translated from a set of three nucleobases, a codon.

7: Combinations of two nucleotides could encode at most $4^2 = 16$ amino acids and combinations of three nucleotides up to $4^3 = 64$ amino acids.

8: Some of the codons also serve as start and stop codons, which serve as initiation site and endpoint for translation. The start codon also encodes for an amino acid, but the stop codon not.

9: Folding can already start during the translation process. It happens rather fast, in the order of milliseconds (Kubelka et al., 2004).

processes, such that appropriate time delays can be attributed in the model and separation of timescales can be used where fit.

## 2.2 Tree of life

An important facet of biology is *classification*. We want to know how different species relate to each other based on shared characteristics (taxonomy) or from an evolutionary perspective (phylogenetics). For the latter, Carl Woese, Otto Kandler and Mark Wheelis proposed the *phylogenetic tree* in 1990 (Figure 2.5). They concluded that life can be separated into *three domains*: the eukaryotes, the bacteria, and the archaea, of which the two last ones can be grouped as prokaryotes.

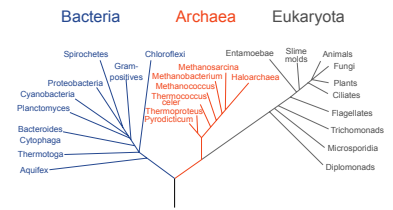

**Figure 2.5:** Phylogenetic tree: life can be classified into three domains: the bacteria, the archaea and the eukaryotes.

The main differences between prokaryotes and eukaryotes are:

- ▶ Eukaryotes have a *nucleus* and prokaryotes do not.
- ▶ Eukaryotes have membraneous *organelles* such as the endoplasmic reticulum, Golgi complex, and mitochondria. Prokaryotes do not.
- ▶ The *ribosomes* are specific to the domain of life.
- ▶ Eukaryotes have *linear chromosomes*, whereas prokaryotes have a single *cyclic DNA* and *plasmids*.
- ▶ Prokaryotes reproduce asexually with *binary fission*. After replication of the DNA, the cell divides into two daughter cells that can grow to be as large as the mother cell. Eukaryotes can also reproduce sexually, where two haploid[10] cells, called the gametes (such as the sperm and egg cells) combine.
- ▶ Prokaryotes can share genetic information between unrelated organisms through a process called *horizontal or lateral gene transfer*. This mechanism opposes vertical gene transfer where parents transfer their genes to their offspring.

10: Haploid cells only contain a single set of chromosomes, whereas diploid cells contain two sets.

As already mentioned in the section about transcription, there is a difference between pro- and eukaryotic transcription. Because of the spatial separation of the transcription machinery (DNA and RNAPs in nucleus) and translation machinery (ribosomes in cytoplasm) in eukaryotes, there is a time separation between the transcription and translation events. After the transcription, the mRNA is transported from the nucleus to the cytoplasm to the ribosomes for translation. In prokaryotes, there is no nucleus and both processes can happen simultaneously.

## Archaea

Archaea often only get limited attention in biology textbooks. Here, we will go more in depth into this domain of life because we studied a gene regulatory network of an archaeon, the *Sulfolobus solfataricus*.

Carl Woese, a renowned American microbiologist of the 20th century, one of the co-founders of the phylogenetic tree, was one of the first to study organisms by observing their DNA sequence. By considering the sequence, he found that a whole lot of species that were previously classified as bacteria were in fact biologically quite different. He called these the archaebacteria, later shortened to archaea (Woese & Fox, 1977).

The difference between archaea and bacteria is very fundamental (Pace, 2006). At the genetic level, it is larger than the difference between humans and plants, but it can only be noticed at molecular scales. It comes down to the presence or absence of molecules in the cell wall (bacteria use peptidoglycan, archaea do not) and the chemistry of the cell membrane (bacteria have a phospholipid bilayer with ester links, archaea use ether links). Another distinction is the presence of introns, which are non-coding sequences in genes, in archaea. Introns can also be found in eukaryotes, but not in bacteria. Also regarding the transcription machinery, many parallels can be drawn between archaea and eukaryotes and most archaeal transcription factors have homologs in bacteria (Kyrpides & Ouzounis, 1999). Likewise, the translation process in archaea shares features with the bacterial and archaeal translation processes (Bell & Jackson, 1998).

Archaea include many *extremophiles*, organisms that thrive in environments that are considered extreme for most forms of life we know. Such environments encompass extreme high and low temperatures, high (alkaline) and low (acidic) pH values, the presence of a large amount of radiation, pressure, or high salt concentrations. The particular archaeon that we use as a case study in Chapter 5, the *Sulfolobus solfataricus*, is a thermoacidophile: it grows best at 80°C and a pH level between 2 and 4. They were first discovered in the Solfatara volcano near Naples in Italy (Figure 2.6), hence their name.

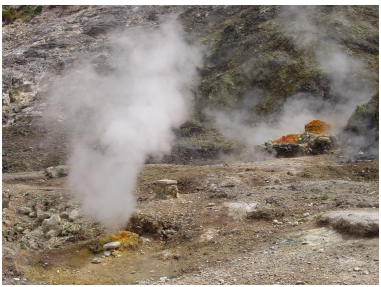

**Figure 2.6:** The Solfatara volcano. The archaeon *Sulfolobus solfataricus* was first discovered in these extreme conditions.

Their remarkable living conditions often make them difficult to study, but all the more interesting for both scientific and industrial purposes. They may be more suitable to grow on an industrial scale than the common lab organisms. I quote Christopher Voigt:

> More and more labs are taking on arcane organisms – I think the *S. cerevisiae* and *E. coli* dominance is dropping. (Eisenstein, 2016)

The enzymes and proteins of extremophilic archaea are unusually stable and can, therefore, be used in extreme industrial environments. Archaea have the potential for being used as biomass, for biogas production, in waste treatment processes, etc. (Schiraldi et al., 2002). Furthermore, some gene regulatory systems of archaea are not found in bacteria. Synthetic biologists can transpose these modules to bacteria, e.g. *E. Coli*, to reduce interference with the original genome of the cell. Using modules of different domains of life, a toolbox with orthogonal modules can be constructed. For all these reasons, the development of synthetic biology with archaea is very promising.

Although archaea include many extremophiles, they are commonly mesophilic and appear in and on human bodies (Lurie-Weinberger & Gophna, 2015). No archaeon has yet been identified as a pathogen, however, archaea have definitely the possibility to cause illness (Cavicchioli et al., 2003; Eckburg et al., 2003). The absence of identified pathogens is probably due to a lack of investment in research on archaea.

### Definition of species

The number of species is large, estimates run between 8 and 8.7 million and about 90% of all species still needs to be described (Mora et al., 2011). One interesting aspect of ecosystems is their diversity of species. In order to study diversity, we need to classify all individuals into categories and subsequently, we can compare the numbers of individuals for all groups. We already discussed the three domains of life, but many smaller classification units exist (Figure 2.7). The unit at the lowest level is the species. But what exactly is a species? This question is known as the *species problem* and its answer is not as simple as it may seem. There are in fact multiple definitions for species, each one having their proper challenges. Here follows a quote from Charles Darwin from his famous book *On the origin of species*:

> No one definition has as yet satisfied all naturalists; yet every naturalist knows vaguely what he means when he speaks of a species. Generally the term includes the unknown element of a distinct act of creation. (Darwin & Mayr, 2003)

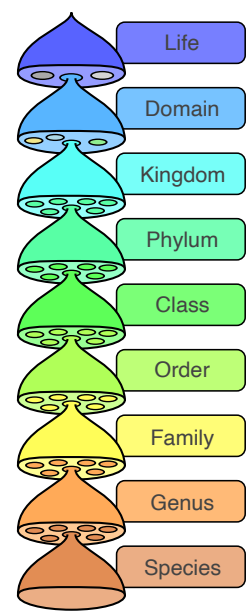

**Figure 2.7:** Biological classification scheme: life is classified at different taxonomic levels.

One of the older definitions is based on reproduction. The *reproductive species* is defined as the group of organisms in which two individuals of the required sexes[11] can produce fertile offspring[12]. Other definitions are for example based on the morphology of the individuals.

11: Not all species have binary physical sexuality, some organisms have more sexes or mating types, such as the *Tetrahymena thermophila* that has seven.

12: *Hybrids* are the offspring of two different species such as mules from a male donkey and a female horse, ligers from a male lion and a female tiger, or tigons from a male tiger and a female lion. *Haldane's rule* states that if one of the sexes of this hybrid is absent, rare or sterile, it is the heterozygous sex. For the given examples, the male is the sterile sex.

A problem with the first definition becomes immediately apparent in the case of asexual reproduction such as for bacteria. One could base the definition of species on its DNA sequence, but due to mutations in the DNA, there will be variations between individuals. Furthermore, prokaryotes can exchange DNA through horizontal gene transfer, making it impossible to use the complete genome as a classification tool.

13: Next to mRNA, there exist other types of RNA such as ribosomal RNA, or rRNA. The 16S and 18S rRNA are small subunits of the ribosome. Their genes evolve very slowly and are therefore excellent candidates for classification.

There is no theoretical definition of a species, but operational definitions exist such as the widely used *operational taxonomic unit* (OTU). The 16S or 18S rRNA[13] marker gene sequences are read for prokaryotes or eukaryotes respectively, and when the similarity between individuals is higher than a particular threshold (often 97%) they are classified as the same OTU (Chen et al., 2013). This method of classification was pioneered by Carl Woese and maintains ecological consistency when hierarchical complete linkage clustering is used for OTU clustering (Schmidt et al., 2014). In our study of the composition of microbial communities (Part 'Microbial communities'), the OTU is used as the classification unit.

## 2.3 Synthetic biology

Besides studying the dynamics of natural systems, there is a large field in biology which engineers new cell biological systems: *synthetic biology* (Cameron et al., 2014; Khalil & Collins, 2010; Mukherji & van Oudenaarden, 2009; Purnick & Weiss, 2009). This field was pioneered in 1961 by François Jacob and Jacques Monod with their study of the *lac* operon in *E. Coli* (Monod & Jacob, 1961), but the "nucleic acid nanotechnology" has developed tremendously, as a result of the growing catalog of cellular components and their interactions, the cloning, and the recombination of existing DNA. Two of the most influential synthetic gene regulatory networks are the toggle switch of Collins (Gardner et al., 2000)—a bistable network— and the repressilator of Elowitz and Leibler, 2000—an oscillator. The goal of synthetic biology is to forward-engineer cellular behavior and design new gene-regulatory networks.

The development of biotechnology contributes to many health applications. Cells are engineered to produce high-valuable chemicals, such as drugs for malaria (Martin et al., 2003; Ro et al., 2006) or cancer[14]. Synthetic biology has also many applications in other industries such as biofuel production, waste treatment, or commodity chemical production.

[14]: The Kid-Kis system in *E. Coli* kill a predetermined population of human cells. They can be used to induce apoptosis in cancer cells (Preston et al., 2016).

### Toolbox and modeling

Synthetic biology needs to have an extended toolbox and good designs for assembly of the components. Inspired by electronics, biological equivalents of electronic components to keep time, provide memory, or logic gate behavior have been developed (Stanton et al., 2014; Tamsir et al., 2011). These and other functions are obtained by combining different genes and the resulting networks—toggle switches, oscillators, cascades, pulse generators, sensors, actuators—serve as fundamental blocks of the toolbox of synthetic biology (Atkinson et al., 2003). An accurate understanding of all the modules is required to facilitate network building with predictable behavior (Le Novère, 2015). The parameters of the modules, such as delay, binding rates, or degradation rates can often be tuned. For example, the transcription rate is defined by the strength of the promoter, which is determined by the position of the transcription factors. Combinatorial promoter libraries collect information of different promoter designs and serve as a guide for the construction of networks (Cox et al., 2007; Rydenfelt et al., 2014). Mathematical models can help predict the dynamical behavior for different parameter sets. Specified functions can also be found by the aid of evolutionary procedures in silico (François & Hakim, 2004), such as oscillating gene regulatory network motifs (van Dorp et al., 2013).

Compared to synthetic biology in eukaryotes and bacteria, synthetic biology in archaea is still very much in its infancy (Atomi et al., 2012). In Chapter 5, we study a gene regulatory system of the archaeon *Sulfolobus solfataricus*. In addition to attempt to predict the natural behavior, we study parameter space to determine parameters sets that lead to interesting dynamical behavior. The ultimate goal is to manipulate the system such that it can perform the desired function in synthetic networks.

# Introduction to nonlinear dynamics | 3

MANY NATURAL SYSTEMS BEHAVE IN NONLINEAR WAYS, they contain many variables that work together in a multiplicative way or via other more complicated relations. In this chapter, we present a short introduction to nonlinear dynamics applied to biological systems. We describe the concepts used in our work about gene regulation and microbial communities. The field of nonlinear dynamics is, however, much wider and extensive introductions into nonlinear dynamics and systems biology can be found in Ingalls, 2013; Strogatz, 2015. Before delving into the realm of nonlinear dynamics, let us shortly revisit what are linear relationships and why we like them so much in physics. Linear maps $f$ are defined such that they preserve vector addition[1] and scalar multiplication[2]. Because of this first property, linear systems can be separated into their elemental components, which can each be solved independently and in the end, the components can be recombined by the *superposition principle*. This scheme makes it easy to keep track of the influence of all components and to intuitively predict the outcome for changing input parameters. However, in nature, many systems do not behave linearly and the nonlinearities that are created by interference, cooperation, and competition, make it very difficult to maintain an intuitive feeling about the resulting behaviors. The advantage of mathematical models is that they can quantitatively predict the behavior of the system even when many of the components show nonlinear responses. Biologists, who have long relied on conceptual models with verbal descriptions of the mechanisms, make use of nonlinear dynamics to offer an understanding of the increasing amount of discovered components and nonlinearities.

1: $f(\vec{x} + \vec{y}) = f(\vec{x}) + f(\vec{y})$
2: $f(a\vec{x}) = a f(\vec{x})$

## 3.1 Mass-action kinetics

Biological systems consist of many parts that are interacting. To write down the mechanisms as sets of equations, we can use the framework that was originally developed for chemical reaction networks. An important principle of this technique is the law of *mass-action kinetics* (Chou & Voit, 2009). This law states that the reaction rate, *i.e.* the probability of a reaction happening, is proportional to the concentrations of the participating components. This idea is based on the fact that a reaction takes place when the reactants collide and that the probability of collision depends on the concentrations of all reactants.

Using the law of mass-action kinetics, all gene regulatory networks can be translated to ordinary differential equations (ODEs). We will here give the example of dimer formation[3]. Consider two proteins $A$ and $B$ which together can form the dimer $AB$. Using the notation of chemical equations we write

$$A + B \underset{q}{\overset{k}{\rightleftharpoons}} AB \qquad (3.1)$$

3: Dimers are complexes formed by two proteins. These can either be identical or different, in which cases they are called *homodimers* or *heterodimers* respectively. More generally, complexes of two or more monomers are called *multimers* where $k$ is the dimerization rate and $q$ the dissociation rate. The corresponding differential equation reads

$$[\dot{AB}] = \underbrace{-q[AB]}_{\text{dimer dissociation}} + \underbrace{k[A][B]}_{\text{dimerization}} \qquad (3.2)$$

where the bracket notation $[X]$ was used to denote the concentration of protein $X$. When writing this equation, we relied on two important assumptions. The first one, dictated by the law of mass action, is that the system needs to be *well mixed*. The reaction rates need to be independent of space. The second assumption is that the numbers of the reactants (i.c. $A$, $B$, and $AB$) are large. This is needed to approximate the concentrations of the reactants ($[A]$, $[B]$, and $[AB]$) as continuous variables instead of a set of discrete numbers.

4: Dilution of proteins in a cell is a direct result of cell growth. Because we do not consider active degradation that depends on the protein concentration, both the degradation and dilution processes are treated in the same way in our equation. They both reduce the amount of proteins in a linear fashion.

We showed the example of dimer formation. But, using mass-action kinetics, any other biological process can be written as a set of differential equations. For example, the degradation and dilution[4] process of proteins,

$$x \xrightarrow{\gamma} \emptyset, \qquad (3.3)$$

can be resumed to

$$\dot{x} = -\gamma x \tag{3.4}$$

with $\gamma$ the combined degradation and dilution rate. The solution of this ODE is the exponential decreasing function $x_0 \exp\{-\gamma t\}$. For many proteins in a healthy cell, the concentration does not go to zero because there is the production of the protein from gene regulation:

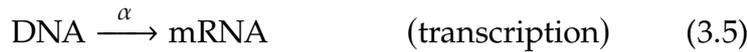
$$\text{DNA} \xrightarrow{\alpha} \text{mRNA} \qquad \text{(transcription)} \tag{3.5}$$

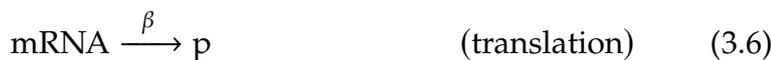
$$\text{mRNA} \xrightarrow{\beta} \text{p} \qquad \text{(translation)} \tag{3.6}$$

where $\alpha$ and $\beta$ represent the transcription and translation rate, respectively. Notice that for these processes, we only wrote DNA and mRNA on the left-hand side of our mass-action process. The transcription and translation rates definitely depend on the number of RNAPs and ribosomes in addition to the availability of nucleotides and amino acids, but we assume that the abundance of the aforementioned molecules is high so that the limiting factor in the cell for these processes is the amount of DNA and mRNA[5]. However, one needs to be aware of this assumption because, in some cases, the limiting factor for translation is not the number of mRNAs but the number of ribosomes and likewise for transcription and the number of RNAPs (Kafri et al., 2015).

5: A model including the RNAPs is provided by (Bintu et al., 2005). Even if all the mentioned elements were to be included in the model, there are still many aspects to transcription and translation such as the initiation, elongation, and termination which are omitted from the model.

The total rate of change of mRNA and protein is equal to the difference of the production rate ($\alpha$ for mRNA and $\beta$ for protein) and the degradation/dilution rate ($\gamma_{\text{mRNA}}$ and $\gamma_p$). The resulting equations are:

$$\dot{\text{mRNA}} = \alpha \, \text{DNA} - \gamma_{\text{mRNA}} \, \text{mRNA}, \tag{3.7}$$
$$\dot{p} = \beta \, \text{mRNA} - \gamma_p \, p. \tag{3.8}$$

Once the set of equations is written down, we can determine the dynamics by analytical or numerical solutions. Such techniques are discussed in the following section.

## 3.2 Solving the equations

Once the biological system is summarized to differential equations, we can solve them to predict the time evolution of the variables given any initial condition. We first study the steady states of the equation. A *steady state* is defined as a state that is in equilibrium, *i.e.* for which the rate of change is zero. Obtaining the steady states can thus be done by equalling the differential equations to zero. For the earlier example of the dimer (Equation 3.2), we obtain

$$[AB]_{ss} = \frac{k}{q}[A]_{ss}[B]_{ss}. \tag{3.9}$$

Being in steady state does not mean that no proteins are dimerizing or that no dimers are dissociating, it means that both these back and forth reactions happen at the same rate. This concept is called *detailed balance*. It also means that, in this steady state, there is time-reversal symmetry.

We now calculate the steady state of the protein concentration (Equation 3.8),

$$p_{ss} = \frac{\beta}{\gamma_p} \text{mRNA}_{ss} \tag{3.10}$$

where the subscript $_{ss}$ denotes the steady-state concentration. Notice that for this process, there is no time-reversal symmetry as the process (Equation 3.6) only goes in one direction as dictated by the central dogma of molecular biology.

An explicit expression for the variables in steady state cannot always be found, because nonlinearities make the equations hard to solve analytically. In this case, numerical methods can be used to determine the steady state for a given set of parameters.

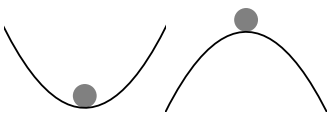

6: The classical textbook example to picture the stability of equilibria is a ball in a landscape. A ball in a valley or on a hilltop correspond to stable and unstable states, respectively. Both will remain in same position if no perturbation is added. This is what makes it equilibria. But a small perturbation makes clear the difference between both states: a ball on the hilltop will roll away from its equilibrium position and a ball in a valley will return to its position.

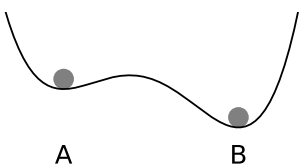

7: Using the analogy of the ball again, we can imagine a landscape with multiple valleys. The ball is in equilibrium in valley *A*, but given a big enough perturbation over the hilltop the ball will end up in a new equilibrium *B*. Therefore equilibrium *A* is not globally stable.

## Stability of fixed points

Steady states are the equilibria of the system. These can be either stable or unstable[6]. The question of stability can be addressed in a local or global context[7]. Because proving global stability can be very challenging, we will consider local stability. To verify whether a state is locally stable or unstable, one needs to look at what happens if any infinitesimaly small perturbation is added to the state. If all trajectories return to the state it is said to be locally stable or *attractive*, if any trajectory goes away from this point, the state is said to be unstable or *repulsive*. These repulsive trajectories could go to another steady state which is stable, to a limit cycle where the state oscillates, or when there are at least three dimensions, to a chaotic attractor. The latter two dynamic solutions are discussed in more details in the next section. There are two other cases for which the distance of the trajectory to the steady state does not grow smaller or larger in time: there is a line of steady states[8] or the steady state is a center.

8: A line of stable states can be represented by a ball on flat terrain.

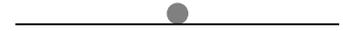

We first consider a one-dimensional system. With the ODE defined as $\dot{x} = f(x)$, the Taylor expansion around a point $x_0$ can be written as

$$f(x) = f(x_0) + \frac{df}{dx}(x_0) \cdot (x - x_0) + \mathcal{O}((x - x_0)^2). \quad (3.11)$$

For a small perturbation $|x - x_0| \ll 1$, the higher-order terms $\mathcal{O}((x - x_0)^2)$ can be safely neglected and $f(x)$ is approximated by a linear function. This is why this technique to study local stability is known as *linear stability analysis*. By evaluating this equation around the steady state ($x_0 = x_{ss}$ and $f(x_0) = 0$), we obtain

$$f(x) = \frac{df}{dx}(x_{ss}) \cdot (x - x_{ss}). \quad (3.12)$$

This equation dictates that the derivative $\frac{df}{dx}(x_0)$ needs to be negative to obtain stability. In that way any small positive (negative) perturbation, results in a negative (positive) derivative $f(x)$ which leads the system back to the steady state. When the derivative $\frac{df}{dx}(x_0)$ is exactly zero and the second derivative $\frac{d^2 f}{dx^2}(x_0)$ is non-zero, the fixed point is called a *saddle node fixed point*. A saddle-node is half-stable: it attracts trajectories from one side and repels it from the other.

For a one-dimensional system, one can represent the velocity of the state as a function of the concentration in a *rate portrait*. In Figure 3.1, an example is give for $f(x) = ax - bx^2$. The points where the rate vanishes ($f(x) = 0$), are $x = 0$ and $x = a/b$. These are the steady states. For concentrations with a positive (negative) rate, the concentration increases (decreases), which we denote by an arrow to the right (left). It is thus clear that the steady state at $x = 0$ is unstable, as a small perturbation applied to this steady state would drive the system away from this point towards another steady state that is stable. In this case, that would be the other steady state $x = a/b$. Notice that the derivative of the rate in the unstable

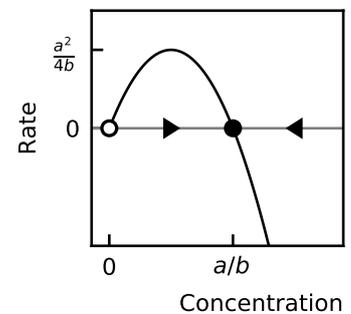

**Figure 3.1:** Rate portrait of $f(x) = ax - bx^2$

and stable states are respectively positive and negative, as we derived in the paragraph above.

A second example includes a half-stable steady state or saddle node. For the rate $f(x) = -x^3 + 2ax^2 - a^2x$, there are two steady states. The first $x = 0$ is stable and the second $x = a$ is half-stable. It attracts trajectories from the right and repels trajectories from the left. These rate portraits prove to be very useful in studying the steady states and the rate at which a system goes towards a steady state. We, therefore, use them in Chapter 4 and Chapter 5.

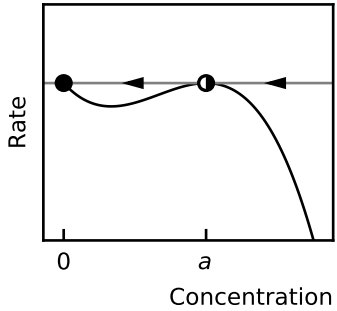

**Figure 3.2:** Rate portrait of $f(x) = -x^3 + 2ax^2 - a^2x$. The system has two steady states $x = 0$ and $x = a$. The derivative $df/dx$ is negative in $x = 0$, therefore, $x = 0$ is a stable steady state. The derivative of $df/dx$ is zero in $x = a$, it is stable from the right, but unstable from the left.

The stability analysis of steady states can be generalized to higher dimensions. In the $N$-dimensional case, we define a derivative for every variable $x_i$

$$\frac{dx_i(t)}{dt} = f_i(x(t)) \tag{3.13}$$

and the steady state $x_{ss}$ is defined such that $f_i(x_{ss}) = 0$ for every $i$.

The Taylor expansions of the derivative functions are

$$f(x) = f(x_{ss}) + J|_{x_{ss}}(x - x_{ss}) + \mathcal{O}((x - x_{ss})^2) \tag{3.14}$$

with $J$ the Jacobian, an $N \times N$-matrix with the derivatives of all $f_i$ with respect to all $x_j$,

$$J_{ij} = \frac{\partial f_i(x)}{\partial x_j}. \tag{3.15}$$

In the context of community dynamics, this Jacobian evaluated in a fixed point is referred to as the *community matrix* $M = J|_{x_{ss}}$ (Levins, 1968). A system with multiple equilibria will, therefore, have multiple community matrices. Furthermore, different systems can lead to the same community matrix.

Using the definition of the community matrix $M$, it follows from Equation 3.14 that

$$\frac{d\Delta x(t)}{dt} = \frac{dx(t)}{dt} - \frac{dx_{ss}}{dt} = \frac{dx(t)}{dt} = f(x(t)) \approx M\Delta(x) \tag{3.16}$$

where we defined $\Delta x = x - x_{ss}$ and used that $x_{ss}$ is a constant and $f(x_{ss}) = 0$ by definition. The solution of this differential

equation is

$$\Delta x(t) = \Delta x(0) \exp\{Mt\} = \Delta x(0) V \exp\{\Lambda t\} V^{-1} \quad (3.17)$$

with $V$ a matrix with the eigenvectors and $\Lambda$ a diagonal matrix with the eigenvalues of $M$. It follows that the eigenvalues of Jacobian determine the stability given they are not zero. This result is referred to as the *Hartman-Grobman theorem*, which states that

> the local phase portrait near a hyperbolic fixed point is 'topologically equivalent' to the phase portrait of the linearization . . . [and] topologically equivalent means that there is a homeomorphism (a continuous deformation with a continuous inverse) that maps one local phase portrait onto the other such that trajectories map onto trajectories and the sense of time . . . is preserved.
> 
> (Strogatz, 2015)

Separating the eigenvalues $\lambda$ of the community matrix $M$ into their real and imaginary part,

$$\text{Re}\{\exp \lambda t\} = \cos(\text{Im}(\lambda)t) \exp(\text{Re}(\lambda)t),$$

we see that the steady state is only stabilizing when all real parts of the eigenvalues are negative. Any positive real part will drive the trajectories away from equilibrium after a perturbation. If additionally, the imaginary part in non-zero, the trajectories show oscillatory behavior.

### Limit cycles and chaotic attractors

In one-dimensional systems, the important features of the system are the fixed points and their stability. Trajectories can only lead to the fixed point or infinity[9]. In two and three dimensions, two other types of qualitative behavior of solutions exist: limit cycles and chaotic attractors. Limit cycles are isolated closed trajectories. Like fixed points, they can be stable, unstable or exceptionally, half-stable. Stable limit cycles are used in systems biology to describe oscillatory dynamics. The *Poincaré-Bendixson theorem* states that two-dimensional systems can only exhibit fixed points and limit cycles. In three or more dimensions, chaos can appear. Chaos is deterministic behavior that is aperiodic and very sensitive

[9]: Periodic solutions can be obtained if the chosen one-dimensional system is a circle (Strogatz, 2015)

to the initial conditions. Because chaotic systems are very difficult to predict, we do not expect them to appear in biological systems. Surprisingly, occasionally chaotic systems are found in nature, they remain an active field of research (Benincà et al., 2008; Lloyd & Lloyd, 1995; Scharf, 2017; Skinner, 1994; Toker et al., 2020).

**Bifurcation theory**

Through the variation of parameters the phase portrait of the system changes. When the phase portrait changes into a topologically nonequivalent state, we say that a *bifurcation* has happened. This is for example the case when by variation of the parameters, the eigenvalues of the Jacobian evaluated in the steady state cross the imaginary axis: at the critical value a limit cycle appears. This particular bifurcation is called the *Hopf bifurcation*. It is a local bifurcation because the limit cycle emerges from the fixed point. Other global bifurcations of cycles exist[10], but in our search for oscillatory behavior in the genetic circuit of Ss-LrpB (Chapter 5), we rely on the Hopf bifurcation. Additionally, there is a whole range of bifurcations that do not involve limit cycles, bifurcations that change the stability of fixed points or include chaotic attractors. We will not discuss these, but an overview can be found in Strogatz, 2015.

10: We refer to Strogatz, 2015 for an overview.

## 3.3 Nonlinear response curves

Many biological processes are complicated and require intermediate steps. To construct a model, we extract a limited number of variables to describe the system of interest and relate the outputs to the inputs, typically with a nonlinear response function. To reduce the model complexity, we can apply *separation of timescales*. In this reduction method, intermediate steps are removed. We first elaborate on this technique and subsequently use it to derive nonlinear response curves in enzyme kinetics.

## Separation of timescales

In biology, the timescales of different mechanisms can vary over orders of magnitude. They go from (milli)seconds for binding of signaling molecules, hours for circadian rhythms, and months for seasonal changes. An example of the timescales for gene regulation in *E. Coli* is given in Table 3.1.

| process | timescale |
| --- | --- |
| (Un)binding of a small molecule to a transcription factor to change its activity | 1 msec |
| (Un)binding of a transcription factor to DNA | 1 sec |
| Transcription and translation of a gene | minutes |
| 50% change in concentration of a stable protein | 1 hour (cell generation) |

**Table 3.1:** Timescales of gene regulation in *E. Coli* (Alon, 2019)

When multiple processes occur at different timescales, we can consider the variables with the fast reactions to be in their equilibrium concentration with respect to the slow variables. This assumption of rapid equilibrium is also referred to as the *quasi-steady-state approximation*. Just like the method to calculate a regular steady state, the quasi-steady state is found by equalling the time derivative of the variable in the fast reactions to zero. The idea of timescale separation was first introduced by Michaelis and Menten (Gunawardena, 2014).

## Michaelis-Menten kinetics and the Hill curve

In 1913, Leonor Michaelis and Maud Menten proposed a reduced model for enzyme-catalyzed reactions. The individual chemical reactions to go from the substrate S to the product P are

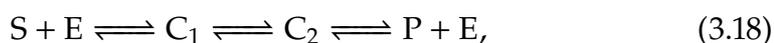

$$S + E \rightleftharpoons C_1 \rightleftharpoons C_2 \rightleftharpoons P + E, \quad (3.18)$$

with E the enzyme, $C_1$ the enzyme-substrate complex and $C_2$ the enzyme-product complex. Using time-scale separation and the assumption that the back reaction from product and enzyme to $C_2$ never occurs[11], the rate to go from substrate to

11: The complete derivation can be found in Ingalls, 2013.

product can be written as

$$\frac{V_{\max} s}{K_M + s}$$

with $s$ the concentration of the substrate, $V_{\max}$ the maximal rate and $K_M$ the half-saturation concentration. Because of the shape of this rate, it is called a hyperbolic rate. This rate law is known as *Michaelis-Menten kinetics*.

Archibald Hill generalized this function such that it can fit empiric responses without knowledge of the underlying networks. The expression of this *Hill function* is

$$\frac{V_{\max} x^n}{K^n + x^n} \qquad (3.19)$$

where $n$ is the *Hill coefficient*. It proves particularly useful when there is cooperativity. In this case, the Hill coefficient $n$ dictates whether there is positive cooperation ($n > 1$), negative cooperation ($n < 1$), or no cooperation ($n = 1$). This function is increasing for an increasing variable $x$ and the Hill coefficient $n$ denotes the slope of the curve: the higher $n$, the steeper the slope. It is therefore well suited to model a gene that is activated by a transcription factor[12]. A similar expression exists for a down-regulated gene

$$\frac{V_{\max} K^n}{K^n + x^n}.$$

This function is decreasing. Examples of both Hill curves are given in Figure 3.3.

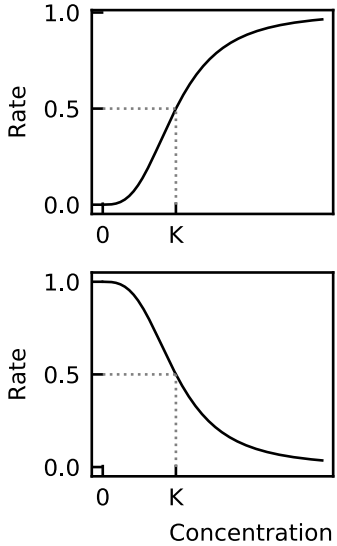

**Figure 3.3:** Increasing and decreasing Hill curves.

12: The ODE of a protein $X$ up-regulated by transcription factor $Y$, would be

$$\frac{d[X]}{dt} = \frac{V_{\max}[Y]^n}{K^n + [Y]^n}.$$

Typically $n$ represents the number of transcription factors that need to bind for maximal transcription rate. Therefore, multiple binding sites enhance the nonlinear effect.

## 3.4 Stochastic equations

We mentioned earlier that one of the constraints of using mass-action kinetics to write a set of ODEs is that the numbers of proteins need to be large. For large numbers, the randomness of collision events between the different components are averaged out. For small numbers, other equations are needed.

There are two approaches: one based on probability distributions and the other on numerical simulations of trajectories. The probabilistic method was born with the *Kolmogorov equations* which model the time evolution of the probability distributions of discrete stochastic processes. A simplified version, the *master equation*, is a set of differential equations dictated by a transition rate matrix that specifies the rate at which the system goes from one state to another.

In this thesis, we use the other, numerical method which is a direct simulation of different possible trajectories by dynamic *Monte Carlo processes*. To reduce computer time for stochastic simulations, a technique, called the *Gillespie stochastic simulation algorithm*[13] was developed. Its idea is very simple: after initialization, an iteration of calculating the Monte Carlo steps is started until the desired simulation time is covered. Every Monte Carlo step is defined by two stochastic variables: the next reaction $R_i$ and the time it takes until this reaction $t_i$. The probability for every reaction $R_i$ is proportional to the propensity of the reaction, a variable that according to the law of mass-action kinetics, is proportional to the abundance of the involved reactants and the reaction rate. As an example, we state the propensities of the processes discussed in this chapter in Table 3.2.

13: The algorithm was constructed by Doob, 1945, but popularized by Dan Gillespie thirty years later (Gillespie, 1976, 1977).

The time until this reaction is an exponential random variable, for this is the distribution of times between two Poisson events. After every Monte Carlo step is calculated, the time series is updated, *i.e.* the species abundances are adapted according to the reaction and the time is increased. Because stochastic simulations are computationally expensive, multiple modifications have been developed to reduce the simulation time such as the K-leap and tau-leap methods (Cai & Xu, 2007; Gillespie, 2001).

**Table 3.2:** Propensities of the rates in this chapter. A variable with a hat $\hat{x}$ denotes the number of molecules of the variable $x$.

| Equation | Rate |
|---|---|
| $A + B \xrightarrow{k} AB$ (3.1) | $k \cdot \hat{A} \cdot \hat{B}$ |
| $AB \xrightarrow{q} A + B$ (3.1) | $q \cdot \hat{AB}$ |
| $x \xrightarrow{\gamma} \emptyset$ (3.3) | $\gamma \cdot \hat{x}$ |
| $DNA \xrightarrow{\alpha} mRNA$ (3.5) | $\alpha \cdot \hat{DNA}$ |
| $mRNA \xrightarrow{\beta} p$ (3.6) | $\beta \cdot \hat{mRNA}$ |

For systems with larger numbers of elements, stochasticity can be implemented by adding a noise term to the deterministic equations. Such a noise term is based on Brownian motion and Ito calculus (Appendix B.6). Stochastic differential equations are also known as *Langevin equations*, after Paul Langevin who used them to describe the Brownian motion (Langevin, 1908). Different types of noise, intrinsic and extrinsic, are expressed through different noise terms. These will be listed and discussed in detail in the context of generalized Lotka-Volterra equations (Section 7.1).

Stochastic simulations lack predictability but are closer to reality for systems with a relatively small number of components. It was long believed that stochasticity is disturbing the dynamics, but more recent theories acknowledge the beneficial effects of stochasticity in gene expression (Kaern et al., 2005). We used stochastic simulations in the modeling of a gene regulatory network in Chapter 5.

# Gene regulatory systems

# Mathematics of gene regulation | 4

To respond to changing environments, cells regulate the concentrations of their proteins. As discussed in Chapter 2, regulations happen at multiple stages. In this chapter, we discuss the regulation of the transcription. Much theoretical work on gene regulatory circuits has been performed by the lab of Uri Alon. A more elaborate introduction to gene regulation can be found in his textbook (Alon, 2019).

Some proteins can regulate the transcription rate of defined genes by binding to a specific DNA sequence. Such proteins are called *transcription factors*. They can either enhance or suppress the transcription of the gene, called activators and repressors, respectively. The widely used notation is

$$X \rightarrow Y \qquad \text{(read 'X activates Y')},$$
$$X \dashv Y \qquad \text{(read 'X represses Y')}.$$

In some cases a protein acts as the transcription factor of its own gene, this process is called *autoregulation*. There are two possibilities, self-activation and self-repression representing positive and negative feedback, respectively.

The gene of a transcription factor is called a regulatory gene, it controls the expression of one or multiple other genes. All these interacting genes together form a (large) gene regulatory network. Understanding and predicting these large networks seems like a daunting task. Fortunately, these complex networks are combinations of simple circuits. These subunits can be studied independently. Some of the possible theoretical circuits are found in nature more often than what is expected from random networks[1], they are referred to as *network motifs* (Milo et al., 2002; Shen-Orr et al., 2002). Whenever a circuit is found in nature more often than expected from random networks, we expect it to have an advantage over other circuits because it was preserved over evolutionary timescales without being replaced by mutations.

[1]: Random networks here are *Erdös-Rényi graphs*. In these graphs, the set of vertices is fixed, as is the number of edges. The edges are uniformly distributed, which means that the probability for all possible edges is equal.

We will use mass-action kinetics to describe gene regulation. As discussed in Section 3.1, this technique relies on several assumptions. Can we assume that the cell is a well-mixed environment? Molecules move around in cells through *diffusion* or through active transport in the case of eukaryotes. Diffusion in cells is slower than in water. The time for molecules to diffuse across cells depends on the cell size and the molecule's size (von Hippel & Berg, 1989). The smaller the molecule the faster the diffusion. The diffusion time for macromolecules is in the order of 10-100 milliseconds in a bacterial cell and up to minutes in animal cells (Phillips, 2013; Schnitzer et al., 2000; Zhao et al., 2008). Noise due to diffusion is only important for small numbers of molecules. After repressive transcription factors have dissociated from the operator, they can rebind much faster than RNAPs can be acquired. But if effective (un)binding rates are used, the assumption of a well-mixed environment is valid (van Zon et al., 2006).

The second assumption that the number of reactants is large is difficult to meet for gene regulation. Every cell contains at most two copies of its DNA after duplication. The DNA can contain multiple copies of a gene, but this number remains low (between 1 and 5) (Bremer & Dennis, 2008). The number of mRNAs per gene is in the order of 0 to 10 (Taniguchi et al., 2010), and the copy number for a particular protein ranges between 1 and $10^4$ (Xie et al., 2008). Because the numbers of reactants in gene regulation are small, deterministic solutions that use mass-action kinetics will only be an approximation of the real dynamics and stochastic solutions should be performed.

Autoregulation is an example of a common network motif and in this chapter, we discuss the functions of positive and negative autoregulation.

## 4.1 Negative feedback

Negative feedback is a ubiquitous process found in many natural as well as engineered applications. It is the main design principle of a thermostat: the actual temperature is compared to the desired temperature and a response opposite to the difference between both temperatures is returned[2]. It is important that the feedback happens fast because a delay in the feedback could lead to oscillations[3]. An example of a

---

2: If the actual temperature is too high, the difference is positive and as a response the heater will be turned lower. On the other side, if the actual temperature is too low, the difference is negative and the response will increase the power of the heater.

3: This phenomenon is well known to people who take a shower where you manually need to mix the hot and cold water. If there is a long delay between the water actually arriving and the change at the faucet, one will overcorrect for too cold or too hot water resulting in oscillations between hot and cold water.

natural system is the Earth's atmosphere, where plant growth, solar radiation, clouds, and the planet's temperature are kept in balance by negative feedback (Charlson et al., 1987). There exist many more examples, but in this section, we focus on negative feedback in gene regulation.

Before we describe negative autoregulation, we shortly revisit the case of protein production without regulation. Consider a protein with a constant production rate $\alpha$ and a constant degradation/dilution rate $\beta$, the time evolution of the protein concentration $x$ can be summarized as

$$\dot{x} = \alpha - \beta x. \tag{4.1}$$

The steady state equals $x_{ss} = \alpha/\beta$ and the response time, *i.e.* the time to reach half of the steady state starting from zero is $T_{1/2} = \ln 2/\beta^4$. The higher the degradation rate, the faster steady state is attained.

4: The response time is found by

$$T_{1/2} = \int_0^{x_{ss}} \frac{dx}{\alpha - \beta x}$$

$$= \left. \frac{-\ln(\alpha - \beta x)}{\beta} \right|_0^{x_{ss}/2} = \frac{\ln(2)}{\beta}.$$

## Negative feedback speeds up the response

In the case of negative autoregulation (NAR), the protein production rate is not a mere constant function but decreases for increasing concentrations of the protein. An example of such a function which is a good approximation for a multistep process is the Hill function

$$f(x) = \frac{\alpha_{\text{NAR}} K^n}{K^n + x^n}, \tag{4.2}$$

as was discussed in Section 3.3. An illustration of this function is given by the dashed line in Figure 4.1. When there are no proteins ($x = 0$), they cannot bind to the promoter to repress transcription and, therefore, the transcription rate is maximal ($f(x) = \alpha_{\text{NAR}}$). For a large concentration of proteins ($x > \alpha_{\text{NAR}}/\beta$ with $\beta$ the degradation/dilution rate of the protein), the probability for the promoter to be bound by a repressor is high and the transcription rate drops to zero.

The time evolution of a self-repressing protein can thus be summarized as

$$\dot{x} = \frac{\alpha_{\text{NAR}} K^n}{K^n + x^n} - \beta x. \tag{4.3}$$

5: The steady states of Equation 4.3 with $n = 1$ are

$$x_{ss} = \frac{K}{2}(-1 \pm t)$$

with $t = \sqrt{1 + 4\alpha_{\text{NAR}}/(K\beta)} > 1$ of which we only consider the solution with the plus sign because it is the only positive case. The response time is

$$T_{1/2} = \int_0^{x_{ss}/2} \frac{dx}{\frac{\alpha_{\text{NAR}} K^n}{K^n + x^n} - \beta x}$$
$$= \frac{\ln(2) - (1 - t^{-1})\ln\left(\frac{3t+1}{t+1}\right)}{\beta}.$$

Because $0 < t < 1$, we obtain that $T_{1/2} < \ln(2)/\beta$.

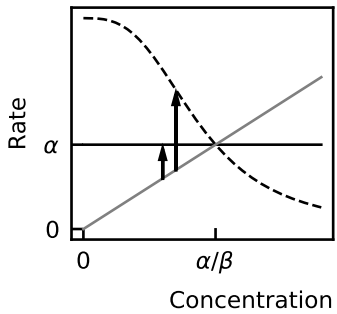

**Figure 4.1:** Negative feedback speeds up response.

6: It is important to keep as many parameters equal as possible to make a fair comparison. This is called a mathematically controlled comparison (Savageau, 2009).

This equation cannot be solved analytically in the general case because of the nonlinearities. However, it is possible for the case of $n = 1$, and the response time is smaller than the response time without regulation[5]. Hence this negative feedback has accelerated the response. The speedup can in fact be obtained with any negative feedback curve (Rosenfeld et al., 2002). Due to the nonlinearities, it is difficult to prove this analytically in the general case. However, using a graphical method, we can argue that the shape of the negative response curve is unimportant as long as it is monotonically decreasing.

We draw the production and degradation/dilution rates as a function of the protein concentration (Figure 4.1), this is called a *rate plot*. The speed $\dot{x}$, which is the difference between the production and degradation/dilution rate, is visualized by the distance between the two curves. The steady state, which is defined as the point where the speed is 0, is thus obviously the point where both curves intersect. It is now easy to see how the distance between the curves increases by using negative feedback and maintaining the same steady state and degradation/dilution rate[6] (Figure 4.1). Hence, negative feedback increases the speed to steady state as well from values lower as from values higher than the steady state, independent of the shape of the curve.

Another benefit from negative feedback is the increased robustness of the steady state against fluctuating production rate $\alpha_{\text{NAR}}$ and degradation/dilution rate $\beta$. A graphical argument for theoretical systems can be found in Alon, 2019. This concept is also proven experimentally (Becskei & Serrano, 2000; Klumpp et al., 2009).

**Negative feedback and delay give rise to oscillations**

As was mentioned in the introduction, to take advantage of the speedup of the negative response, it is important that the feedback is fast, because a delay can lead to *oscillatory behavior* (Novák & Tyson, 2008). This delay, of which the biological nature can be the time that is needed for transcription, translation, post-transcriptional or post-translational modification, or transport of the mRNA from the nucleus to the cytoplasm, can be modeled explicitly by delay differential equations or by modeling these intermediate steps. Here, we present a short example of how oscillations can arise by modeling a protein, its mRNA and and its transcriptional repressor. This model is known as the *Goodwin model* (Gonze & Abou-Jaoudé, 2013). The differential equations of this system read:

$$\frac{dp}{dt} = \alpha_p m - \beta_p p, \tag{4.4}$$

$$\frac{dm}{dt} = \frac{\alpha_m}{1 + (r/K)^n} - \beta_m m \tag{4.5}$$

$$\frac{dr}{dt} = \alpha_r p - \beta_r r \tag{4.6}$$

with $p$, $m$, and $r$ the number of proteins, mRNA, and repressors, $\alpha_x$ and $\beta_x$ their respective maximal production and degradation rates. A decreasing Hill function with Hill coefficient $n$ is used for the production rate of the mRNA to represent the negative feedback. Given that the nonlinearity in the system is sufficiently high, i.c. the Hill coefficient $n$ is at least 10, the system exhibits oscillations (Figure 4.2).

Oscillations can also be obtained in models that include explicit delay for the transcription and translation process, such as the model of the genetic oscillator of somitogenesis in zebrafish (Lewis, 2003) and the oscillatory gene expression of the Hes1 gene (Monk, 2003). Mather et al., 2009 have shown that the period of the oscillation can be substantially longer than the total delay, as long as there are strong nonlinearities.

Oscillations play an important role in many biological systems: most life on earth is dictated by a circadian rhythm, the cell division process of some bacteria is regulated by the oscillation of Min proteins, oscillations govern vertebrate somitogenesis and neural progenitor cell differentiation, etc. (Kruse & Jülicher, 2005; Uriu, 2016). Also in the context of synthetic biology do we want to build robust genetic oscillators with a specific frequency and amplitude. Novak and Tyson summarized the design principles of biochemical oscillations (Novák & Tyson, 2008): time delays can be integrated into the model explicitly, by a series of intermediates or by positive feedback. The first robust and tunable synthetic oscillator was built by the lab of Hasty in 2008 (Stricker et al., 2008).

Using an evolutionary algorithm, van Dorp et al., 2013 searched for small theoretical gene regulatory networks that

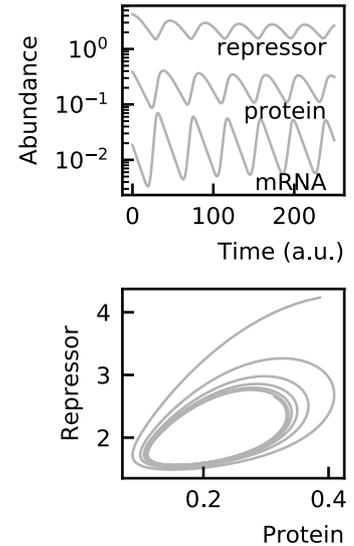

**Figure 4.2:** The Goodwin oscillator

Time series of the Goodwin oscillator (top panel) and its representation in phase space (bottom panel).

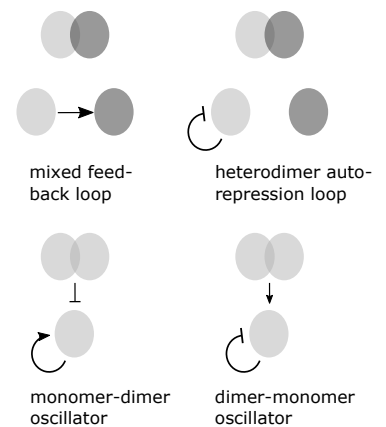

**Figure 4.3:** The four smallest genetic oscillators involving at most two proteins (van Dorp et al., 2013). Both the monomers (single ellipses) and homo/heterodimers (two bound ellipses) are active. Through activation (arrows) and repression (blunt arrows), oscillations can be obtained.

result in oscillatory behavior. They reported four oscillators of which the number of components and nonlinearities are minimal (Figure 4.3). Here, we explain the monomer-dimer oscillator because it is used in our study of the Ss-LrpB network. The protein of this oscillator appears in two active forms: a monomeric and a dimeric form. They regulate their proper gene positively and negatively, respectively. Essentially, at low protein concentrations there are more monomeric proteins that will bind to the DNA and upregulate the gene expression until a high concentration is obtained. At the high concentration, there are more dimeric proteins which, in their turn, downregulate the gene expression. In this way, oscillatory behavior is obtained.

The emergence of oscillations in the Goodwin oscillator could be explained deterministically. For insufficient nonlinearities ($n < 10$), the deterministic solution may only show damped oscillations. However, noise can help sustain the oscillations (Figure 4.4) (Li & Lang, 2008; Scott et al., 2007), especially when the number of repressor binding sites is high (Lengyel & Morelli, 2017). A self-repressing gene can exhibit regular oscillations when the mRNA and protein half-lives are approximately equal to the gene response time (Wang et al., 2014). Moreover, multiple binding sites increase the nonlinearity, which results in enhanced oscillations, *i.e.* larger amplitudes (Lengyel et al., 2014) and cooperativity between the autorepressor molecules is not mandatory when the number of binding sites is high (Guisoni et al., 2016). These oscillations are called *noise-induced oscillations*. The role of noise in synthetic oscillations is, however, two-edged: it can be both constructive and destructive for the generation of oscillations (Purcell et al., 2010).

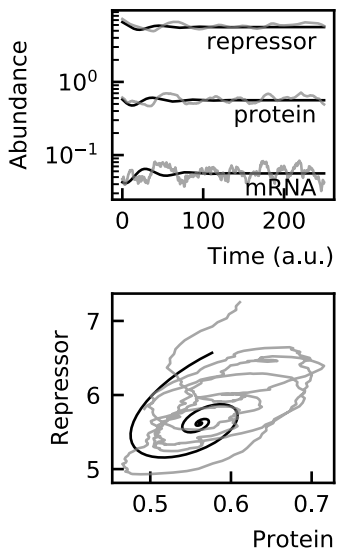

**Figure 4.4:** The damped Goodwin oscillator and noise-induced oscillations.

Time series of the damped Goodwin oscillator (top panel) and its representation in phase space (bottom panel). Noise can maintained oscillatory behavior (grey lines).

## 4.2 Positive feedback

Just like for negative feedback, positive feedback is not a unique concept of biological problems. We recognize it when a microphone picks up the sound of the amplifiers resulting in a screech, when people start to panick when other people are panicking, etc. We will here discuss the function of positive feedback in gene regulation.

We have seen that negative autoregulation speeds up the response and increases the robustness of the steady state against changing parameters, but which benefits can be attributed to positive feedback? First of all, positive feedback works in the opposite way as negative feedback, and using the same graphical arguments as for negative feedback, we can argue that positive feedback slows down the response time and amplifies the noise in parameters.

**Positive feedback can give rise to bistability**

A characteristic that can emerge from positive feedback is *bistability* (Ferrell & Xiong, 2001). Bistable systems have two distinct stable steady states, which can be denoted as the low and high state, or even the "on" and "off" state if the "off" state is sufficiently low. Projecting this theory onto our examples, we see that for the microphone the volume remains zero as long as the input noise is low, keeping it in the "off" state; for a loud input noise, the positive feedback will keep the screech going even when the input noise is not present anymore, thus keeping it in the "on" state.

Just like the microphone, when a protein concentration exceeds the threshold, the protein can keep itself at a high concentration. Once the protein concentration drops below the threshold, the feedback is insufficient to keep the production up and the concentration drops to a low value. These two states, the "on" and "off" state define the bistable system. In some systems, random transitions between these two states can happen because of molecular noise.

Using a rate plot, we can show how bistability may arise from positive feedback. Without feedback, the production rate (horizontal line) and the degradation/dilution rate (diagonal line) have only one intersection point (Figure 4.5). The concentration of the intersection is the steady state. For positive feedback (the dashed curve), multiple intersections and, therefore, multiple steady states are possible. In Figure 4.5, we discern a low (L), intermediate (I), and high (H) steady state. Because the slope of the degradation rate is larger than the slope of the production rate in the low and high states, these states are stable. Using a similar reasoning, we conclude that the intermediate state is unstable. The high and low states can be interpreted as the "on" and "off" states, respectively and the intermediate state represents the threshold value. The existence of bistability depends on the parameters. One can see that if the degradation/dilution rate is increased, only the low steady state remains and on the contrary, if this rate is decreased, only the high state remains. A right balance of the parameters is necessary to exhibit bistability. In some systems, one wants to reduce the parameter region where bistability exists, to obtain a unique low or high state depending on the parameters, without being in a state where both solutions are allowed and where the system could fluctuate between both because of noise. Extra binding sites (decoy binding sites) will reduce the parameter region of bistability (Burger et al., 2010). Cooperativity in the binding of transcription factors to the DNA also promotes bimodal gene expression (Gutierrez et al., 2012).

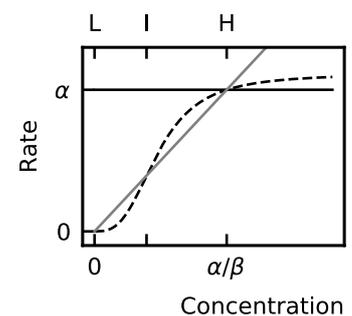

**Figure 4.5:** Positive feedback can give rise to bistability.

Positive and negative feedback are the main single gene motifs. Other motifs consisting of two or more genes can be found too. These are known as *feed-forward loops*. An overview can be found in Alon, 2019.

## 4.3 Non-monotonic feedback in Ss-LrpB

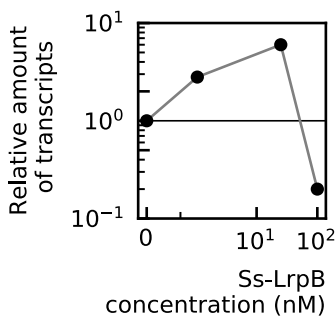

**Figure 4.6:** The response of Ss-LrpB is non-monotonic (Data from (Peeters et al., 2013)). It is increasing for small concentrations and decreasing for large concentrations.

7: An illustration of the different types of feedback—monotonic and non-monotonic—can be found in Figure 4.7.

Some years ago, it was discovered by Eveline Peeters' lab that the leucine responsive protein B of the *Sulfolobus solfataricus* (Ss-LrpB) has a non-monotonic[7] feedback regulation (Figure 4.6) (Peeters et al., 2013). Although positive and negative feedback are ubiquitous, non-monotonic autoregulation is rare. Therefore, it had not been studied extensively. The dynamical behavior of Ss-LrpB has not yet been observed and many parameters of the system have not been measured. However, through mathematical models we can predict the qualitative behavior of the protein for different parameter sets. In the following chapter, we present our work about non-monotonic autoregulation.

To solve this problem of non-monotonic autoregulation, we propose two hypotheses. The first one combines the advantages of both positive and negative autoregulation. We describe the parameter regions for which non-monotonic feedback results in a bistable solution for which the response to the high state is faster than for regular positive feedback. Although many of the parameters of the natural Ss-LrpB

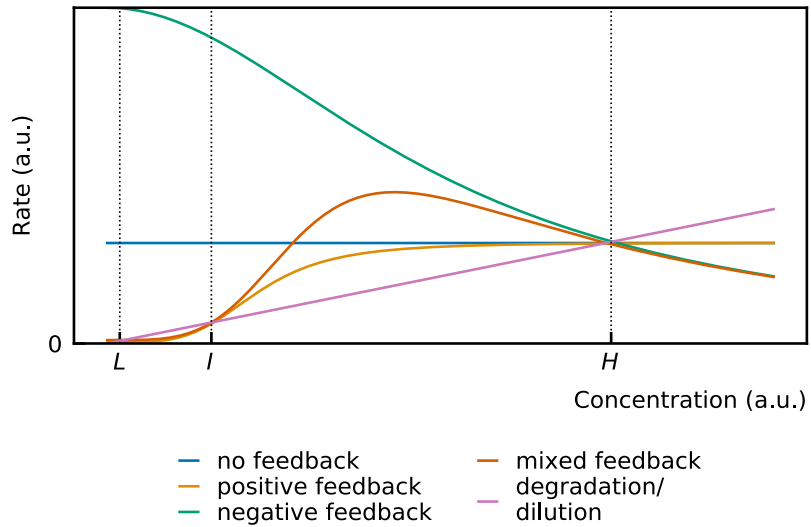

**Figure 4.7:** Monotonic and non-monotonic response curves.

system are unknown, we conclude that the natural system can not be bistable.

Our second hypothesis is oscillatory behavior. This was inspired by the monomer-dimer oscillator proposed by van Dorp et al., 2013[8]. Although no proper deterministic oscillations could be discovered for the Ss-LrpB system, spiky behavior was obtained through stochastic modeling. A possible explanation for pulsing genetic circuits is proposed in (Levine et al., 2013). They argue that frequency modulated pulsing allows for "bang bang" control. This provides regulation of the time a molecule such as a transcription factor is active, instead of a regulation of the concentration at which it is present in the cell. We suggest that the stochastic spiking of the Ss-LrpB protein can be a sensing mechanism with a reduced burden on the cell, in the sense that the concentration of the protein is not maintained at a high threshold concentration but reaches this concentration at irregular time intervals.

8: We discuss this oscillator in Section 4.1.

# Non-monotonic autoregulation in single gene circuits | 5

WE THEORETICALLY STUDY THE EFFECTS OF NON-MONOTONIC RESPONSE CURVES in genetic autoregulation by exploring the possible dynamical behaviors for such systems. Our motivation is twofold : we aim at conceiving the simplest genetic circuits for synthetic biology and at understanding the natural autoregulation of the LrpB protein of the *Sulfolobus solfataricus* archaeon which exhibits non-monotonicity.

We analyzed three toy models, based on mass-action kinetics, with increasing complexity and sought for oscillations and (fast) bistable switching. We performed large parameter scans and sensitivity analyses, and quantified the quality of the oscillators and switches by computing relative volumes in parameter space reproducing the sought dynamical behavior. All single gene systems need finely tuned parameters in order to oscillate, but bistable switches are more robust against parameter changes. We expected non-monotonic switches to be faster than monotonic ones, however solutions combining both autoactivation and -repression in the physiological range to obtain fast switches are scarce. Our analysis shows that the Ss-LrpB system can not provide a bistable switch and that robust oscillations are unlikely. Gillespie simulations suggest that the function of the natural Ss-LrpB system is sensing via a spiking behavior, which is in line with the fact that this protein has a metabolic regulatory function and binds to a ligand.

## 5.1 Introduction

Synthetic biology aims at building an extended toolbox of elementary genetic circuits and efficient designs for assembling them. These building blocks are inspired by electronics. The biological equivalent of many circuits have been built for timekeeping, electronic memory storage, toggle switches, oscillators, cascades, pulse generators, time-delayed circuits, spatial patterning and logic gate behavior (Cameron et al., 2014; Guet et al., 2002; Khalil & Collins, 2010; Mukherji & van Oudenaarden, 2009; Purnick & Weiss, 2009; Stanton et al., 2014). In order to construct predictable complex circuits, each building block must itself be predictable. In this work, we searched for the simplest genetic networks consisting of a single gene that produce, at the deterministic level, a dynamical behavior other than a stable steady state, *i.e.* oscillations or bistable switching. We also assessed the importance of molecular noise by performing Gillespie simulations. Our motivation is twofold. First, conceiving the simplest building blocks with only one gene is of interest to synthetic biology as it can potentially reduce undesired interference with other modules and facilitate the construction of complex circuits based on orthogonal compounds. Second, we want to understand the possible functions that can be fulfilled by single gene circuits. We are in particular interested in the function of the protein Ss-LrpB in the archaeon *Sulfolobus solfataricus*, both in the natural context and for its potential utility to develop simple building blocks for synthetic biology with Archaea, a territory almost unexplored. The relevance of non-monotonic regulation is broader than the Ss-LrpB system and is of importance for instance for toxin-antitoxin systems (Cataudella et al., 2013; Gelens et al., 2013).

Ss-LrpB is a protein forming dimers that regulates positively or negatively its own production via binding to three sites in front of the promoter. This system can be considered as a one gene mixed feedback loop: at low concentrations the protein activates itself, and at high concentrations it represses its own production (Peeters, Peixeiro, et al., 2013). We wonder what could be the role of such a complicated gene architecture, and under which circumstances this non-monotonic autoregulation generates oscillations, bistability, bursting behavior or simply leads to a steady state. Since the Ss-LrpB protein has a metabolic regulatory function and binds to a ligand, we formulated two hypothetic dynamical behaviors relevant for sensing. Oscillations can provide a sensing mechanism which measures the input signal at regular time intervals. Alternatively, bistability of this protein could provide a switch to maintain a high concentration of Ss-LrpB when the ligand is present (absent) and low concentration of the protein otherwise.

It is well known that bistability can be obtained through autoactivation, while autorepression is known to speed up the reaction time (Rosenfeld et al., 2002). We hypothesize that the dual feedback can result in a faster switch, see Figure 5.1B. In order to go from one steady state to the other an external trigger needs to decrease/increase the concentration past the intermediate unstable state. Although this switch is reversible, the main advantage of the mixed feedback is the time gain to switch from the intermediate state to the high steady state. Fast switching provides increased fitness at the individual level, contrarily to the bet-hedging strategy operating at the population level. Notice that bet-hedging also relies on a bistable switch, however the switching is stochastic (Cohen, 1966). Another hypothesis for the Ss-LrpB dynamics is related to the possibility that the threshold concentration of the protein needed to sense the presence of the ligand is too high to be maintained in steady state. Oscillatory dynamics or irregular spiking can produce a high enough concentration only at certain time intervals, thereby possibly reducing the burden on the cell. Noise could also be useful to facilitate the evolution of gene regulation (Wolf et al., 2015). Like negative feedback, non-monotonic mixed feedback can give rise to oscillations when combined with implicit delay (Figure 5.1A).

The broader question we addressed in this work is: what are the possible dynamical behaviors for single gene circuits? More specifically, we considered deterministic models consisting of ordinary differential equations (ODEs) based on mass-action kinetics and without explicit time delays, similarly to Karapetyan and Buchler, 2015 where slow DNA unbinding kinetics promotes oscillations. Introducing explicit time delays can lead to oscillations or even chaotic behavior, as is the case for Mackey-Glass systems (Mackey & Glass, 1977). However they are not considered here because we are mainly interested in prokaryotes. For these organisms, time delays necessary to generate oscillations are typically 5 to 20 minutes which is considerably longer than the time to transcribe and translate a gene, even if in some cases delay-induced degrade-and-fire oscillations can be obtained for delays as short as 3 to 5 minutes (Mather et al., 2009). The Ss-LrpB protein which inspired this work is a small protein and is expected to be produced rapidly (more details in Section 5.4). Implicit time delays can be provided by positive feedback (Alon, 2019).

In order to produce nonlinearities and negative feedback, which are crucial for oscillations without explicit time delays and to obtain interesting dynamics in general, we only allowed for dimerization and multiple binding sites. We did not include phenomenologically high Hill coefficients to obtain oscillations such as in the famous Goodwin model. This single gene oscillator without explicit time delay describes oscillatory dynamics with only three variables and only one source of nonlinearity, a Hill function. The Hill coefficient needs to be considerably high ($n > 8$) in order to obtain sustained oscillations through a Hopf bifurcation (Griffith, 1968). Similar limit-cycle oscillations can be obtained by fast phosphorylation and dephosphorylation of the protein (Gonze & Abou-Jaoudé, 2013).

To seek minimal requirements to generate oscillations or a fast bistable switch based on a single gene with multiple binding sites and dimerization only, we considered toy models of increasing complexity. We first considered a single binding site. In that case, a monomer is known to be insufficient to generate oscillations without explicit delay, we therefore allowed for dimerization. The protein can bind to the binding site in both its monomeric and dimeric form and the binding site occupancy determines the up- or down-regulation of the transcription rate. We call this system the monomer dimer system (MDS). This network was based on the theoretical monomer dimer oscillator of which van Dorp discovered its oscillatory potential (van Dorp et al., 2013). He considered the case of transcriptional repression by the dimer and activation by the monomer, which is a simple conceptual analog of the complete Ss-LrpB system as it is based on positive and negative feedback provided by a single gene. To our knowledge, this monomer dimer oscillator has not yet been observed in nature. We explored next the two-dimer (2DS) model which can be regulated only in dimeric form but by binding to two separate binding sites. If autoactivation occurs when one dimer is bound and autorepression when two dimers are bound, we again obtain a simplification of the three binding site Ss-LrpB system. For prokaryotic transcription factors, dimers are generally more stable than their respective monomers. Therefore we expect that a synthetic implementation of this regulatory network in prokaryotes will be easier than an implementation of the monomer dimer system. Finally, we turned to systems with three binding sites with regulations inspired by the Ss-LrpB system (3DS) and analyzed their ability to generate oscillations, bursty behavior, or function as bistable switches. Figure 5.2 illustrates the questions we addressed together with our strategy to tackle them.

The paper is organized as follows, we first present our toys models, the MDS, 2DS and 3DS. We then explain our search strategy for oscillators and bistable switches. The results of these explorations in high dimensional parameter spaces are presented for each toy model, first on the oscillatory dynamics and then on the ability to serve as a fast bistable switch. We conclude by a discussion about the natural Ss-LrpB system, both at the deterministic and stochastic level.

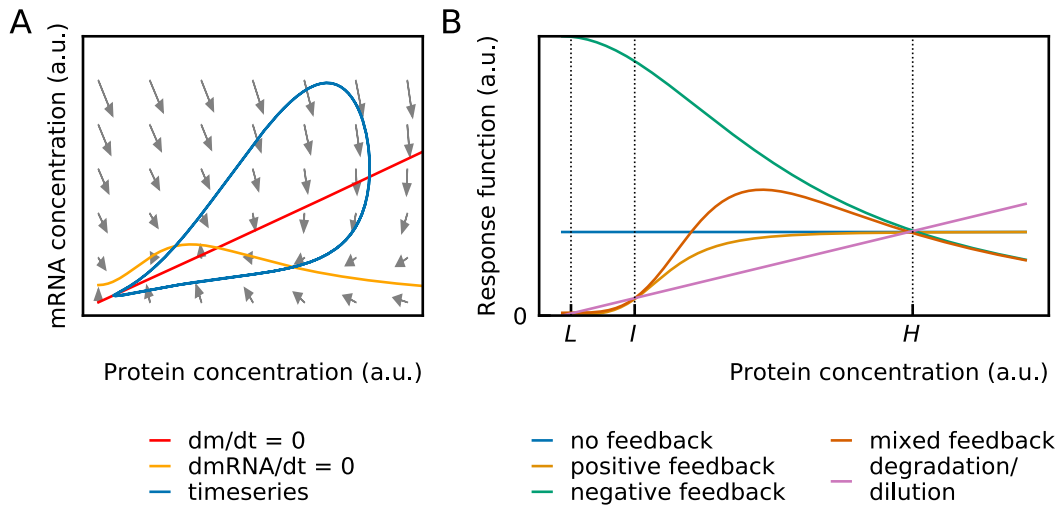

**Figure 5.1:** Conceptual representation of the two hypotheses. (A) Oscillations: a non-monotonic response curve, together with implicit delay can give rise to oscillatory behavior. A non-monotonic response, *i.e.* the mRNA production is non-monotonic with respect to the protein concentration ($m$), results in a non-monotonic nullcline (dmRNA/dt = 0). With sufficient delay (implicit or explicit), the steady state which is the intersection of both nullclines (dm/dt = 0 and dmRNA/dt = 0) becomes unstable and the time series becomes a clockwise oscillation in the phase plane. The grey arrows denote local velocities under the assumption of quasi-steady state for the remaining variables, *i.e.* the DNA states and dimer concentration. (B) A fast bistable switch : Steady state is obtained when the production rate (feedback) and removal rate (degradation/dilution) are equal. For concentrations where the feedback is bigger than the degradation/dilution, the concentration will increase, for concentrations where the feedback is smaller than the degradation/dilution, the concentration will decrease. Positive feedback can lead to bistability, *i.e.* there is a low ($L$) and high stable steady state ($H$), the intermediate state ($I$) is unstable. Because the instantaneous speed of the reaction is the difference of the response function and the degradation/dilution, the induction time to the steady state ($H$) will be smaller for systems with negative feedback than for systems with no feedback. Similarly, mixed feedback can speed up the induction time of a bistable switch, *i.e.* the time it takes to go from the unstable intermediate state $I$ to the stable high steady state $H$.

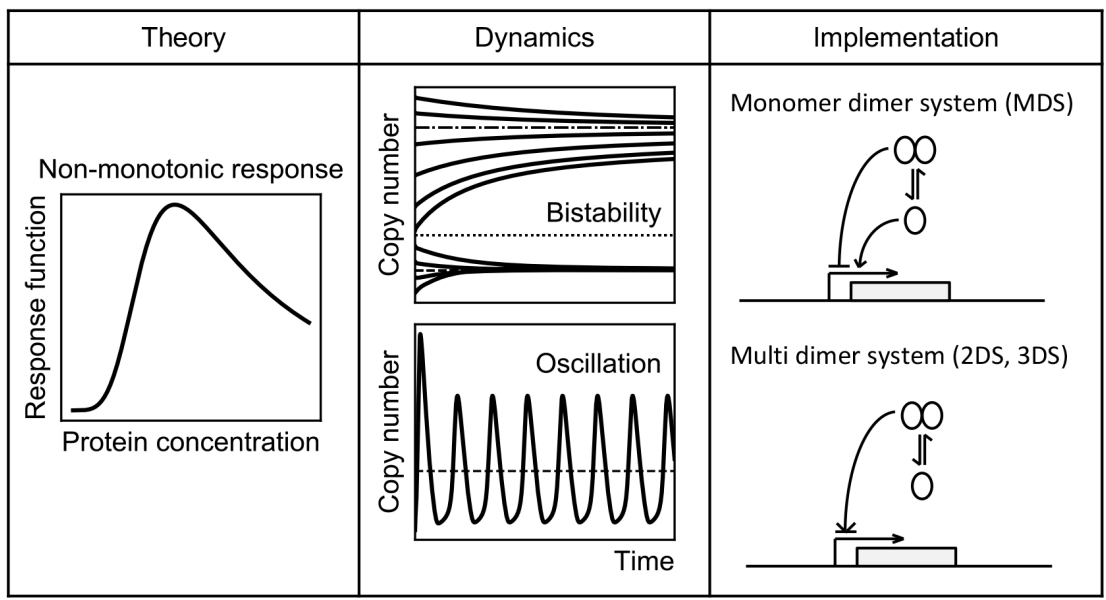

**Figure 5.2:** Theory, dynamics and implementation of non-monotonic response curves. Non-monotonic response curves with a maximum can lead to different types of dynamics such as bistability and oscillations. The most simple one gene implementations for such a non-monotonic response are the monomer dimer system (MDS) and multi dimer systems (details in Materials and methods).

## 5.2 Materials and methods

**Toy models**

We considered three families of toy models of increasing complexity which are based on mass-action kinetics. A set of ODEs describes the time derivatives of the concentrations of the different variables (DNA, mRNA and proteins). These concentrations are expressed in number per cell (instead of mol per liter), but can be fractional, for example a concentration of 1 nM in a cell of 4 fL corresponds to 2.4 molecules per cell. Expression in number per cell facilitates the comparison of deterministic and stochastic simulations. An overview of the physiological ranges of the parameters we selected based on literature for our models is provided in Table C.4. All python codes used in this work are available on github.

**Monomer dimer system (MDS)**: The MDS is a single gene system consisting of a protein that can regulate its production via one binding site in the regulatory region, both in its monomeric and dimeric versions. When the transcriptional fold changes $f_i$ (either activating, $f_i > 1$, or repressing $f_i < 1$) of DNA bound by a monomer or dimer(s) are suitably chosen, a non-monotonic response curve can be obtained. Using mass-action kinetics, this system can be described by a deterministic model of five ODEs and 13 parameters (we refer to Appendix C for details, and a list of parameters is given in Table 5.1). It was first discovered by van Dorp that the MDS can oscillate (van Dorp et al., 2013). However no wide search for oscillating parameter regions had been performed.

**Two dimer system (2DS)**: The two dimer system (2DS) we propose is again a self-regulatory gene, *i.e.* the gene transcribes for a protein which in dimer-form is the transcription factor of this gene. The regulatory region of the gene contains two binding sites for the transcription factor. When choosing adequately the activation/repression folds for all DNA configurations, non-monotonic curves can be obtained. A possible design to implement a promoter with a non-monotonic response synthetically is the following: when placing the first binding site for an activating transcription factor before and one binding site after the initiation site for transcription the former will activate transcription and the latter will repress transcription by steric hindrance. The deterministic model of the 2DS is slightly larger than the MDS model and has six ODEs and 16 parameters (details in Appendix C.1).

**Table 5.1:** Overview of all variables and all parameters of the different models.

| variable | dimension | |
|---|---|---|
| $DNA_i$ | number per cell | concentration of DNA with no bound proteins ($i = 0$), one bound monomer ($i = m$), one bound dimer ($i = d$) or dimers bound to site(s) i ($i \in \{1, 2, 3\}$) |
| mRNA | number per cell | mRNA concentration |
| $m$ | number per cell | monomer concentration |
| $d$ | number per cell | dimer concentration |
| $k_{bi}$ | min$^{-1}$ | binding rate of monomer ($i = m$), dimer ($i = d$) or dimer to site i ($i \in \{1, 2, 3\}$) |
| $k_{ui}$ | min$^{-1}$ | unbinding rate of monomer ($i = m$), dimer ($i = d$) or dimer to site i ($i \in 1, 2, 3$) |
| $K_i$ | min$^{-1}$ | binding constant of monomer ($i = m$), dimer ($i = d$) or dimer to site i ($i \in 1, 2, 3$), $K_i = k_{bi}/k_{ui}$ |
| $\phi_0$ | min$^{-1}$ | transcription rate |
| $f_i$ | n.a. | transcriptional fold change when monomer ($i = m$) or dimer ($i = d$) is bound or dimers are bound to site(s) $i$ ($i \in \{1, 2, 3\}$) with respect to no protein bound to the DNA |
| $\beta$ | min$^{-1}$ | translation rate |
| $\gamma_i$ | min$^{-1}$ | degradation rate of monomer ($i = m$), dimer ($i = d$) or mRNA ($i = $ mRNA) |
| $\alpha_{ass}$ | min$^{-1}$ | association rate of monomers to dimers |
| $\alpha_{diss}$ | min$^{-1}$ | dissociation rate of dimers to monomers |
| $co_{b,uij(k)}$ | n.a. | cooperativity factor for binding ($b$) or unbinding ($u$) between sites $i$ and $j$ (and $k$) ($i, j, k \in \{1, 2, 3\}$) |
| $\omega_{ij(k)}$ | n.a. | cooperativity factor between sites $i$ and $j$ (and $k$) ($i, j, k \in \{1, 2, 3\}$) $\omega_{ij(k)} = co_{bij(k)}/co_{uij(k)}$ |

**Three dimer system (3DS)**: The last system we considered is the three dimer system (3DS), a three binding site version of the 2DS. This system can describe Ss-LrpB autoregulation without modeling the DNA loop. It is described by ten ODEs and 28 parameters (details in Appendix C.1).

## (Non-)monotonicity types

To analyze our results we made a division between different types of (non-)monotonicity. Response curves can be classified into monotonic and non-monotonic curves. We furthermore divided the non-monotic class in curves with one maximum (type 1), curves with one minimum (type 2) and curves with both a minimum and a maximum (type 3), see Figure 5.3.

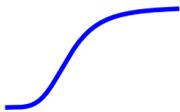

**Figure 5.3:** Classification of (non-)monotonic curves.

## Search for single gene oscillators

To generate oscillations, the following four requirements need to be met: negative feedback, nonlinearity, proper balance of timescales and time delay (Novák & Tyson, 2008). The latter can be inserted explicitly in the model to account for instance for the time of transcription, translation, splicing, transportation between nucleus and cytoplasm or implicitly by modeling intermediate states such as phosphorylation processes or via positive feedback. Although explicit transcriptional delay can transform stable gene networks into oscillators (Atkinson et al., 2003; Bratsun et al., 2005; Lewis, 2003; Monk, 2003), we focused as mentioned above on single gene oscillators that generate oscillations *in the absence of explicit time delay*.

Given the large number of parameters in all our models, we performed a random search through parameter space to seek oscillations. We focused on oscillations that arise through a *Hopf bifurcation*. In practice, we look for systems which have one fixed point whose eigenvalues have a nonzero imaginary part and a positive real part while all other fixed points have eigenvalues with negative real parts. We do not consider more complex solutions with multiple fixed points whose eigenvalues have a positive real part, which results in the coexistence of both a locally stable steady state and locally stable oscillations. Subsequently, we imposed physical selection criteria. Given the maximum value max($x$) and the amplitude $A(x) = \max(x) - \min(x)$ of the oscillation of variable $x$ (mRNA for the mRNA copy number, m for the monomer copy number and d for the dimer copy number), we require that

$$
\begin{aligned}
&\max(\text{mRNA}) < 20, & &A(\text{mRNA}) > 1, & &A(\text{mRNA}) > \max(\text{mRNA})/3, \\
&\max(m) < 5000, & &A(m) > 1, & &A(m) > \max(m)/3, \\
&\max(d) < 5000, & &A(d) > 10 \text{ and} & &A(d) > \max(d)/3.
\end{aligned}
\tag{5.1}
$$

The criteria on the maxima ensure that the copy number of mRNA remains low enough and that the copy numbers of the protein stay within their physiological bounds. For deterministic oscillations to be robust against molecular noise, we suspect that the changes in copy number of the mRNA and proteins in the deterministic model must be at least comparable to one copy number. We even set the lower limit on the amplitude to 10 for the dimer count, to obtain significant oscillations. The last criterion forces the oscillation amplitudes to be high enough with respect to the maximum values in order to distinguish the oscillation from the stochastic noise.

For every solution we performed a bifurcation analysis to assess the stability of the oscillatory behavior against parameter changes. We defined a logarithmic parameter volume $V$ of a solution $s$,

$$
V(s) = \prod_{p} \log_{10}\left(\frac{p_r(s)}{p_l(s)}\right),
\tag{5.2}
$$

where the product is over all $p$ parameters, and $p_l(s)$ and $p_r(s)$ are the left and right bifurcation points of this parameter for solution $s$ keeping all other parameters fixed (the latter are set equal to the borders of the physiological ranges when the bifurcation point exceeds these). This volume is used to estimate the coverage of parameter space of our random scan, and the relative volume taken by oscillatory solutions. The logarithm of the values is chosen over the values themselves as fold differences are more important in biology than absolute differences.

## Search for single gene bistable switches

To find bistable switches, we seek dynamical systems with three fixed points: a low concentration stable fixed point $L$, an intermediate unstable fixed point $I$ and a high concentration steady state $H$. The dynamics of a dimeric protein is regulated by four key processes: production, degradation/dilution, dimerization and dimer dissociation. In our three toy models, the time evolution of the monomer concentration is given by the following ODE, with aforementioned processes in order:

$$\frac{\mathrm{d}m}{\mathrm{d}t} = \frac{\beta \mathrm{DNA}_{\mathrm{tot}} \phi_0 f(m,d)}{\gamma_{\mathrm{mRNA}}} - \gamma_m m - 2\alpha_{\mathrm{ass}} m^2 + 2\alpha_{\mathrm{diss}} d. \tag{5.3}$$

This equation is obtained by assuming quasi-steady state for the DNA configurations, a more elaborate derivation of this equation can be found in Appendix C.4. Using the quasi-steady-state approximation for the dimer concentration, $d = \frac{\alpha_{\mathrm{ass}}}{\alpha_{\mathrm{diss}} + \gamma_d} m^2$ and assuming the dissociation rate of the dimer is much faster than degradation rate of the dimer ($\gamma_d \ll \alpha_{\mathrm{diss}}$), the last two terms of Equation 5.3 cancel out and the time derivative of the monomer concentration is proportional to the difference of the transcription and degradation term,

$$\frac{\mathrm{d}m}{\mathrm{d}t} \propto f(d(m)) - \gamma m, \tag{5.4}$$

with $\gamma = \gamma_m \gamma_{\mathrm{mRNA}} \beta^{-1} \mathrm{DNA}_{\mathrm{tot}}^{-1} \phi_0^{-1}$. Bistability arises when three positive values exist for $m$ such that $dm/dt$ vanishes ($dm/dt = 0$), *i.e.* the feedback term $f(d(m))$ and the degradation term $\gamma m$ are equal (Figure 5.1B). The shape of the transcription function $f$ and number of parameters involved depend on the specific system considered. Those functions are given here under together with the ranges of the grid over which we performed a parameter scan. More details can be found in Appendix C.4.

**MDS** In the quasi-steady-state approximation, the transcription function can be written as

$$f_{MDS}(m) = \frac{f_d K_d^* m^2 + f_m K_m m + 1}{K_d^* m^2 + K_m m + 1} \tag{5.5}$$

with $K_d^* = K_d \frac{\alpha_{\mathrm{ass}}}{\alpha_{\mathrm{diss}} + \gamma_d}$. There are only four parameters in this equation. We performed a scan over these parameters in the following ranges:

$$f_d : 10^{-3} \to 10^2, \qquad K_d^* : 10^{-10} \to 10^3,$$
$$f_m : 10^{-3} \to 10^2 \text{ and} \qquad K_m : 10^{-7} \to 10^3$$

with 30 values logarithmically spaced for every range.

**2DS** For a two dimer model the transcription function can be written as

$$f_{2DS}(d) = \frac{Ad^2 + Bd + 1}{Cd^2 + Dd + 1} \tag{5.6}$$

with

$$C = K_{d1}K_{d2}\omega, \qquad A = f_{12}C,$$
$$D = K_{d1} + K_{d2} \text{ and} \qquad B = f_1 K_{d1} + f_2 K_{d2}.$$

We perform a scan over parameter space analogously to the MDS case but adapting ranges to the combined variables:

$$D : 2 \cdot 10^{-4} \to 2 \cdot 10^4, \qquad B : 10^{-3}D \to 10^2 D,$$
$$C : 0.05/2010^{-4}(D - 10^{-4}) \to 20/0.05(D/2)^2 \text{ and} \quad A : 10^{-3}C \to 10^2 C$$

with 20 values logarithmically spaced for every range.

**3DS** For a system with three binding sites, the transcription function can be written as

$$f_{3DS}(d) = \frac{Ad^3 + Bd^2 + Cd + 1}{Dd^3 + Ed^2 + Fd + 1} \tag{5.7}$$

with

$$D = K_{d1}K_{d2}K_{d3}\omega_{12}\omega_{13}\omega_{23}\omega_{123} \quad A = f_{123}D,$$
$$E = K_{d1}K_{d2}\omega_{12} + K_{d1}K_{d3}\omega_{13} \quad B = f_{12}K_{d1}K_{d2}\omega_{12} + f_{13}K_{d1}K_{d3}\omega_{13}$$
$$\quad + K_{d2}K_{d3}\omega_{23} \qquad\qquad + f_{23}K_{d2}K_{d3}\omega_{23},$$
$$F = K_{d1} + K_{d2} + K_{d3} \qquad C = f_1 K_{d1} + f_2 K_{d2} + f_3 K_{d3}.$$

and an analogous scan was performed with 8 values logarithmically spaced for every range.

When performing the scan on the grids defined above, we imposed physical constraints:

(a) $I - L \geq 10$,
(b) $500 < H < 600$ and  (5.8)
(c) $I < 100$

where constraint (a) prohibits the lower state $L$ and the intermediate state $I$ from lying too close together to avoid stochastic noise to switch constantly between these states, constraint (b) fixes an interval for the high steady state $H$ because the induction time depends on the level of this state, and constraint (c) prohibits the intermediate state $I$ from being too high.

If the system is bistable, we can calculate the time to reach the high steady state when starting from the intermediate one. In the reduced system (Equation 5.4), the time it takes for the system to go from the unstable intermediate $I$ to the stable high steady state $H$ is proportional to the following expression

$$\tilde{\Delta} t \propto \int_I^H \frac{\mathrm{d}m}{f(d(m)) - \gamma m}. \tag{5.9}$$

Although this is the time obtained for the reduced system using the quasi-steady-state assumption for the dimer and DNA concentrations, it provides a good estimate for the time obtained from the complete system. For every set of parameters that dictate the response curve $f$ in the scan, we choose the last remaining free parameter $\gamma$ such that the approximated induction time $\tilde{\Delta} t$ is minimized. Then we compute the induction time by doing a deterministic simulation (details in Appendix C.4).

## 5.3 Results

**Single gene oscillators** For the MDS very few oscillating solutions were found. Out of $2.4 \cdot 10^8$ searched parameter sets, only 18 met the conditions for a Hopf bifurcation. Moreover, parameters of these oscillatory solutions need to be finely tuned, which is consistent with the fact that we found only very few solutions. The mean and mean logarithmic range for the different parameters are represented in Figure 5.4A. More detailed figures of the distribution of the ranges for the different parameters can be found in Appendix C.3. Oscillators with finely tuned parameters are not expected to be realistic due to the unavoidable noise in cellular processes. Moreover, when the system oscillates, the amplitudes of oscillations are typically very small. Amplitudes of oscillations of the monomer versus dimer are represented in Figure 5.5. We conclude therefore that the MDS cannot provide a realistic genetic oscillator. For the 2DS more oscillating solutions can be found. However they are still very rare. Out of the $2.4 \cdot 10^8$ examined sets 1894 were oscillating and only 170 meet the selection criteria (Equation 5.1). Around 58% of the solutions are non-monotonic. Oscillatory parameter ranges are considerably wider than in the MDS case, as illustrated in Figure 5.4.

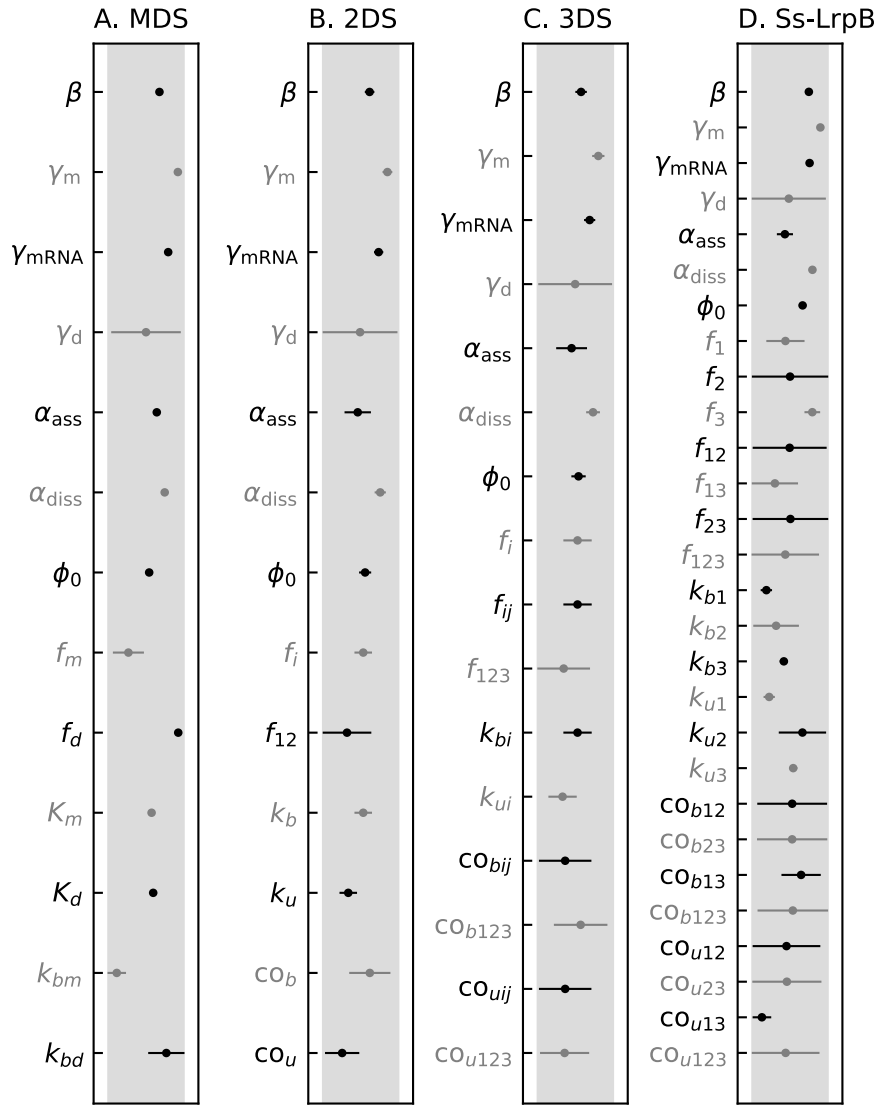

**Figure 5.4:** Oscillatory solutions for the different toy models. The shaded region represents the physiological range. The black and gray lines represent the mean oscillatory ranges for the different parameters. The axis is logarithmically scaled, the length of the lines thus represent fold ratios. The line for each parameter is scaled according to the physiological range of this parameter. Ranges are very small for the MDS and become wider for the 2DS and 3DS. Individual solutions can be found in Appendix C.3.

Even more solutions can be found for the 3DS : 2999 out of $0.8 \cdot 10^8$ parameter sets have a Hopf bifurcation. 291 solutions remain after imposing the selection criteria. 56% of the solutions are non-monotonic. For oscillating solutions the monomer copy number is most often higher than the dimer copy number, this is the case for all toy models (Figure 5.5).

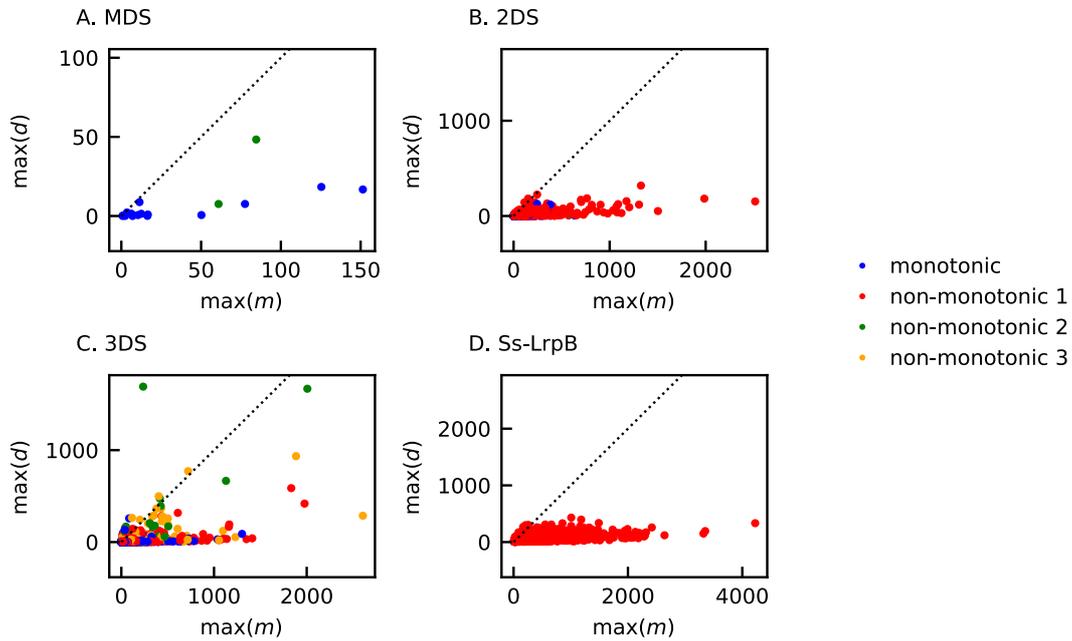

**Figure 5.5:** Amplitudes of oscillation in the monomer (m)-dimer (d) plane for the different toy models. The dashed line represents same copy number of monomer and dimer. Oscillations for the MDS are small: the dimer copy number does not exceed 50. The copy number of the monomer is higher than the copy number of the dimer for the majority of the solutions for all systems, contrarily to what is observed in prokaryotic systems in nature.

The results of our random scans are summarized in Table 5.2. To quantify the likelihood of a model to provide oscillatory solutions, we computed the logarithmic parameter volume of the scanned parameter space (definition by Equation 5.2), the mean volume of the found oscillatory solutions and the estimated relative volume of the found oscillatory solutions as a proxy for the capacity of the model to generate oscillations. It needs to be mentioned that the distribution of the logarithmic volumes of the different found oscillatory solutions is non-Gaussian (more details in Figure C.4). The ratio of the total logarithmic volume of parameter space to the mean logarithmic parameter volume of an oscillating region gives an estimate of the number of parameter sets that needs to be studied. Except for the MDS, where the amount is orders of magnitude higher than what is computationally manageable, the number of studied sets is in the same order of magnitude as this ratio. Considering the relative oscillating volumes, we conclude that it is 5 times more likely to find an oscillatory solution in the 3DS model than in the 2DS model, and 100 times more likely to find an oscillatory solution in the 2DS model than in the MDS model.

**Bistable switches** Our results show that the typical bistable switches are monotonically increasing or non-monotonic of type 2 for the MDS and 2DS, while for the 3DS all types of (non-)monotonicity are possible (Table 5.2). In Figure C.7 the regions within parameter space providing bistable switches for the different systems are represented. For 2DS and 3DS, a large proportion of parameter space leads to bistability while the MDS needs to be finely tuned to provide this dynamical property.

To assess the potential of the different models to provide bistable switches, we represented the relative logarithmic parameter volume in parameter space which they occupy (Table 5.2). The proportion of parameter space occupied by bistable switches clearly increases considerably from MDS to 2DS and from 2DS to 3DS. The complexity of the model increases its ability to provide switches. We also investigated the type of non-monotonicity favoring bistability. For the simpler MDS and 2DS, bistability can be obtained more easily with a monotonic or non-monotonic type 2 response curve, for the more complex 3DS, other non-monotonic types can also lead to bistability. Performing bifurcation analysis also shows that the parameter ranges providing bistability becomes larger for 3DS with respect to 2DS and MDS (details in Appendix C.4).

The fastest responses are found for non-monotonic type 1 curves as expected (Figure 5.6D), but only a small selection of the solutions is faster than any monotonic curve. The speed advantage of the non-monotonicity is thus only effective when in the most optimized case. For most of the non-monotonic solutions, a monotonic solution can be found that is as fast as the considered non-monotonic one (Figure 5.6C).

**Table 5.2:** Summary of the search for oscillations and the scan for bistability

|  |  | Type | MDS | 2DS | 3DS | SsLrpB |
|---|---|---|---|---|---|---|
|  | Number of parameters |  | 13 | 16 | 28 | 21 |
|  | Logarithmic volume of parameter space |  | $2 \cdot 10^7$ | $2 \cdot 10^9$ | $10^{16}$ | $3 \cdot 10^{11}$ |
| Oscillations | Number of random sets |  | $2.4 \cdot 10^8$ | $2.4 \cdot 10^8$ | $0.8 \cdot 10^8$ | $0.8 \cdot 10^8$ |
| Oscillations | Hopf solutions (meeting selection criteria) | M | 16 (2) | 789 (9) | 1322 (36) | - |
|  |  | N1 | 0 (0) | 1103 (161) | 973 (171) | 2294 (1158) |
|  |  | N2 | 2 (0) | 2 (0) | 379 (19) | - |
|  |  | N3 | 0 (0) | 0 (0) | 325 (65) | - |
|  |  | T | 18 (2) | 1894 (170) | 2999 (291) | 2294 (1158) |
|  | Mean logarithmic volume of oscillating region (Equation 5.2) | M | $5.4 \cdot 10^{-9}$ | $4.4 \cdot 10^0$ | $4.8 \cdot 10^7$ | - |
|  |  | N1 | - | $3.1 \cdot 10^0$ | $4.8 \cdot 10^6$ | $8.4 \cdot 10^2$ |
|  |  | N2 | $3.4 \cdot 10^{-10}$ | $1.8 \cdot 10^{-3}$ | $2.6 \cdot 10^7$ | - |
|  |  | N3 | - | - | $2.4 \cdot 10^6$ | - |
|  |  | T | $4.8 \cdot 10^{-9}$ | $3.6 \cdot 10^0$ | $2.3 \cdot 10^7$ | $8.4 \cdot 10^2$ |
|  | Ratio of the total logarithmic volume of parameter space to the mean logarithmic volume of an oscillating region |  | $5 \cdot 10^{15}$ | $6 \cdot 10^8$ | $4 \cdot 10^8$ | $4 \cdot 10^8$ |
|  | Total relative oscillating volume |  | $7.5 \cdot 10^{-8}$ | $7.9 \cdot 10^{-6}$ | $3.7 \cdot 10^{-5}$ | $2.8 \cdot 10^{-5}$ |
| Bistability | Number of sets |  | $8.1 \cdot 10^5$ | $1.6 \cdot 10^5$ | $2.6 \cdot 10^5$ | $2.6 \cdot 10^6$ |
| Bistability | Bistable solutions | M | 4474 | 2334 | 6101 | - |
|  |  | N1 | 0 | 0 | 789 | 0 |
|  |  | N2 | 76 | 1195 | 2336 | - |
|  |  | N3 | 0 | 0 | 434 | - |
|  |  | T | 4550 | 3529 | 9660 | 0 |
|  | Total relative bistable volume |  | $5.6 \cdot 10^{-3}$ | $2.2 \cdot 10^{-2}$ | $3.7 \cdot 10^{-2}$ | 0 |

M: monotonic, N1: non-monotonic type 1, N2: non-monotonic type 2,
N3: non-monotonic type 3, T: total (M + N1 + N2 + N3)

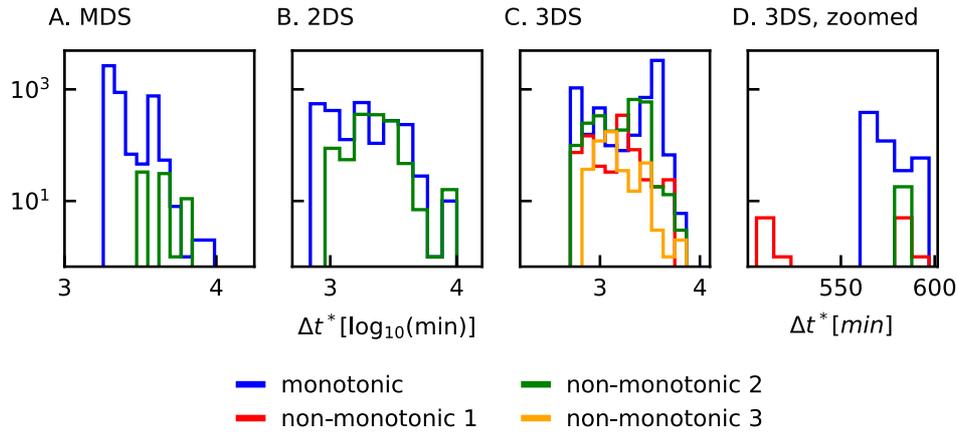

**Figure 5.6:** Distribution of induction times for the different toy models. For the MDS and 2DS no bistable non-monotonic type 1 solutions were found. For the 3DS many bistable solutions are found and the lowest inductions times are found for non-monotonic type 1 as we expected (panel D). Non-monotonic responses are not in general faster than monotonic responses (panel C): the induction time depends on the actual shape of the response curve and therefore on the parameters.

## 5.4 Discussion about the dynamics of the Ss-LrpB natural system

We conclude by a discussion on the possible dynamical behavior of the leucine responsive protein B of the archaeon *Sulfolobus solfataricus* (Ss-LrpB). This protein regulates itself in a unique way. The regulation site of this protein contains three binding sites to which the protein can bind in dimeric form. The outer sites of this regulation site have the highest affinity and will be occupied before the middle site (Peeters, van Oeffelen, et al., 2013). Experimental results suggest that transcription is activated when one or both outer sites are occupied. Due to cooperativity the middle site gets bound when the outer sites are occupied and subsequently, DNA undergoes a conformational change and loops on itself. In this configuration the transcription is repressed (Peeters, Peixeiro, et al., 2013). The mechanism of unlooping is unknown. Possible manners include unlooping when the proteins in the loop are degraded (Stricker et al., 2008) or via faster direct unlooping (Manzo et al., 2012). Both paths are denoted by dashed arrows in Figure 5.7. The temporal evolution of this system has not yet been observed experimentally, and most parameters have not been measured. As mentioned above, one hypothesis is that the Ss-LrpB system is a (possibly bursty) oscillator.

This system contains all necessary elements for oscillations: (1) negative feedback loop by the autorepression in the state where all dimers are bound, (2) nonlinearities are provided by dimerization, cooperative binding (Buchler et al., 2005), and (3) delay by modeling the intermediate states of the DNA. The sequence of gene Ss-LrpB is only 468nt long (data from KEGG library (Kanehisa & Goto, 2000; Kanehisa et al., 2017, 2016)) and with a transcription rate of minimum 40-80nt/s (Phillips, 2013), the transcriptional delay would only be in the order of a couple of seconds. Translation goes equally fast at a rate of at least 20aa/s (Phillips, 2013). Because the Ss-LrpB protein only contains 155aa (data from KEGG library (Kanehisa & Goto, 2000; Kanehisa et al., 2017, 2016)) and transcription and translation are coupled in prokaryotes, the total delay due to transcription and translation remains in the order of several seconds. This motivated us to not include an explicit time delay. Furthermore, it was put forward by Novak and Tyson (Novák & Tyson, 2008) that positive feedback can be used to add implicit delay to a negative-feedback system. The fourth element—balanced time-scales—can be obtained by fine tuning the parameters.

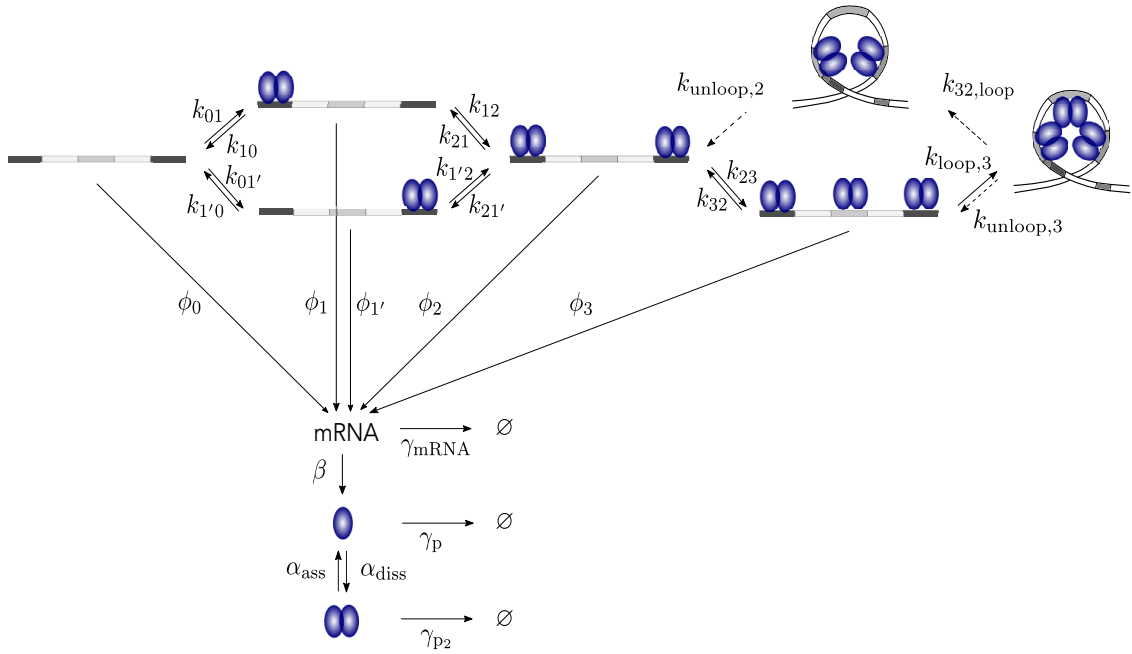

**Figure 5.7:** Schematic of the Ss-LrpB system. Ss-LrpB is a dimeric autoregulative transcription factor. The control region of its own gene contains three binding sites of which the first and third one have a high affinity. Due to cooperativity the middle site will get bound and the DNA subsequently loops on itself. The system of unbinding is unknown and two hypotheses, either through direct unlooping or either by first unbinding the protein of the middle site, are denoted by dashed arrows. All DNA configurations have their own transcription rates, which lead to a non-monotonic response curve. mRNA will be translated to monomers which can dimerize. The model also takes into account degradation/dilution of all proteins (mRNA, monomers and dimers).

To find oscillatory behavior, we performed the same analysis for the Ss-LrpB

system as for the 3DSs but with parameters fixed to their experimental values when known (see Appendix C.5). Furthermore it is imposed that the response is non-monotonic of type 1 and that the maximum of the response curve is at least twice the basal response and the minimum at most half the basal rate in accordance to the experimental response curve (Peeters, Peixeiro, et al., 2013). We found that some parameters, such as the degradation rates $\gamma_m$ and $\gamma_{mRNA}$, need to be finely tuned for oscillations. The sensitivity of each parameter is represented in Figure 5.4. And, as for general 3DS, many solutions have higher monomer copy numbers than dimer copy numbers (Figure 5.5). We conclude that the relative volume in parameter space leading to an oscillatory behavior is very small, $2.8 \, 10^{-5}$. It is therefore highly unlikely that the natural Ss-LrpB system shows oscillatory behavior.

To assess the role of noise in oscillatory systems, we selected two solutions which are compatible with the measured response curve of Ss-LrpB and performed stochastic simulations. Time series for different copy numbers of the DNA are shown in Figure 5.8. Stochastic models can differ much from the deterministic behavior, especially when considering a wild type *Sulfolobus solfataricus* cell which only contains at most two copies of the genome. In synthetic biology, target genes can be inserted in vector plasmids which can be injected in the cell in higher numbers. This would reduce the stochasticity in mRNA and protein concentrations (Figure 5.8).

We next envisaged the possibility to have a bistable switch. We undertook the same analysis as for the 3DS bistable switch analysis but again with fewer free parameters. For the Ss-LrpB system, seven parameters have been measured and we estimated two more based on literature, as detailed in Appendix C.5. Five free parameters remain: $f_1$, $f_2$, $f_3$, $f_{12}$, $f_{23}$ over which we performed a scan with the different $f$ ranging from $10^{-3}$ to $10^2$. The bistable region in the $f_{13}$-$f_{123}$-plane is shown in yellow and green in Figure 5.9. Outside this region, no bistable systems were found. The color represents the induction time of the fastest solution found for each $f_{13}$-$f_{123}$-combination. The smallest induction times can be found when both the DNA$_{13}$ and the DNA$_{123}$ are activating ($f_{13} > 1$ and $f_{123} > 1$) and the response curve is monotonically increasing. In red the percentages of Ss-LrpB compatible solutions are given, *i.e.* solutions with response curves of non-monotonicity type 1 and a maximal transcription rate of at least twice the basal rate and a minimal rate lower than half the basal rate. Such response curves are found when DNA$_{123}$ is repressing ($f_{123} < 1$) and DNA$_{13}$ either somewhat activating or repressing. The regions of bistable systems (yellow-green) and Ss-LrpB compatible solutions (red) are not overlapping, we can therefore conclude that the natural Ss-LrpB system cannot exhibit bistability.

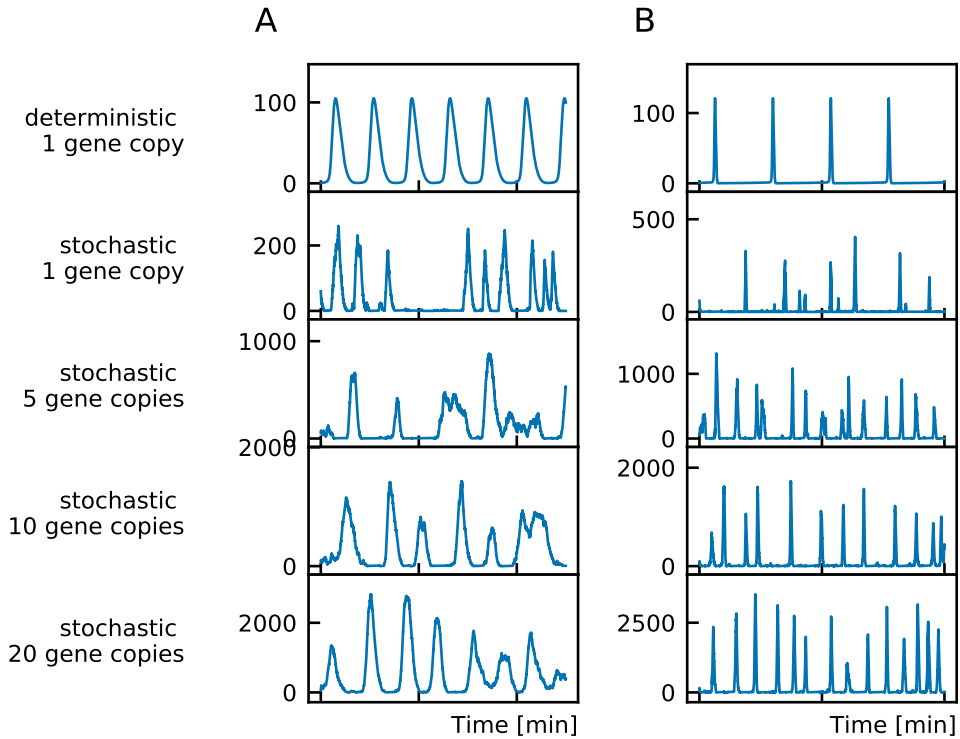

**Figure 5.8:** Stochastic time series. Time series of two examples of Ss-LrpB compatible systems are shown (A and B), in a deterministic and stochastic models from 1 to 20 DNA copy numbers. The oscillations become more regular for higher copy numbers. Note that the frequency is higher in the stochastic models with respect to the deterministic model for the example on the right.

We conclude that it is difficult to obtain deterministic solutions other than a stable steady state with the architecture and parameters of Ss-LrpB. Since a stable steady state can be obtained by the simplest self-regulations, we speculate that the choice for the complicated architecture of the natural Ss-LrpB system roots in its stochastic properties. To test whether the natural system has evolved toward parameters leading to a noisier, and more spiky behavior, we ran stochastic timeseries for random 3DS configurations as well as for random Ss-LrpB compatible solutions. After comparison of the Fano factors, a measure of stochastic spiking, of the different sets, we concluded that the distribution of Fano factors is similar for the Ss-LrpB system and the 3DS, as illustrated in Figure 5.10. Since the shape of the Fano factor histograms will be affected by fixing additional values for the parameters, we cannot conclude on our hypothesis. Without extra experimental measurements, it would be time intensive to obtain a better comparison of the 3DS and the SsLrpB compatible systems. We hypothesize that the function of spiky dynamics, besides the ones proposed by Levine et al., 2013, could be a sensing mechanism with a reduced burden on the cell. The idea is that the time averaged mean concentration of the protein is lower than the threshold concentration for sensing. Although it is slightly more frequent to exhibit rich dynamical behavior for systems with three binding sites than with two, two and three binding sites systems essentially allow for similar dynamical behaviors. We therefore expect that additional properties of systems with three binding sites, such as looping of DNA when three dimers are bound, should explain their functional role. However, without experimental evidences for a specific property, we did not elaborate more on this possibility.

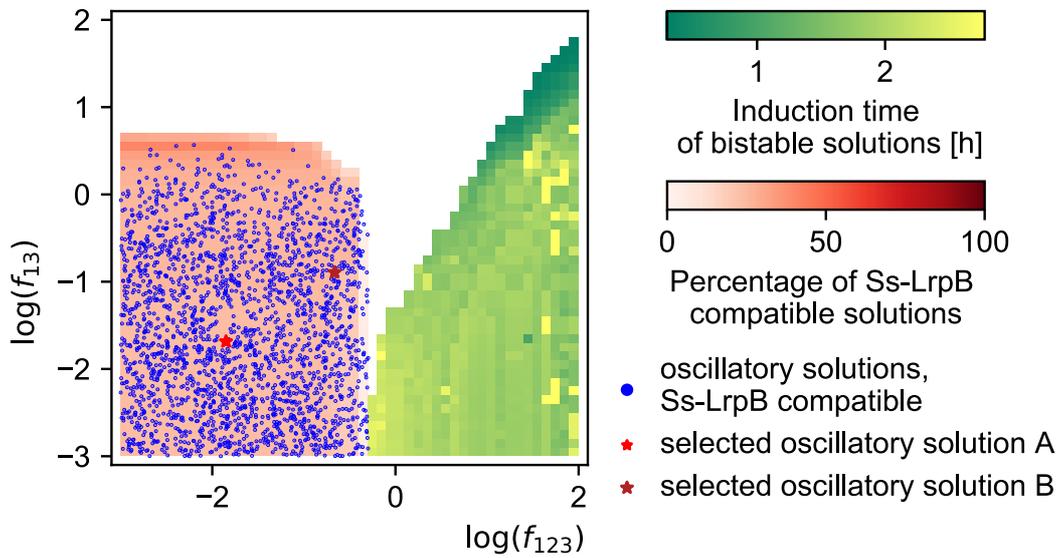

**Figure 5.9:** Bistable region with induction times in the $f_{123}$-$f_{13}$-plane. The bistable region is shown in yellow and green where the color represents the induction time of the fastest solution found for every $f_{13}$-$f_{123}$-combination. In red the percentages of Ss-LrpB compatible solutions are given. The regions of bistable systems (yellow-green) and Ss-LrpB compatible solutions (red) are not overlapping, we can therefore conclude that the natural Ss-LrpB system cannot exhibit bistability.

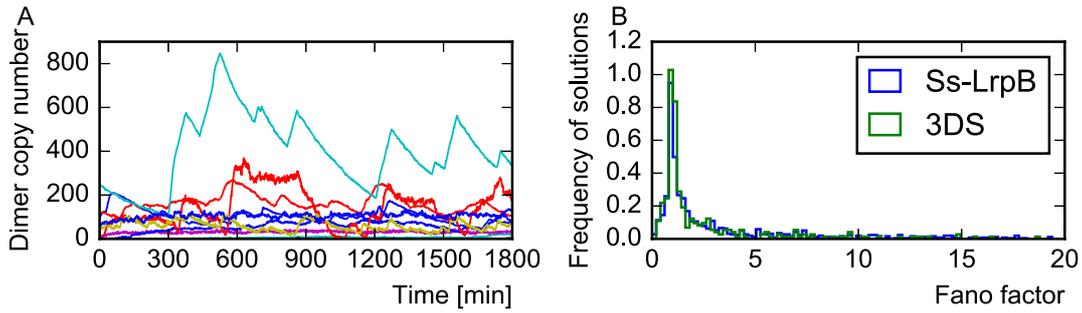

**Figure 5.10:** Bursting behavior of the 3DS systems: (A) 15 time traces compatible with the Ss-LrpB system (B) Histogram of Fano factors distributions for 3DS and 3DS systems compatible with Ss-LrpB systems.

## 5.5 Conclusion

The question we addressed is the function of non-monotonic autoregulation. A system with such autoregulation exists in nature and we therefore assume it should have an advantage over other gene regulatory networks. We seek whether the non-monotonicity could result in dynamics other than a stable steady state, in particular oscillations and bistability. We studied three different single gene networks with increasing complexity that can exhibit non-monotonicity.

Despite the fact that oscillations through Hopf bifurcations become more abundant when the number of binding sites increases, they stay very scarce in parameter space. Moreover the oscillating range is finely tuned for multiple parameters. We also found that the amplitudes of monomers often exceed the amplitudes of the dimers even though the dimer is typically the active form of the protein. Because of the counterintuitive amplitudes and the finely tuned parameters, we conclude that it is highly unlikely that a natural system with a non-monotonic self response would be oscillating.

Besides oscillations, a non-monotonic type 1 response can, just as a monotonically increasing response, give rise to bistability. Repression for high concentrations can theoretically shorten the induction time to go from the unstable intermediate state to the high steady state. If the high steady state is a response on external stress, the fast reaction would give an advantage for survival for each individual cell in contrast to the stochastic switching of the bet-hedging strategy which assures survival on population level. We find indeed that the fastest solutions have a non-monotonic response type 1 curve, but only a small fraction of the non-monotonic curves is faster than monotonic implementations. The parameters dictating the response curve need to be adjusted carefully in order to be faster than monotonic responses.

For synthetic circuits, we would advise to build a fast bistable switch out of the two or three binding sites models. To build an oscillator, our sensitivity analysis provides a list of parameters that should be finely tuned and should be screened over to find a working system, however we expect it to be challenging to succeed in that enterprise.

The question of the function of the natural Ss-LrpB system is discussed in the previous section. To sum up, a scan over parameter space revealed that the response curves compatible with the natural Ss-LrpB system cannot give rise to bistability. Oscillations would need finely tuned parameters. Ruling out switches and oscillators, we consider simply steady states and fluctuations due to intrinsic noise around these steady states. We concluded by proposing that the natural Ss-LrpB system dynamics has spiking behavior around a steady state. A possible explanation for such behavior is that the concentration of Ss-LrpB needed for sensing is too high to be maintained at steady state.

# Microbial communities

# Microbes and bacteria | 6

A WHOLE NEW WORLD WAS DISCOVERED by Antoni van Leeuwenhoek in the second half of the 17th century when he looked into his self-made microscope and saw a multitude of small living entities. He called them "diertjes" in his mother tongue, Dutch, which translates literally as small animals and is translated in the literature as "animalcules". To demonstrate his astonishment, I quote van Leeuwenhoek

> I saw ... incredibly many of the very little animalcules ... and the whole water seemed to be alive[1]         (van Leeuwenhoek, 1676)

[1]: The original Dutch version goes: "ik ... ongelooflijk veel seer kleijne diertgens ... sag ... en scheen het gansche water ... als te leven".

These omnipresent "animalcules" are now referred to as microorganisms of which bacteria are probably the most familiar. Microorganisms are classified as such through their size, but this group encompasses species of different domains in the phylogenetic tree such as archaea, micro-eukaryotes, and bacteria. It is no surprise that van Leeuwenhoek reports large numbers of microorganisms, because estimates of the total number of bacteria on earth mount up to $10^{30}$. Microorganisms are also found at almost any place on earth, including in and on our bodies. Over 5000 operational taxonomic units (OTUs)[2] have been identified in the human gastrointestinal tract (Nayfach et al., 2019) and the number of bacteria in our gut is about $4 \cdot 10^{13}$, which is as high as the number of human cells in our body (Sender et al., 2016).

[2]: OTUs can be interpreted as species. The exact definition is given in Section 2.2.

Soon after their discovery, bacteria were associated with *pathogens*. They were believed to make us sick, but in reality, most of them are harmless and we even need large bacterial communities to survive. The importance of bacteria and other microbes cannot be underestimated. Not only are they involved in digestion, but they also play large roles in processes in many domains such as food processing, agriculture, medicine, biotechnology, etc.

Microbial life is important in many aspects of our lives. We here list some of them, but the list is certainly not exhaustive and plenty of other examples can be found.

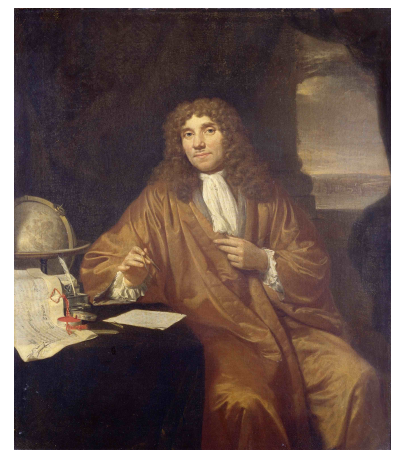

**Figure 6.1:** A portrait of Antoni van Leeuwenhoek (1632-1723).

▶ In natural ecosystems, microbes play an important role in *biogeochemical transformations*. They help with the decomposition of dead organisms. Hence, they release the stored $CO_2$ in the air, where it can be absorbed by plants during photosynthesis. Without microorganisms the carbon cycle would come to rest.

▶ For thousands of years humans have exploited microorganisms in the preparation of food to improve the taste and achieve fermentation for the *preservation of food*. Yoghurt, cheese, beer, bread, and chocolate are only some examples of products of which the preparation relies on microbial cells.

▶ Microorganisms are also deployed in the *chemical industry* to produce a vast number of chemicals for the energy industry, dietary supplements such as vitamins, pharmaceutical products, or insulin.

▶ Some bacteria can be used as insecticides (Chattopadhyay et al., 2017). This alternative to chemical pesticides is increasingly explored as the latter raise many ecological questions.

▶ Some bacteria which are used to treat wastewater, produce at the same time electricity. This concept is extensively used in *microbial fuel cell technology* (El-Naggar et al., 2010; Pandey et al., 2016).

## 6.1 Microbial species live in stable communities[3]

3: The difference between the words "population" and "community" is that populations denote a group of individuals of all the same species and communities are a combination of multiple populations.

Microbial species are often found in communities. Different species live together in the same environment in specific compositions: the relative abundances of the species depend on the ecosystem. Experiments have shown that species interact with each other. These interactions are often mediated by the environment, *i.e.* microbes chemically modify their surroundings. The interactions can be classified into different categories. Some species share a mutualistic relationship through metabolic cross-feeding. Others compete for the same resources. Evidence suggests that competition for shared resources happens particularly between closely related taxa (David et al., 2014). The large number of species and interspecies interactions in microbial communities make them fairly complex. However, many microbial communities are sta-ble[4]. It is not yet clear how many species can coexist and what the underlying mechanisms to guarantee the stability of large and complex communities are. This is why they are studied with techniques from theoretical population dynamics.

We can distinguish two forms of bacterial growth: a unicellular and a multicellular life stage. In the first, cells are planktonic, *i.e.* free-swimming in their environment. In the latter, cells become sessile by attaching to a surface. They do so in many-species communities and by building extracellular matrices out of exopolysaccharides, nucleic acids, proteins, etc. These matrices, which are also called biofilms (Figure 6.2), have characteristics that planktonic cells do not possess. Their physiology is different, they show better resistance to antibiotics and molecules are shared between the metabolic pathways of various species. Moreover, biofilms create spatial structures, which are important for their dynamics (Nadell et al., 2016). Biofilms have been reported in the human appendix. After a harmful shift of the gut microbiota, the appendix may have a crucial role to reinoculate the gut (Bollinger et al., 2007). Whether biofilms are also present in other parts of the gastrointestinal tract is still subject to debate (W. M. de Vos, 2015).

4: Interestingly, in experimental setups with constant functionality of the species, it is found that bacterial populations can show chaotic fluctuations, whereas populations of archaea have stable abundances (Fernández et al., 1999).

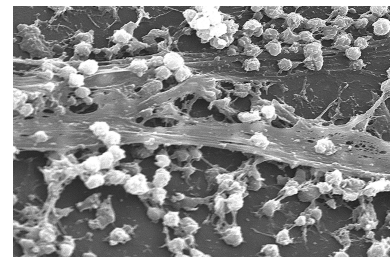

**Figure 6.2:** Staphylococcus aureus biofilm on the surface of a catheder.

### Experimental microbial communities show high diversity and heavy-tailed abundance distributions

In recent years, biodiversity has become a hotly debated subject in the context of global warming and human overexploitation of nature. Pollution ruins the habitats of species that subsequently go extinct. This loss of biodiversity is a real catastrophe, but why is biodiversity this important? First of all, a large portion of the world's industry and economy depends on biological resources. Nobody knows which biological products that can be used in the medical, food, or other industries still remain to be discovered. The extinction of such unexplored treasures would be a huge loss for society. Also from other perspectives than the anthropologic one, diversity is important for keeping stability. Decreasing diversity leaves more room for invasive species (Groom et al., 2006). For the aforementioned reasons, it is important to study how high diversity is obtained and maintained.

Several concepts are related to diversity:

▶ The *species richness* is simply the number of different species in the ecosystem.
▶ The *species evenness* measures the similarity of the abundances of all species. It can be described by *Pielou's evenness index*. The mathematical definition can be found in Appendix A.2.
▶ The *effective number of species* is the total number of individuals divided the mean abundance of all species.

Just like other ecosystems on earth, microbial communities show high diversity. Some mechanisms that promote or maintain diversity have been studied. One such mechanism is cooperation or altruism. These types of interactions are favored because colonizing species are selected for by the residing composition: species that are not contributing to the overall welfare of the community have a smaller probability of establishing in the new community (Griffin et al., 2004; Travisano & Velicer, 2004). For example, in the gut microbiome Bacteroides and Firmicutes probably coexist because of their complementary functions. However, some species are functionally redundant, especially within the same family, but they can coexist by preferring different substrates (Sonnenburg et al., 2005). One theory also states that diversity in communities can grow through a positive feedback loop where diversity creates new niches and, therefore, even more species can join the community. This theory has been used to demonstrate the plant and animal diversity on islands (Emerson & Kolm, 2005).

5: rRNA tags are used to define OTUs, which classify individuals (see Section 2.2).

Microbial communities, like many ecological systems, exhibit heavy-tailed distributions, which means that there are few species in high abundance and many species in low abundance. For example, in microbial communities of agricultural soil, 99% of unique rRNA tags[5] is from less than 1% of the population (Ashby et al., 2007). The abundance distribution can often be described by a lognormal or power law function (Preston, 1948). The unevenness can also be represented by a *rank abundance distribution*, a plot with the rank—the position of the species when all species are sorted from most to least abundant—on the *x*-axis and the abundance on the *y*-axis. Because of the high variance, a log-log scale is often used. For heavy-tailed distributions, the rank abundance distribution is steep.

## 6.2 Gut microbiota

We live together with a multitude of microbes, but most of them are harmless or even necessary for our survival (Grice & Segre, 2012). These microorganisms are not only bacteria but also archaea, viruses, phages[6], yeast, fungi, and other micro-eukaryotes. The whole body is colonized by microbes, but the gut is the most densely populated region. As mentioned earlier, the number of bacteria in our gut is about $4 \cdot 10^{13}$ which is as high as the number of human cells in our body (Sender et al., 2016). Not only are we almost outnumbered in cell count, but the combined genome of the human intestinal microbiota exceeds the human genome by more than two orders of magnitude[7].

The microbial *composition* of the gut microbiome plays an important role in human health. An imbalance of the composition has already been related to cardiometabolic disorders, inflammatory bowel disease, diabetes, neuropsychiatric diseases[8], autism, and colorectal cancer (Cani, 2018; W. M. de Vos & de Vos, 2012). For example, the ratio of Firmicutes over Bacteroides has been associated with metabolic disorders (Cani, 2018). This shows that even the composition at the phylum level has a large impact. Even though the relative abundances of all species are crucial for good health, the absolute quantities of the microbes seem to be even more important (Vandeputte et al., 2017).

An unhealthy, ill microbiome is often associated with an unstable microbiome (Lozupone et al., 2012) which shows large fluctuations over time or which has a small diversity, *i.e.* a low number of different species. Despite the large turnover rate—species are growing, dividing, and dying incessantly—healthy people show a *stable gut composition* over long periods of time which is resistant to invasions (Caporaso et al., 2011; David et al., 2014; Faith et al., 2013; Zoetendal et al., 1998).

The relationship between the human microbiome and its host is not merely commensal or symbiotic, but rather *mutualistic*. The human body and its associated microbial communities benefit from each other and even depend on each other. They make up for 10% of the calorie intake. They provide a thousand different digestive enzymes, whereas human cells only produce 20. They take care of the fermentation of non-digestible carbohydrates through the production of

6: Phage, short for bacteriophage, is a virus of bacteria and archaea. In the human gut, phages outnumber bacteria by 10 to 1 (Cani, 2018).

7: The combined genome of the human intestinal microbiota contains over 10 million genes (J. Li et al., 2014) and the human genome contains only 35.000 genes.

8: Researchers have proven that gut microbiota are connected to the brain, this connection is called the *microbiome-gut-brain axis*. Via biochemical signaling, the gastrointestinal tract (GIT) can influence the central nervous system (CNS). It is not yet known whether the microbiome is either the cause, or the consequence of specific neurological conditions or a combination through multiple complex feedback-loops in the signaling between the GIT and the CNS.

short-chain fatty acids and the synthesis of K, B2, and B12 vitamins.

On evolutionary scales, the distributions of the gut microbiota have co-evolved with their hosts. To be able to establish a population in the gut, microbes need enzymes that consume nutrients, molecular structures to attach to the gut wall and a fast growth rate to avoid washout (Ley et al., 2006). In the next section, we discuss the evolution of a specific microbiome at a smaller timescale, the timescale of a life.

## Development of individual gut microbiome

9: This is still a little controversial. Theories of an in utero microbiome can be found in Aagaard et al., 2014; Perez-Muñoz et al., 2017.

Fetuses are sterile[9]. The very first contact with bacteria happens during the delivery, this is why the gut microbial composition of a newborn resembles the vaginal or skin microbial composition of the mother depending on the method of delivery (Adlerberth & Wold, 2009; Dominguez-Bello et al., 2010). Soon the microbiome is influenced through breastfeeding and the environment of the child (Rothschild et al., 2018). For example, there is a correlation between the ownership of pets and the microbiome (Azad et al., 2013; Kates et al., 2020; Tun et al., 2017). Also, it was observed that people who live together have similar microbiomes (Dill-McFarland et al., 2019).

In early life, the community has low diversity and is highly dynamic. The composition shifts while the child grows until it attains a steady composition at the age of 2.5 years (Koenig et al., 2011). The mature microbiome is both more diverse and stable. The *immune system* plays a critical role in shaping our microbiome and vice-versa the microbiome trains the immune system. At old age, the diversity gets lost again (Claesson et al., 2011, 2012).

## Spatial and temporal structure of the gut microbiome

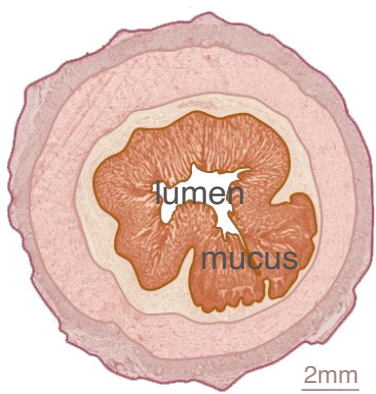

**Figure 6.3:** Transversal cross-section of the small intestine.

Two major axes determine the directions in the intestines: the longitudinal axis from proximal to distal and the radial axis from the lumen to the mucus. Figure 6.3 shows the cross-section of the small intestine. A thick layer of mucus separates the microbiome and the epithelial cells, acting as a barrier. In this way, an immune response from our body towards the microbiome is prevented (Johansson & Hansson, 2016).

Because the gastrointestinal tract is not homogeneous, the gut microbiome has also a spatial structure. Studying the longitudinal axis, the microbial density and diversity increases along the gastrointestinal tract (Donaldson et al., 2016). For the radial axis, the microbial composition in the mucus is very different from the composition in the lumen (Zoetendal et al., 2002). For example in mouse and rat intestines, differences in the morphology of the bacteria can be found along the radial axis: long spiral rod-shaped bacteria thrive in the mucosal areas whereas coccus-shaped microbes have a higher density in the central lumen (Swidsinski et al., 2007).

Next to spatial dependence, the microbial composition depends also on time. On the timescale of years, the composition changes because of the development of the host as discussed above, but also on shorter timescales, over the course of a day, the composition can show variations. This is not surprising because the host lives according to a circadian rhythm and eats at specific timings. It has been demonstrated that these fluctuations are reflected in the microbiome (Liang & FitzGerald, 2017). Apart from changing the community's environment, food and water also introduce *allochthonous microbes*, which are discerned from the *autochthonous microbes*, or the so-called residents. It is still unknown if these two groups of bacteria are functionally different (Nava & Stappenbeck, 2011).

**Experimental analysis**

There is undoubtedly a spatial structure in the gut microbiome, but peristalsis will homogenize the microbial populations and nutrients in the colon to some extent. Zoetendal et al., 2002 found that the bacteria in the mucus are uniformly distributed, but that they differ from the bacteria in the lumen. Therefore the community of the colon can be considered well-mixed with respect to the longitudinal axis. It is very difficult to make direct observations of the composition at different places in the gut. For this reason, most analyses happen by studying the *stool*[10]. The composition of fecal matter is, however, significantly different from the composition of mucosa-associated bacterial communities (Zoetendal et al.,

10: This technique is rather old. In the end of the 19th century, Theodor Escherich studied infant feces. In this work, he described the *Bacterium coli commune*, or the common colon bacillus. This bacterium later was named after this researcher, the *Escherichia Coli*, and became one of the most studied organisms (Shulman et al., 2007).

11: Cross-sectional means taking samples of different individuals at one time point and longitudinal means samples of one individual for different time points.

2002). In fecal matter, on average, 30% of the cells are dead and 20% are damaged (Ben-Amor et al., 2005). Even without sequencing stool samples, certain assumptions can be made by looking at the hardness of the stool, or equivalently the water content of the feces. The latter variable has been correlated to the microbial composition. Higher scores on the Bristol stool scale (diarrhea end of the spectrum) correspond to a higher relative abundance of the Bacteroidetes (Vandeputte et al., 2016). Multiple cross-sectional and longitudinal studies[11] of stool have been performed. Their results are discussed in the subsequent section.

## Gut microbial composition cluster in different groups

All humans have their own unique gut microbial composition (Huttenhower et al., 2012). The variation of microbial composition between different individuals is much larger than the variation over time of the microbial composition of one individual (David et al., 2014). The variation of microbiomes of different locations in and on the body of a single individual is also larger than the variation between microbiomes of the same location in different individuals. Obviously, the microbial compositions are influenced by the lifestyle of the host, for example, fiber is the crucial element of your diet that determines the gut microbiome (David et al., 2014). The question remains how large the role of the host's genetic material is in determining the gut composition. Tests have been performed on people with varying genetic relationships (from mono- and dizygotic twins[12] to unrelated individuals) and the degree of resemblance of their microbiota is correlated with their genetic relatedness (Zoetendal et al., 2001), suggesting that the host's DNA plays a significant role in determining the microbial composition.

12: Monozygotic twins are also called identical and dizygotic non-identical.

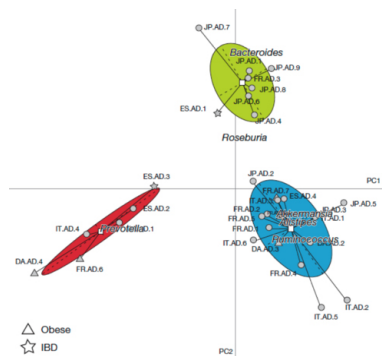

**Figure 6.4:** The human gut microbiome of different individuals clusters in three groups, called the enterotypes.

Although the compositions of many individuals are different, they seem to form clusters. Arumugam et al., 2011 discovered three community types, referred to as *enterotypes* which are independent of race, gender, age or body mass index (Figure 6.4). Each of the three types can be identified by the abundance of three genera: *Bacteroides*, *Prevotella*, and *Ruminococcus*. Whether we can divide the composition space into three distinct types, if there are peaks in a landscape (Falony et al., 2016), or if all variation is closer to a gradient (Gorvitovskaia et al., 2016; Knights et al., 2014), is still subject to debate. The human gut microbiome is not the only ecosystem that shows clustering: enterotypes were also discovered in the cynomolgus macaque microbiota (X. Li et al., 2018).

**Studying the gut microbiome**

In this chapter, we have explained the importance of microbial communities in many domains, focusing on the importance of the gut microbial composition for a human's health. We have shown that the number of involved species is high and that understanding how they all work together is not straightforward. For the above reasons, it is important to propose mathematical models that can predict the community's composition. Especially in the context of individualized medicine, we would like to be able to understand how communities lose stability or diversity and make their host ill. Therefore, the goal is to build predictive models that provide us with strategies that can be used to prevent and cure illness.

# Modeling microbial communities | 7

CULTURING BACTERIA IN THE LAB OFTEN PROVES TO BE VERY DIFFICULT, especially for anaerobic species. However, the ultimate goal of microbial ecology is to have predictive power over the community dynamics. Therefore, theoretical models and simulations are useful tools. One specific field that would profit from predictive models is *individualized medicine*: there is a growing demand for personalized models of the gut microbiome. Such models can predict the microbial composition after administering drugs or a change in diet, they can offer a custom solution for every dysbiosis. Also, other microbial systems, natural or synthetic ones, will benefit from predictive models. But there exist many types of models and using the words of Wennekes:

> Science is an abstraction and it is subjective; scientists leave out elements that they consider to be unimportant, and keep the remainder, but different scientists will have different opinions on what the important elements are.
> 
> (Wennekes et al., 2012)

This quote emphasizes that different approaches exist to model a system depending on which aspects are considered important. This is not different for models of microbial communities where a distinction is made between two types of models: individual-based versus population-level models. These models, respectively following a bottom-up and top-down approach, complement each other (Nisbet et al., 2016).

▶ *Individual-based models* (IBMs), pioneered by de Angelis and Gross (DeAngelis & Gross, 2018)[1], describe all separate individuals or small groups of individuals. Although IBMs lack a general framework because of the individuality of the agents, they account for the variability between individuals and can track the life cycle of (small groups of) individuals. Even though these

1: IBMs were first used in the context of forest canopies (Botkin et al., 1972), but later extended to all kind of problems in ecology (DeAngelis & Gross, 2018). The history of IBMs can be found in (DeAngelis & Grimm, 2014).

models are computationally very demanding, they have often a conceptually understandable implementation: a simple set of rules for individuals determines the evolution of the community from one time point to the next which gives rise to patterns at the population-level. IBMs encompass patch and urn models. Patch models assume that all considered species live on the same patch close to each other, such that there are no spatial effects. In urn models, one or more individuals are subsequently picked like balls from an urn after which they have a certain probability to perform an "action" like growth, death, or interaction. These automaton models can also be described by a master equation[2] (McKane et al., 2000).

2: This stochastic description was discussed in Section 3.4.

▶ On the other hand, there are the *population-level models* (PLMs). They use a mean-field-approximation to describe communities by the abundances of their composing species. This technique results in an elegant mathematical formulation. A drawback is that general interactions between species do not always directly map to microscopic processes.

Other aspects of the problem that can be included or excluded from the model are the spatial structure, the community's environment, and the metabolites through which species interact. These elements can be implemented in both IBMs and PLMs. For spatial modeling, PLMs use partial differential equations (Mimura & Tohma, 2015; Reichenbach et al., 2008). In IBMs, spatial dependence can be implemented through a 2 or 3 dimensional grid, or continuous space (Daly et al., 2019; McCauley et al., 1993). In models with localized interactions, global spatial patterns arise. In our research, we assume that the microbial content of fecal matter is homogeneous and use models that do not have spatial dependence.

The inclusion of metabolites, instead of a direct description of species-species interactions, improves the ability to capture the dynamics of microbial communities (Brunner & Chia, 2019). Besides the addition of the nutrients, Smith, 2002 also examines the effect of the elemental stoichiometry of the nutrients and concludes that the resource availability and the resource supply ratios have a significant impact on plant communities.

One of the very popular models is the generalized Lotka-Volterra model, which is a PLM for interacting species without spatial dependence or metabolites. This model is presented in the following section.

## 7.1 Lotka-Volterra equations

In this section, we introduce the generalized Lotka-Volterra (gLV) equations which we use to model communities of interacting species. A more complete introduction on the gLV equations can be found in the lecture notes of Allesina which can be consulted online (Allesina, 2020).

The organisms for which population growth is most easily modeled, are bacteria in a perfect environment, *i.e.* with an unlimited amount of nutrients. In such conditions, bacterial cells will grow and divide after every time interval $\tau$, which is called the doubling time. Another simplification comes from modeling bacterial growth at the population level. If the number of bacteria in the colony is large, this number can be approximated by a real number instead of an integer one. In ideal conditions, the dynamics of a bacterial population can be described as

$$\dot{N} = gN \tag{7.1}$$

with $N$ the number of bacteria and $g$ the replication rate of the bacterium. It follows that bacterial growth is *exponential* $N(t) = N(0)\exp\{gt\}$[*]. This model is also known as the *Malthusian growth model* named after Thomas Robert Malthus, an English economist of the 18th century.

Of course, no real-life system has unlimited resources, and models that include saturation effects are needed. One such model was introduced by the Belgian mathematician, Pierre François Verhulst. He published an equation for the growth of a population in 1838 (Verhulst, 1838). This equation, referred to as the *Verhulst equation* or the *logistic equation*, is still used today and can describe the growth of non-interacting species. As for the exponential Malthusian model, the growth term is linear with the number of species. The saturation effect is implemented by the carrying capacity $K$, which will prevent endless exponential growth.

[*]: There is of course stochasticity, leading to fluctuations in growth rate and generation times of single cells. The population growth rate depends on the distributions of the single-cell growth rates and their correlations (Lin & Amir, 2017, 2020).

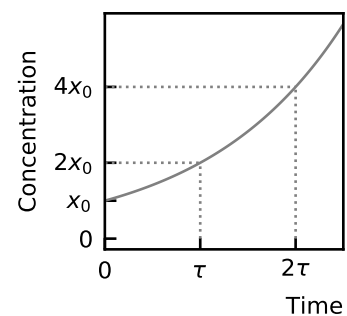

**Figure 7.1:** In perfect conditions, bacterial colonies show exponential growth.

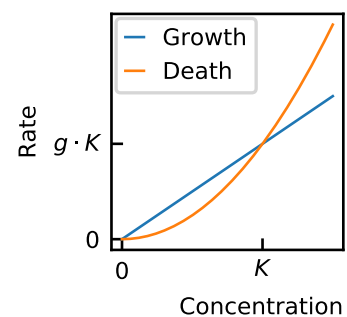

**Figure 7.2:** The growth and death rate of the logistic equation are linear and quadratic to the concentration. The intersections of both curves denote steady states.

The Verhulst equation reads

$$\dot{N} = gN(1 - \frac{N}{K}) \quad (7.2)$$

with $N$ the number of individuals, $g$ the intrinsic per-capita growth rate and $K$ the carrying capacity. The carrying capacity slows down the growth rate for an increasing number of individuals by a factor $(1 - N/K)$, it embodies a crowding effect. In view of terminology which is also used later for the Lotka-Volterra equations, the product of the growth rate and the negative inverse of the carrying capacity, $-gK^{-1}$, can also be interpreted as the self-interaction, the effect an individual receives from interacting with other individuals of the same species. We assume that this interaction effect between individuals of the same species is negative because these species compete for the same nutrients and produce the same toxic waste products. Therefore the mathematical representation of the effect is negative $-K^{-1}$.

The ordinary differential equation Equation 7.2 can be solved analytically, and the solution is

$$N(t) = \frac{KN(0)\exp(gt)}{K + N(0)(\exp(gt) - 1)}. \quad (7.3)$$

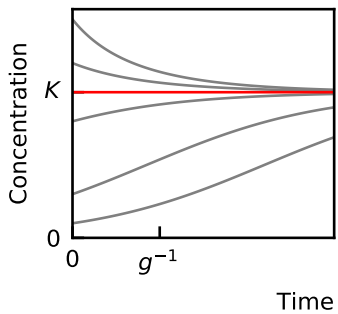

**Figure 7.3:** All trajectories approach the non-trivial steady state $K$.

The steady states of Equation 7.2 can be found by equalling the time derivative to zero ($\dot{N} = 0$), which results in the trivial solution $N^* = 0$ and $N^* = K$.

Experiments show that the logistic equation is indeed a good approximation for modelling bacterial growth in optimal conditions, *i.e.* a constant environment with a continuous supply of nutrients (Zwietering et al., 1990). The growth curves of more complex life forms that have a reproduction scheme with more intermediate steps such as eggs, larvae, etc. or that require multiple individuals for reproduction do not fit well to logistic equations and require more elaborate models.

Logistic models only consider one species at a time. In the next paragraphs, generalized Lotka-Volterra models, which describe the time evolution of a community of interacting species are discussed. The original Lotka-Volterra equations were proposed independently by Alfred J. Lotka (Lotka, 1920) in 1920 and Vito Volterra (Volterra, 1926) in 1926. They describe the dynamics of predator and prey populations. This set of nonlinear differential equations results in deterministic and continuous solutions.

| $\omega_{ij}$ | $\omega_{ji}$ | Name of the interaction |
|---|---|---|
| 0 | 0 | neutralism |
| - | - | competition |
| + | - | predation/parasitism * |
| + | + | mutualism/protocooperation † |
| 0 | - | amensalism |
| 0 | + | commensalism |

**Table 7.1:** Types of two-species interaction, adapted from Odum, 2005. The first two columns denote if the interaction $\omega_{kl}$ is zero (0), has a positive sign (+) or a negative sign (-).

The predator-prey equations can be generalized to all sorts of interactions between species such that they can represent large and complex food webs. Species abundances can increase through immigration, explicit growth, and growth mediated by the interaction with other species. Similarly, the abundances can decrease by emigration, diffusion, explicit death, and death mediated by the presence of other species. The *generalized Lotka-Volterra equations* dictate the time derivative of the abundance of species $i$, $x_i$

$$\dot{x}_i = \lambda_i + x_i \left( g_i + \sum_j \omega_{ij} x_j \right) \qquad (7.4)$$

where $\lambda_i$ is the immigration term, $g_i$ the growth rate of species $x_i$ and $\omega$ the interaction matrix. The growth rate $g_i$ is not necessarily positive. When it is, it means that the species can grow in the absence of other species. Zero or negative growth rates indicate that the species needs other species for survival. For example, the predator in the original Lotka-Volterra equations has a negative growth rate, because it can only survive in the presence of its prey.

Every element of the interaction matrix $\omega_{ij}$ dictates the influence of a species $j$ on species $i$. All types of interactions between two species $i$ and $j$ can be described by combining positive, negative, or zero values for $\omega_{ij}$ and $\omega_{ji}$. They are all listed in Table 7.1.

In microbial communities, competition can arise through fights for shared resources, but also through toxin production (Chao & Levin, 1981; James et al., 2013). Furthermore, in microbial communities, competition dominates over cooperation

*: The difference between parasitism and predation is the relative size of the species causing the harm with respect to the species that is being harmed, parasites are generally smaller than their host, whilst predators are usually larger than their prey.

†: The difference between protocooperation and mutualism is the need for this interaction to the survival, in a protocooperation the interaction is favorable to both species but not obligatory, whereas in mutualism the interaction is needed for the survival of both species

(Foster & Bell, 2012).

## Assumptions and limitations

The generalized Lotka-Volterra model (Equation 7.4) is based on three assumptions:

1. The environment is *well mixed* and stable. Therefore, the parameters cannot be space dependent.
2. The growth of a species is linear with the number of species. This assumption holds for populations of bacteria that reproduce asexually.
3. Breeding is continuous and generations are overlapping. Also this assumption works well for bacteria or other unicellular organisms. Subtleties arise for larger organisms that reproduce seasonally.

The generalized Lotka-Volterra model is an *empiric* model where only effective interactions are modeled between species. Although one obtains better results through modeling both the species and all metabolites (Brunner & Chia, 2019), gLV are still widely used for their simplicity. Another drawback is that gLV only considers pairwise interactions that are not influenced by third parties. Such interactions could potentially be added as higher-order terms and give rise to nonlinear functional responses (Drossel et al., 2004; Gonze et al., 2018; Sidhom & Galla, 2020).

## Dynamics and stability

If the interaction matrix $\omega$ is not singular, we find a non-trivial steady-state solution of Equation 7.4: $\vec{x}^* = -\omega^{-1} \cdot \vec{g}$. Because $\vec{x}$ represents the vector with the species abundances, we are looking for solutions in the positive orthant, which we call feasible equilibria. If a system has no feasible equilibrium, all trajectories starting in $\mathbb{R}^n_{\geq 0}$ will reach a boundary of $\mathbb{R}^n_{0+}$ [3]. In 1976, Smale showed that for systems of five species or more, any asymptotic behavior can be obtained—in particular stable steady states, limit cycles, and chaos (Hirsch, 1982; Smale, 1976).

Let us consider a system with a feasible equilibrium. What can we tell about the stability of such a point? The stability of a steady state can be determined through linear stability

3: The proof of theorem can be found in 5.2.1 of Hofbauer and Sigmund, 1998.

analysis, which was introduced in Section 3.2. An equilibrium is said to be stable if all of the eigenvalues of the Jacobian of the system evaluated in this equilibrium have negative real parts.

To obtain a valid model and keep the solutions bounded, the self-interaction terms $\omega_{ii}$ need to be negative[4]. Complex systems with many and strong interactions are generally not stable. The stability-complexity debate will be discussed in Section 7.3 where different constraints on the interaction matrix that lead to stability are presented.

## Application of Lotka-Volterra models

The Lotka-Volterra model was used to explain classical predator-prey systems such as the lynx-hare dynamics in Canada (Gilpin, 1973; Nedorezov, 2016), but also other population dynamics such as the one of marine phages (Gavin et al., 2006). Aside from population dynamics, these equations were also used in the modeling of cancer and the immune system (Babbs, 2012; Sotolongo-Costa et al., 2003). These models even find applications outside biology: in business (Hung et al., 2017), social sciences (Epstein & Axtell, 1996), network communication (Zhang et al., 2019), economics (Apedaille et al., 1994), etc.

## Fluctuations in time series

Experimental time series show large fluctuations over time which are not all the result of the deterministic dynamics. Stochasticity, which, in this context is also referred to as noise, plays a large role and can be introduced in dynamical systems through different processes. In simulations of microbial communities at the species level, we distinguish intrinsic and extrinsic noise. *Intrinsic noise* is a result of the stochastic nature of cell division and cell death together with the discreteness of individuals. *Extrinsic noise* is caused by environmental fluctuations, *e.g.* changing nutrients, pH, temperature. To implement this extrinsic noise in a system of gLV equations, the changing environments are often interpreted as changing parameters such as the growth rate. Because this parameter

4: If we would imagine a system with a positive self-interaction $\omega_{kk}$ for one $k$, there exists a threshold abundance $\tilde{x}_k > 0$ such that the highest order term is dominating for all abundances equal or bigger than the threshold one,

$$\omega_{kk}x_k^2 > |\lambda_k + x_k(g_i + \sum_{j \neq k}\omega_{jk}x_jx_k)|$$

with $x_k \geq \tilde{x}_k$. Because the time derivative of the species with the positive self-abundance, $\dot{x}_k$, woud be positive for all $x_k \geq \tilde{x}_k$, all trajectories starting at or above this threshold abundance would go towards an inifinite value for $x_k$. Because infinite amounts of species are not physical, any system with positive self-interaction is ill defined.

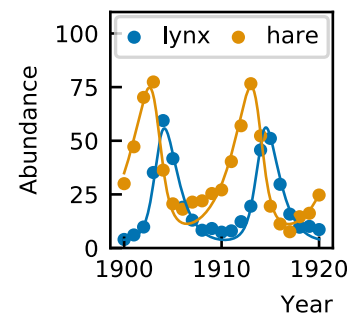

**Figure 7.4:** The dynamics of the lynx and hare populations can be described by the predator-prey model of Lotka and Volterra. Data from (MacLulich, 1937)

This paragraph is based on the supplemental material of our publication (Descheemaeker & de Buyl, 2020).

is multiplied by the species abundance, the resulting noise is linear:

$$dx_i(t) = \lambda_i dt + g_i x_i(t) dt + \sum_j \omega_{ij} x_i(t) x_j(t) dt + x_i(t) \sigma_i dW(t), \quad (7.5)$$

where $dW$ is an infinitesimal element of a Brownian motion defined by a variance of $dt$ ($dW \sim \sqrt{dt}\mathcal{N}(0,1)$). The remaining parameters, inter- and intra-specific interactions, can also change depending on the environment. The formulation of this noise is more subtle (Mao et al., 2003; Zhu & Yin, 2009), and is not used in our work.

The second type of extrinsic noise arises because of fluctuations in the immigration rate of species. This noise is constant and therefore additive:

$$dx_i = \lambda_i dt + g_i x_i dt + \sum_j \omega_{ij} x_i x_j dt + \sigma_{i,\text{const}} dW. \quad (7.6)$$

To derive the form of intrinsic noise in gLV equations, we can consider every species abundance making a random walk in one dimension. The average displacement is zero, and the variance of displacement is the sum of the rate of growth (jumping to the right) and the rate of death (jumping to the left). For the gLV equations, this results in a noise term

$$\langle n_i(t) n_i(t') \rangle = (\text{growth rate}_i + \text{death rate}_i) x_i \quad (7.7)$$
$$= (f(g_i) + h(\omega, \vec{x})) x_i \delta(t - t'), \quad (7.8)$$

with $\omega$ the interaction matrix, and where functions $f$ and $h$ each decouple the growth and death terms. In the gLV model, no difference is made between negative interactions as a result from slowing down the growth rate or increasing the death rate, only the resulting net rates are used. This distinction is however important for the implementation of the intrinsic noise for gLV. For a model without interactions—the logistic equations—the resulting variance of the noise is proportional to the square root of the abundance of a species $\sqrt{x}$ (Walczak et al., 2012). One must be careful not to use this noise for values that are smaller than 1, because this derivation relies on Poisson statistics which is defined for integer numbers. Intrinsic noise can thus be approximated by noise proportional to the square root of the species abundance:

$$dx_i(t) = \lambda_i dt + g_i x_i(t) dt + \sum_j \omega_{ij} x_i(t) x_j(t) dt + \sqrt{x_i(t)} \sigma_{i,\text{sqrt}} dW_{\text{sqrt}}.$$

(7.9)

We here described how one can simulate stochastic time series. As mentioned in Section 3.4, probabilistic models offer another way to study stochasticity. For the gLV model, a description using Fokker-Planck equations can be used (Goel et al., 1971). This technique was not used in this thesis, but an elaborate introduction can be found in Walczak et al., 2012.

## 7.2 Neutral and niche theory

There is a longstanding debate in ecology about the nature of diversity in communities. Two opposing theories have been developed: the neutral and the niche theory. The niche of a species embodies all elements it needs to grow and reproduce, from nutrients to the environment. Many organisms rely on the same elements, but are adapted for particular conditions. These niche differences have been used to explain diversity, but also theories without niche differences, the so called neutral theories, allow diversity and heavy-tailed distributions.

The *neutral theory* assumes that every species is ecologically equivalent. The species' niches, therefore, overlap perfectly (Figure 7.5) and all birth, death, and migration processes are governed by stochasticity (Harpole, 2010). Because all species have equal niches, a neutral model cannot combine species of widely varying sizes or at different trophic levels. For example, plants and microbes cannot be described as one community by a neutral model. The community is studied in the context of a larger meta-community from which species can immigrate. An example is a field with many plants in which seeds from surrounding plants can immigrate and settle. A neutral model can, by definition, not include any specific interactions between species. The focus is not at all on the differences between species but on stochasticity. Individuals are treated equally—in our example, all individual seeds have an equal probability of growing to a plant—and the abundances of species fluctuate over time due to randomness.

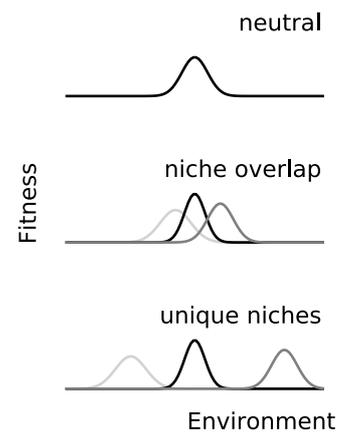

**Figure 7.5:** In a neutral theory, the niches of different species are perfectly overlapping and they all have an equal fitness level in a given environment. In a niche theory, the niches of the different species are overlapping or unique.

Depending on the strength of immigration, the community will either resemble its meta-community (for high immigration rates) or develop into a community different from the meta-community (for low immigration rates). The latter process is called *ecological drift*. In some neutral models, a maximum number of individuals is defined. This is also known as the *zero sum assumption*. Limited capacity results in indirect interactions between species: effective competition emerges. Although the neutral assumption received much criticism for being unrealistic, the theory succeeds in reproducing certain characteristics of ecological systems, such as the rank abundance profile which is heavy-tailed (Bell, 2001; Hubbell, 2001; Volkov et al., 2003). However, it fails to predict which species will have the high abundances and which the low, and how the environmental conditions change the dynamics. The neutral model, being very simple, can be proposed as a null model—an alternative to the niche model which is presented next (Gaston & Chown, 2005; Harte, 2004).

The *niche theory*, on the contrary, assumes that all species have different traits, or in other words their niches are unique or only partly overlapping (Figure 7.5). These differences are the reason why distinct species can coexist. No universal definition of a niche exists, but a mathematical formulation was proposed by Hutchinson, 1967: a *niche* is a (hyper-)volume in a set of dimensions that represents the species' capacity to use resources. Overlapping niches are modeled by interspecies interactions in gLV models, logistic models represent unique niches.

Of course, the degree of difference between all species can range from low to high values. In gLV models, this is expressed through varying growth rates and (self-)interactions. The neutral character is brought by stochasticity. Considering the relative importance of competitive interactions and stochasticity, any system can be situated somewhere along the neutral-niche spectrum (Fisher & Mehta, 2014; Gravel et al., 2006; He, 2005). Although systems at the opposing ends of this spectrum seem unreconcilable, Wennekes et al., 2012 argues that the whole debate is merely philosophical and that a system can be both in the neutral and niche regime depending on the scale at which you are considering the processes. They state that from the perspective of the small scales, all interactions are visible in detail and the dynamics will reside in the niche regime, whereas from the perspective of the large scales, these details are not described and the dynamics seems more stochastic and in the neutral regime.

The degree of neutrality can be described by multiple measures, based on the steady state—Bray-Curtis dissimilarity, Kullback-Leibler divergence, and Jensen-Shannon distance—or on the time series—the neutral covariance test. Their definitions are listed in Appendix A.2.

## 7.3 Stability of models with interactions

We state that many microbial communities are stable, but how exactly do we define stability? There are multiple aspects to dynamic stability McCann, 2000:

- ▶ *Equilibrium stability* is the stability that we discussed in Section 3.2. It is a discrete measures that describes whether a system returns to the equilibrium after a small perturbation.
- ▶ *General stability* measures the extreme values of the population densities. Both the minimal and maximal numbers are taken into account and a system is labelled more stable the further away the abundances are from the extreme values. This definition can be used in non-equilibrium dynamics.
- ▶ *Equilibrium resilience* takes into account the time it takes for the system to return to a stable equilibrium after a perturbation. The faster the response, the more resilient the system is.
- ▶ *General resilience* takes into account the time it takes for the system to return to an equilibrium or non-equilibrium solution. Typical perturbations that are studied in the context of resilience are ecological disasters such as fires, floodings, and storms, but also insect pests and human pollution.
- ▶ *Resistance* measures how populations change after a perturbation, it is the inverse of sensitivity. This measure is often encountered in the context of invasions: when a new species is introduced into a community and it fails to settle, the community is said to resist.

When two species are competing, the one who has an advantage over the other, however small, will eventually outcompete the other. This principle has several names such as *competitive exclusion*, *winner-take-all principle*, or *Gause's law* (Gause & Witt, 1935). It is unknown to what extent ecological principles can be applied to microbial communities. Dynamics are probably more subtle than the winner-takes-all principle where only the most fit species survives. A generalization of this principle is predicted by resource-competition models which state that the number of resources cannot be smaller than the number of coexisting species. In nature, however, we find many systems with a large number of coexisting species and only a small number of resources. One mechanism that allows to maintain diversity is a high immigration rate (Posfai et al., 2017).

### Stability-complexity debate

In the late 1950s, it was thought that stability was created by complexity (Elton, 1972; R. MacArthur, 1955). But almost fifty years ago, May showed that the probability of stability for large ecosystems is close to zero (May, 1972). He proposed an empirical law connecting complexity and stability. Before stating the theorem, we first need to introduce two concepts. We already defined the *community matrix* in Section 3.2, it is the Jacobian of the system evaluated in its fixed point. We define the percentage of non-zero elements in the community matrix as Gardner and Ashby's *connectivity C*—in short the connectivity, or sometimes in the context of ecology the connectance. Furthermore, the non-zero off-diagonal elements of the community matrix have zero mean and variance $\alpha^2$. In this context, we call $\alpha$ the "average interaction strength". For community matrices with -1 on the diagonal[5], May proposed an inequality that separates stable from unstable solutions.

5: One needs to be careful not to confuse the interaction matrix $\omega$ of generalized Lotka-Volterra models and the community matrix
$$M = \mathrm{diag}(\vec{x}^*) \cdot \omega$$
with $\vec{x}^* = -\omega^{-1} \cdot \vec{g}$ the steady state of Equation 7.4 without immigration ($\lambda = 0$).

**May's instability criterion** With $S$ the number of species, $\alpha$ the average interaction strength and $C$ the connectivity, the probability for stability will go to 1 for increasing number of species ($S \to \infty$) if and only if $\alpha^2 S C < 1$, and, the probability for stability will go to 0 for increasing number of species ($S \to \infty$) if $\alpha^2 S C > 1$.

This stability criterion was later derived by Allesina and Tang, 2015 using random matrix theory. May's criterion is sometimes called a paradox as ecologists find that for many systems biodiversity—read complexity—brings stability and that ecosystem with low diversity are more vulnerable to invasions (Ives & Carpenter, 2007; Ptacnik et al., 2008; Tilman & Downing, 1994). However, recently a negative correlation between diversity in microbial communities and interaction strength between its species has been observed (Ratzke et al., 2020). It remains unclear how stability and biodiversity are connected in ecosystems.

As a result of May's criterion, many gLV systems with random parameters are unstable. A possible approach to obtain stable systems is to start from a large pool of species and let the system evolve towards a stable state with a smaller number of species. This smaller subset, also called the non-invasible solution or the saturated rest point, contains a number of species with zero abundance. Using this method, it has been shown that the network structure has little influence on the biodiversity (Serván et al., 2018).

Multiple theoretical studies have proposed restrictions to the interaction parameters to obtain stable systems. We summarize a few of the results in the list below. Some of the conclusions seem contradictory. This only demonstrates how difficult the question of stability is and how it depends on specific choices for the parameters and the model implementation.

- Emmerson and Yearsley, 2004 study Lotka-Volterra systems with an interaction matrix of which the strengths follow an asymmetric distribution skewed to low values– a distribution that is found in real ecosystems. They conclude that compared to uniform interaction distributions, skewed distributions enhance the local and global stability, but only in the presence of omnivory. However, the return time to equilibrium from a non-equilibrium initial condition is larger for the skewed systems.
- Rock-paper-scissors competition can assure coexistence of species (May & Leonard, 1975). The consequence of this intransitive competition is that the population dynamics are periodic limit cycles. For communities with more than three species, multiple definitions of an *intransitivity index* are defined which are related to the stability of the system (Laird & Schamp, 2018). The intransitivities have a stabilizing effect which roots in the network structure instead of the one-on-one competition of the winner-takes-all hypothesis. Allesina and Levine, 2011 show that such the network can only consist of an odd number of species, as a competitive network with maximal connection—all species are interacting with all other species— and an even number of species will have a subnetwork with an odd number species, such that subnetwork wins against others.
▶ Predator-prey interactions are stabilizing, whereas mutualistic and competitive interactions are destabilizing. Predator-prey networks lose stability for more realistic food web structures and a large number of weak interactions (Allesina & Tang, 2012).
▶ The stability decreases when cooperative interactions are added to a competitive system (Coyte et al., 2015).
▶ Stability increases for a growing number of species in the presence of mediated cooperation (Tu et al., 2019).
▶ Mutualistic interactions can be modeled through crossfeeding. Butler and O'Dwyer, 2018 shows that if species consume, but do not produce resources, all positive equilibria will be locally stable. Crossfeeding can disrupt the stability if the mutualistic interactions are not sufficiently weak.

A different mechanism to ensure stability independent of the interaction matrix is spatial structure, *i.e.* coexistence can be maintained if the system is not well-mixed (Frean & Abraham, 2001; Kerr et al., 2002; Laird, 2014).

Another question with respect to diversity is why multiple species that perform the same function coexist (Rosenfeld, 2002; Wohl et al., 2004). The insurance hypothesis explains this functional redundancy (Yachi & Loreau, 1999): there is a buffering effect for the total productivity which counteracts against stochastic fluctuations in the productivity of individual species and there is an increase of the mean productivity.

**Self-organized criticality**

*Self-organized criticality* was proposed by Bak et al., 1987. They argue that systems which have processes on different spatial scales, such as fractals, will organize themselves into barely stable states. This leads to a power law distribution of the temporal fluctuations. Bak et al., 1987 proposed the dynamics of a sandpile to illustrate this concept. The slope of the sandpile will self-organize at a critical value. When grains are added, the slope becomes higher than the critical value, which results in avalanches of sand which decrease the slope of the pile. The distribution of these avalanches is a power law (Bak et al., 1987)—a heavy-tailed distribution[6] with many small and few large avalanches. This sand pile model can be compared to community dynamics, which maintain a critical diversity, or number of distinct species. The result of adding new species to a community is extinction of some or many of the species such that for a constant inflow of species the diversity is maintained constant. Besides ecological networks, many other biological systems are poised at criticality comprising neural networks, bird flocks, etc. (Mora & Bialek, 2011).

[6]: See the next section about heavy-tailed distributions in natural systems.

Self-organized models have been implemented as individual-based models. The community is represented by an automaton where a simple set of rules for the individuals gives rise to patterns at the population level, a phenomenon called *emergence* (Brown, 1995; R. H. MacArthur & Wilson, 1967; Maurer, 1999; May, 2001; Solé et al., 2002).

## 7.4 Heavy-tailed distributions

It is well known that the distribution of the abundances of all species in a community is heavy-tailed, implying that there are few abundant species and many rare. Biological systems have been described by lognormal, power law, exponential and log series distributions. Mathematical descriptions of these different distributions can be found in Appendix B.1. The exact distribution will depend on the sampling intensity and the size of the community (Preston, 1948).

Explaining heavy-tailed distributions from first principles is challenging, and different mechanisms are needed to describe the different heavy-tailed distributions. A model that exhibits a power law distribution is the individual-based model (Solé et al., 2002). Most of the mechanisms that result in power laws are based on a rich-grow-richer scheme (Hoffmann et al., 2007), but power laws can also arise through the mixing of lognormal distributions (Allen et al., 2001). In Chapter 9, we explain how heavy-tailed distributions can be obtained with gLV models.

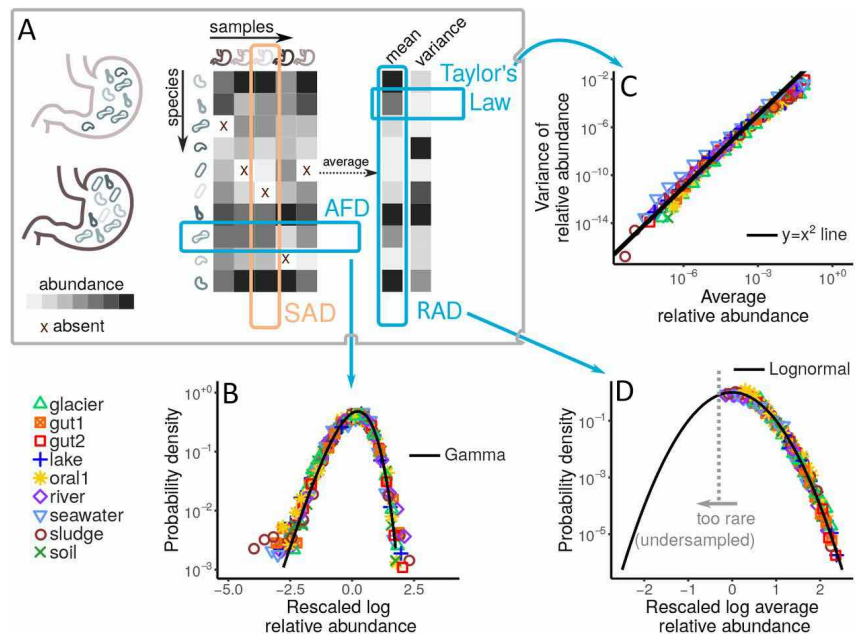

**Figure 7.6:** Macroecological laws from Grilli, 2019. Taken from Grilli, 2019.

## 7.5 Laws of diversity and variation in microbial communities

The title of this section was borrowed from the article of Grilli, 2019. Using a thermodynamic framework, the author obtains three macroecological laws for microbial communities:

1. The probability density of the abundance of a species (*abundance fluctuation distribution*) is well described by a Gamma distribution. An important consequence is that the abundance of a species is never truly zero, zeros in experimental data are only artefacts of sampling and a finite detection threshold[7]. The exclusion of real species absence eliminates strongly interacting gLV models with multiple steady states as these states have zero abundance for a number of species.
2. The coefficient of variation of a species is constant with respect to the mean abundance, thus the variance scales quadratically with the mean. This power law relation between the mean and variance is also known as *Taylor's law*.
3. The *relative abundance distribution* is lognormal. This observation excludes neutral theories which would result in a normal distribution.

Grilli asks the question whether the variability of abundances is a result of *heterogeneity* or *stochasticity*, where the former represents different environmental conditions for different

[7]: The author tested the Gamma distribution against a zero-inflated Gamma distribution to make sure the data followed a distribution for which the probability vanishes at zero.

communities and the latter environmental fluctuations in time. Comparing cross-sectional and longitudinal data[8], he claims that most of the variability can be explained by stochasticity and heterogeneity contributes only little to variability. He then proposes a simple model without interactions, but with linear multiplicative noise to account for the environmental fluctuations and discovers that such a stochastic logistic model can reproduce the three laws mentioned above. Although we know that species interact albeit only limited because most correlation coefficients are close to zero, interactions are not required to explain the three macroecological laws.

8: Cross-sectional means taking samples of different individuals at one time point and longitudinal means samples of one individual for different time points.

We obtained independently a similar conclusion stating that interactions are not required to explain some of the stochastic characteristics of microbial time series. This work is presented in Chapter 8.

## 7.6 Modeling the gut microbiome

In our analysis, we use generalized Lotka Volterra equations to describe the human gut microbiome. This model is simple, and consequently coarse-grained. We here discuss some concerns regarding the choice of the model:

- ▶ The *environment* of the microbial community is not described in gLV descriptions. To build a more realistic model, one should incorporate mechanistic interactions among cells including both the microbiome and the human epithelial cell wall of the gut. The gut environment is relatively similar among individuals, unlike the environment of the skin biome, of which the dynamics are much more host-dependent (Bashan et al., 2016).
- ▶ Another aspect that is lacking in gLV models is the *spatial structure*. Not only will spatial structure in the model give rise to patterns in the solutions, but it can also increase the number of coexisting species or the diversity, as explained in Section 7.3.
- ▶ *Metabolites* play a large role in microbial communities. Fecal water contains at least 133 metabolic compounds (Gao et al., 2010). Modeling metabolites instead of or together with microbial species might reveal the energy flows in communities.

- ▶ The gLV model considers only living bacterial cells. However, dead cells might have a large effect since they release many resources after lysis.
- ▶ Many models exclude immigration of species into the system because immigration is negligible and the number of parameters is reduced by omitting it. However, immigration may play a significant role, because as discussed in Section 6.2, it was observed that people who live together have similar microbiomes and that owning pets changes the composition.

# Stochastic logistic models reproduce experimental time series of microbial communities

MICROBIAL COMMUNITIES ARE FOUND EVERYWHERE ON EARTH, from oceans and soils to gastrointestinal tracts of animals, and play a key role in shaping ecological systems. Because of their importance for our health, human-associated microbial communities have recently received a lot of attention. According to the latest estimates, for each human cell in our body, we count one microbe (Sender et al., 2016). Dysbiosis in the gut microbiome is associated with many diseases from obesity, chronic inflammatory diseases, some types of cancer to autism spectrum disorder (Gilbert et al., 2016). It is therefore crucial to recognize what a healthy composition is, and if unbalanced, be able to shift the composition to a healthy state. This asks for an understanding of the ecological processes shaping the community and dynamical modeling.

The dynamics of complex ecosystems can be studied by considering the number of individuals of each species, referred to as abundances, at subsequent time points. There are several ways to characterize experimental time series properties. Models typically focus on one specific aspect such as the stability of the community (Coyte et al., 2015; Gavina et al., 2018; Gibbs et al., 2018; Grilli et al., 2017; Levine et al., 2017; May, 1972), the neutrality (Fisher & Mehta, 2014; Washburne et al., 2016), or mechanisms leading to long-tailed rank abundance distributions (Brown et al., 2002; Matthews & Whittaker, 2015; McGill et al., 2007; Solé et al., 2002). Different types of dynamical models have been proposed. A first distinction can be made between neutral and non-neutral models. Neutral models assume that species are ecologically equivalent and that all variation between species is caused by randomness. In such models, no competitive or other interactions are included. A second distinction is made between population-level and individual-based models. Generalized Lotka-Volterra (gLV) models describe the system at the population level and assume that the interactions between species dictate the community's time evolution. Both deterministic and stochastic implementations exist for gLV models. Stochastic models include a noise term. There are multiple origins of the noise: intrinsic noise captures the fluctuations due to small numbers, extrinsic noise models external factors such as changing immigration rates of species or changing growth rates mediated by a varying flux of nutrients. Individual-based or agent-based models include self-organized instability models (Solé et al., 2002) and the controversial neutral model of Hubbell (Hubbell, 2001; Rosindell et al., 2011). A classification scheme that assesses the relative importance of different ecological processes from time series has been proposed in Faust et al., 2018. The scheme is based on a test for temporal structure in the time series via an analysis of the noise color and neutrality. Applied to the time series of human stool microbiota, it tells us that stochastic gLV or self-organized instability models are more realistic than deterministic gLV and Dirichlet-multinomial models. Here, we will however only focus on stochastic gLV models. The reason for this is twofold. First, one can encompass the whole spectrum of ecosystems from neutral to niche with gLV models (Fisher & Mehta, 2014). Second, we aim at describing dense ecosystems and even though an individual-based model might be more accurate, in the large number limit it will be captured by a Langevin approximation, *i.e.* by the stochastic gLV model.

Our goal is to compare time series generated by stochastic gLV models with experimental time series of microbial communities. We aim at capturing all observed properties mentioned above—the rank abundance profile, the noise color, and the niche character—as well as the statistical properties of the differences between abundances at successive time points with one model. As is shown in Appendix D.2, the abundance distribution is heavy-tailed, which means that few species are highly abundant and many species have low abundances. Despite the large differences in abundances, the ratios of abundances at successive time points and the noise color are independent of these abundances and although the fluctuations are large, the results of the neutrality tests indicate that the experimental time series are in the niche regime. To sum up, we seek growth rates, interaction matrices, immigration rates, and an implementation of the noise in stochastic gLV models to obtain the experimental characteristics.

We simulated time series using gLV equations. The interaction matrices are random as was introduced by May, 1972. The growth rates are determined by the choice of the steady state, which is set to either equal abundances for all species or abundances according to the rank abundance profiles found for experimental data. For the noise, we consider different implementations corresponding to different sources of intrinsic and extrinsic noise.

Our analysis constrains the type of stochastic gLV models able to grasp the properties of experimental time series. First, we show that there is a correlation between the noise color and the product of the mean abundance and the self-interaction of a species. The noise color profile for such models will, therefore, depend on the steady state. This implies that imposing equal self-interaction strengths for all species, what can be done to ensure stability (Fisher & Mehta, 2014; Gibson et al., 2016), is incompatible with the properties of experimental time series. Second, from the differences between abundances at successive time points, we conclude that a model with mostly extrinsic (linear) noise agrees best with the experimental time series. Third, neutrality tests often result in the niche regime for time series generated by noninteracting species with noise. We, therefore, conclude that all stochastic properties of experimental time series are captured by a logistic model with large linear noise. However, interactions are not incompatible with those properties. This suggests using stochastic logistic models as null models to test for interactions. Our results go along the lines of the ones obtained by Grilli, 2019 which state that the stochastic logistic model can be interpreted as an effective model capturing the statistics of individual species fluctuations.

## 8.1 Properties of experimental time series

We study time series from different microbial systems: the human gut microbiome (David et al., 2014), marine plankton (Martin-Platero et al., 2018), and diverse body sites (hand palm, tongue, fecal) (Caporaso et al., 2011) (Figure 8.1A). A study of the different characteristics for a selection of this data is represented in Figure 8.1. The complete study of all time series can be found in Appendix D.2. We propose a detailed description of the properties of the experimental time series. They fall essentially into two categories. The stability and rank abundance are tightly connected to the deterministic part of the equations while the differences between abundances at successive time points and noise color explain the stochastic behavior. The neutrality is more subtle and depends on the complete system.

**The time series show fluctuations over time.**

The experimental time series show large fluctuations over time. We can ask the question whether the origin of this variation is biological or technical, and assume that most of the variation can be contributed to biological processes. This hypothesis is supported by the results of Silverman et al., 2018 for microbial communities of an artificial gut. Here, the biological variation becomes five to six times more important than the technical variation for the sampling interval of a day. Also, Grilli, 2019 shows the time correlation of experimental time series which is non-zero. In the case where the variation is mostly due technical errors, we expect to see no correlation. Because no experimental errorbars are available for most of the data and because we assume most variation has a biological origin, we did not consider the errors on the species abundances.

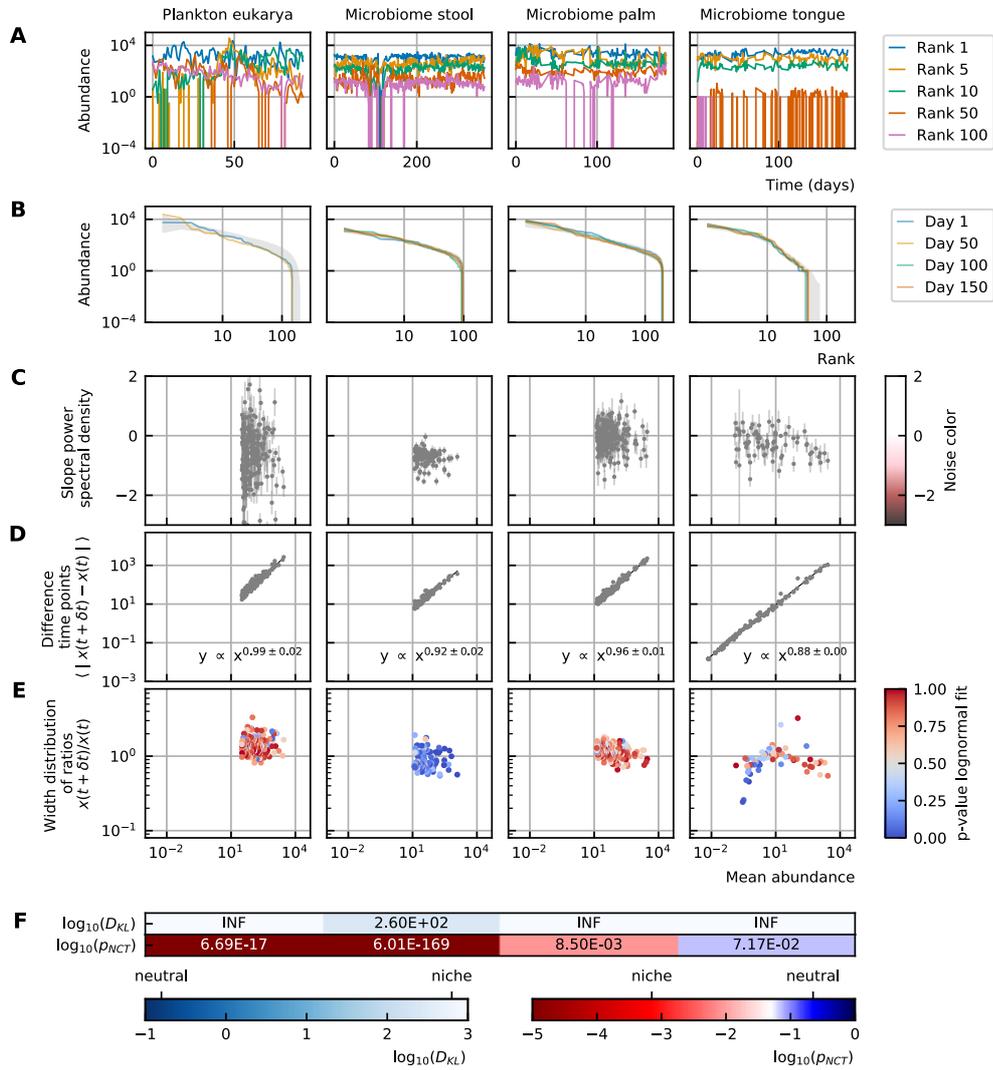

**Figure 8.1:** Characteristics of experimental data (A) Time series. (B) Rank abundance profile. The abundance distribution is heavy-tailed and the rank abundance remains stable over time. (C) Noise color: No clear correlation between the slope of the power spectral density and the mean abundance of the species can be seen. The noise colors corresponding to the slope of the power spectral density are shown in the colorbar (white, pink, brown, black). (D) Absolute difference between abundances at successive time points: There is a linear correspondence (in log-log scale) between the mean absolute difference between abundances at successive time points and the mean abundance of the species. Because the slope is almost one, this hints at the linear nature of the noise. (E) Width of the distribution of the ratios of abundances at successive time points: The width of the distribution of successive time points is large (order 1) and does not depend on the mean abundance of the species. Most of the species fit well a lognormal distribution: the p-values of the Kolmogorov-Smirnov test are high. (F) Neutrality: The values of the Kullback-Leibler divergence ($D_{KL}$) and the neutral covariance test ($p_{NCT}$) are explicitly given. Additionally, we use color codes for both tests with the neutral regime represented by dark blue. White and red indicate the niche regime for the KL test and NCT respectively. We conclude that most experimental time series are in the niche regime.

### The abundance distribution is heavy-tailed.

The first aspect of community modeling that has been widely studied during the last years is the stability of the steady states. Large random networks tend to be unstable (May, 1972). This problem is often solved by considering only weak interactions, sparse interaction matrices (May, 2001) or by introducing higher-order interactions (Gavina et al., 2018; Grilli et al., 2017; Sidhom & Galla, 2020). Although the stability of gLV models decreases with an increasing number of participating species, the stability only depends on the interaction matrix and not on the abundances (Gibbs et al., 2018). The abundance distribution of the experimental data is heavy-tailed. This means that there are few common and many rare species. The distribution of the steady-state values can also be represented by a rank abundance curve. The **rank abundance distribution** describes the commonness and rarity of all species. It can be represented by a rank abundance plot, in which the abundances of the species are given as a function of the rank of the species, where the rank is determined by sorting the species from high to low abundance. These curves can generally be fitted with power law, lognormal, or logarithmic series functions (Brown et al., 2002; Limpert et al., 2001; McGill et al., 2007). Although the abundances show large fluctuations over time, the rank abundance remains stable (Figure 8.1B).

### The differences between abundances at successive time points are large and linear with respect to the species abundance.

Time series can be described by the differences between abundances at successive time points. We propose to focus on two specific representations of the information contained in those differences. First, we consider the mean absolute difference between abundances at successive time points $\langle |\, x(t+\delta t) - x(t)\, | \rangle$ as a function of the mean abundance $\langle x(t) \rangle$. These values represent the jumps of the abundances from one time point to the next. For the experimental data, the relation between these variables is a monomial—this means that it is linear on the log-log scale (Figure 8.1D). The fact that the slope of this line is almost one hints at a linear nature of the noise.

Second, we examine the distribution of the ratios of the abundances at two successive time points $x(t + \delta t)/x(t)$. The advantage of this method is that the direction of the jump between two time points is captured: for ratios higher than 1 the jump is positive, for ratios lower than 1 the jump is negative. The width of this distribution tells how large the fluctuations are. To measure this width, we fit the distribution with a lognormal curve for which the mean is fixed to be one as the fluctuations occur around steady state. For most of the species of experimental data (except for the stool data), the fit of the distribution to a lognormal curve is good (Figure 8.1E). Furthermore, we notice that the distribution is wide—in the order of 1—and that the width does not depend on the mean abundance of the species (Figure 8.1E).

## The noise color is independent of the mean abundance of the species.

The noise of a time series can be studied by considering the distribution of the frequencies of the fluctuations. This distribution can be defined by its slope, which is interpreted as the noise color. White, pink, brown and black noise correspond to slopes around 0, -1, -2 and -3 respectively. The more negative the slope is—this corresponds to darker noise—the more structure there is in the time series (Faust et al., 2018). We notice that there is no correlation between the noise color and the mean abundance of the species for experimental time series (Figure 8.1C). More details about the estimation of the noise color can be found in Appendix D.1.

## Experimental time series are in the niche regime.

In neutral theory, it is assumed that all species or individuals are functionally equivalent. It is challenging to test whether a given time series was generated by neutral or niche dynamics. We use two definitions of neutrality measures: the Kullback-Leibler divergence as used in Fisher and Mehta, 2014 and the neutral covariance test as proposed by Washburne et al., 2016. The **Kullback-Leibler divergence** measures how different the multivariate distribution of species abundances is from a distribution constructed under the assumption of ecological neutrality. The idea of the **neutral covariance test** is to compare the time series with a Wright-Fisher process. A Wright-Fisher process is a continuous approximation of Hubbell's neutral model for a large and finite community. In particular, it tests the invariance with respect to grouping. More about the validity of these neutrality measures can be found in Appendix A.2.

Both neutrality measures indicate that most experimental time series are in the niche regime (Figure 8.1F).

## 8.2 Modeling generalized Lotka-Volterra equations

In a microbial community different species interact because they compete for the same resources. Moreover, they produce byproducts that can affect the growth of other species. Depending on the nature of the byproducts, harmful, beneficial, or even essential, the interaction strength will be either negative or positive. To describe the dynamics of interacting species, one can use the generalized Lotka-Volterra equations:

$$\dot{x}_i = \lambda_i + g_i x_i + \sum_j \omega_{ij} x_i x_j, \tag{8.1}$$

where $x_i$, $\lambda_i$ and $g_i$ are the abundance, the immigration rate, and the growth rate of species $i$ respectively, and $\omega_{ij}$ is the interaction coefficient that represents the effect of species $j$ on species $i$. The diagonal elements of the interaction matrix $\omega_{ii}$, the so-called self-interactions, are negative to ensure stable steady states. The off-diagonal elements of the interaction matrix $\omega_{ij}$ are drawn from a normal distribution with standard deviation $\alpha$ ($\omega_{ij} \sim \mathcal{N}(0, \alpha^2)$). The gLV equations only consider pairwise effects and no saturation terms, or other higher-order terms. Due to this drawback, these models sometimes fail to predict microbial dynamics (Levine et al., 2017; Momeni et al., 2017). However, they are among the most simple models for interacting species and therefore widely studied and used. Noninteracting species can be described by the logistic model, which is a special case of the gLV model obtained by setting all off-diagonal elements of the interaction matrix to zero.

### Implementations of the noise

There exist two principal types of noise: intrinsic and extrinsic noise. *Extrinsic noise* arises due to external sources that can alter the values of the different variables: the immigration rate and growth rate fluctuate in time through colonization of species or a changing flux of nutrients. These processes give rise to additive and linear multiplicative noise respectively. The remaining parameters, inter- and intra-specific interactions can also, change depending on the environment. The formulation of this noise is more subtle (used in Zhu and Yin, 2009). *Intrinsic noise* is due to the discrete nature of individual microbial cells. Thermal fluctuations at the molecular level determine the fitness of the individual cells. Therefore, cell growth, cell division, and cell death can be considered as stochastic Poisson processes. For large numbers of microbes, these fluctuations will be averaged out.

We first consider the extrinsic noise. If the time series is calculated by $x_i(t + dt) = x_i(t) + dx_i(t)$, the implementation of the linear multiplicative noise is as follows,

$$dx_i(t) = \lambda_i dt + g_i x_i(t) dt + \sum_j \omega_{ij} x_i(t) x_j(t) dt + x_i(t) \sigma_i dW(t), \quad (8.2)$$

where $dW$ is an infinitesimal element of a Brownian motion defined by a variance of $dt$ ($dW \sim \sqrt{dt} \mathcal{N}(0, 1)$). Changes in immigration rates of microbial species can be modeled with additive noise,

$$dx_i = \lambda_i dt + g_i x_i dt + \sum_j \omega_{ij} x_i x_j dt + \sigma_{i,\text{const}} dW_{\text{const}}, \quad (8.3)$$

with $dW_{\text{const}} \sim \sqrt{dt} \mathcal{N}(0, 1)$. Our main motivation is to model the gut microbiome in the colon. Here, we ignore the immigration of species for two reasons. First, the number of microbes in the colon is orders of magnitude larger than the number of microbes in the other parts of the gut (Gorbach et al., 1967; Marteau et al., 2001)—therefore, the flux of incoming microbes in the colon is small. Second, we only consider systems around steady state, for which we assume immigration does not play an important role. For perturbed systems, which are far from equilibrium, immigration rates cannot be ignored. Ignoring immigration may be too restrictive for some microbial systems such as the skin microbiome or plankton.

We implement the intrinsic noise by a term that scales with the square root of the species abundance (Fisher & Mehta, 2014; Walczak et al., 2012),

$$dx_i(t) = \lambda_i dt + g_i x_i(t) dt + \sum_j \omega_{ij} x_i(t) x_j(t) dt + \sqrt{x_i(t)} \sigma_{i,\text{sqrt}} dW_{\text{sqrt}}, \quad (8.4)$$

with $dW_{\text{sqrt}}$ again an infinitesimal element of a Brownian motion defined by a variance of $dt$ ($dW_{\text{sqrt}} \sim \sqrt{dt} \mathcal{N}(0, 1)$). The size of this noise $\sigma_{i,\text{sqrt}}$ is determined by the cell division ($g^+$) and death rates ($g^-$) separately, which are in our model com-bined to one growth vector ($g = g^+ - g^-$, $\sigma_{i,\text{sqrt}} = \sqrt{g^+ + g^-}$), for large division and death rates the intrinsic noise will be larger.

To sum up, we focus on linear multiplicative noise because: (a) extrinsic noise is dominant as microbial communities contain a very large number of individuals and (b) we ignore the immigration of individuals in our analysis.

We verified that our analysis is robust with respect to the multiple possibilities for the discretization of these models. We also compare our population-level approach with individual-based modeling approaches. Details can be found in Appendix D.1.

## 8.3 Reproducing properties of experimental time series from stochastic generalized Lotka-Volterra models

We find that the aforementioned characteristics of experimental time series can be reproduced by stochastic logistic equations. We first explain how to choose the growth rate to obtain the heavy-tailed experimental abundance distribution. Next, we discuss how the noise color determines the self-interaction of a species given its abundance and how the implementation of the noise determines the slope of the mean absolute increment $\langle |\, x(t + \delta t) - x(t)\, | \rangle$ and the mean abundance $\langle x(t) \rangle$ (such as in Figure 8.1D). In the end, by using the appropriate choice for the self-interactions, growth rates, and noise implementation, we conclude that a stochastic logistic model can reproduce all the stochastic properties, including the niche regime for the neutrality tests although the model does not include any interactions.

### The rank abundance distribution can be imposed by fixing the growth rate.

Random matrix models do typically not give rise to heavy-tailed abundance distributions. Neither is it known which properties of the interaction matrix and growth rates are required to obtain a realistic rank abundance distribution. We can however enforce the desired rank abundance artificially by solving the steady state of the gLV equations. Given the steady-state abundance vector $\vec{x}^*$ and interaction matrix $\omega$, we impose the growth rate $\hat{g} = -\omega \vec{x}^*$. One model that results in heavy-tailed distributions is the self-organized instability model proposed by Solé et al., 2002 (see Chapter 9).

For logistic models, the growth rate is equal to the product of the self-interaction and mean abundance. The noise color and the width of the distribution of ratios $x(t + \delta t)/x(t)$ depend on this product. To obtain given characteristics—a predefined noise color and width of the distribution of ratios $x(t + \delta t)/x(t)$—the choice of the growth rate will dictate the choice of the remaining free parameters, the sampling time step $\delta t$ and the noise strength $\sigma$.

### The noise color is determined by the mean abundance and the self-interaction of the species.

To study the noise color, we first consider a model where the species are not interacting. The noise color is independent of the implementation of the noise but depends on the product of the mean abundance and the self-interaction of the species (Figure 8.2A). For noninteracting species, the growth rate equals the product of the self-interaction and the steady-state abundance. Because we consider fluctuations around steady state, the mean and the steady-state abundance are nearly equal and the $x$-axis of Figure 8.2A-C can be interpreted as the growth rate. Also, the strength of the noise does not change its color (Figure 8.2C). A parameter that is important for the noise color is the sampling rate: the higher the sampling frequency the darker the noise becomes (Figure 8.2B). This is in agreement with the results of Faust et al., 2018. Darker noise corresponds to more structure in the time series. The more frequent the abundances are sampled the more details are visible and the underlying interactions become more visible. We conclude that the noise color is only dependent on the mean abundance, the self-interactions, and the sampling rate. Figures of the dependence on the mean abundance and self-interaction separately can be found in Appendix D.3.

For interacting species, increasing the strength of the interactions makes the color of the noise darker in the high mean abundance range (Figure 8.2D-E). Importantly, for interacting species with a lognormal rank abundance, the correlation between the noise color and mean abundance is preserved (Figure 8.2E). The data can be fit to obtain a bijective function between the product of the mean abundance and the self-interaction, and the noise color. Assuming this model is correct, we can obtain an estimate for the self-interaction coefficients given the mean abundance and noise color by fixing the sampling rate and the interaction strength. The uncertainty on the estimates is larger where the fitted curve is more flat (slopes of the power spectral density around -1.7 and 0), but many experimental values of the stool microbiome data lie in the pink region where the self-interaction can be estimated for this model.

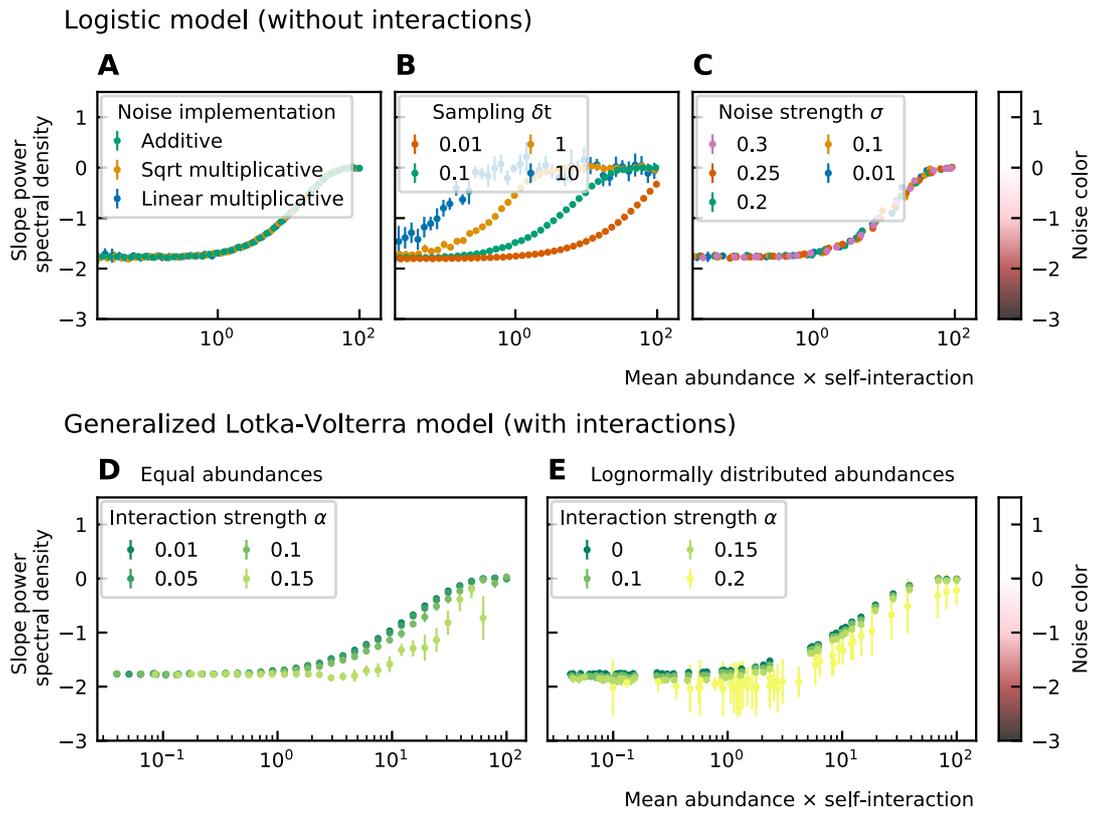

**Figure 8.2:** Noise color as a function of the mean abundance and self-interaction for stochastic logistic and gLV equations. The noise colors corresponding to the slope of the power spectral density are shown in the colorbar (white, pink, brown, black).
The mean abundance determines the noise color when there is no interaction, the implementation method (A) and the strength of the noise (C) have no influence. A smaller sampling time interval $\delta t$, which is equivalent to a higher sampling rate, makes the noise darker (B). For gLV models with interactions, larger interaction strengths make the noise colors darker for systems with equal abundances (D) as well as systems with heavy-tailed abundance distributions (E).

## The implementation of the noise determines the correlation between the mean absolute increment $\langle |\, x(t + \delta t) - x(t)\, | \rangle$ and the mean abundance $\langle x(t) \rangle$.

Next, we study the differences between abundances at successive time points. From the results of the noise color, we can estimate the self-interaction for the dynamics of the experimental data. We use the rank abundance and the self-interaction inferred from noise color of the microbiome data of the human stool to perform simulations and calculate the characteristics of the distribution of differences between abundances at successive time points. We, here, assume that there are no interactions. More results for dynamics with interactions are in Appendix D.3. We first study the correlation between the mean absolute difference between abundances at successive time points $\langle |\, x(t + \delta t) - x(t)\, | \rangle$ and the mean abundance $\langle x(t) \rangle$. For linear multiplicative noise, the slope of the curve of the logarithm of the mean absolute difference between abundances at successive time points $\log_{10}(\langle |\, x(t + \delta t) - x(t)\, | \rangle)$ as a function of the logarithm of the mean abundance $\log_{10}(\langle x(t) \rangle)$ is one. For multiplicative noise that scales with the square root of the abundance, the slope is around 0.66 and for additive noise, the slope is zero. By combining both linear noise and noise that scales with the square root of the abundance, slopes with values between 0.6 and 1 can be obtained (Figure 8.3A). The slopes of experimental data range between 0.84 and 0.99, we therefore conclude that linear noise is a relatively good approximation to perform stochastic modeling of microbial communities.

## The strength of the noise determines the width of the distribution of ratios $x(t + \delta t)/x(t)$.

Next, we examine the distribution of the ratios of abundances at successive time points. As expected, for significant noise, this distribution can be approximated by a lognormal curve and the width of the distribution becomes larger for increasing noise strength (Figure 8.3B). In order to have widths that are of the same order of magnitude as the ones of the experimental data, the noise must be sufficiently strong. Another way of increasing the width is through interactions, this effect is only moderate. These results are presented in the Appendix D.3.

## Stochastic logistic models capture the properties of experimental time series.

By using all previous results and imposing the steady state of experimental data, we find that it is possible to generate time series with identical characteristics to the ones seen in the experimental time series (Figure 8.4). Furthermore, these time series can be generated without introducing any interaction between the different species, but their neutrality measures can still be in the niche regime (Figure 8.4F). Out of 100 simulations, 62 had a p-value smaller than 0.05 for the neutral covariance test which means they are in the niche regime. The colors of the noise fix the self-interaction values (Figure 8.4C), next the rank abundance distribution is imposed by calculating the growth vector $\hat{g}$ (Figure 8.4B). The slope of the curve of the mean absolute difference between abundances at successive time points as a function of the mean abundance is one by using linear multiplicative noise (Figure 8.4D) and the width of the fluctuations is tuned by choosing a large noise size $\sigma$ (Figure 8.4E). In most experimental time series, only the fractional abundances of species can be measured per time point and not the absolute ones. Because the total abundance of all species remains nearly constant in time series generated by a stochastic logistic equation, our results still hold for time series with fractional abundances (see Appendix D.3). Similar results can be obtained for models with interactions (see Appendix D.3), but we want to stress that interactions are not needed to reproduce the properties of experimental time series.

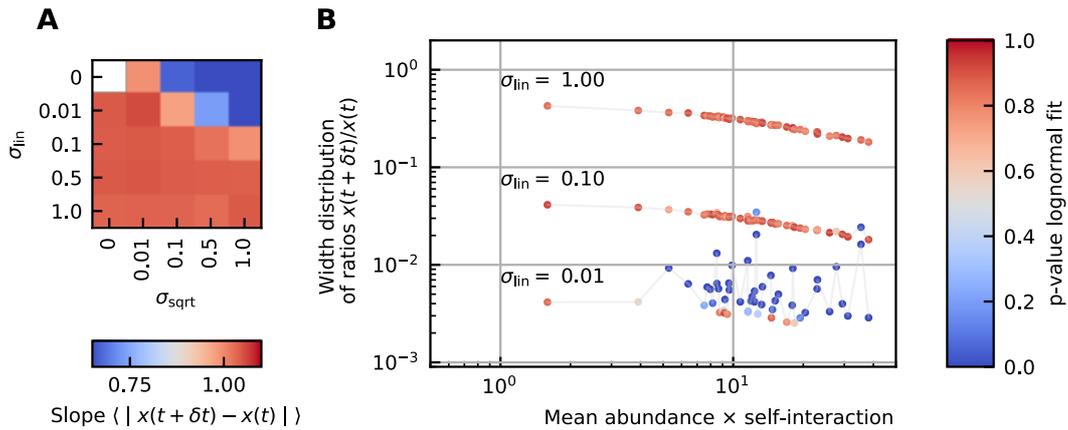

**Figure 8.3:** (A) Correlation between the mean absolute differences between abundances at successive time points and the mean abundance for different strengths of the linear noise ($\sigma_{\text{lin}}$) and multiplicative noise that scales with the square root of the abundances ($\sigma_{\text{sqrt}}$). More specifically, the parameter represents the slope of the logarithm of the mean absolute difference between abundances at successive time points as a function of the logarithm of the mean abundance. Examples of such slopes are given by Figure 8.1D. Here, the slope ranges from 0.66 for noise that scales with the square root to 1 for linear noise. (B) The width of the distribution of the ratios of abundances at successive time points increases for increasing strength of the noise. For sufficiently strong noise the distribution is well fitted by a lognormal function (high p-values for the Kolmogorov-Smirnov test).

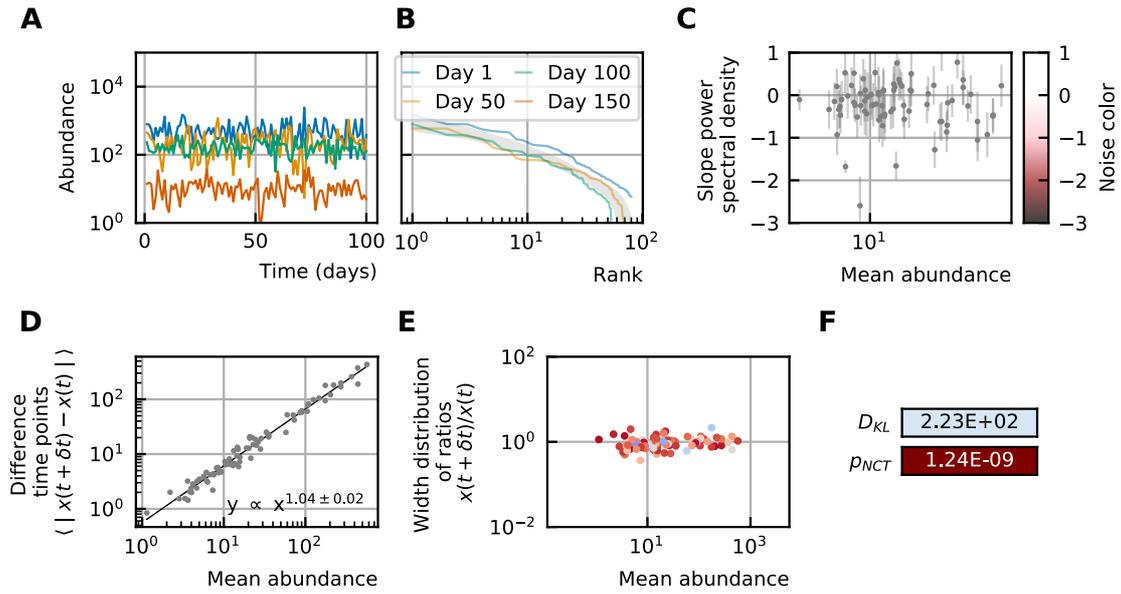

**Figure 8.4:** A stochastic logistic model is able to reproduce the different characteristics of the noise: (A) Time series. (B) A rank abundance that remains stable over time. (C) Results of the neutrality test in the niche regime. (D) Noise color in the white-pink region with no dependence on the mean abundance. (E) The slope of the mean absolute difference between abundances at successive time points is around 1. (F) The width of the distribution of the ratios of abundances at successive time points is in the order of 1 and independent of the mean abundance.

## 8.4 Discussion

Recent research has focused on different aspects of experimental time series of microbial dynamics, in particular the rank abundance distribution, the noise color, the stability, and neutrality. Within the framework of stochastic generalized Lotka-Volterra models, we studied the influence of growth rates, interactions between species, and the different sources of stochasticity on the observed characteristics of the noise and on neutrality. Our observations are:

▶ Even when we consider the case without interactions between species, the result of the neutrality test on the time series is often niche. We should, therefore, be careful in the interpretation of the results of neutrality tests.

▶ For a given sampling step $\delta t$, the noise color depends on the product of the self-interaction and the mean abundance, which for noninteracting species reduces to a dependence on the growth rate. Assuming the model can be used for microbial communities, the self-interaction coefficients can be estimated given the mean abundance, noise color, and sampling rate. Low sampling rates result in larger errors (Figure 8.2B). For sparsely sampled experimental data, the standard deviation of the self-interaction inferred using the noise color will be larger. For the experimental time series (plankton, gut, and human microbiome) the self-interaction strengths range over several orders of magnitude. The convention of equalling all self-interactions to -1 used in several studies (Fisher & Mehta, 2014; Gibson et al., 2016), cannot be adopted for stochastic models of communities with a heavy-tailed abundance distribution.

▶ The exponent of the mean absolute differences between abundances at successive time points with respect to the mean abundances is slightly smaller than one for experimental time series. Linear multiplicative noise results in a value of one, square root noise results in lower values (0.6). A mix of linear and square root noise can result in slopes with intermediate values.

▶ A large multiplicative linear noise is in agreement with both the distribution of the ratios of abundances at successive time points and the relation between the differences between abundances at successive time points and mean values.

To conclude, characteristics of experimental time series, from plankton to gut microbiota, can be reproduced by stochastic logistic models with a dominant linear noise. We expect, however, that for higher sampling rates, modeling the interactions between microbes would be necessary to explain the properties of the time series. For gut microbial time series, the system is sampled only once a day and therefore dominated by the noise in the growth terms corresponding to a linear noise.

Predictive models for the dynamics of microbial communities will certainly require a more in-depth description of the system. Nutrients and spatial distribution of microbes should play a role to dictate the evolution of the community, as well as the interaction with the environment. Synthetic microbial communities are currently being developed and will hopefully provide a more comprehensive view on the complexity of microbial communities (Vrancken et al., 2019).

# Heavy-tailed abundance distributions with stochastic generalized Lotka-Volterra models  |  9

The abundance distribution of experimental data is heavy-tailed. This means that there are many rare species and few abundant ones. Heavy-tailed distributions include the power law and lognormal distribution, but most experimental data of microbial communities fit best a lognormal distribution. All mathematical details about heavy-tailed distributions can be found in Appendix B.1.

The *origin* of heavy-tailed abundance distributions is not fully understood yet. Here, we will discuss how they arise as an emergent property from large communities with many interacting species or through specific distributions of the parameters. For the latter case, we show that non-interacting models lead to lognormal abundance distributions when the parameters follow lognormal distributions and that power law distributions can be obtained by uniformly distributed parameters.

If species abundances would be the result of random walks—the process only involving constant noise, for example through immigration and abundance-independent emigration—the central limit theorem tells us that the abundances follow a normal distribution. However, the evolution of a community is not solely governed by random events: next to immigration, it involves growth and death of species and interactions between species, all processes that depend on the abundances of the species. When variables are multiplied instead of summed, lognormal or other heavy-tailed distributions arise instead of normal distributions (Nair et al., 2020).

The emergence of heavy-tailed distribution has been reported as the result of self-organized critical systems (Bak et al., 1987). Using individual-based models (IBMs), Solé et al., 2002 reproduce heavy tails and three other characteristics of microbial communities:

1. *long-tailed abundance distributions*,
2. the *species-connectivity relation*: there is a power law relation between the number of species and the connectivity,
3. fluctuations and chaotic time series, and,
4. the *species-area relation*: there is a power law relation between the number of species observed in an ecosystem and the area of the ecosystem.

The models of Solé et al., 2002 include interactions between individuals and immigration of individuals. All individuals of a species were equivalent, *i.e.* differences between individuals of the same species are not taken into account. In this respect, the model is similar to a stochastic simulation, such as the Gillespie algorithm which was introduced in Section 3.4. The details about the model of Solé et al., 2002 can be found in Appendix E.1. Because IBMs are computationally expensive, we propose population-level models (PLMs) with the same characteristics. Moreover, deterministic models are more amenable to study analytically.

Let us first examine the third and fourth characteristics. It was already proven that all dynamical behaviors including chaos can be obtained for models with minimally five species (Smale, 1976)—an example of chaos through generalized Lotka-Volterra with four species can be found in (Allesina, 2020). Furthermore, the fluctuations and stochastic characteristics of experimental time series can be simulated by sgLV models. This matter was discussed extensively in Chapter 8. Therefore, we will not study the third characteristic here. The last characteristic necessitates spatial modeling and hence, we will not consider it.

Here, we want to focus mainly on the first characteristic: reproducing long-tailed abundance distributions. Long-tailed distributions are a subclass of *heavy-tailed distributions* and many of the common heavy-tailed distributions such as the Pareto and lognormal distribution are long-tailed (Nair et al., 2020). We are mostly interested in these distributions and we will use the terminology of heavy-tailed distributions. Although many abundance distributions in nature are heavy-tailed, in the context of population models, the abundance distribution has often been ignored or imposed. In this research, we want to study how heavy-tailed distributions can arise from stochastic generalized Lotka-Volterra (sgLV) equations when we introduce a maximal capacity. Additionally, we study whether such models can reproduce the second characteristic, *i.e.* the species-connectivity relation.

In this chapter, we will first present an analysis of experimental data of microbial communities and propose different observables to characterize the abundance distribution and time series. Next, we reproduce and generalize the results of Solé et al., 2002: heavy-tailed distributions can be obtained by IBMs. Afterward, we show, inspired by the set-up of the IBM, how heavy-tailed distributions can also be obtained by sgLV models by including a maximal capacity. We finish by approaching the problem from another, simpler perspective. We consider the logistic model without a maximal capacity and show that even without complex interactions heavy-tailed abundance distribution can be obtained by choosing parameters wisely.

In parallel, we look at the species-connectivity relation. In the context of ecological systems, the connectivity of a network, defined as the ratio of links between different components over all possible links, is often called the connectance. This translates in the number of non-zero (off-diagonal) elements in the interaction matrix. In the 70's, May presented his seminal work about the stability of large ecosystems (May, 1972). He proposed an empirical boundary that separates stable from unstable distributions. This inequality, known as *May's instability criterion*, has been confirmed by random matrix theory (Allesina & Tang, 2012) and generalized for more parameter settings (Allesina & Tang, 2015). The instability criteria describe the relation between

1. the number of species, also called the *diversity S*,
2. the *complexity* of the network, which depends on the connectance $C$ and a parameter that reflects the strength of the interactions, and
3. the stability of the system.

The shape of the boundary between stable and unstable systems is defined by these parameters and assumptions about the interaction matrix and abundance distribution. For the assumptions made by May, the border of instability is a hyperbolic relation between the number of species $S$ and the connectance $C$, $S \propto C^{-1}$.

May defined the boundary of the stable and unstable regions. In general, stability is guaranteed for systems that are situated below the boundary, *i.e.* the number of species is smaller than the maximal number of species allowed for the given complexity or inversely, the complexity is smaller than the maximal allowed complexity for the given number of species. Therefore, one expects to find solutions scattered all over the region of stable solutions, but surprisingly, when the connectance $C$ and diversity $S$ of ecological systems are compared, they form a boundary instead of being randomly spread under a boundary. Different empirical proposals for the species-connectivity relation have been proposed, such as the link-species scaling law and constant connectance hypothesis (Cohen et al., 1990; Martinez, 1992). The species-connectivity relation studied by Solé et al., 2002 considers $S \propto C^{-0.5}$, a relation obtained from experimental data from many different habitats (freshwater, marine, desert, rainforest, etc.) (Montoya & Solé, 2003).

Solé et al., 2002 are able to explain the particular expression for the boundary and the fact that the communities are poised at this boundary by adding an immigration rate for the species. They propose a mean-field model that results in $S \propto C^{-1+\epsilon}$ where $\epsilon$ depends on the immigration rate. This model has only two parameters: the diversity $S$ and connectance $C$. They demonstrate these results with their IBM. Immigration increases the number of species until the border of instability is attained where self-organization of the system keeps the system poised at criticality. This behavior is found not only in community dynamics, but in many biological systems (Mora & Bialek, 2011).

We find a power law relation for high connectance values but the diversity $S$ saturates for small connectance because there is a finite number of species and a maximal number of individuals in the model. We show that similar results are also obtained with a PLM.

## 9.1 Experimental data has a lognormal abundance distribution

We use experimental data of plankton and microbial communities related to the human body. The sources of all datasets can be found in Table D.1. The abundance distributions of the microbial communities under consideration are heavy-tailed and the best fit is obtained by a lognormal function rather than a power law function (Figure 9.1). We will, here, use the terminology of a power law distribution, which is also referred to as the Pareto distribution. More about the definitions of the heavy-tailed distributions can be found in Appendix B.1.

In Figure 9.1, we show the fit of a normal (dotted line), lognormal (solid line) and power law distribution (dot-dashed line). The color of the line represents the p-value of the Kolmogorov-Smirnov (KS) test (details about this test can be found in Appendix B.3). This means that higher values (darker red) correspond to a better fit, and that fits with values smaller than 0.05 can be rejected (blue). In particular, the normal distribution is rejected for all considered microbial communities and the power law distribution for most of them. In general, a lognormal distribution fits the data best. In the legend, the most important parameter of the different distributions are given, *i.e.* for the normal distribution, the standard deviation $\sigma$, for the lognormal distribution, the width $s$ as defined in Equation 9.1 and for the power law, the exponent $\alpha$.

The lognormal distribution can be described by

$$f_X(x) = \frac{1}{s(x-\mu)\sqrt{2\pi}} \exp\left\{\frac{-1}{2}\left(\frac{\ln((x-\mu)/x_0)}{2s}\right)^2\right\}, \tag{9.1}$$

where $\mu$ defines the location of the distribution, $x_0$ the scale, and $s$ a measure for the width of the distribution. In experimental data, the latter ranges from 1.42 to 2.78 (Figure 9.1). Although the species abundances show large fluctuations over time, the width of the abundance distribution remains nearly stable (Figure 9.2). The colors again denote the p-value of the KS test and points in blue can be rejected. The width of the lognormal distribution thus remains between 1 and 3 (grey area).

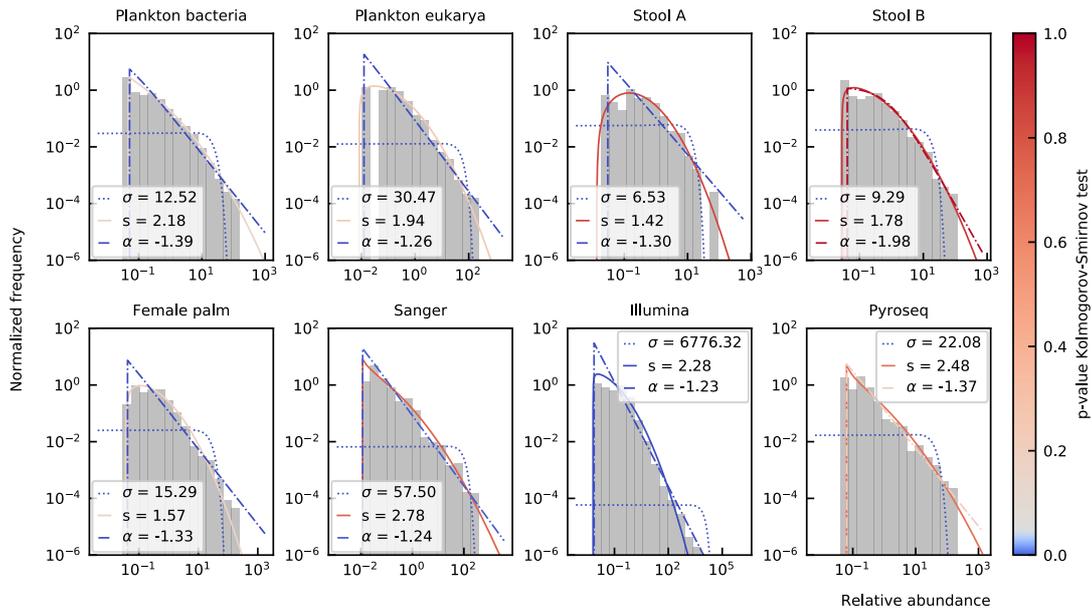

**Figure 9.1:** Experimental data has a heavy-tailed distribution. The lognormal distribution (full line) fits best. The power law (dash-dotted) and exponential (dotted) distribution are rejected for most communities because the p-values of the KS test are lower than 0.05 (lines are colored in blue). The widths *s* of the lognormal distribution range between 1.42 and 2.78. For data coming from a time series, we considered the first time point.

An alternative for studying the abundances of communities is the rank abundance curve. For power law abundances, the rank abundance is also a power law (see Appendix B.2). For the lognormal abundance distribution, the width *s* reflects the steepness of the rank abundance distribution. Larger values of the width *s* result in steeper curves (Figure 9.3).

However, it is not enough to find models that lead to heavy-tailed distributions. As we will see, larger amounts of noise make the abundance distributions wider, but we want the fluctuations of the modeled time series not to be larger than fluctuations in experimental data. We can describe the fluctuations of separate species by studying the coefficient of variation for the species. Additionally, we can look at the fluctuations of the composition as a whole by considering the Jensen-Shannon (JS) distance between microbial compositions at different time points.

The definition of the JS distance can be found in Appendix 1. Basically, it is a metric that is used to describe the distance between community compositions.

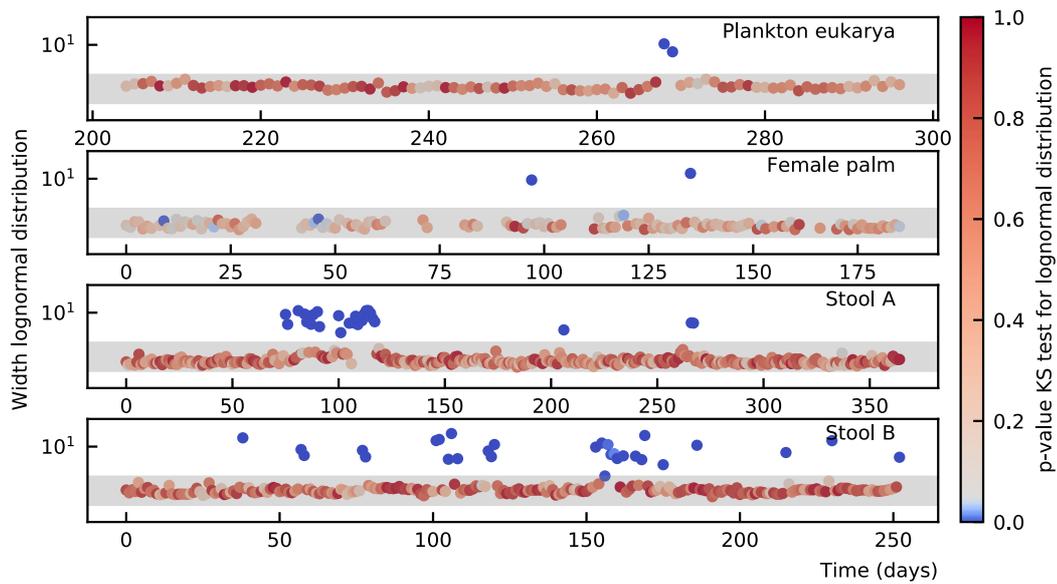

**Figure 9.2:** The species abundances of microbial communities show large fluctuations over time, but the width of the abundance distribution remains nearly stable over time. The grey horizontal band extends from 1 to 3. The color of the dots shows the p-value of the KS test to a lognormal distributions, blue values are poor fits. The lognormal fits for which the width is higher than 3, can all be rejected.

The median JS distance between states at different time points in experimental data ranges from 0.3 to 0.4 (Figure 9.4). This is for longitudinal data (multiple time points of one community). For comparison, in cross-sectional data of the gut (different communities at one time point) the median distance ranges from 0.4 to 0.5 (Appendix E.4). We calculated the Pearson correlation coefficient and the p-value for the hypothesis of non-correlation. For the planktonic eukaryotes, we reject the hypothesis that there is no correlation between the time interval and JS distance. The community is not stable and is slowly evolving which can be explained by a seasonal change in the seawater (Martin-Platero et al., 2018). On the other hand, for the stool data of subject A, this hypothesis cannot be rejected. The community fluctuates around a steady state. In our simulations, we want to mimic the results of communities such as the latter: fluctuating around a steady state with a median JS distance of 0.3 to 0.4 and a lognormal abundance distribution with a width *s* between 1 and 3.

Another way of assessing the strength of the fluctuations is by considering fluctuations of separate species. According to Grilli, 2019, the distribution of relative species abundance at different time points is a Gamma distribution. The coefficient of variation—the ratio of the standard deviation to the mean—is a variable that determines this distribution (see Appendix E.2). We find that some of the species indeed follow a Gamma distribution and that this is particularly true when the coefficient of variation is small. In Figure 9.5, the coefficient of variation of all species is shown and the color of the dots represents the p-value of the KS test for the Gamma distribution. This means that we can reject the hypothesis that the abundance distribution over time for a species follows a Gamma distribution for the species with blue points and that the species with dark red points are likely to follow a Gamma distribution. The median coefficient of variation is between 0.7 and 2. We expect the median coefficient of variation to correlate with the median JS distance, but this is not necessarily true as one can imagine a community where all species populations grow and decrease maintaining the same relative abundances. In this case, the JS distance would remain zero while the coefficients of variation do not.

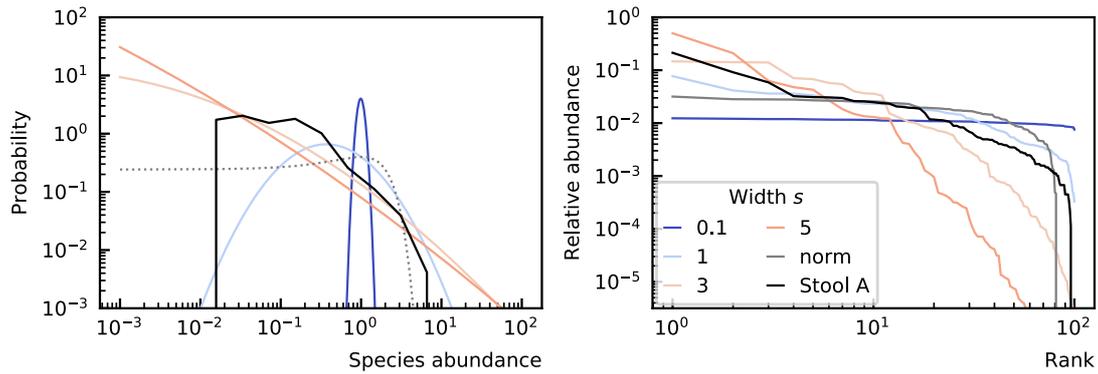

**Figure 9.3:** The width of the lognormal distribution *s* reflects the steepness of the rank abundance curve. The larger the width of the abundance distribution, the steeper the rank abundance curve.

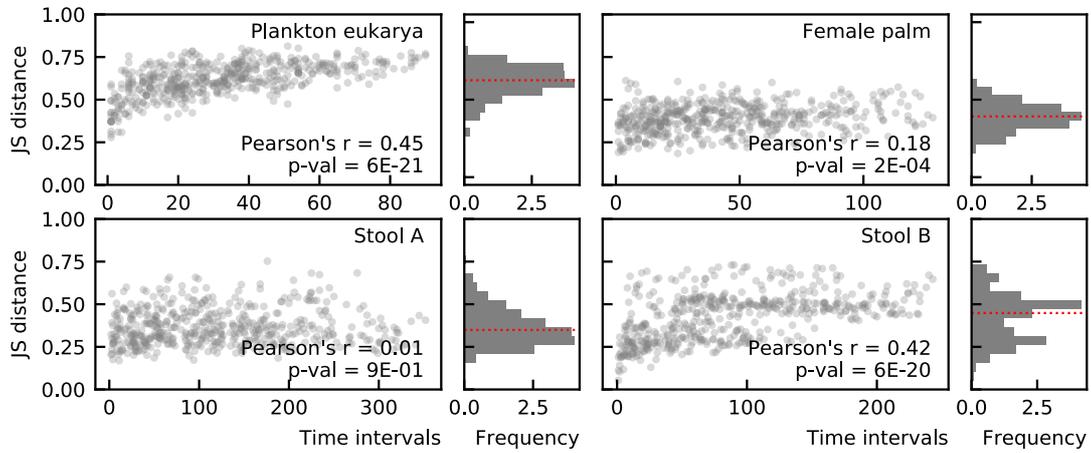

**Figure 9.4:** Jensen-Shannon (JS) distance of longitudinal experimental data as a function of the time interval between both compositions. The median value (red line) ranges from 0.3 to 0.4. The Pearson correlation function and p-value for non-correlation are given.

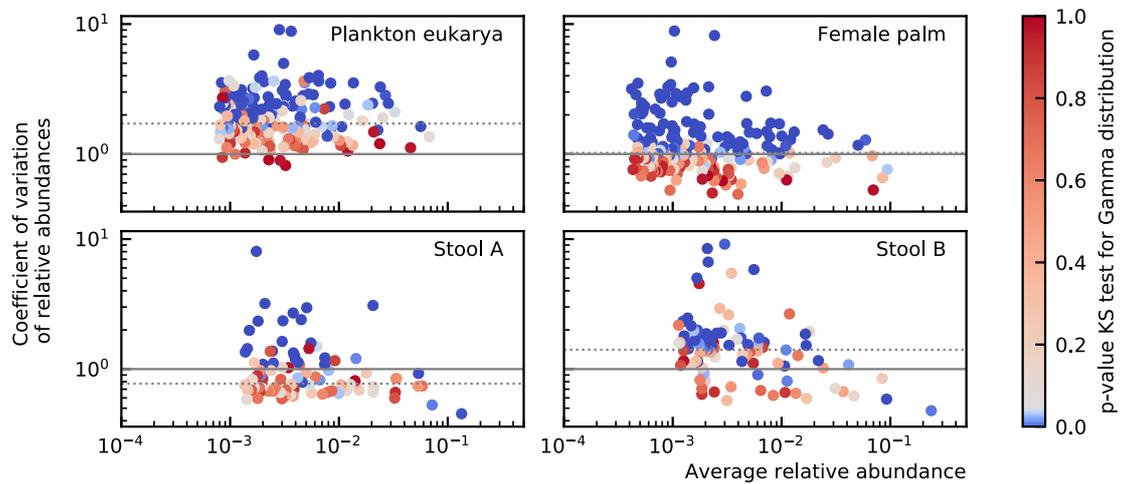

**Figure 9.5:** Longitudinal experimental data of microbial communities has a coefficient of variation that is around one. For species with a large coefficient of variation, we can reject the hypothesis that the population follows a Gamma distribution. The dotted horizontal lines denote the median coefficient of variation per community.

## 9.2 Species-connectivity relation

There is a trade-off between the complexity of a network and its stability. The complexity is related to the number of species $S$, number of interactions or connectance $C$ and the strength of the interactions. Consider a community matrix—defined as the Jacobian of a system evaluated in the fixed point—with -1 on the diagonal and mean 0 and variance $\xi^2$ of the off-diagonal elements. In this case, May found that systems for which

$$\xi\sqrt{SC} < 1 \tag{9.2}$$

have a high probability to be stable. Using random matrix theory Allesina and Tang, 2015 generalized this inequality to community matrices with random elements on the diagonal. In that case the inequality becomes

$$\xi\sqrt{SC} < \bar{d} \tag{9.3}$$

where $-\bar{d}$ is the average of the diagonal elements of the community matrix. Remarkably, the distribution of the off-diagonal elements in the community matrix is unimportant (Allesina & Tang, 2015) and although the normal distribution is often used, uniform distributions lead to similar results provided the appropriate variance is used and the mean is zero. Furthermore, they defined the criterion in the case that the mean of the off-diagonal elements is $\mu \neq 0$, but assuming the diagonal elements $-d$ are all identical,

$$\max\left(\sqrt{SC(\xi^2 + (1-C)\mu^2)} - C\mu, (S-1)C\mu\right) < d. \tag{9.4}$$

A result of May's instability criterion (Equation 9.2) is that the instabilty border is a hyperbolic relationship between the number of species, or diversity $S$ and the connectance $C$: $S \propto C^{-1}$. Without the assumption of May that the mean of the off-diagonal elements of the community matrix are zero, the relationship becomes more complicated (Equation 9.4).

For gLV models without immigration (Equation 9.6 with $\lambda_i = 0$), the community matrix is

$$M_{ij} = x_i^* \omega_{ij} \tag{9.5}$$

where $x_i^*$ is the steady state given by $x_i^* = \sum_j (\omega^{-1})_{ij} g_j$. From the definition of the community matrix, one sees that its connectance is equal to the one of the interaction, but that the distribution of elements of the community matrix is distorted by the abundance distribution. Also in the presence of immigration, the steady state is altered which changes the community matrix. How these effects influence the stability is still an open question (Allesina & Tang, 2015).

Montoya and Solé, 2003 reported experimental data that deviated from May's hyperbolic species-connectivity relation, $S \propto C^{-1+\epsilon}$ with $0 \leq \epsilon \leq 0.5$. Solé et al., 2002 shows that an IBM with immigration can explain this relationship. In the next sections, we repeat the analysis of this work and reproduce the power law relation with IBM for large complexity. After, we show that similar results are obtained with PLM by using an sgLV with a maximal capacity.

## 9.3 Heavy-tailed distributions through dynamical modelling

There exist many models for microbial community dynamics. We can distinguish population-level (PLMs) and individual-based models (IBMs). Solé et al., 2002 reported that heavy-tailed distributions can be obtained with an IBM with a fixed maximum number of individuals. They argue that ecological systems might be organized at a critical point, balancing immigration of new species with interaction between species. We will first show which criteria need to be met for IBMs to exhibit heavy tails and next, we propose sgLV and logistic models that result in heavy-tailed distributions.

We will use the IBM proposed in Heyvaert, 2017 which is an explicit interpretation and generalization of the model of Solé et al., 2002 that allows for all types of interactions, *i.e.* not only predation, but also mutualism, competition, amensalism and commensalism (see Table 7.1 for the definition of the different interaction types). This model exhibits heavy-tailed distributions even when all species have equal growth rates, self-interaction, and immigration probability. Every species unevenness is, therefore, a result of stochastic fluctuations or the interaction between individuals of distinct species. Heyvaert, 2017 and Solé et al., 2002 show that depending on the immigration rate, both power law and lognormal abundance distributions can be obtained. We repeat this analysis. To quantify which distribution fits the abundance distribution best we use the KS test. More details about the fit methods can be found in Appendix E.2. Furthermore, when the lognormal distribution has the best fit and is not rejected (p-value of KS test higher than 0.05), we calculate the width of the lognormal distribution. We also look at the average JS distance between different time points, and the average coefficient of variation of the relative abundances of species. Notice that all these measures are defined by the relative abundances of species and not the absolute abundances. The width of the lognormal distribution is independent of the scale, the JS distance is defined for fractional abundances and we choose to measure the coefficient of variance of the relative abundances. As many experimental data is measured as fractional abundances, this facilitates comparing theoretical with experimental results.

We will study the influence of the immigration rate and interaction matrix. The interaction matrix is characterized by two parameters: the interaction strength and the connectance. The interaction strength $a$ determines the distribution from which the interaction parameters are drawn: we use the uniform distribution $\mathcal{U}(-a, a)$. The connectance $C$ represents the ratio of interactions (off-diagonal elements of the interaction matrix) that are non-zero. In practice, given the interaction strength $a$ and connectance $C$, we first draw all elements of the interaction matrix from the uniform distribution $\mathcal{U}(-a, a)$ and subsequently put random elements to zero with a probability $1 - C$. This is similar to the construction of an Erdős-Rényi graph. Finally, to study the effects of interactions and stochasticity, we impose all remaining parameters to be equal for the different species: the self-interactions $\omega_{ii} = -1$, the growth and death or emigration rates $g_i^+ = 1.5$ and $g_i^- = 0.5$ and immigration rate $\lambda_i = \lambda$.

We first study IBMs with equal growth rates and self-interactions, and subsequently the analogous PLM. To mimic the maximum number of individuals in the IBM, we introduce a maximal capacity in gLV and show that the same results—heavy-tailed abundance distributions and the species-connectivity relation—can be obtained. At the end of this section, we show how heavy-tailed distributions also arise from logistic equations with uniformly and lognormally distributed parameters.

## 9.4 Heavy-tailed distributions with individual-based models

We used the model of Heyvaert, 2017, which is based on the model of Solé et al., 2002. It is a stochastic cellular automaton in which immigration, growth, and extinction events can happen at every time step as well as the interaction between species. More details about the exact implementation of this model can be found in Appendix E.1. The abundance distribution shifts from a lognormal to a power law distribution for a decreasing immigration rate and increasing connectance (Figure 9.6A). This result is in agreement with the results of Solé et al., 2002. The transition from lognormal to power law distributions is gradual with increasing width of the lognormal distribution (Figure 9.6B). The number of species is high when there is immigration but low without immigration because species that go extinct cannot reappear in the community. Both measures of fluctuation—the average coefficient of variation (Figure 9.6C) and average JS distance (Figure 9.6D)—increase for decreasing immigration. The results of Figure 9.6 are for a maximum number of individuals $N_{max} = 10^4$. For smaller maximum numbers of individuals, the diversity decreases and the width of the lognormal abundance distribution and the measures of fluctuation increase. This is discussed in Appendix E.4.

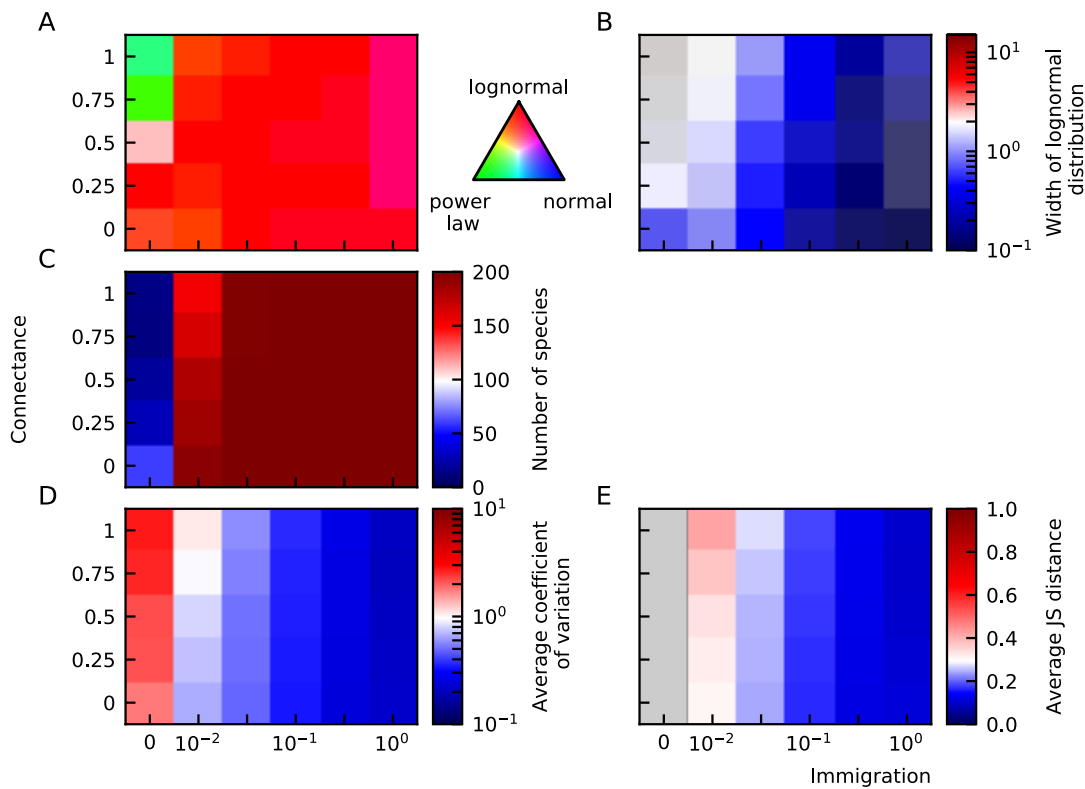

**Figure 9.6:** Individual-based models result in heavy-tailed abundance distributions. (A) Abundances mostly fit a lognormal distribution best, with a tendency to power law distributions for small immigration and high connectance. (B) The width of the lognormal distribution increases for decreasing immigration and increasing connectance. (C) The species diversity is maintained except for small immigration rates. The average coefficient of variation (D) and average JS distance (E), increase for decreasing immigration. The fixed parameters of these plots are the number of species $N_{spec} = 200$, the interaction strength $a = 0.75$ and the maximum number of individuals $N_{max} = 10000$.

Solé et al., 2002 reported that the diversity—defined as the number of (non-zero) species—and connectance are connected through a power law relation. We also include the interaction strength in our definition of complexity based on the definition of May. However, we cannot compare the complexity defined here as $a^2C$ with the one of May defined as $\xi^2C$ as $a^2$ is the variance of the interaction matrix and $\xi^2$ the variance of the community matrix. The latter depends on the abundance distribution (Equation 9.5). Although there is a difference between both definitions, we still see that the definition of complexity creates a boundary for stability (Figure 9.7). For small complexity, we see saturation because the number of species in the immigration pool is finite ($N_{\text{spec}} = 500$) as well as the maximum number of individuals ($N_{\text{max}}$). The exponent of the power law is more negative for smaller immigration rates (Figure 9.7). This quantitatively agrees with the mean-field model of Solé et al., 2002. We also show how the width of the abundance distribution grows with the complexity (Figure 9.7).

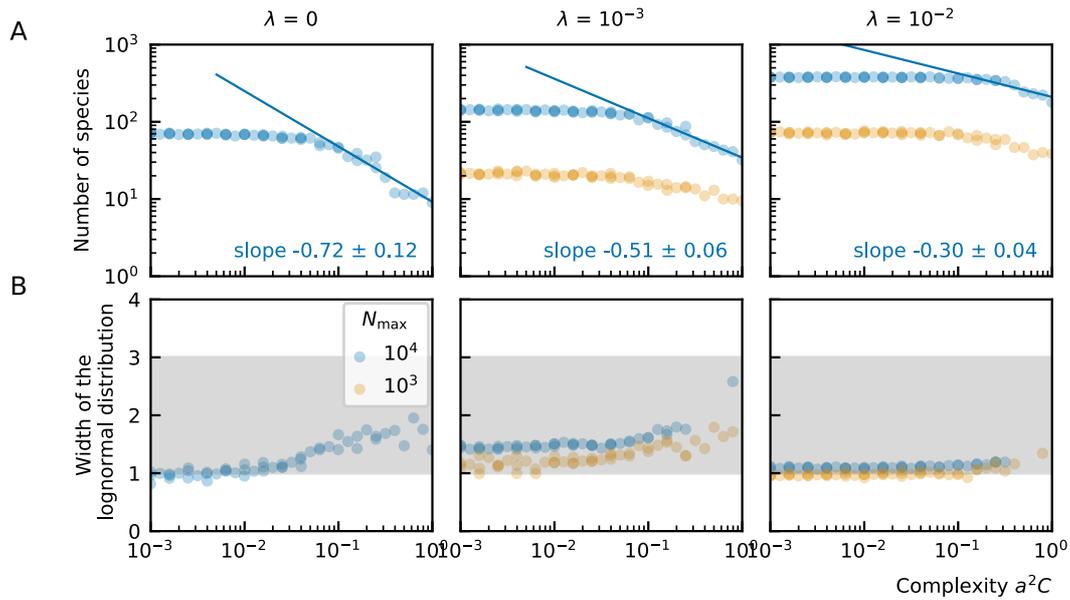

**Figure 9.7:** (A) Diversity-complexity relation for IBM for different immigration rates ($\lambda = 0$, $10^{-3}$ and $10^{-2}$). The solutions form a border. Because of immigration, solutions under the curve are unstable and because of interactions and a limited number of lattice sites any solutions above the curve are unstable. There is relation between the complexity and number of species can be approximated by a power law. For smaller complexity, we see saturation because the number of species in the immigration pool is finite ($N_{\text{spec}} = 500$). The exponent of the power law is more negative for smaller immigration rates. (B) The width of the abundance distribution fitted with a lognormal. The width increases for increasing complexity.

## 9.5 Heavy-tailed distributions with stochastic Lotka-Volterra models with a maximal capacity

We have shown that IBMs can result in heavy-tailed abundance distributions. The question remains whether one obtains qualitatively similar behavior for PLM. We study the widely used generalized Lotka-Volterra (gLV) equations. In addition to the immigration term $\lambda_i$, the growth rate $g_i$, and the self-interaction term $\omega_{ii}$ for every species, this model includes pairwise interactions between the different species: $\omega_{ij}$ represents the effect of species $j$ on species $i$ which can be beneficial ($\omega_{ij} > 0$), harmful ($\omega_{ij} < 0$) or non-existent ($\omega_{ij} = 0$). We will consider a linear noise strength $\eta_i$ as argued in Chapter 8. The stochastic gLV (sgLV) equation of species $i$ reads

$$dx_i(t) = \left(\lambda_i + g_i x_i(t) + \sum_j \omega_{ij} x_i(t) x_j(t)\right) dt + \eta_i x_i dW \quad (9.6)$$

with $dW$ an infinitesimal element of a Brownian motion which is defined by a variance of $dt$ ($dW \sim \sqrt{dt}\mathcal{N}(0,1)$). The values for all these parameters were discussed in Section 9.3.

The emergence of heavy-tailed distributions for IBMs happens for low immigration rates and high interaction rates, at the boundary of instability. The IBM restricts solutions that go to infinity by construction: there is a maximum number of individuals. In gLV models, there is no such restriction and high interaction rates lead to a loss of diversity (blue region in Figure 9.8A for $N_{\max} = \infty$). Therefore, we do not find the same qualitative behavior as for IBM. To still obtain heavy-tailed distributions with gLV, we want to mimic the IBM. Hence, we first study the differences between the IBM and sgLV model.

For gLV models, one subtlety regarding the extinct species needs to be taken into account. In IBMs species can be absent, in which case they are not considered for the abundance distribution. In gLV models without noise, species that go extinct will take an infinite amount of time to reach zero. Noise or finite integration steps can lead a population to obtain zero or negative abundances (in the latter case, the abundance is put to zero), but otherwise, the species have a very small abundance at the end of the time series. We will consider all species that have an abundance that is six or more orders of magnitude smaller than the maximum abundance in the community as if they were extinct. For the IBMs we performed simulations for a maximum number of individuals up to 10,000 which leads to a difference of maximally four orders of magnitude between populations. This constraint is important for fitting the heavy-tailed distributions, but even more so in the study of diversity-complexity.

The simulation rules of the IBM are based on the gLV model which describes the abundances at the population level. The immigration, extinction, and interaction rates can be chosen to correspond to the parameters of a gLV model, but one important difference is the fixed maximum number of individuals. The growth rate of species, therefore, depends on the number of empty sites, whereas the growth rate in the gLV model is a scalar.

Biologically, the maximal capacity can be interpreted as limits by fundamental resources such as water, nutrients, and space. In the next paragraph, we propose a PLM with a maximal capacity.

A maximal capacity can be added to logistic and sgLV models by multiplying all positive growth terms with a growth probability $\gamma$ that depends on the total number of individuals $N_{\max}$, $\gamma = 1 - \sum_j x_j/N_{\max}$. The growth vector $g$ can be split into its positive and negative components $g = g^+ + g^-$. Species for which the net growth term is positive can grow in the absence of other species, species with a strictly negative growth term will need other species to interact positively with them to have a non-zero positive steady-state abundance. Analogously, the interaction matrix can be split into its positive and negative parts: $\omega = \omega^+ + \omega^-$. Only the positive growth terms and immigration rate are multiplied by the probability for growth $\gamma$ that depends on the fraction of "empty space". Furthermore, we limit the probability for growth $\gamma$ such that it has a value between 0 and 1. This constraint is important because we could imagine a steady state starting with more individuals than the maximal capacity. Without a limit at 0, $\gamma$ could become negative and the growth term becomes a death term. The implementation of the maximal capacity $N_{\max}$ into gLV models (Equation 9.6) thus resumes:

$$dx_i(t) = \left( g_i^- x_i(t) + \sum_j \omega_{ij}^- x_i(t) x_j(t) + \gamma(t) \left( \lambda_i + g_i^+ x_i(t) + \sum_j \omega_{ij}^+ x_i x_j \right) \right) dt,$$

$$\gamma(t) = \max\left(0, 1 - \frac{\sum_k x_k(t)}{N_{\max}}\right).$$

(9.7)

Noise can be implemented by adding a linear random fluctuation $\eta_i x_i dW$ to the right-hand side of Equation 9.7 similar to the noise in Equation 9.6. Notice that the total number of individuals can exceed the total capacity because noise term is not bounded by the maximal capacity. The maximal capacity is, therefore, not a rigid constraint. One could also implement the maximal capacity effect for both the deterministic and stochastic parts, *i.e.* multiply the noise term by $\gamma$. This gives a bias to negative values for the noise term.

We will now consider the simplest case of our model: without interactions, *i.e.* the logistic equations.

## Logistic equations with a maximal capacity are generalized Lotka-Volterra equations

The logistic equations are similar to the gLV equations (Equation 9.6), but the immigration and interaction terms are omitted:

$$dx_i(t) = \left(g_i x_i(t) - \beta_i x_i^2(t)\right) dt \tag{9.8}$$

with $\beta_i$ the self-interaction of species $i$. We do not include the noise term in the next derivations, but one can easily add it to these equations. The self-interaction is related to the carrying capacity $K_i = g_i/\beta_i$. After adding a maximal capacity to the logistic equations

$$\begin{aligned} dx_i(t) &= \left(g_i^- x_i(t) + \left(1 - \frac{\sum_j x_j(t)}{N_{\max}}\right) g_i^+ x_i(t) - \beta_i x_i^2\right) dt, \\ &= \left((g_i^- + g_i^+) x_i(t) - \sum_j \left(\frac{g_i^+}{N_{\max}} + \beta_i \delta_{ij}\right) x_i(t) x_j(t)\right) dt, \end{aligned} \tag{9.9}$$

we can identify an interaction matrix $\omega_{ij} = -\frac{g_i^+}{N_{\max}} - \beta_i \delta_{ij}$. Logistic equations with a maximal capacity can thus be interpreted as gLV equations as long as we assume that the self-interaction has only a negative component ($\beta_i = \beta_i^-$, $\beta_i^+ = 0$). The reduction to a gLV model is valid as long as the sum of the abundances is smaller than the maximal capacity. One needs to be careful to not use initial conditions where $\sum_j x_j(0) > N_{\max}$. A smaller maximal capacity translates into more competitive interactions in the gLV model.

Logistic equations with a maximal capacity have thus effective interactions between all species. However, in this chapter, we use the terminology of interactions only for the specific interactions and not the indirect interactions through the maximal capacity.

## The width of the lognormal distribution is determined by the maximal capacity and noise level.

We perform simulations for the system of logistic equations with maximal capacity (all growth rates equal to 1, all self-interactions equal to -1, no specific interactions). Both increasing the noise and decreasing the maximal capacity make the lognormal distribution wider (Figure 9.9). To obtain a width of around 2 (white values), we need a balanced combination of maximal capacity and noise. The smaller the maximal capacity, the less noise is needed to obtain a similar width. We see that decreasing the maximal capacity increases the JS distance (Figure 9.9D). As expected increasing the noise also increases the JS distance.

In sgLV models, we can impose the strength of the noise. In IBMs, the level of stochasticity is determined by the maximum number of individuals. For decreasing maximum number of individuals the fluctuations increase (Figure E.2C-D). Similarly, for sgLV models, the parameters of the fluctuations increase with increasing noise and decreasing maximal capacity (Figure 9.9).

### Interactions make the lognormal abundance distribution wider.

We previously showed that logistic equations with a maximal capacity simply reduce to gLV equations. When including interactions (Equation 9.7), higher-order interaction terms remain:

$$
\begin{aligned}
dx_i(t) &= \left( \lambda_i \left(1 - \frac{\sum_k x_k(t)}{N_{\max}}\right) + g_i x_i(t) + \sum_j \omega_{ij} x_i(t) x_j(t) \right. \\
&\quad \left. - \frac{\sum_k x_k(t)}{N_{\max}} \left( g_i^+ x_i(t) + \sum_j \omega_{ij}^+ x_i x_j \right) \right) dt \\
&= \left( \lambda_i \left(1 - \frac{\sum_k x_k(t)}{N_{\max}}\right) + g_i x_i(t) + \sum_j \left( \omega_{ij} - \frac{g_i^+}{N_{\max}} \right) x_i(t) x_j(t) \right. \\
&\quad \left. - \sum_j \sum_k \frac{\omega_{ij}^+}{N_{\max}} x_i(t) x_j(t) x_k(t) \right) dt
\end{aligned}
\tag{9.10}
$$

Such higher-order terms already appear in some models and are often interpreted in a phenomenological way and not as distinct mechanistic processes (Letten & Stouffer, 2019). The strength of the interaction is still only dependent on two species ($\omega_{ijk} = -\omega_{ij}^+/N_{\max}$) because the term originated from a pairwise interaction description.

The interaction matrix is both determined by the interaction strength and connectance. Both parameters have similar effects on the abundance distribution and fluctuations. An analysis of these parameters can be found in Appendix E.4.

Without maximal capacity, gLV equations are unstable for a large number of species and a large number of interactions. When a maximal capacity is imposed, the diversity does not decrease for larger connectances and interaction strengths (Figure E.6 A). This leads to lognormal distributions (Figure E.6B). Furthermore, we see that increasing the immigration rate, decreases the width of the abundance distribution (Figure E.6C). By the implementation of a maximal capacity, we get results that are qualitatively similar to the ones of the IBM.

We again look at the relation of the diversity and complexity using the same definition $a^2C$. For small complexity, there is again saturation because the number of species in the equations is finite ($N_{\text{spec}} = 500$). Similar to the results for the IBM, the slope of the power law becomes flatter for increasing immigration rates when there is a maximal capacity (Figure 9.10). We also show how the width of the abundance distribution grows with the complexity (Figure 9.7). The width of the abundance distribution increases for increasing complexity. However, for increasing complexity $a^2C > 0.05$ the hypothesis that the data follow a lognormal function is rejected (Figure 9.10).

As a conclusion, immigration allows for more diversity in models with a maximal capacity. The diversity-complexity relation is a power law. The width of the abundance distribution increases for increasing complexity but the distribution deviates from a lognormal distribution. Smaller maximal capacities lead to larger widths.

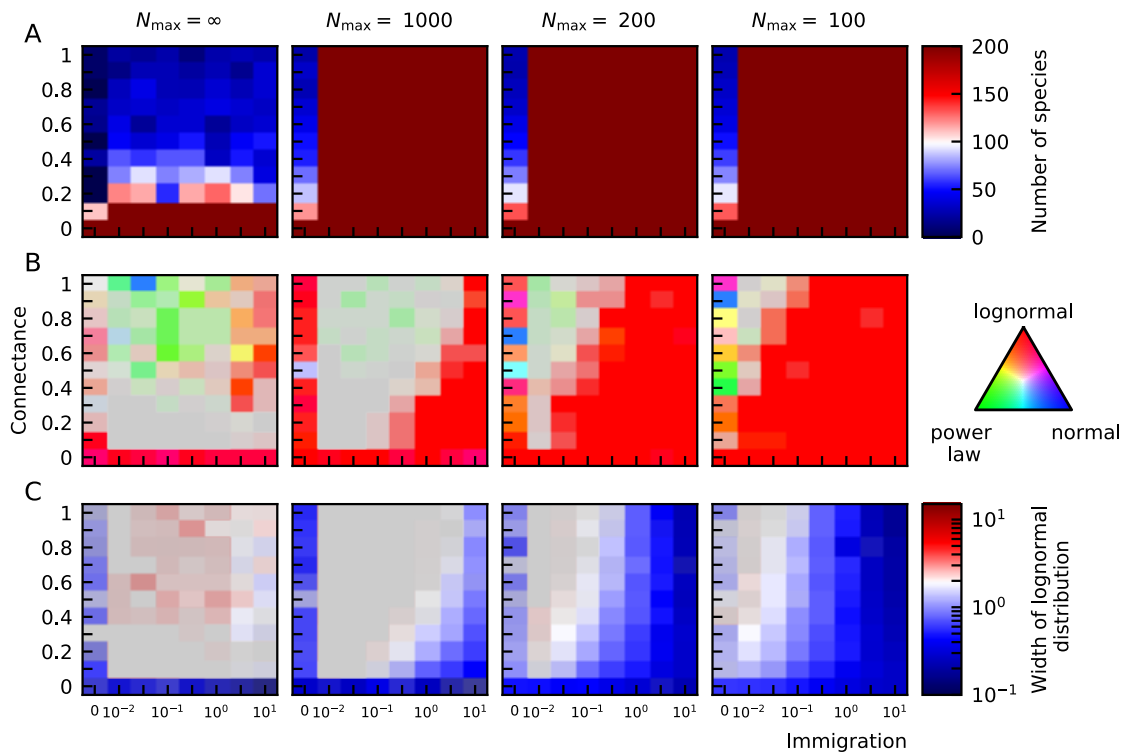

**Figure 9.8:** A finite maximal capacity and non-zero immigration rate allow for high diversity in the presence of interactions (A). Notice that the maximal capacity does not limit the number of species, *e.g.* there can be 200 species even when $N_{max} = 100$ because the species abundances can be smaller than 1 in population-level models. For decreasing maximal capacities the abundance distribution becomes lognormal (B). The grey area represents distributions that are neither lognormal, power law or normal. The width of this lognormal distribution increases for decreasing immigration and increasing connectance (C).

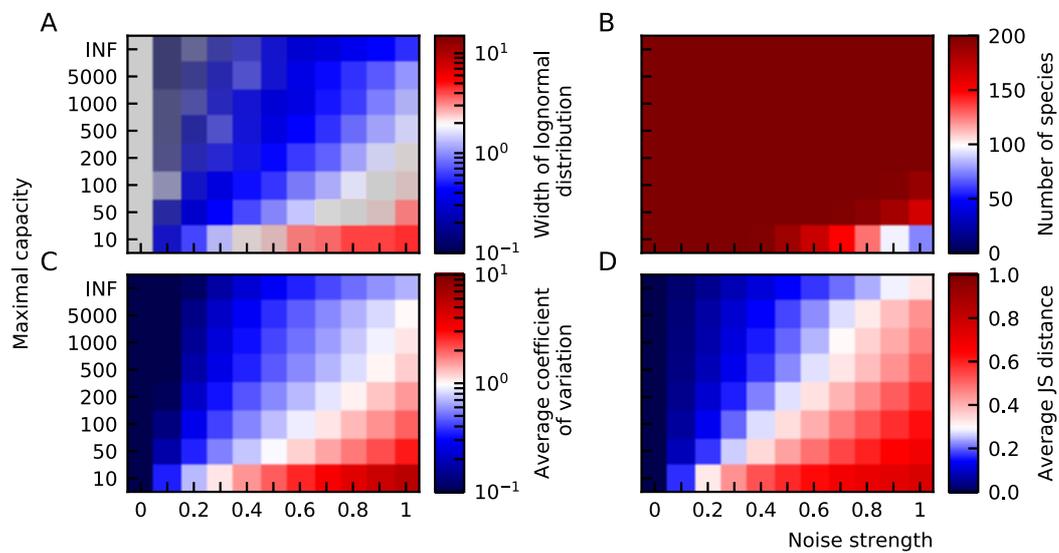

**Figure 9.9:** Heavy-tailed distributions for logistic equations, the influence of the maximal capacity and noise strength. (A) Abundances mostly fit a lognormal distribution best, with a tendency to normal distributions for small noise. This transition is gradual as the width of the lognormal distribution decreases in this direction (B). The species diversity is maintained except for small maximal capacities and high noise (C). The average coefficient of variation (E) and average JS distance (F), increases for decreasing maximal capacities and increasing levels of noise. Because logistic equations are used for these plots, the fixed parameters are the interaction strength $a = 0$, the connectance $C = 0$ and immigration rate $\lambda = 0$.

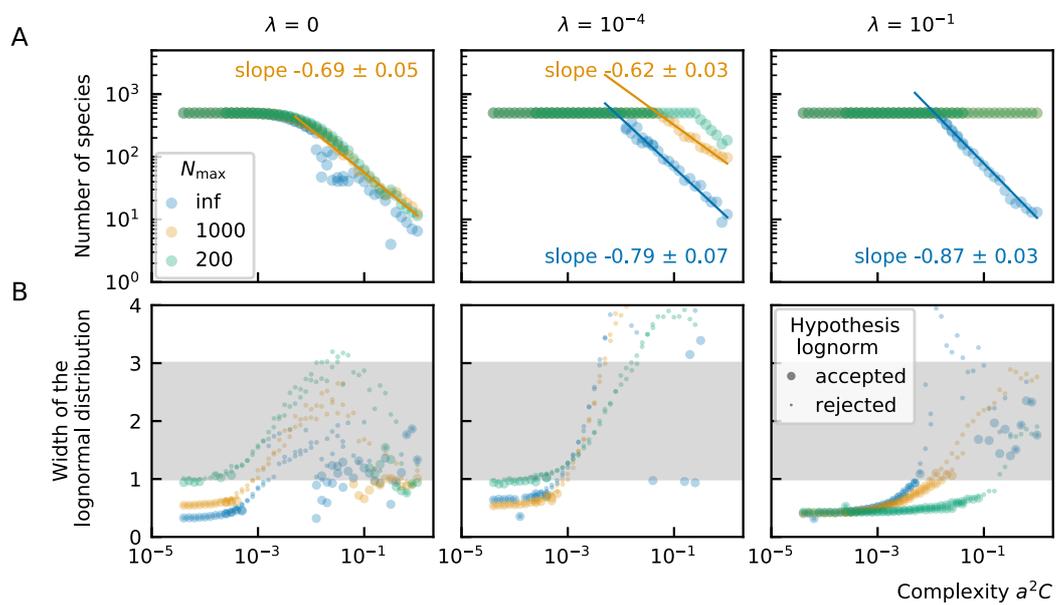

**Figure 9.10:** (A) Diversity-complexity relation for sgLV for different immigration rates ($\lambda = 0$, $10^{-4}$ and $10^{-1}$). There is relation between the complexity and number of species can be approximated by a power law. For smaller complexity, we see saturation because the number of species in the immigration pool is finite ($N_{\text{spec}} = 500$). The exponent of the power law is more negative for smaller immigration rates. (B) The width of the abundance distribution fitted with a lognormal. The width increases for increasing complexity, but the abundance distribution deviates from a lognormal distribution.

## 9.6 Heavy-tailed distributions with logistic equations

In the absence of interactions (Equation 9.8), the steady-state abundance of a species $x_i$ is given by the ratio of the growth rate to the self-interaction, $-g_i/\beta_i$.

If the self-interaction is equal for all species—like the common convention of using $\beta_i = -1$—the distribution of the steady-state abundances is obviously the distribution of the growth rates. On the other hand, if the growth rates are equal for all species, then the distribution of the steady-state abundances is the inverse distribution of the self-interactions.

### Uniformly distributed parameters result in power law abundance distributions

If the self-interactions would be uniformly distributed between $a$ and $b$, the distribution of the steady states would be

$$f_X(x) = \begin{cases} \frac{g}{x^2(b-a)} & \text{for } g/b \leq x \leq g/a, \\ 0 & \text{elsewhere.} \end{cases} \quad (9.11)$$

This is a power law distribution with power -2 on the bounded region $[g/b, g/a]$. The derivation of this PDF can be found in Appendix E.3.

If the distribution of growth rates and the distribution of self-interactions were uniform over $[0, g_{\max}]$ and $[-\beta_{\max}, 0]$, respectively, then the distribution of the steady states would be a ratio of uniforms distribution (RUD). By defining $\phi = g_{\max}/\beta_{\max}$, we write the specific distribution of the steady states as

$$f_X(x) = \begin{cases} \frac{1}{2\phi} & \text{for } 0 \leq x \leq \phi, \\ \frac{\phi}{2x^2} & \text{for } x < \phi, \\ 0 & \text{for } x < 0. \end{cases} \quad (9.12)$$

The derivation of this PDF can be found in Appendix E.3. The distribution is shown in Figure 9.11 and its median is $\phi$. Experimental data corresponds well to a lognormal distribution, but can the data also be approximated by a RUD? We can calculate $\phi$ for experimental data as the median abundance value and calculate the p-value of the corresponding RUD, the results can be rejected for most of the experimental distributions (see Appendix E.4). It is important to keep in mind that to use this RUD, *the growth rate and self-interaction must be uncorrelated*.

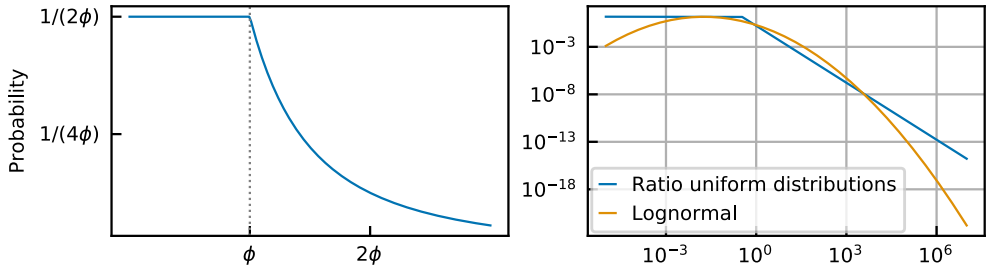

**Figure 9.11:** The ratio of uniforms distribution in a linear scale on the left and in a log-log scale on the right. In the right plot a lognormal distribution is drawn for comparison.

## Lognormal abundance distributions when parameters follow lognormal distributions

If the distributions of the growth rates, and self-interactions follow a lognormal distribution, $\ln(g) \sim \mathcal{N}(\mu_g, \sigma_g^2)$ and $\ln(\beta) \sim \mathcal{N}(\mu_\beta, \sigma_\beta^2)$, it follows that the ratio of the growth rate to the self-interaction is also a lognormal distribution,

$$\ln(g/\beta) \sim \mathcal{N}(\mu_g - \mu_\beta, \sigma_g^2 + \sigma_\beta^2 - 2\sigma_{g\beta}). \tag{9.13}$$

Notice that this also holds when both distributions are not independent. The derivation of this PDF can be found in Appendix E.3.

Biological growth rates and self-interactions are empiric parameters to which many microscopic processes contribute. Their effects are often multiplicative, and where multiplicative effects reign, lognormal distributions appear (Limpert et al., 2001; Zhang & Popp, 1994). Moreover, the distribution of doubling times of bacteria in the wild has been estimated to be lognormal (Gibson et al., 2018). For abundances much smaller than the carrying capacity, the growth is exponential and the doubling time is inversely proportional to the growth rate. Because the inverse distribution of a lognormal distribution is again a lognormal distribution, this implies that the bacterial growth rates follow a lognormal distribution. Furthermore, lognormal distributions of the self-interactions are not incompatible with the results of the noise color in Chapter 8 (see Section E.4).

From Equation 9.13, we conclude that the width of the abundance distribution depends on the widths of the growth rate $\sigma_g$ and self-interaction $\sigma_\beta$. Fixing the width of the abundance distribution $s$, we can find the width of the growth rate given the width of the self-interaction, $\sigma_\beta = \sqrt{s^2 - \sigma_g^2}$ where we assumed that the distributions of the self-interaction and growth rate are independent. In this way, we can scan over different proportions of both parameters. In Figure 9.12, we scan over the width of the growth rates and the strength of the noise. We choose the width of the abundance $s = 2$ and, as expected, the width of the distribution remains around this value (Figure 9.12A). For large noise, and a large width of the growth rate distribution, a number of species goes extinct and diversity is lost (Figure 9.12B). In the analysis above where the growth rate and self-interactions were identical for all species, the two parameters describing the fluctuations—average coefficient of variation and average JS distance—were positively correlated. Here, with the growth rate and the self-interactions following a lognormal distribution, the parameters of fluctuations are negatively correlated with respect to the proportion of variability in the self-interaction and growth rates. For increasing widths of the growth rate and, therefore, self-interactions that become more similar for all species, the coefficient of variation increases, but the average JS distance decreases. The effect of the noise is the same as for the gLV case, it increases the parameters of the fluctuations.

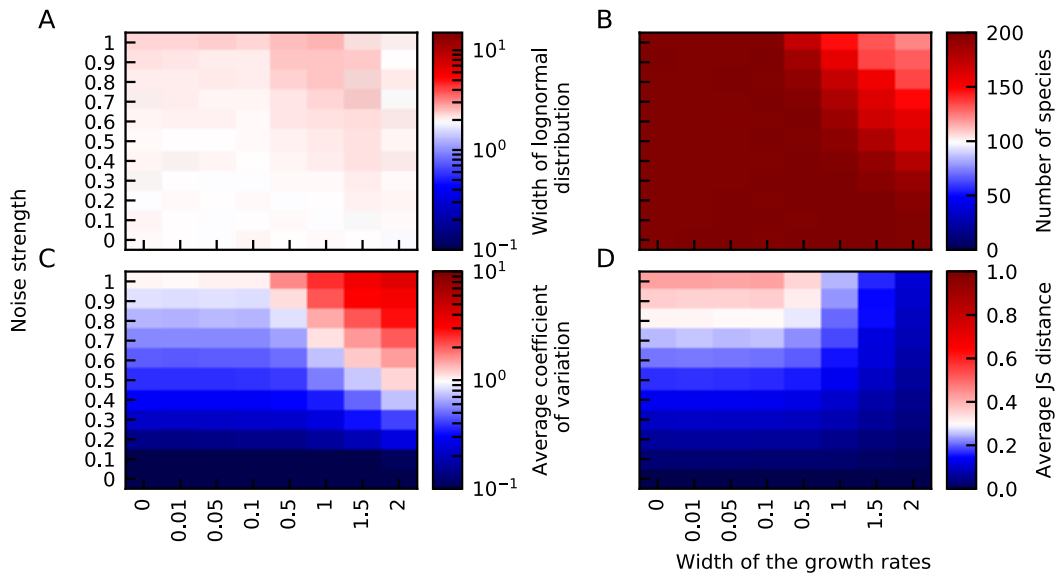

**Figure 9.12:** Lognormal parameter distributions lead to lognormal abundance distributions. (A) The width of the abundance distribution $s$ does not depend on the width of the growth rate $\sigma_g$. (B) For large noise and large width of the growth rate, species go extinct. The parameters of the fluctuation, average coefficient of variation (C) and average JS distance (D) are negatively correlated with respect to the width of the growth rate distribution. They both increase for increasing noise. The fixed parameters of these plots are the imposed width of the abundance distribution $s = 2$, the immigration $\lambda = 0$, the interaction strength $a = 0$ and connectance $C = 0$.

## 9.7 Discussion

The abundance distribution of microbial communities is heavy-tailed and typically fits a lognormal distribution. Heavy-tailed distributions can be obtained by differ-ent modeling approaches. In IBMs, heavy tails are the result of the system being self-organized at boundary of instability (Solé et al., 2002). Immigration increases the number of species, and interactions between species cause individuals to die. The resulting abundance distribution is heavy-tailed and generally a lognormal distribution. The width of the latter is determined by the interactions and immigration rate: larger interactions—interaction strength and connectance—and smaller immigration rates result in larger widths of the abundance distribution. In gLV models, strong interactions lead to unstable solutions and many extinct species. We propose a maximal capacity in the gLV model that mimics the maximum number of individuals in IBMs. We show that with this maximal capacity lognormal abundance distributions are obtained with increasing width for increasing connectance and interaction strength.

In the IBM and gLV model with maximal capacity, the growth rate was equal for all species and likewise for the self-interaction. The heavy-tailed rank abundance distributions were the result of interactions between species and stochasticity. Heavy tails can also be obtained by the appropriate choice of the parameters. We have shown that for logistic equations, a uniform distribution of the self-interaction leads to a power law abundance distribution, but more importantly, lognormal distributions of the growth rate and the self-interactions give rise to lognormal abundance distributions. This is in line with our expectations of such distributions. The majority of these parameters have not been measured as most of the species cannot be cultured in the lab. *In vivo* experiments are even more challenging, but data of other microbes is consistent with lognormal distributed parameters (Gibson et al., 2018). The width of the distribution of growth rates and the width of the distribution of self-interactions contribute equally to the width of the final abundance distribution. However, the fluctuations in the presence of noise have a different character depending on which distribution is widest.

We expect that the heavy-tailed abundance distribution of experimental data is the result of both heavy-tailed parameter distributions and self-organization through interactions. We show that one does not need to go to IBM to obtain heavy-tailed distributions. They can be obtained with stochastic gLV equations with a maximal capacity which are computationally less demanding. Moreover, the number of individuals in the gut is $4 \cdot 10^{13}$ which is orders of magnitude higher than what realistically can be modeled by an IBM. Furthermore, in IBMs, species with low abundance go extinct and reappear through immigration. The distribution of a species is thought to be Gamma, where the probability of the species actually having an abundance of zero is zero (Grilli, 2019). Without considering major perturbations, we do not expect species to go extinct. In experimental data, the abundance of species can become zero, but the cause is most probably the abundance being lower than the detection experimental threshold.

# Conclusions and outlook | 10

Biological systems are complex. They consist of many parts that interact in non-trivial ways. The large number of components, nonlinear responses, and stochasticity make it challenging to study these systems quantitatively. Biological systems exist at different scales: from the small scale of DNA to the large scale of ecologies. In this thesis, we study two systems: the autoregulation of a gene at the single-cell level and microbial communities at the population level.

In the first part, we study non-monotonic autoregulation of genes. Inspired by the natural system of the Leucine responsive protein B of the *Sulfolobus solfataricus* (Ss-LrpB) (Peeters et al., 2013), we look at the possible dynamics of genes that show autoactivation for low concentrations and autorepression for high concentrations. As this genetic network has persisted through evolution, we expect it to have an advantage over other simpler networks.

Positive autoregulation allows bistability (Ferrell & Xiong, 2001), negative autoregulation speeds up the response to the steady state (Rosenfeld et al., 2002) or can give rise to oscillations when there is delay which can arise from strong nonlinearities (Novák & Tyson, 2008). Based on these ideas, we propose two hypotheses for the dynamical behavior of proteins with non-monotonic autoregulation: oscillations and bistability with an increased speed toward the high steady state with respect to monotonic positive feedback. We approach the problem with three increasingly complex single gene networks with one to three binding sites for the protein.

We show that in theory, *non-monotonic autoregulation* can result in oscillations or bistable switches with a fast response to the high concentration. However, the parameter regions for which these dynamical behaviors are present are small. The proposed theoretical single gene networks are nonetheless relevant for synthetic biology as they are small yet result in interesting dynamical behavior. Concerning the natural Ss-LrpB system, the dynamical behavior of the protein has never been measure in vitro or in vivo because of technical challenges. Consequently, we can only make predictions about its dynamics. We conclude that when our model is compatible with the Ss-LrpB system, it does not allow bistability and needs finely tuned parameters for oscillations. We propose a third dynamical behavior for the network: stochastic spiking. We speculate that the protein is a sensing molecule and that the threshold concentration for sensing is high. Stochastic spiking of the protein could allow the concentration to exceed the threshold, but the burden on the cell is reduced by not maintaining the protein at the high concentration. However, to confirm this idea a more elaborate analysis of the energy in the system should be performed.

Our study does not reveal a clear advantage for the specific architecture of the natural Ss-LrpB system. Whether this is due to the simplicity of the model or the fact that we only search for interesting dynamical behavior and did not consider cost, efficiency, etc., remains to be uncovered. Maybe the system needs to be studied in a larger context or the mechanisms of the DNA looping when three dimers are bound are important for the function of the protein (Peeters et al., 2006). In order to know which hypothesis should be studied in more depth, more of the experimental parameters should be measured as well as the concentration of the protein as a function of time. Our findings of which of the parameters are critical because they need fine-tuning provides a guideline for experimentalists which parameters will reveal more about the dynamical behavior and are preferably measured first.

The analysis of this gene regulatory network contributes to the *development of a genetic toolbox for archaeal networks*, but the techniques that we use to study this system are no different from techniques used in bacteria or eukaryotes. Therefore, the results of non-monotonic autoregulation are directly applicable to systems in these organisms as well.

From the study of a small genetic network, we made a large leap in biology to the dynamics of *microbial communities*. Although these biological fields are separate, the mathematical methods we use to study them are similar. Population-level models are constructed along the lines of mass-action kinetics, like the models of gene regulatory networks.

Microbes such as bacteria, archaea, micro-eukaryotes are ubiquitous (Martiny et al., 2006). Apart from vastly outnumbering other organisms on earth, they are indispensable for ecological systems and human health (Graham et al., 2016). Microbes live in communities and the composition of these communities determines their properties. In this way, the composition of the human gut microbiome dictates the health of its host (de Vos & de Vos, 2012). Having a good understanding of community dynamics is therefore crucial.

Two actual topics of research in the field of community dynamics are the *neutral-niche debate* and the stability-complexity relation. Neutral theories assume that all species are ecologically equivalent and that community dynamics are dominated by random birth, death, and immigration processes (Hubbell, 2001). Niche theories, in contrast, consider all species to be adapted for a specific environment, and depending on the environment certain species will have an advantage over others (Harpole, 2010). As neutral theories assume that no individual has an advantage over another individual, interactions such as competition are excluded. In this context, we associate neutral and niche regimes with non-interacting and interacting models. Which of both hypotheses should be used in which context is still under discussion.

The other subject of debate is the *stability-complexity criterion*. In the '70s, May provided a criterion that separates stable from unstable systems and concluded that, theoretically, stability and high complexity—defined by the number of species and interactions—are mutually exclusive (May, 1972). This surprising fact is called a paradox by ecologists as biodiversity and complexity in ecosystems are thought to be a stabilizing factor (Ives & Carpenter, 2007; Ptacnik et al., 2008; Tilman & Downing, 1994). Furthermore, apart from the existence of a boundary between stable and unstable solutions, natural ecosystems seemed to live at this critical border. Such critical behavior is also observed in multiple other biological systems (Mora & Bialek, 2011). How ecosystems remain in this position is still an active field of research.

In this thesis, we addressed both questions with stochastic *generalized Lotka-Volterra models*. These models consider communities at the population level and include different types of interactions between species like competition and mutualism. We compared these models with *logistic models* that do not include interspecies interactions.

Our first project related to microbial communities studies the question of the neutral-niche debate. Experimental time series of microbial communities are characterized by a heavy-tailed abundance distribution, fluctuations, and correlations between different species. We studied the latter two characteristics through the color of the noise of the different species, the differences between time points, and two neutrality measures. We found that the color of the noise depends on the product of the self-interaction and the mean abundance such that the self-interaction can be estimated given the noise color. Furthermore, we show that to obtain fluctuations similar to the ones in experimental time series strong linear multiplicative noise is needed. A comparison of the properties in logistic models allows us to conclude that all stochastic properties of experimental time series can be reproduced by *stochastic logistic equations*. This conclusion is in agreement with the results of Grilli, 2019. The need for interacting models to study microbial communities is sometimes justified by neutrality tests such as the neutral covariance test indicating a niche regime. We show that also non-interacting (logistic) models can be classified in the niche regime by these tests. Hence, given the characteristics we consider, we cannot conclude that interactions are compulsory to model microbial communities at the level and sampling rate of the experiment. We do not question the existence of interactions between species in a community but rather whether these interactions are important in a noisy environment at a low sampling rate. It has been shown that almost none of the interaction parameters of a generalized Lotka-Volterra model inferred from experimental time series of the microbiome of stool with a least-square estimator can be inferred with a significant non-zero value (Tuijnder, 2020). This strengthens our hypothesis that interactions are not crucial for the communities at the measured timescales. As a result, we propose stochastic logistic models as a null model.

Our second project related to microbial communities focusses on the question of how *heavy-tailed abundance distributions* can be obtained by mechanistic models. Many theoretical studies have either imposed or ignored the rank abundance curve of the models. Solé et al., 2002 propose an individual-based model for which heavy-tailed abundance distributions emerge. This emergent property is the result of the system being self-organized at the border of instability. An immigration term increases the number of individuals and interactions between individuals decrease this number again. As a result of the pushing and pulling on the abundances by these two opposing mechanisms heavy-tailed abundance distributions arise. We propose a population-level model that obtains the heavy-tailed distributions by adding a maximal capacity term to the generalized Lotka-Volterra equations.

Additionally, we propose a simpler mechanism to obtain heavy-tailed abundance distributions without interspecies interactions. In a model without such interactions, the steady state of the abundance is determined by the distributions of its parameters: the growth rate and self-interaction. We derive that uniform distributions of these parameters lead to power law abundance distribution with exponent -2 and that lognormal distributions of these parameters lead to lognormal abundance distributions. The hypothesis of a lognormal parameter distribution is reasonable and lognormal distributions have already been found for bacterial doubling times (Gibson et al., 2018). Furthermore, the self-interactions and growth rates that we estimate from the noise color assuming the model follows a logistic equation in the previous project are compatible with this assumption.

The ultimate goal is to build predictive models for microbial communities. These are the first step towards personalized medicine. These models are a large puzzle with many pieces. We have studied the relation between stability, complexity and heavy-tailed abundance distributions, but also the influence of environmental factors, the availability of nutrients, and spatial effects need to be understood. Both theoretical and experimental research is building towards a global understanding of microbial communities. With our research, we hope to have contributed to solving a small piece of the puzzle.

# Appendices

# Dissimilarity and neutrality measures | A

## A.1 Dissimilarity measures

The composition of distinct communities is different and the composition of a given community changes over time. To quantify the difference between different communities, there are multiple measures. In this section, we present three of them.

**Bray-Curtis dissimilarity**

The *Bray-Curtis dissimilarity* is named after J. Roger Bray and John T. Curtis (Bray & Curtis, 1957) who used it to study plant communities in forests. It is defined as

$$BC = \frac{\sum \left| \vec{x}_1 - \vec{x}_2 \right|}{\sum (\vec{x}_1 + \vec{x}_2)}. \tag{A.1}$$

Importantly, this measure is not a distance because it does not satisfy the triangle inequality (Greenacre & Primicerio, 2014).

An important assumption made by this measure is that both samples were taken from similar areas or volumes. Because the absolute counts and not the relative counts of species are used. The Bray-Curtis dissimilarity is a value between 0 and 1, where 0 denotes that all species have equal abundances and 1 means that the samples are completely disjoint, *i.e.* no species are shared.

The opposite measure of the dissimilarity is the similarity or *Bray-Curtis index* $1 - BC$.

The Bray-Curtis index is sometimes defined differently, as the index that is also known as the *Czekanowski's quantitative*

*index* (Bloom, 1981)

$$CZ_{ij} = 1 - \frac{2C_{ij}}{S_i + S_j} \qquad (A.2)$$

where $i$ and $j$ denote the two communities, $S_i$ the number of species counted in $i$ and $C_{ij}$ the sum of the lesser count for each species in both communities[1] Also this value ranges between 0 and 1, where 0 denotes that all species are shared and 1 means that no species are shared.

1: For example, consider communities at site 1 and 2 with the following species abundances (sp denotes species)

|        | sp 1 | sp 2 | sp 3 |
|--------|------|------|------|
| site 1 | 3    | 7    | 2    |
| site 2 | 5    | 0    | 4    |

The values for $C_{12}$, $S_1$ and $S_2$ would be $C_{12} = 3 + 0 + 2 = 5$, $S_1 = 3 + 7 + 2 = 12$ and $S_2 = 5 + 0 + 4 = 9$. This results in a Czekanowski's quantitative index dissimilarity of $CZ_{12} = 1 - (2 \cdot 5)(12 + 9) = 0.41$.

## Kullback-Leibler divergence

The *Kullback-Leibler (KL) divergence* is a measure that compares two probability distributions introduced by Solomon Kullback and Richard Leibler (Kullback & Leibler, 1951). It is also known as *relative entropy* and it is used in information theory, neuroscience, machine learning, etc. In the context of community dynamics, it is defined by

$$KL(x|y) = \sum_i x_i \ln \frac{x_i}{y_i} \qquad (A.3)$$

where $x$ and $y$ represent the relative abundance vectors of the community composition. This definition is only defined if all elements of the $x$ and $y$ vector are strictly positive. In experimental data, we nevertheless have zero values for some of the abundances. There are two possibilities for this to occur: some species are absent in a certain environment, or some species are extremely rare with respect to other species in the same environment such that their abundance is too low for the detection threshold. In order to calculate the KL divergence with zero abundance, a pseudocount ($\epsilon \ll 1$) is added to the vectors. Like the Bray-Curtis divergence, the KL divergence is not a distance, because it does not satisfy the triangle inequality. Notice also that the KL divergence is not symmetrical for $x$ and $y$. The minimal value for the KL divergence is zero, this happens when both communities have exactly the same composition. The measure is, however, not bounded above, and can become infinite for some distributions. Multiple symmetric and bounded versions of this measure exist, we introduce the Jensen-Shannon divergence in the next section.

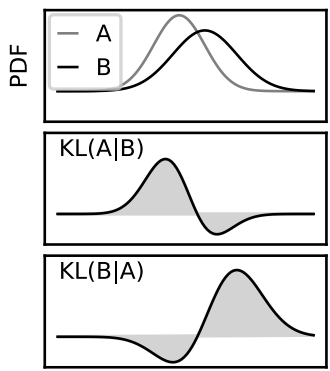

**Figure A.1:** The Kullback-Leibler divergences between two normal distributions $A$ and $B$ (top) are defined as the integral under the curves in middle and bottom. The measure is not symmetric and $KL(A|B) \neq KL(B|A)$.

### Jensen-Shannon divergence and distance

The *Jensen-Shannon (JS) divergence* is a measure that compares two probability distributions which is derived from the KL divergence, but which is bounded and symmetric (Lin, 1991). It is used in bioinformatics, machine learning, social sciences, etc. In the context of community dynamics, it is defined by

$$JS(x|y) = \frac{1}{2}\left(KL\left(x\bigg|\frac{x+y}{2}\right) + KL\left(y\bigg|\frac{x+y}{2}\right)\right). \quad (A.4)$$

where $x$ and $y$ represent the relative abundance vectors of the community composition. The value of the JS divergence ranges between 0 and 1, denoting two equal community compositions and two very different compositions. The square root of the JS divergence is a metric and is also called the *JS distance*.

## A.2 Neutrality measures

There is no consensus on the definition of neutrality. In general, ecosystems are considered neutral if the dominating cause of fluctuations are random birth and death processes and not fitness advantages of species.

> This paragraph is based on the methods section of our publication (Descheemaeker & de Buyl, 2020).

Different neutrality measures focus on different aspects of neutrality. The Kullback-Leibler divergence and Pielou's evenness index verify whether all species are equal (equal abundances and equal covariances). The neutrality covariance test studies the grouping invariance of species in time series (Washburne et al., 2016).

### Pielou's eveness index

For a system with $S$ species with frequencies $p_i$, this index is

$$J = \frac{\sum_{i=1}^{S} p_i \ln p_i}{\ln S},$$

where numerator represents the Shannon diversity index and the denominator this same index for the case all species had the same abundance. The evenness index ranges from 0 to 1 with low values representing very uneven communities and high values more even communities.

**Kullback-Leibler neutrality**

The density function of a multivariate Gaussian distribution is

$$P(x) = \frac{1}{(2\pi)^{N/2}\sqrt{\det K}} \exp\left(-\frac{1}{2}(x-\mu)^T K^{-1}(x-\mu)\right) \quad \text{(A.5)}$$

where $\mu$ and $K$ are the mean and covariance matrix of the distribution respectively. The Kullback-Leibler divergence for two multivariate Gaussian distributions in $\mathbb{R}^n$ is (Duchi, 2007)

$$\begin{aligned}D_{KL}(P|Q) = \frac{1}{2}\Bigg(&\ln\frac{\det K_Q}{\det K_P} - n + \text{Tr}\left(K_Q^{-1}K_P\right) \\ &+ (\mu_Q - \mu_P)^T K_Q^{-1}(\mu_Q - \mu_P)\Bigg).\end{aligned} \quad \text{(A.6)}$$

For every time series, we can calculate the $\mu$ and $K$ and we define values $\mu_N$ and $K_N$ for a corresponding neutral time series in which all species are equal (Fisher & Mehta, 2014). The distance to neutrality $D_{KL}(P|P_N)$ can thus be calculated by computing the probability distribution of the original time series $P$ and the associated neutral distribution $P_N$ with mean values $\mu_N = S^{-1}\sum_{i=1}^{S}\mu_i$ and $K_{P,ii} = S^{-1}\sum_{i=1}^{S}K_{ii}$ and $K_{P,ij} = S^{-1}(S-1)^{-1}\sum_{i=1}^{S}\sum_{j=1,i\neq j}^{S}K_{ij}$ with $S$ the number of species.

**Validity of neutrality measures**

One way of generating a neutral time series is considering a lattice of $N$ individuals on which every time step one individual is replaced by another. Each individual of the lattice has an equal probability of being replaced (probability is $N^{-1}$). The disappearance of the first species can be interpreted as the result of either death or emigration. The replacing individual is either the result of immigration or growth. The probability of immigration depends on the immigration rate $\lambda$ ($0 \leq \lambda \leq 1$). In case of an immigration event, all species of the external species pool $S$ have an equal probability of immigrating. The probability of a growth event is thus given by the remaining $1-\lambda$. In case of growth, every individual has an equal probability of growing. Time series generated in this way depend on three variables: the length of the simulation time $T$, the immigration probability $\lambda$ and the number of individuals $N$. We study the effect of these three variables on both neutrality measures (Figure A.2).

As discussed in Wennekes et al., 2012, the neutral-niche debate is a scale problem. Looking at a process from close-by and describing all the details, results in the niche regime. Dynamics of large systems where the details are not modeled appear to be dominated by random processes. The questions are: at which scale is one studying the process and which of the deterministic or stochastic processes is dominating? We can see this in the upper row of Figure A.2A. Larger immigration gives rise to a neutral regime, however for short time series and large communities one will only see the "details" because the time series is not yet in steady state. Consequently this results in niche values. In the absence of immigration, the "winner-take-all" principle holds. During a transient period all species but one go extinct in a stochastic manner. In the end, the steady state consists of only one species surviving. The duration of the transient depends on the size of the community. This is why the time series are in the niche regime with respect to grouping in the absence of immigration when the time series is long enough (Figure A.2B). For small immigration and small communities, the "winner" cannot take over due to the immigration of other species, but the most fit species dominate the community one by one in a random order. For short time series, there are only a few changes of dominating species and this apparent structure in the time series gives rise to a niche regime (Figure A.2B). For longer time series however, there are many random changes of the dominating species, which results in a neutral character (Figure A.2B).

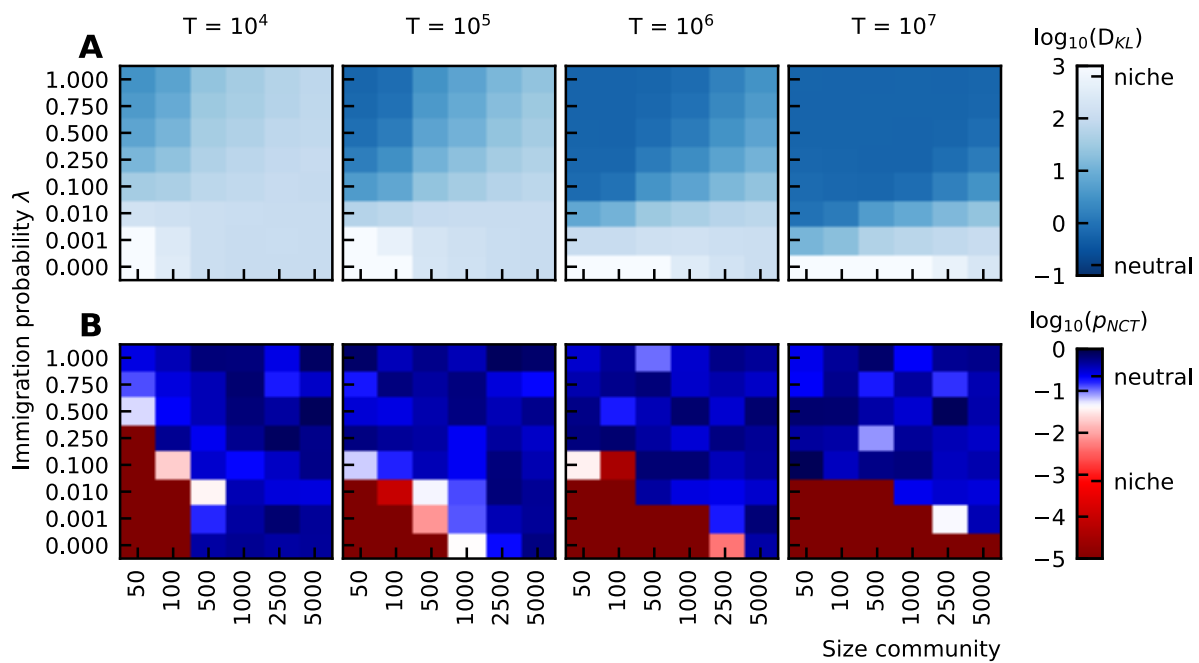

**Figure A.2:** Neutrality tests, (A) Kullback-Leibler divergence and (B) the p-value of the neutral covariance test. We expect to see the neutral regime for all time series because they are generated by random birth, death and immigration events. For short time series and large size of the community, only the details are seen and the result is niche. For small immigration, there is a "winner-take-all" effect, for which the neutrality covariance test, which tests for invariance with respect to grouping gives a niche result.

# Statistical methods | B

## B.1 Heavy-tailed distributions

There is the misconception that most variables follow a Gaussian distribution. We even call it the "normal" distribution. The *central limit theorem* states that sums of independent random variables tend towards Gaussian distributions as the number of terms increases, even when these individual random variables do not follow a Gaussian distribution[1]. Although this theorem is very useful and the normal distribution is omnipresent, there are as many natural processes that do not combine variables through sums, but rather through products, extremum functions, or other more complicated nonlinear functions. In that case, we observe the emergence of heavy-tailed distributions rather than normal distributions. Such distributions are ubiquitous. In this section, we first define the concept of heavy-tailed distributions mathematically. Next, we present three of the most common heavy-tailed distributions and some of their applications.

Heavy-tailed distributions are defined as distributions with tails that are heavier than tails of exponential distributions. We can define them by considering their cumulative distribution function $F$ (CDF) and more specifically the complementary cumulative distribution function $1 - F$ (Nair et al., 2020)

> **Heavy-tailed distribution** A distribution is called heavy-tailed if and only if for its cumulative distribution function $F$, for all $\mu > 0$,
> $$\lim_{x \to \infty} \sup \frac{1 - F(x)}{e^{-\mu x}} = \infty.$$

This means that there exists a threshold value for which all higher values have a larger probability than an exponential distribution. The tails of heavy-tailed distributions are said to decay more slowly than the distribution of an exponential function (Figure B.1).

1: There is a more extensive version of the central limit theorem in which additive processes can lead to stable distributions with heavy tails given that the random variables have an infinite variance (Nair et al., 2020).

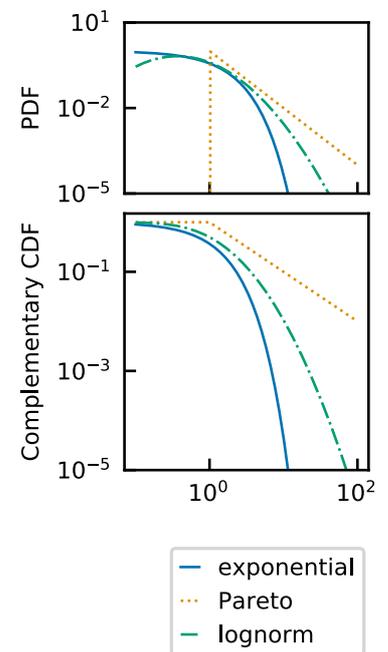

**Figure B.1:** The probability density function (top) and complementary cumulative distribution function (bottom) of an exponential and two heavy-tailed distributions (Pareto and lognormal). The tails of the exponential function decay faster.

The opposite of heavy-tailed distributions are light-tailed distributions. The intuitive conceptions we have about both types of distributions are quite different. This is manifested by the catastrophe and conspiracy principles. A probability density function (PDF) $f$ defined over the positive real numbers satisfies the *catastrophe principle* if, for $X_1, ..., X_n$, independent random variables with distribution $f$ and $n > 1$,

$$\lim_{t \to \infty} P(\max(X_1, ..., X_n) > t \mid X_1 + ... + X_n > t) = 1.$$

In other words, if the sum of random variables is large, it is because one of the samples is large and not because multiple samples are large. The heavy-tailed distributions we present here (lognormal and power law) satisfy this principle.

In contrast, light-tailed distributions tend to follow the conspiracy principle. A PDF $f$ defined over the positive real numbers satisfies the *conspiracy principle* if, for $X_1, ..., X_n$, independent random variables with distribution $f$ and $n > 1$,

$$\lim_{t \to \infty} \frac{P(\max(X_1, ..., X_n) > t)}{P(X_1 + ... + X_n > t)} = 0.$$

In other words, if the sum of the random variables is larger than its expectation value, it is most probably because many individual samples had a value higher than the expectation value. Common light-tailed distributions—the Gaussian and exponential distribution—satisfy this principle.

### Exponential distribution

The exponential distribution serves as a boundary to distinguish light- from heavy-tailed distributions. The random variable of the exponential distribution describes the time between two events in a Poisson process. Its PDF is

$$f_{\exp}(x, \lambda) = \begin{cases} \lambda e^{-\lambda x} & \text{if } 0 \geq x, \\ 0 & x < 0, \end{cases} \quad (B.1)$$

where $\lambda$ is the rate parameter. The mean of this distribution is $\lambda^{-1}$ and the variance is $\lambda^{-2}$. The distribution describes time-intervals of Poisson processes, such as decay times of radioactive particles, but also length-intervals for processes with a constant probability per unit length (Jorgensen, 1987),

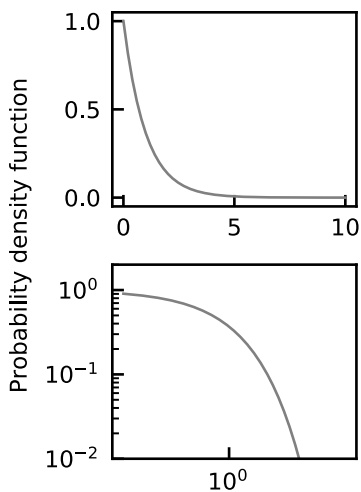

**Figure B.2:** Exponential distribution in linear (top) and log-log (bottom) scale.

such as the distance between mutations on the DNA (Kendal, 2003).

## Lognormal distribution

A lognormal distribution, is a continuous PDF and the logarithm of its random variable is normally distributed. Many natural phenomena are characterized by a lognormal distribution. These distributions often occur for values that cannot be negative with low mean values and high variance (Limpert et al., 2001). Growth processes are usually linear multiplicative[2] and can be described by percentage changes of the variable. An accumulation of many of such changes is multiplicative in a linear scale, but becomes linear in a logarithmic scale. The *central limit theorem* tells us that normal distributions arise by sums of random variables. Multiplication of random variables results in lognormal distributions.

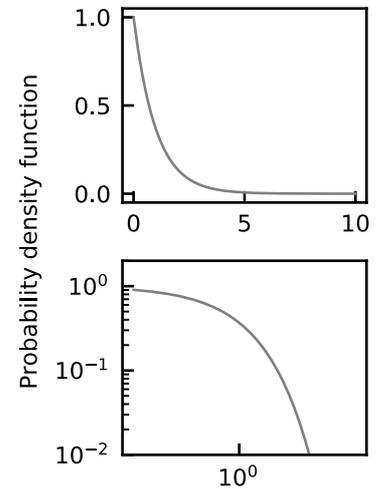

**Figure B.3:** Lognormal distribution in linear (top) and log-log (bottom) scale.

[2]: This assumption was discussed in the context of mass-action kinetics in Section 3.1.

The lognormal PDF is

$$f_{\log}(x, \lambda) = \begin{cases} \dfrac{1}{sx\sqrt{2\pi}} \exp\left(\dfrac{-\ln^2(x)}{2s^2}\right) & \text{if } 0 < x, \\ 0 & \text{otherwise,} \end{cases} \quad (B.2)$$

The theorem introduced in the previous paragraph, which states that the multiplication of random variables is a lognormal distribution, is universal and can be applied in many fields. In econometrics, it is known as *Gibrat's law*, which stipulates that the growth rate of companies is independent of their actual size, and that as a result the distribution of company sizes is lognormal (Gibrat, 1930). Other examples of lognormal distributions in nature and social behavior, are the distribution of the firing rate of neurons (Buzsáki & Mizuseki, 2014), blood pressure (Gaddum, 1945), and the size of the length of posts on internet fora (Sobkowicz et al., 2013).

## Pareto or power law distribution

Power laws appear in a wide variety of areas such as ecology, economics, seismology, and demography (Newman, 2005).

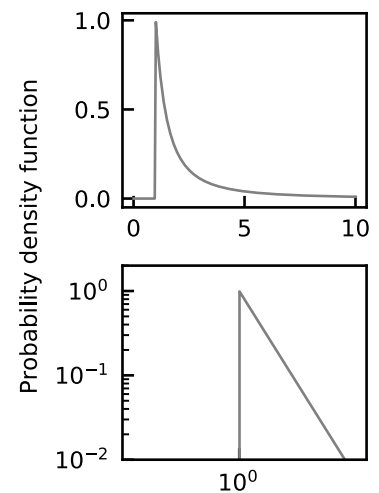

**Figure B.4:** Pareto distribution in linear (top) and log-log (bottom) scale.

One unique and useful characteristic of the power law function is its *scale invariance*. After function variable $x$ is rescaled, the function remains proportional to the original function:

$$f_{\text{pow}}(cx) = (cx/x_m)^{-\alpha} = c^{-\alpha} f_{\text{pow}}(x) \propto f_{\text{pow}}(x).$$

Because the characteristics of this function remain the same for any scale, they are self-similar, fractal-like. The signature of a power law is the straight line in a log-log scale with a slope equal to the exponent $\alpha$ of the power law.

A power law distribution is not bounded for arbitrary exponents and domain of the variable. The standardized power law distribution is defined for $0 \leq x \leq 1$ and $\alpha > -1$[3]

$$f_{\text{pow}}(x, \lambda) = \begin{cases} \alpha x^\alpha & \text{if } 0 \leq x \leq 1, \\ 0 & \text{otherwise.} \end{cases} \quad (B.3)$$

3: In python, this distribution is available in the scipy package as scipy.stats.powerlaw.

To allow for exponents smaller than $-1$, the domain of the power law distribution needs to have a lower bound larger than 0. The power law distribution that is defined for $x \in \mathbb{R}_{\geq x_m}$, is also known as the *Pareto distribution* after Vilfredo Pareto who used it to describe phenomena in scientific as a well as social fields. In the context of economics, he noticed that 80% of the wealth of society belongs to 20% of the population. This observation became known as the *80-20 rule* or the *Pareto principle*. Some other of the many examples of power law distributed quantities are the sizes of craters (Neukum & Ivanov, 1994), solar flares (Lu & Hamilton, 1991), cities (Gabaix, 1999), and the length of protein sequences (Jain & Ramakumar, 1999).

The Pareto distribution is a power law PDF[4]

$$f_{\text{pareto}}(x, \lambda) = \begin{cases} \left(\dfrac{x}{x_m}\right)^\alpha & \text{if } 0 < x_m \leq x, \\ 0 & \text{otherwise,} \end{cases} \quad (B.4)$$

4: In python, this distribution is available in the scipy package as scipy.stats.pareto.

where $\alpha < 0$.

One can also truncate the domain by giving it both a lower and an upper bound: $f_{\text{tpow}}(x) \propto x^\alpha$ for $x_m < x < x_M$[5].

5: In python, this function is available in the powerlaw package (Alstott et al., 2014) which was developed using methods of Clauset et al., 2009 and Klaus et al., 2011.

Discrete versions of this distribution are the Zipf and the zeta distributions. The Zipf distribution is discussed in the next section about the rank abundance distribution.

## B.2  Rank abundance distribution

Next to abundance distributions, rank abundance curves are much used in population studies. However, one must be careful not to confound both. In rank abundance plots, the rank of all abundances is determined by ordering them from high to low. Such plots can have a power law shape. This was for example observed by George Zipf for the frequency of words in a language—the frequency of a word is inversely proportional to the rank of the word. This empirical characteristic is known as *Zipf's law*.

Power law rank abundance distributions are associated with power law PDFs. Here follows a short demonstration (Adamic, 2000). If the rank abundance curve follows a power law, the expected value $E[x_r]$ of the variable $x$ at rank $r$ is

$$E[x_r] = c_1 r^\beta \tag{B.5}$$

This means that there are $r$ variables that have at least this value:

$$P[x \geq c_1 r^\beta] = c_2 r. \tag{B.6}$$

After changing variables, $y = c_1 r^\beta$, we obtain the expression of the CDF $F$,

$$F = P[x \geq y] = \frac{c_2}{c_1^{1/\beta}} y^{(1/\beta)}. \tag{B.7}$$

The PDF $f$ is recovered by taking the derivative of the CDF $F$

$$f = \frac{c_2}{\beta c_1^{1/\beta}} y^{(1/\beta - 1)} \propto y^\alpha, \tag{B.8}$$

the PDF is thus power law with a coefficient $\alpha = 1/\beta - 1$.

One must be careful when drawing conclusions, because the rank abundance distribution is sometimes erroneously represented in a log-linear scale instead of a log-log scale (Wierman, 2020).

## B.3 Kolmogorov-Smirnov test

The *Kolmogorov-Smirnov test* can be used to compare either two samples and asses how likely it is that they share an underlying distribution or it can compare a sample and a reference PDF directly and tell how likely it is that the underlying distribution of the sample is the reference distribution.

The test compares the two cumulative distribution functions $F_i$ and $F_j$ of the samples or distributions. The test statistic $D$ is defined as the maximal distance between the two curves,

$$D_{ij} = \sup_x \left| F_i(x) - F_j(x) \right|. \tag{B.9}$$

Using the Kolmogorov distribution, one can attribute a p-value to the statistic (Marsaglia et al., 2003), or equivalently to the hypothesis that the compared distributions are the same.

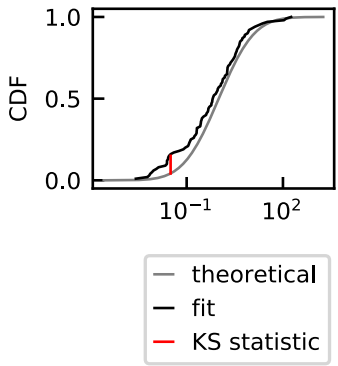

**Figure B.5:** The Kolmogorov-Smirnov statistic is the maximal absolute difference between two distributions. We use it as a goodness-of-fit-test.

## B.4 Pearson correlation coefficient

The Pearson correlation coefficient, also referred to as Pearson's r, is a statistic that measures the linear correlation between two variables $X$ and $Y$. It is defined as the ratio of the covariance between both variables over the product of their standard deviations $\sigma_X$ and $\sigma_Y$,

$$\rho_{X,Y} = \frac{\text{cov}(X,Y)}{\sigma_X \sigma_Y}. \tag{B.10}$$

The value ranges between -1 and 1, with -1 and 1 denoting an exact linear relationship with negative or positive correlation respectively. A value of 0 means no correlation.

## B.5 Fano factor

The *Fano factor*, named after Ugo Fano, measures the dispersion of a PDF. It is defined as the ratio of the variance $\sigma^2$ to the mean $\mu$,

$$F = \frac{\sigma^2}{\mu}. \tag{B.11}$$

It can be interpreted as a noise-to-signal ratio. It is used in neuroscience, to characterize the variability in neural spikes.

## B.6 Brownian motion and Ito calculus

A *Brownian motion* or *Wiener process* is described by a probability distribution over the set of continuous functions $B : \mathbb{R}_{\geq 0} \to \mathbb{R}$ which is defined by three characteristics:

1. $P(B(0) = 0) = 1$, the motion starts at the origin,
2. the motion is stationary: $\forall 0 \leq s \leq t : B(t) - B(s) \sim \mathcal{N}(0, t-s)$,
3. the increments are independent: if intervals $[s_i, t_i]$ are not overlapping than $B(t_i) - B(s_i)$ are independent.

Some properties of Brownian motion are (Lee, 2013)

- it will cross $y = 0$ infinitely often,
- it does not deviate considerably from $y = \sqrt{t}$,
- it is not differentiable anywhere and
- it is self-similar: the shape of the Brownian motion does not depend on the scale.

This paragraph is based on the supplemental material of our publication (Descheemaeker & de Buyl, 2020).

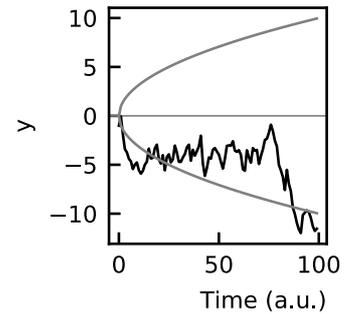

**Figure B.6:** A Brownian motion.

An important feature of such a motion is the quadratic variation, which states that the expectation value of $dW^2$ is $dt$. In a regular derivative, the square and other higher order terms of the infinitesimal $dt$ of the Taylor expansion are ignored. Due to the quadratic variation, the square term of $dW$ cannot be ignored and Ito's lemma states that for a stochastic process $X_t$:

$$dX_t = \mu dt + \sigma dW_t, \tag{B.12}$$

and for $f$ a smooth function, we have that

$$df(t, X_t) = \left(\frac{\partial f}{\partial t} + \mu \frac{\partial f}{\partial x} + \frac{1}{2}\sigma^2 \frac{\partial^2 f}{\partial x^2}\right) dt + \frac{\partial f}{\partial x} dW_t, \tag{B.13}$$

and as a consequence,

$$d(\ln x_i) = \frac{dx_i}{x_i} - \frac{(dx_i)^2}{2x_i^2}. \tag{B.14}$$

These integration techniques are known as *Ito calculus*.

# Supplementary material of "Non-monotonic autoregulation in single gene circuits" C

## C.1 The models: Ordinary differential equations and quasi-steady states

The systems were modeled through mass-action kinetics. The models take into account protein-DNA binding and unbinding, dimerization, transcription, translation and dilution/degradation of mRNA and proteins.

### MDS

The monomer-dimer system contains five ordinary differential equations (ODEs):

$$\frac{\text{dDNA}_0}{\text{d}t} = -\left(k_{bm}m + k_{bd}d\right)\text{DNA}_0 + k_{um}\text{DNA}_\text{m} + k_{ud}\text{DNA}_\text{d},$$

$$\frac{\text{dDNA}_\text{m}}{\text{d}t} = k_{bm}m\text{DNA}_0 - k_{um}\text{DNA}_\text{m},$$

$$\frac{\text{dmRNA}}{\text{d}t} = \phi_0 \left(\text{DNA}_0 + f_m\text{DNA}_\text{m} + f_d\text{DNA}_\text{d}\right) - \gamma_{\text{mRNA}}\text{mRNA},$$

$$\frac{\text{d}m}{\text{d}t} = -k_{bm}m\text{DNA}_0 + k_{um}\text{DNA}_\text{m} + \beta\text{mRNA} - 2\alpha_{\text{ass}}m^2 + 2\alpha_{\text{diss}}d - \gamma_m m \text{ and}$$

$$\frac{\text{d}d}{\text{d}t} = -k_{bd}d\text{DNA}_0 + k_{ud}\text{DNA}_\text{d} + \alpha_{\text{ass}}m^2 - \alpha_{\text{diss}}d - \gamma_d d$$

where $\text{DNA}_\text{d} = \text{DNA}_{\text{tot}} - \text{DNA}_0 - \text{DNA}_\text{m}$ because the total amount of DNA is conserved. A summary of all variables and parameters is given in Table C.1. In case of fast dimerization and fast (un)binding of the proteins to the DNA, the time derivatives of the DNA complexes $\text{DNA}_\text{i}$ and dimer $d$ can be put equal to zero, the quasi-steady state values of these variables are then

$$\text{DNA}_{0,\,\text{qss}} = \frac{\text{DNA}_{\text{tot}}}{K_d d + K_m m + 1},$$

$$\text{DNA}_{\text{m},\,\text{qss}} = \text{DNA}_{0,\,\text{qss}} K_m m \text{ and}$$

$$d_{\text{qss}} = \frac{\alpha_{\text{ass}}}{\alpha_{\text{diss}} + \gamma_d} m^2.$$

**Table C.1:** Overview of all variables and all parameters of the different models.

| variable | dimension | |
|---|---|---|
| $DNA_i$ | number per cell | concentration of DNA with no bound proteins ($i = 0$), one bound monomer ($i = m$), one bound dimer ($i = d$) or dimers bound to site(s) i ($i \in \{1, 2, 3\}$) |
| mRNA | number per cell | mRNA concentration |
| $m$ | number per cell | monomer concentration |
| $d$ | number per cell | dimer concentration |
| $k_{bi}$ | min$^{-1}$ | binding rate of monomer ($i = m$), dimer ($i = d$) or dimer to site i ($i \in \{1, 2, 3\}$) |
| $k_{ui}$ | min$^{-1}$ | unbinding rate of monomer ($i = m$), dimer ($i = d$) or dimer to site i ($i \in 1, 2, 3$) |
| $K_i$ | min$^{-1}$ | binding constant of monomer ($i = m$), dimer ($i = d$) or dimer to site i ($i \in 1, 2, 3$), $K_i = k_{bi}/k_{ui}$ |
| $\phi_0$ | min$^{-1}$ | transcription rate |
| $f_i$ | n.a. | transcriptional fold change when monomer ($i = m$) or dimer ($i = d$) is bound or dimers are bound to site(s) $i$ ($i \in \{1, 2, 3\}$) with respect to no protein bound to the DNA |
| $\beta$ | min$^{-1}$ | translation rate |
| $\gamma_i$ | min$^{-1}$ | degradation rate of monomer ($i = m$), dimer ($i = d$) or mRNA ($i = $ mRNA) |
| $\alpha_{ass}$ | min$^{-1}$ | association rate of monomers to dimers |
| $\alpha_{diss}$ | min$^{-1}$ | dissociation rate of dimers to monomers |
| $co_{b,uij(k)}$ | n.a. | cooperativity factor for binding ($b$) or unbinding ($u$) between sites $i$ and $j$ (and $k$) ($i, j, k \in \{1, 2, 3\}$) |
| $\omega_{ij(k)}$ | n.a. | cooperativity factor between sites $i$ and $j$ (and $k$) ($i, j, k \in \{1, 2, 3\}$) $\omega_{ij(k)} = co_{bij(k)}/co_{uij(k)}$ |

## 2DS

Analogously to the system of ODEs for the MDS, we can write down the system for the 2 dimer system which contains six ODEs

$$\frac{d\text{DNA}_0}{dt} = -(k_{b1} + k_{b2})\,d\text{DNA}_0 + k_{u1}\text{DNA}_1 + k_{u2}\text{DNA}_2,$$

$$\frac{d\text{DNA}_1}{dt} = k_{b1}d\text{DNA}_0 - k_{u1}\text{DNA}_1 + k_{u2}\text{co}_{u12}\text{DNA}_{12} - k_{b2}\text{co}_{b12}d\text{DNA}_1,$$

$$\frac{d\text{DNA}_2}{dt} = k_{b2}d\text{DNA}_0 - k_{u2}\text{DNA}_2 + k_{u1}\text{co}_{u12}\text{DNA}_{12} - k_{b1}\text{co}_{b12}d\text{DNA}_2,$$

$$\frac{d\text{mRNA}}{dt} = \phi_0\left(\text{DNA}_0 + f_1\text{DNA}_1 + f_2\text{DNA}_2 + f_{12}\text{DNA}_{12}\right) - \gamma_{\text{mRNA}}\text{mRNA},$$

$$\frac{dm}{dt} = \beta\text{mRNA} - 2\alpha_{\text{ass}}m^2 + 2\alpha_{\text{diss}}d - \gamma_m m \quad \text{and}$$

$$\begin{aligned}\frac{dd}{dt} = &-(k_{b1} + k_{b2})\,d\text{DNA}_0 - (k_{b2}\text{DNA}_1 + k_{b1}\text{DNA}_2)\text{co}_{b12}d \\ &+ k_{u1}\text{DNA}_1 + k_{u2}\text{DNA}_2 + (k_{u1} + k_{u2})\text{co}_{u12}\text{DNA}_{12} + \alpha_{\text{ass}}m^2 - \alpha_{\text{diss}}d - \gamma_d d\end{aligned}$$

where $\text{DNA}_{12} = \text{DNA}_{\text{tot}} - \text{DNA}_0 - \text{DNA}_1 - \text{DNA}_2$ because the total amount of DNA is conserved. A summary of all variables and parameters is given in Table C.1. The quasi-steady states of the DNA complexes and dimer are:

$$\text{DNA}_{0,\,\text{qss}} = \frac{\text{DNA}_{\text{tot}}}{\omega_{12}K_1K_2d^2 + (K_1 + K_2)d + 1},$$

$$\text{DNA}_{1,\,\text{qss}} = \text{DNA}_{0,\,\text{qss}}K_1d,$$

$$\text{DNA}_{2,\,\text{qss}} = \text{DNA}_{0,\,\text{qss}}K_2d \quad \text{and}$$

$$d_{\text{qss}} = \frac{\alpha_{\text{ass}}}{\alpha_{\text{diss}} + \gamma_d}m^2.$$

## 3DS

The 3DS contains ten ODEs:

$$\frac{d\text{DNA}_0}{dt} = -(k_{b1} + k_{b2} + k_{b3})\,d\text{DNA}_0 + k_{u1}\text{DNA}_1 + k_{u2}\text{DNA}_2 + k_{u3}\text{DNA}_3,$$

$$\frac{d\text{DNA}_1}{dt} = k_{b1}d\text{DNA}_0 - k_{u1}\text{DNA}_1 + k_{u2}\text{co}_{u12}\text{DNA}_{12} + k_{u3}\text{co}_{u13}\text{DNA}_{13}$$
$$- (k_{b2}\text{co}_{b12} + k_{b3}\text{co}_{b13})\,d\text{DNA}_1,$$

$$\frac{d\text{DNA}_2}{dt} = k_{b2}d\text{DNA}_0 - k_{u2}\text{DNA}_2 + k_{u1}\text{co}_{u12}\text{DNA}_{12} + k_{u3}\text{co}_{u23}\text{DNA}_{23}$$
$$- (k_{b1}\text{co}_{b12} + k_{b3}\text{co}_{b23})\,d\text{DNA}_2,$$

$$\frac{d\text{DNA}_3}{dt} = k_{b3}d\text{DNA}_0 - k_{u3}\text{DNA}_3 + k_{u1}\text{co}_{u13}\text{DNA}_{13} + k_{u2}\text{co}_{u23}\text{DNA}_{23}$$
$$- (k_{b1}\text{co}_{b13} + k_{b2}\text{co}_{b23})\,d\text{DNA}_3,$$

$$\frac{d\text{DNA}_{12}}{dt} = k_{b1}\text{co}_{b12}d\text{DNA}_2 + k_{b2}\text{co}_{b12}d\text{DNA}_1 - (k_{u1} + k_{u2})\,\text{co}_{u12}\text{DNA}_{12}$$
$$- k_{b3}\text{co}_{b13}\text{co}_{b23}\text{co}_{b123}d\text{DNA}_{12} + k_{u3}\text{co}_{u13}\text{co}_{u23}\text{co}_{u123}\text{DNA}_{123},$$

$$\frac{d\text{DNA}_{23}}{dt} = k_{b2}\text{co}_{b23}d\text{DNA}_3 + k_{b3}\text{co}_{b23}d\text{DNA}_2 - (k_{u2} + k_{u3})\,\text{co}_{u23}\text{DNA}_{23}$$
$$- k_{b1}\text{co}_{b12}\text{co}_{b13}\text{co}_{b123}d\text{DNA}_{23} + k_{u1}\text{co}_{u12}\text{co}_{u13}\text{co}_{u123}\text{DNA}_{123},$$

$$\frac{d\text{DNA}_{13}}{dt} = k_{b1}\text{co}_{b13}d\text{DNA}_3 + k_{b3}\text{co}_{b13}d\text{DNA}_1 - (k_{u1} + k_{u3})\,\text{co}_{u13}\text{DNA}_{13}$$
$$- k_{b2}\text{co}_{b12}\text{co}_{b23}\text{co}_{b123}\text{DNA}_{13} + k_{u2}\text{co}_{u12}\text{co}_{u23}\text{co}_{u123}\text{DNA}_{123},$$

$$\frac{d\text{mRNA}}{dt} = \phi_0\left(\text{DNA}_0 + f_1\text{DNA}_1 + f_2\text{DNA}_2 + f_3\text{DNA}_3 + f_{12}\text{DNA}_{12} + f_{23}\text{DNA}_{23} + f_{13}\text{DNA}_{13}\right.$$
$$\left. + f_{123}\text{DNA}_{123}\right) - \gamma_{\text{mRNA}}\text{mRNA},$$

$$\frac{dm}{dt} = \beta\text{mRNA} - 2\alpha_{\text{ass}}m^2 + 2\alpha_{\text{diss}}d - \gamma_m m \text{ and}$$

$$\frac{dd}{dt} = -(k_{b1} + k_{b2} + k_{b3})\,d\text{DNA}_0 + k_{u1}\text{DNA}_1 + k_{u2}\text{DNA}_2 + k_{u3}\text{DNA}_3$$
$$- ((k_{b2}\text{co}_{b12} + k_{b3}\text{co}_{b13})\,d\text{DNA}_1 + (k_{b1}\text{co}_{b12} + k_{b3}\text{co}_{b23})\,d\text{DNA}_2 + (k_{b1}\text{co}_{b13} + k_{b2}\text{co}_{b23})\,d\text{DNA}_3)$$
$$+ (k_{u1} + k_{u2})\,\text{co}_{u12}\text{DNA}_{12} + (k_{u2} + k_{u3})\,\text{co}_{u23}\text{DNA}_{23} + (k_{u1} + k_{u3})\,\text{co}_{u13}\text{DNA}_{13}$$
$$- (k_{b1}\text{co}_{b12}\text{co}_{b13}\text{DNA}_{23} + k_{b2}\text{co}_{b12}\text{co}_{b23}\text{DNA}_{13} + k_{b3}\text{co}_{b13}\text{co}_{b23}\text{DNA}_{12})\,\text{co}_{b123}d$$
$$+ (k_{u1}\text{co}_{u12}\text{co}_{u13} + k_{u2}\text{co}_{u12}\text{co}_{u23} + k_{u3}\text{co}_{u13}\text{co}_{u23})\,\text{co}_{u123}\text{DNA}_{123}$$
$$+ \alpha_{\text{ass}}m^2 - \alpha_{\text{diss}}d - \gamma_d d,$$

with $\text{DNA}_{123} = \text{DNA}_{\text{tot}} - \text{DNA}_0 - \text{DNA}_1 - \text{DNA}_2 - \text{DNA}_3 - \text{DNA}_{12} - \text{DNA}_{13} - \text{DNA}_{23}$ because the total amount of DNA is conserved. A summary of all variables and parameters is given in Table C.1. The quasi-steady states of the DNA complexes and dimer are:

$$\begin{aligned}
\text{DNA}_{0,\text{qss}} &= \text{DNA}_{\text{tot}} \left[ \omega_{123}\omega_{12}\omega_{13}\omega_{23}K_1K_2K_3 d^3 \right. \\
&\quad \left. + (\omega_{12}K_1K_2 + \omega_{23}K_2K_3 + \omega_{13}K_1K_3)\,d^2 + (K_1 + K_2 + K_3)\,d + 1 \right]^{-1}, \\
\text{DNA}_{1,\text{qss}} &= \text{DNA}_{0,\text{qss}} K_1 d, \\
\text{DNA}_{2,\text{qss}} &= \text{DNA}_{0,\text{qss}} K_2 d, \\
\text{DNA}_{3,\text{qss}} &= \text{DNA}_{0,\text{qss}} K_3 d, \\
\text{DNA}_{12,\text{qss}} &= \text{DNA}_{0,\text{qss}} \omega_{12} K_1 K_2 d, \\
\text{DNA}_{23,\text{qss}} &= \text{DNA}_{0,\text{qss}} \omega_{23} K_2 K_3 d, \\
\text{DNA}_{13,\text{qss}} &= \text{DNA}_{0,\text{qss}} \omega_{13} K_1 K_3 d, \\
d_{\text{qss}} &= \frac{\alpha_{\text{ass}}}{\alpha_{\text{diss}} + \gamma_d} m^2.
\end{aligned}$$

## C.2 Physiological ranges of the parameters

**Table C.2:** Physiological ranges for the different parameters (Buchler et al., 2005; Karapetyan & Buchler, 2015; Stricker et al., 2008; Zavala & Marquez-Lago, 2014) for a cell with volume $4 - 40$fL ($1$nM $= 2.4 - 24$ molecules)

| Parameter | Boundaries | Parameter | Boundaries |
|---|---|---|---|
| $\beta(\text{min}^{-1})$ | $0.01 - 120$ | $f_{\text{activation}}$ | $1 - 100$ |
| $\gamma_m(\text{min}^{-1})$ | $10^{-3} - 1$ | $f_{\text{repression}}$ | $0.001 - 1$ |
| $\gamma_{\text{mRNA}}(\text{min}^{-1})$ | $10^{-3} - 10$ | $f$ | $0.001 - 100$ |
| $\gamma_d(\text{min}^{-1})$ | $5 \cdot 10^{-4} - 0.1$ | $k_b(\text{min}^{-1})$ | $10^{-3} - 10^2$ |
| $\alpha_{\text{ass}}(\text{min}^{-1})$ | $10^{-2} - 1$ | $k_u(\text{min}^{-1})$ | $0.01 - 1000$ |
| $\alpha_{\text{diss}}(\text{min}^{-1})$ | $10^{-3} - 10^3$ | co$_b$ | $0.05 - 20$ |
| $\phi_0(\text{min}^{-1})$ | $10^{-2} - 10$ | co$_u$ | $0.05 - 20$ |

# C.3 Oscillations

## Bifurcation analysis

In order to check if a system oscillates, we first look if we can find a single steady state which is unstable (at least one of the eigenvalues of the Jacobian evaluated in this steady state has a positive real part). Unstable steady states can lead to solutions that oscillate, chaotic behavior or solutions that go to infinity. To be certain the system is oscillating, we do a time series and look for two characteristics of stable oscillations:

1. Stable amplitude: the difference between consecutive maxima in the time series must be less than 1%, the same holds for consecutive minima.
2. Stable period: the difference between to consecutive distances between local maxima must be small.

We want to find the oscillatory regions around a found oscillating solution for every parameter. This means all parameters are kept constant except for one, the parameter of which we determine the bifurcation points. Therefore we use method based on the bisection method for finding roots. To find the left boundary of the oscillating region, we first check whether the system with the considered parameter equal to the left boundary of the physiological range is oscillating. If this is the case, we have found the left boundary of the oscillating region, if not, we know that the bifurcation happens in the interval $[l_1, l_2]$ where we initialize $l_1$ with the parameter value at the left boundary of the physiological range and $l_2$ with the parameter value of the oscillating solution. Next we will update the boundaries of the interval around the bifurcation until the interval is considerably small and approximate the bifurcation point by the point that is logarithmically halfway this interval. To update the interval we take the point logarithmically halfway the interval

$$l_m = 10^{\frac{\log_{10}(l_1)+\log_{10}(l_2)}{2}} \tag{C.1}$$

and check whether it is oscillating. In the case that it is oscillating we update right bound of the interval, $l_2 = l_m$, otherwise the left bound $l_1 = l_m$. The right boundary of the oscillating region is found in a similar way. Finding these two outer solutions does not guarantee that all solutions in the range are oscillating, since other bifurcations might happen in between. Therefore we computed time series for five logarithmically equally spaced samples to see if they were oscillating (large computation time limits the number of solutions that can be tested).

**Oscillatory ranges distributions**

In Figure 5.4, the mean values and mean ranges are given for the different parameters. Here different bifurcation ranges are given for different solutions (Figure C.1, Figure C.2, Figure C.3, and Figure C.4). We can see that some parameters can have different values in the physiological range but still be finely tuned, for example $\beta$ or $\phi_0$ for the MDS. For other parameters such as $f_i$ of the 2DS the distribution is bimodal. These characteristics are not represented in Figure 5.4, but are unimportant for a global overview of the oscillating regions. Because there is much variation in the ranges, there is also much variation in the volumes (Figure C.5).

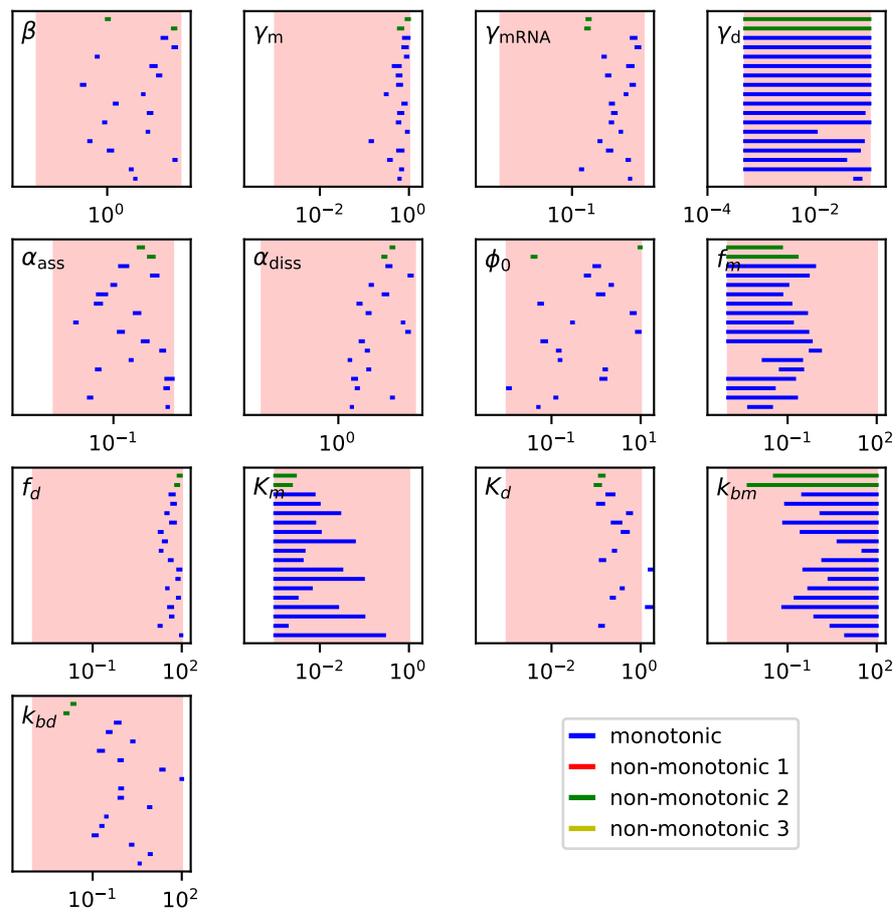

**Figure C.1:** Oscillatory ranges of the different parameters for the different solutions of the MDS. The red region represents the physiological range. The axis is logarithmically scaled, the length of the lines thus represent fold ratios. Different lines represent different solutions.

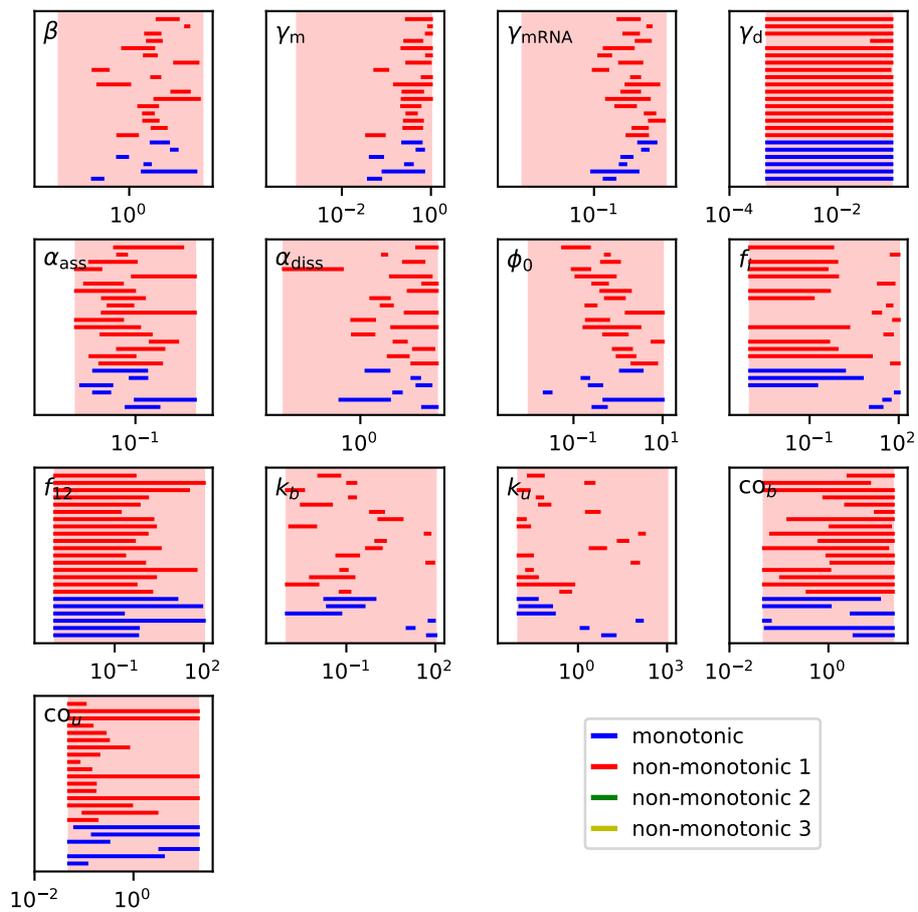

**Figure C.2:** Oscillatory ranges of the different parameters for selection of solutions of the 2DS. The red region represents the physiological range. The axis is logarithmically scaled, the length of the lines thus represent fold ratios. Different lines represent different solutions.

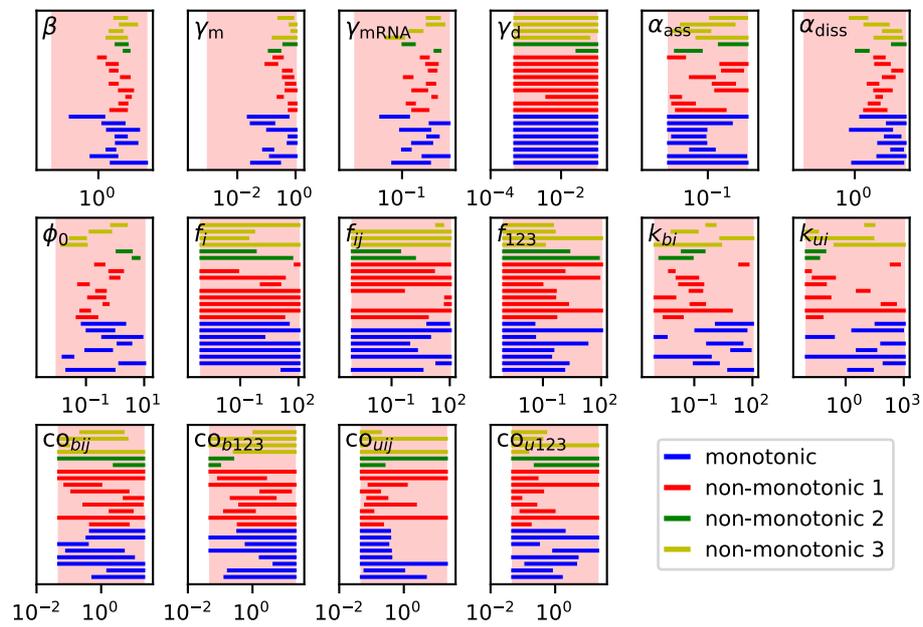

**Figure C.3:** Oscillatory ranges of the different parameters for a selection of solutions of the 3DS. The red region represents the physiological range. The axis is logarithmically scaled, the length of the lines thus represent fold ratios. Different lines represent different solutions.

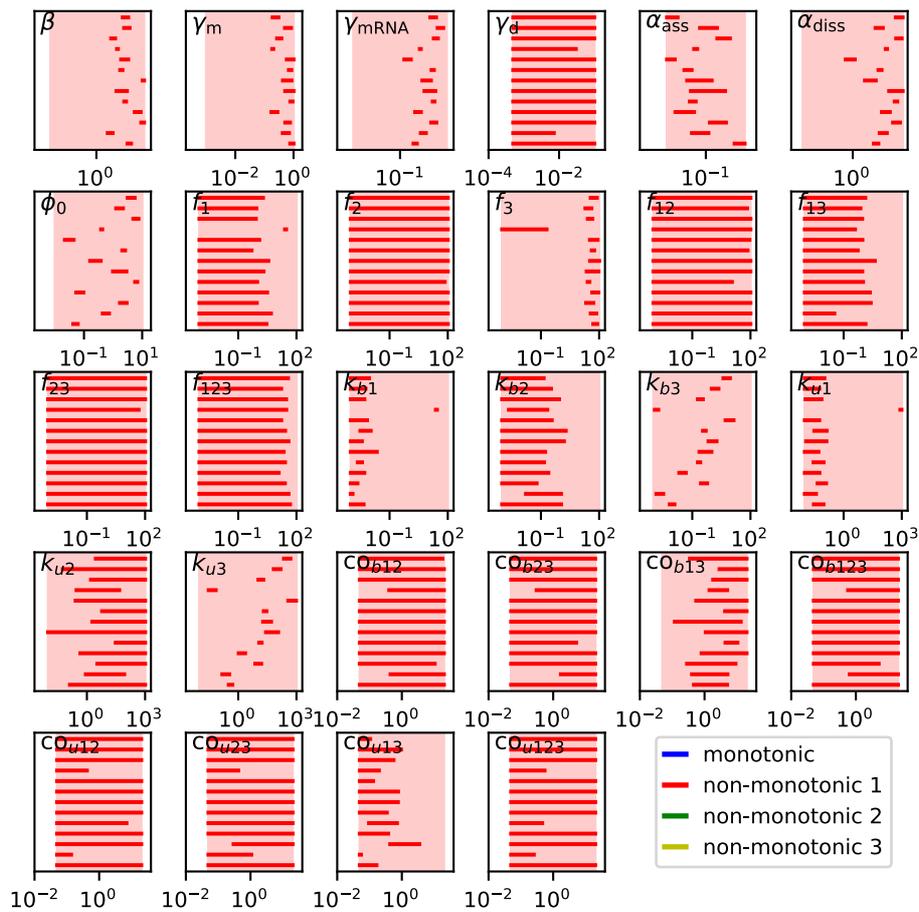

**Figure C.4:** Oscillatory ranges of the different parameters for a selection of solutions of the Ss-LrpB. The red region represents the physiological range. The axis is logarithmically scaled, the length of the lines thus represent fold ratios. Different lines represent different solutions.

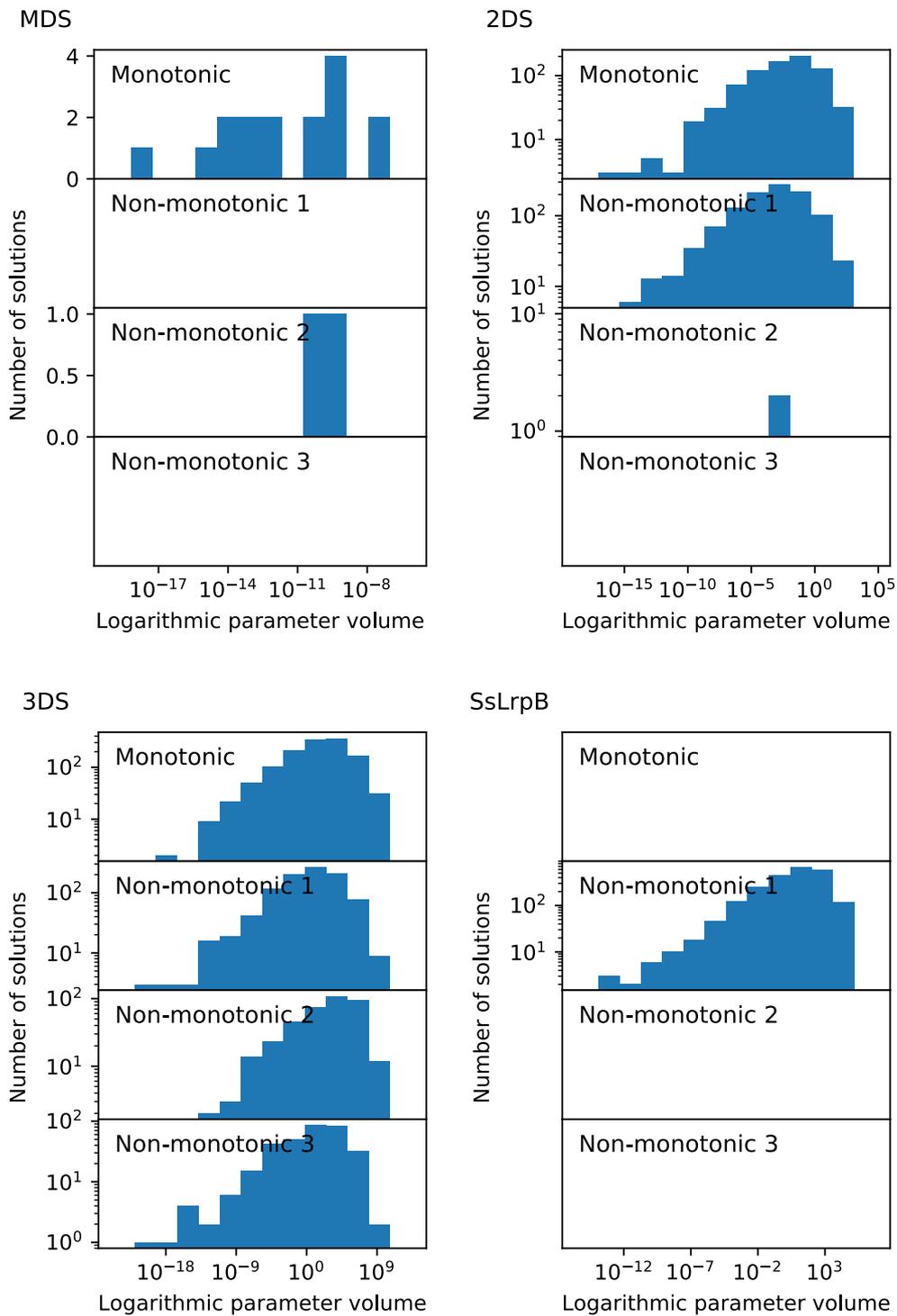

**Figure C.5:** Volume distribution for the different toy models.

## C.4 Bistability

In order to check if a system is bistable, we look if we can find three steady states of which one is unstable (at least one of the eigenvalues of the Jacobian evaluated in this steady state has a positive real part) and the other two stable (all eigenvalues of the Jacobian evaluated in this steady state have a negative real part).

### Quasi-steady-state approximation

In the assumption of fast binding and unbinding of the dimers to the DNA and fast dimer association and dissociation, we can equal the time derivatives of the DNA complexes and the dimer concentration $d$ to zero and use their quasi-steady-state approximation in the equations for mRNA and monomer concentration $m$ (given in Appendix C.1):

$$\frac{dm}{dt} = \beta \text{mRNA} - \gamma_m m - 2\alpha_{\text{ass}} m^2 + 2\alpha_{\text{diss}} d_{qss}, \tag{C.2}$$

$$\frac{d\text{mRNA}}{dt} = \phi_0 \sum_i f_i \text{DNA}_{\text{qss, i}} - \gamma_{\text{mRNA}} \text{mRNA} \tag{C.3}$$

with $i \in [0, m, d]$ for the MDS, $i \in [0, 1, 2, 12]$ for the 2DS and $i \in [0, 1, 2, 3, 12, 23, 13, 123]$ for the 3DS. Assuming quasi-steady state for the mRNA concentration and fast dissociation of the dimer with respect to degradation of the dimer ($\gamma_d \ll \alpha_{\text{diss}}$), we obtain

$$\frac{dm}{dt} = \frac{\beta \phi_0 \text{DNA}_{\text{tot}} f(m, d)}{\gamma_{\text{mRNA}}} - \gamma_m m. \tag{C.4}$$

with $f(m, d) = \sum_i f_i \text{DNA}_{\text{qss, i}} / \text{DNA}_{\text{tot}}$ the transcription function. This function depends on the system, using the steady state expression of Appendix C.1 we obtain

$$f_{MDS}(m, d) = \frac{f_d K_d d + f_m K_m m + 1}{K_d d + K_m m + 1}, \tag{C.5}$$

$$f_{2DS}(d) = \frac{A d^2 + B d + 1}{C d^2 + D d + 1}, \tag{C.6}$$

$$f_{3DS}(d) = \frac{A' d^3 + B' d^2 + C' d + 1}{D' d^3 + E' d^2 + F' d + 1}, \tag{C.7}$$

with

$$C = K_{d1}K_{d2}\omega, \qquad\qquad A = f_{12}C,$$
$$D = K_{d1} + K_{d2}, \qquad\qquad B = f_1 K_{d1} + f_2 K_{d2}$$
$$D' = K_{d1}K_{d2}K_{d3}\omega_{12}\omega_{13}\omega_{23}\omega_{123}, \qquad A' = f_{123}D,$$
$$E' = K_{d1}K_{d2}\omega_{12} + K_{d1}K_{d3}\omega_{13} + K_{d2}K_{d3}\omega_{23}, \quad B' = f_{12}K_{d1}K_{d2}\omega_{12} + f_{13}K_{d1}K_{d3}\omega_{13}$$
$$F' = K_{d1} + K_{d2} + K_{d3} \text{ and} \qquad\qquad + f_{23}K_{d2}K_{d3}\omega_{23},$$
$$C' = f_1 K_{d1} + f_2 K_{d2} + f_3 K_{d3}.$$

### Induction time of bistable systems

The induction time is a representation of the time it takes a bistable system to attain the stable high steady state when starting from the unstable intermediate steady state. Deterministically, systems will remain forever in the unstable intermediate steady state if they are not perturbed, therefore the initial condition is chosen to be a slight perturbation,

$$s_i = \{\text{DNA}_j = \text{DNA}_{j,I} \;\forall j \in S,$$
$$\qquad \text{mRNA} = \text{mRNA}_I,$$
$$\qquad m = m_I,$$
$$\qquad d = 1.1 d_I\}$$

where $S$ is $(0, m, d)$ for the MDS, $(0, 1, 2, 12)$ for the 2DS and $(0, 1, 2, 3, 12, 13, 23, 123)$ for the 3DS case and $x_H$, $x_I$ and $x_L$ are respectively the high, intermediate and low steady state of component $x$. A deterministic simulation will never reach the high steady state but approach this solution as an asymptote. Therefore the final state is defined by a state which has the dimer concentration higher than 90% of the high steady state,

$$s_f = \{\text{DNA}_j \;\forall j \in S,$$
$$\qquad \text{mRNA},$$
$$\qquad m,$$
$$\qquad d > 0.9 d_H\}.$$

The induction time $\Delta t$ is thus defined by the time it takes the system to attain a dimer number higher than $0.9 d_H$ starting from initial condition $s_i$.

Simulations show that different choices for $K_i$, $f_i$, $co_{bi}$ and $co_{ui}$ – when coefficients $A$, $B$, $C$, $D$, $E$ and $F$ are fixed– do not influence the induction time. Other free parameters get the values dictated by Table C.3.

**Table C.3:** Values used for the simulation to measure the induction time of bistable systems

| Parameter | Value |
| --- | --- |
| $\gamma_m$ | $0.1 \text{ min}^{-1}$ |
| $\gamma_{mRNA}$ | $0.01 \text{ min}^{-1}$ |
| $\gamma_d$ | $0.001 \text{ min}^{-1}$ |
| $\alpha_{ass}$ | $0.01 \text{ min}^{-1} \text{ molecule}^{-1}$ |
| $\alpha_{diss}$ | $2 \text{ min}^{-1}$ |
| $\phi_0$ | $1 \text{ min}^{-1}$ |
| $k_{bi}$ | $1 \text{ min}^{-1}$ |
| $\beta$ | $\left(\gamma_m \gamma_{mRNA} \sqrt{(\alpha_{diss} + \gamma_d)/\alpha_{ass}}\right)/(\gamma \phi_0)$ |

We compare the induction time calculated by an explicit simulation as explained in this section with the approximated induction time $\tilde{\Delta t}$ defined in Section 5.2 (Figure C.6). As mentioned earlier in this section, the initial condition and final state need to be defined. These will affect the simulated induction time. Depending on the exact shape of the response curve and the parameters, the approximated time is thus an over- or underestimation of the simulated induction time. This explains why there is no one-on-one match of the approximated time and the simulated time.

## Bistability in parameter space

In Figure C.7 in the regions within parameter space providing bistable switches for the different systems are represented. For 2DS and 3DS, a large proportion of parameter space leads to bistability while the MDS needs to be finely tuned to provide this dynamical property. In this figure, we used parameters $A - F$. In order to study the parameters listed in Table C.1, we selected 100 random solutions in the bistable region and performed a bifurcation analysis. This bifurcation analysis is similar to the one for oscillations, but the boundaries are checked for bistability instead of oscillations. The mean parameter value and mean width of the bistable range is given in Figure C.8.

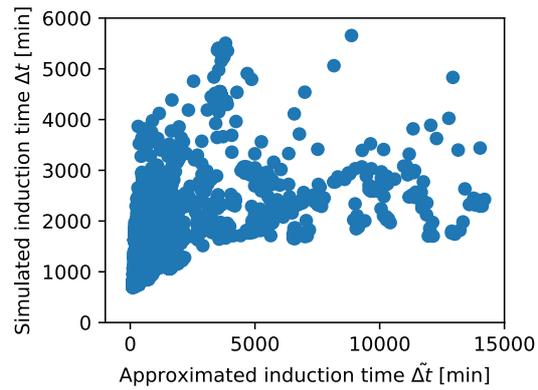

**Figure C.6:** Comparison of the approximated and simulated induction time.

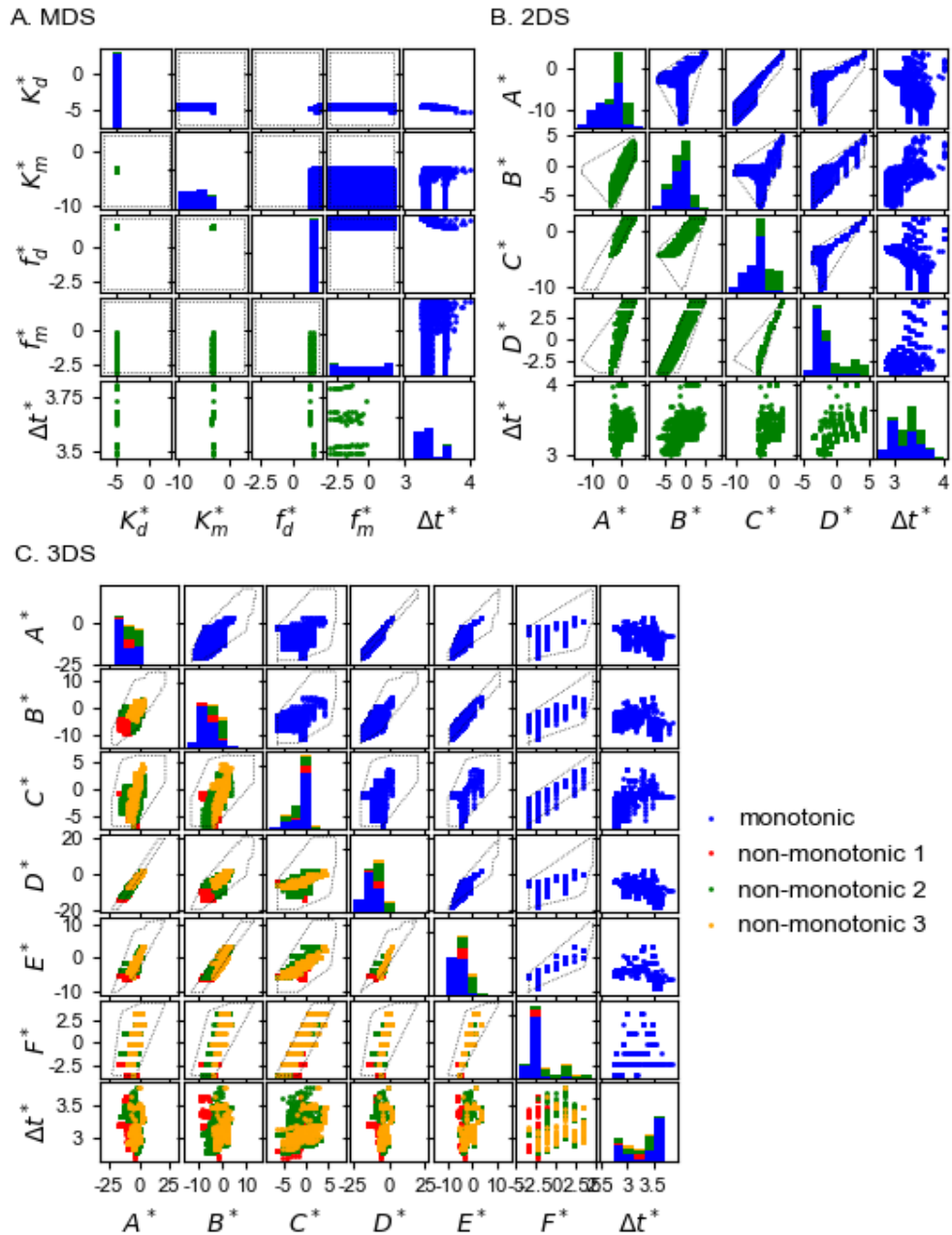

**Figure C.7:** Parameter distribution of bistable solutions for the different toy models. Bistable solutions are shown as a function of the logarithm of the parameters dictating the response curve and the logarithm of the induction time ($X^* = \log_{10}(X)$ with $X \in \{K_d, K_m, f_d, f_m, \Delta t, A, B, C, D, E, F\}$). Monotonic solutions are shown in the upper triangles (blue solutions). Non-monotonic solutions are shown in the lower triangles (red, green and orange). Gray dashed lines indicate physiological ranges.

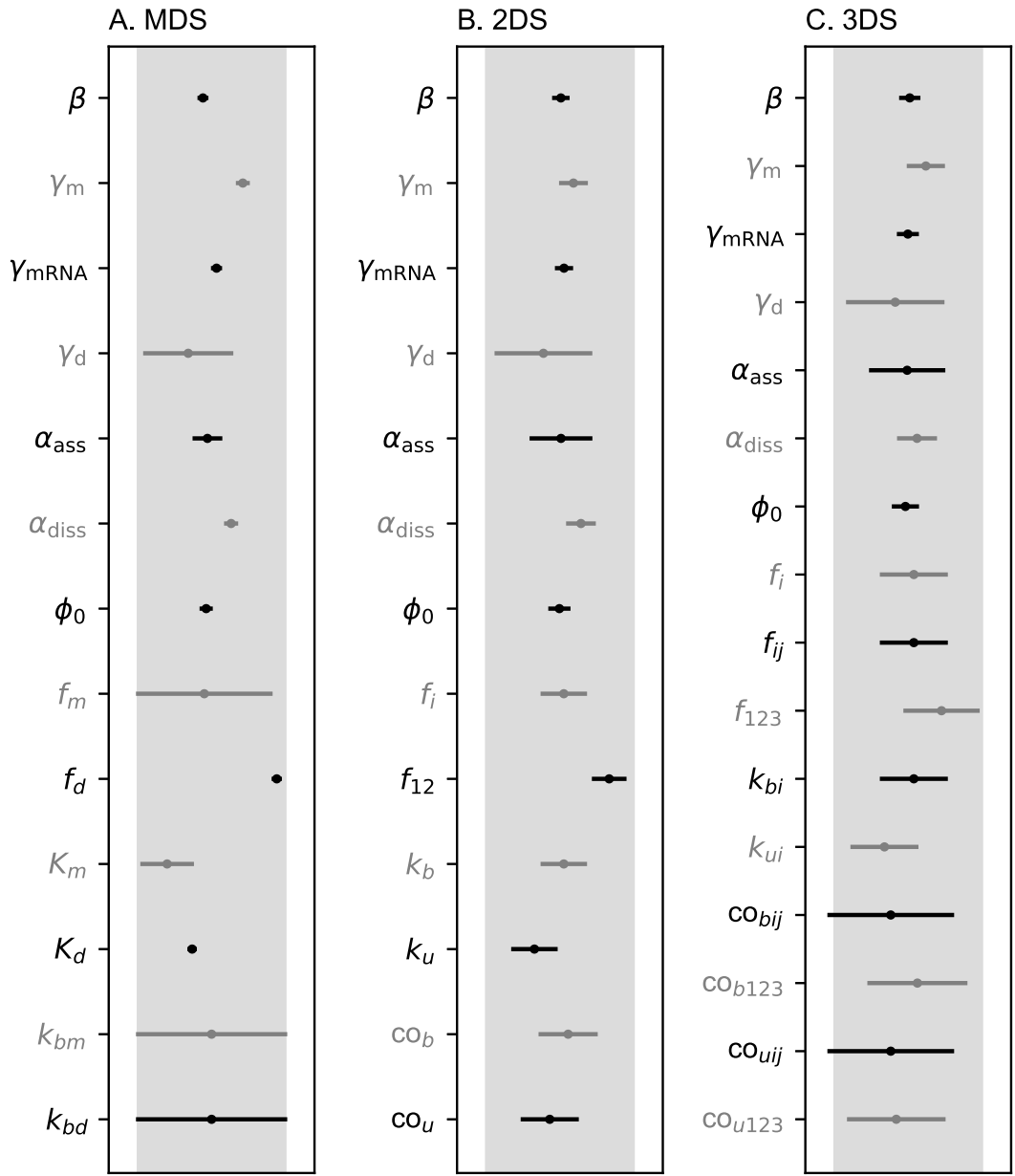

**Figure C.8:** Bistable solutions for the different toy models. The shaded region represents the physiological range. The black and gray lines represent the mean bistable ranges for the different parameters. The axis is logarithmically scaled, the length of the lines thus represent fold ratios. The line for each parameter is scaled according to the physiological range of this parameter. Ranges are very small for the MDS and become wider for the 2DS and 3DS.

## C.5 SsLrpB compatibility

In order to be compatible with the natural SsLrpB system, a 3DS needs to have the same values as measured experimentally (Table C.4) and moreover the response curve needs to be of non-monotonic type 1 with a maximum higher than 2 and a minimum lower than 0.5 (Peeters et al., 2013). The radius of *Sulfolobus solfataricus* is around 1 μm, thus the volume can be estimated at 4 fL. A concentration of 1 μM in one cell corresponds therefore to 2400 molecules. Since the non-monotonicity is experimentally shown between 0 and 100 nM (Peeters et al., 2013), we only considered systems up to 1000 molecules (equivalent to 300 nM), *i.e.* the response curve needed to meet the above constraints on non-monotonicity in the interval from 0 to 1000 dimers, only considering the integer numbers.

The measured parameters of Table C.4 fix parameters, $A$, $D$, $E$ and $F$, which are defined in Section 5.2, such that only 2 free parameters remain for the bistable switch search : $B$ and $C$. The former is determined by $f_{12}$, $f_{23}$ and $f_{13}$ and the latter by $f_{123}$. For each $f_{13} - f_{123}$ combination, we scanned over the only free parameters $f_{12}$ and $f_{23}$, for which we get different response curves. Thsese are shown in grey in Figure C.9. Bistable curves meeting the conditions Equation 5.1 have a dashed red line and curves meeting the Ss-LrpB criteria have a dashed blue line. For every bistable (red) response curve, different degradation rates are tested (green lines) and the one with the smallest approximated induction time is chosen. A simulation is done with this solution to calculate the real induction time, which is shown by the yellow-green coloring in Figure 5.9.

**Table C.4:** Measured values for the Ss-LrpB system

| Parameter | value | | |
|---|---|---|---|
| $K_{d1} = k_{b1}/k_{u1}$ | 73.5 µM$^{-1}$ | $3.1 \times 10^{-2}$ molecule$^{-1}$ | |
| $K_{d2} = k_{b2}/k_{u2}$ | 0.7 µM$^{-1}$ | $2.9 \times 10^{-4}$ molecule$^{-1}$ | |
| $K_{d3} = k_{b3}/k_{u3}$ | 49.1 µM$^{-1}$ | $2.0 \times 10^{-2}$ molecule$^{-1}$ | |
| $\omega_{12} = co_{b12}/co_{u12}$ | 4.1 | | |
| $\omega_{13} = co_{b13}/co_{u13}$ | 2.1 | | |
| $\omega_{23} = co_{b23}/co_{u23}$ | 7.9 | | |
| $\omega_{123} = co_{b123}/co_{u123}$ | 3.1 | | |

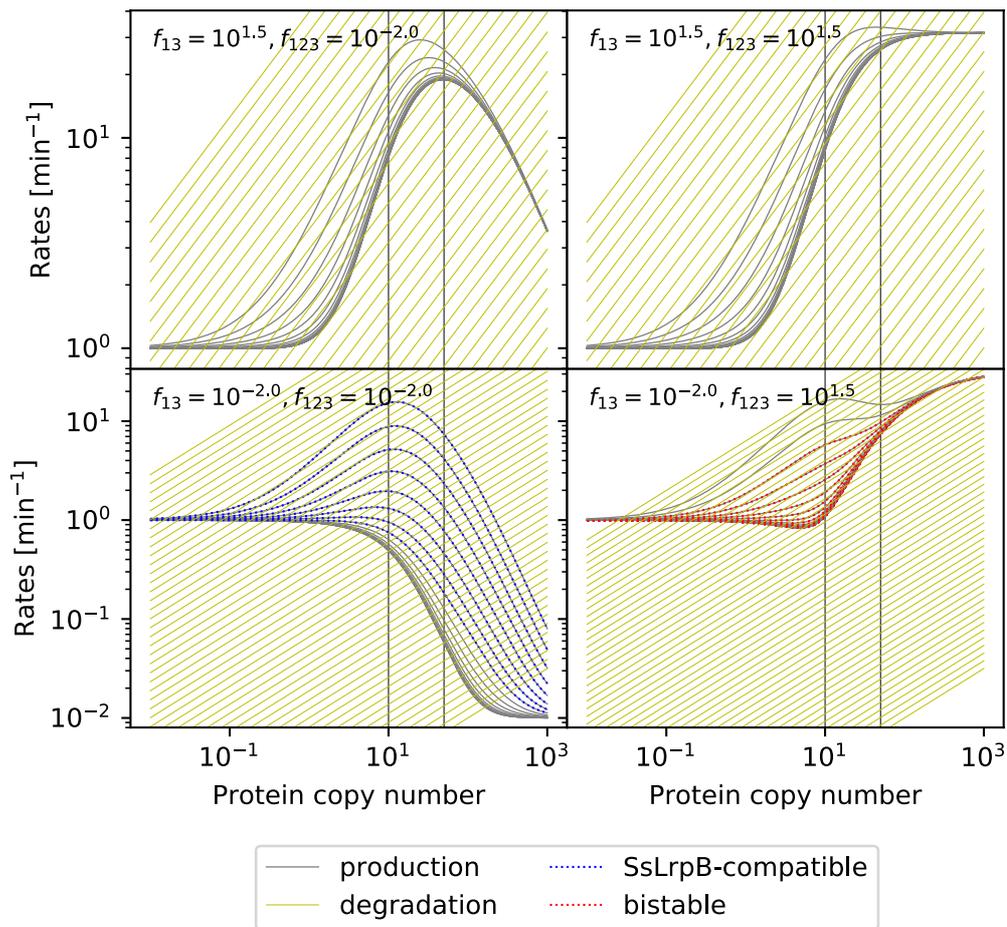

**Figure C.9:** Rates for different $f_{13} - f_{123}$ combinations.

# Supplementary material of "Stochastic logistic models reproduce experimental time series of microbial communities"

D

## D.1 Methods

**Examples of width distribution of ratios of abundances at successive time points**

To study the jumps that the species abundances make over time, we consider the distribution of the ratios of abundances at successive time points $x(t + \delta t)/x(t)$. This distribution can be fitted to a lognormal function:

$$f(x) = \frac{1}{\sqrt{2\pi}sx} \exp\left(-\frac{\ln^2(x)}{2s^2}\right). \tag{D.1}$$

We impose the constraint that the median of this distribution is one because we assume that the time series fluctuates around steady state and that there will be as many steps for which the abundance is increasing as there are steps for which the abundance is decreasing. We define the shape parameter $s$ as the width of the step distribution. Some examples of fits are shown in Figure D.1. The goodness of the fit is obtained through the Kolmogorov-Smirnov test, of which we represent the p-value. A higher p-value corresponds to a better fit of the distribution to a lognormal function. The p-value is represented by the color of the fitted line.

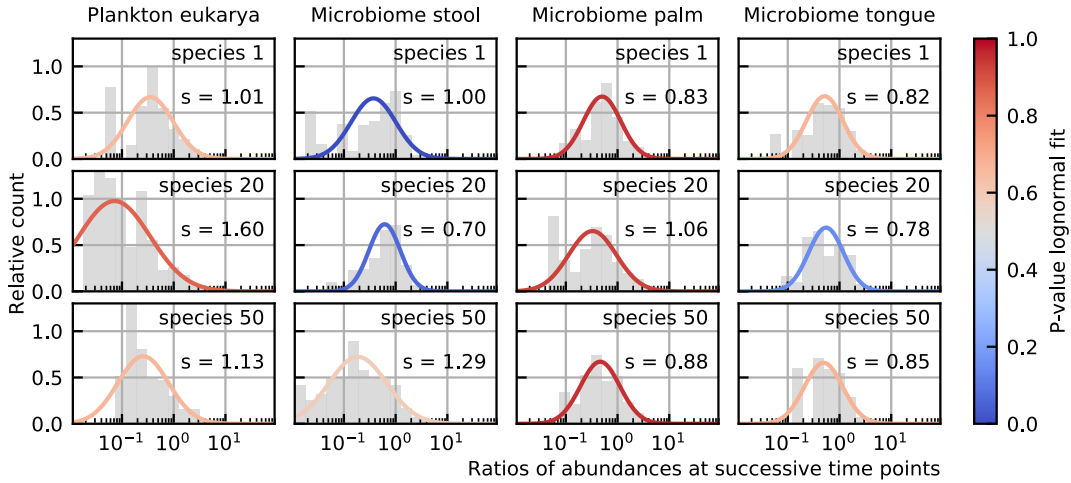

**Figure D.1:** A lognormal function is fitted to the ratios of abundances at successive time points $x(t + \delta t)/x(t)$ for different species. The species number denotes its rank. We interpret the shape parameter $s$ as a measure for the width of the steps. Here we see that this width $s$ is of the order of 1 for all species. The goodness of the fit (p-value of the Kolmogorov-Smirnov test) is represented by the color of the fitted line. For most species the fit is good (high p-values).

## Implementation and interpretation of the noise

Noise is introduced in dynamical systems by different processes. In simulations of microbial communities at the species level we distinguish intrinsic and extrinsic noise. Intrinsic noise is a result of the stochastic nature of cell division and cell death together with the discreteness of individuals. Extrinsic noise is caused by environmental fluctuations, *e.g.* changing nutrients, pH, temperature. To implement this extrinsic noise in a system of generalized Lotka-Volterra equations, the changing environments are often interpreted in changing parameters such as the growth rate. Because this parameter is multiplied by the species abundance, the resulting noise is linear. The remaining parameters, inter- and intra-specific interactions can also change depending on the environment. The formulation of this noise is more subtle (used in Zhu and Yin, 2009).

To derive the form of intrinsic noise in generalized Lotka-Volterra equations, we can consider every species abundance making a random walk in one dimension. The average displacement is zero and the variance of displacement is the sum of the rate of growth (jumping to the right) and the rate of death (jumping to the left). For the generalized Lotka-Volterra equations this results in a noise term

$$\langle n_i(t) n_i(t') \rangle = (\text{growth rate}_i + \text{death rate}_i) x_i = (f(g_i) + h(\omega, \vec{x})) x_i \delta(t-t'), \quad \text{(D.2)}$$

with $\omega$ the interaction matrix and where functions $f$ and $h$ decouple the growth and death terms. In the generalized Lotka-Volterra model no difference is made between negative interactions as a result from slowing down the growth rate or increasing the death rate, only the resulting net rates are used. This distinction must however be made to implement the intrinsic noise for gLV. In our analysis, we use the simpler logistic models where the resulting variance of the noise is proportional to the square root of the abundance $\sqrt{x}$. One must be careful not to use this noise for values that are smaller than 1, because this derivation relies on Poisson statistics which is defined for integer numbers.

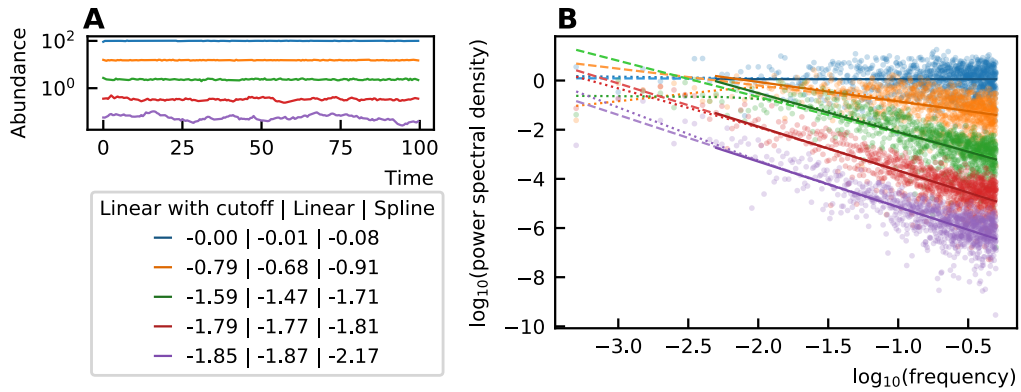

**Figure D.2:** The noise color of time series (A) is determined by the slope of the power spectral density (B). This slope can be measured through a linear fit of all values (dashed), a linear fit through the higher frequency range (solid line) or by performing a spline fit (dotted). A linear fit through all frequencies can be influenced by the windowing effect for low frequencies and the spline fit can make the slope steeper at the low frequencies and result in a darker noise as can be seen for the purple curves. The values of the noise color determined by the different techniques are given in the legend. Therefore, in our work, we opt for the linear fit with a cutoff for low frequencies.

### Noise color

The color of the noise in a time series is determined by the slope of the power spectral density in log-log scale. This slope can be determined by a linear fit through the spectrum. A different technique has been put forward to estimate the slope of the spectral density by Faust et al., 2018. There it is argued that the power spectral density does not have a constant slope and that therefore a nonlinear curve must be fitted. They choose for a spline fit and consider the minimal value of its derivative as the value for the noise color. Because the minimal value of the slope of the fit is taken, the noise color tends to be darker when using this technique. For our time series however we see that the spline fit only deviates from the linear fit for low frequencies (Figure D.2). We ignore the low frequencies for fitting because of the windowing effect. Therefore, we opt for a linear fit after omitting the values for low frequencies (one order of magnitude of the lowest frequencies).

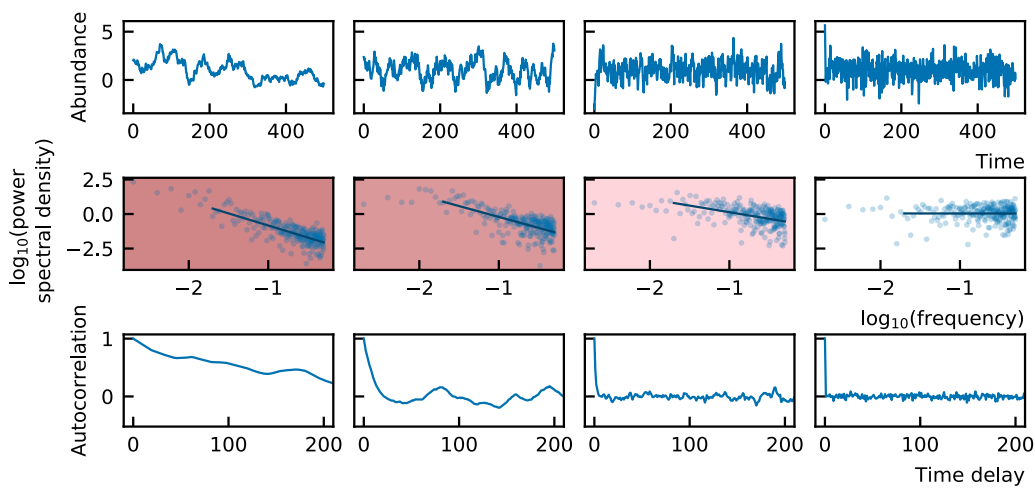

**Figure D.3:** Different stochastic time series (first row) with the corresponding power spectral density (second row) and autocorrelation functions (last row). The background color of the power spectral density corresponds to the color of the noise. When there is more structure present in the time series, the autocorrelation function is less steep at small time delays and the noise is darker.

The correspondence between the colors and slopes is here:

| Slope | Color |
|-------|-------|
| 0     | white |
| -1    | pink  |
| -2    | brown |
| -3    | black |

## Noise color and autocorrelation

For any time series, an autocorrelation function can be calculated. The latter is obtained by calculating the correlation of a time series with the same time series starting at a later time point. The difference in time between the starting point of both time series is the delay. The autocorrelation function is defined as the autocorrelation for all positive delay values. As explained in the supplemental material of Faust et al., 2018, the noise color is related to the autocorrelation of the time series. The lighter the noise color is, the steeper the autocorrelation function is for small time delays. For dark noise, the autocorrelation decreases more slowly and the time series is said to contain more structure. An illustration of this is given in Figure D.3.

## Discretizations of stochastic models with linear noise

The implementation of the linear multiplicative noise is as follows (Equation 8.2),

$$dx_i = \lambda_i dt + g_i x_i dt + \sum_j \omega_{ij} x_i x_j dt + x_i \sigma_i dW, \tag{D.3}$$

with $dW$ an infinitesimal element of a Brownian motion which is defined by a variance of $dt$ ($dW \sim \sqrt{dt}\mathcal{N}(0,1)$).

Because of its nonlinear nature, generalized Lotka-Volterra equations cannot be solved analytically. It is therefore necessary to perform simulations of the stochastic differential equations.

When we consider a linear multiplicative noise, we must first choose a discretization of the deterministic differential equations. Two possibilities can be found in the literature. The first one is the Ricker implementation and the second one we here call the Langevin implementation.

### Langevin discretization

To solve the stochastic equations numerically, we can use the Euler-Maruyama method. Here the discretization becomes

$$x_i(t + \delta t) = x_i(t) + g_i x_i(t)\delta t + \omega_{ij} x_i(t) x_j(t)\delta t + \sigma \delta W x_i(t). \tag{D.4}$$

Due to the choice of discretization $x$ can become zero or negative. Because this is the most straightforward discretization of the Langevin equation, we call it the Langevin discretization.

### Ricker discretization

For this implementation, we assume there is no migration ($\lambda_i = 0$). First, we rewrite Equation D.3 as the derivative of the species abundance divided by the species abundance itself, such that we can replace the latter by a derivative of the logarithm of the species abundance:

$$\frac{\dot{x}_i}{x_i} = g_i + \sum_j \omega_{ij} x_j + \sigma \frac{dW}{dt}, \tag{D.5}$$

$$\dot{\ln x_i} = g_i + \sum_j \omega_{ij} x_j + \sigma \frac{dW}{dt}. \tag{D.6}$$

When dividing by the species abundance, we assumed that $x_i$ can never become zero (or negative for the solution is continuous). This would be true in the theoretically continuous framework and in the absence of noise, but this is not what is observed in experimental time series where the abundances are discrete numbers and can become zero in the event of extinction.

Using the assumption $\delta t \ll 1$, we obtain

$$\ln x_{t+\delta t} - \ln x_t = \delta t(\omega x + g) + \sigma \delta W, \tag{D.7}$$

$$x_{t+\delta t} = x_t \exp\{\delta t(\omega x + g)\} \exp\{\sigma \delta W\} \tag{D.8}$$

The $\delta W$ are normally distributed random elements ($\delta W \sim \sqrt{\delta t} \mathcal{N}(0, 1)$).

Due to the choice of discretization, $x_i$ can never become zero or negative, which is fundamentally different from the previous implementation.

**Arato discretization**

Using Ito calculus (see Appendix B.6 for a short introduction) a third discretization can be obtained:

$$x_i(t) = x_i(0) \exp\left\{\left(g_i t - \frac{\sigma_i^2 t}{2} - \sum_{j=1}^n \omega_{ij} \int_0^t x_j(s) ds + \sigma W_i(t)\right)\right\} \tag{D.9}$$

as calculated in Arató, 2003.

In the limit of small noise or small timesteps, the Langevin, Ricker, and Arato discretizations are equal (Figure D.4A). For large noise, a clear difference can be seen for both implementations (Figure D.4B).

The parameters of the implementation, such as the discretization, length of the simulation, and integration time step $dt$ are not influencing the results of the noise color (Figure D.4C-E).

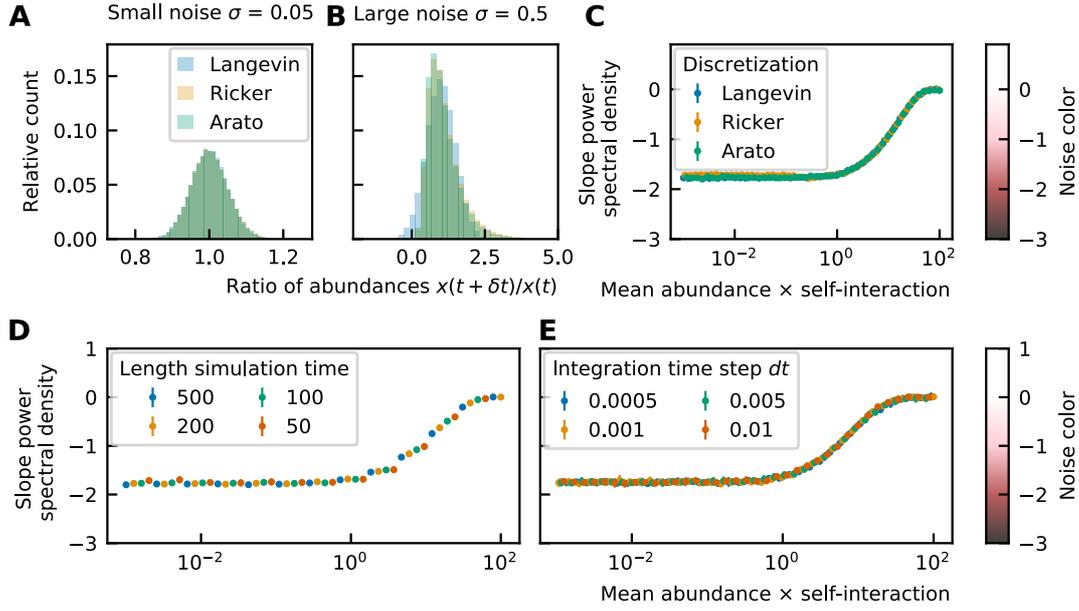

**Figure D.4:** Distributions of $x_{t+dt}$ for $x_t = 1$ for three different discretizations: Langevin $x_{t+dt} = x_t + x_t \sigma dW$, Ricker $x_{t+dt} = x_t \exp\{(\sigma dW)\}$ and Arato $x_{t+dt} = x_t \exp\left\{\left(\frac{-\sigma^2}{2} dt + \sigma dW\right)\right\}$ with $dW \sim \mathcal{N}(0,1)$ and $dt = \sigma^2$ for small noise $\sigma = 0.05$ (A) and large noise $\sigma = 0.5$ (B). For both these figures the sampling time step $\delta t$ is equal to the integration time step $dt$. For small noise, all implementations result in the same abundance distribution. For large noise, the distributions are different. The Langevin implementation allows abundances to become zero and negative (which will be equalled to zero). The abundances of the Ricker and Arato implementation never become zero. The noise color does not depend on the particular implementation: it is independent of the discretization (C), length of the time series (D) and integration time step $dt$ (E).

### Discretizations of stochastic models with non-linear noise

In the main manuscript, we discussed choosing a noise term that is linear in the species abundance because this represents best the external noise on growth and death of the species. Intrinsic noise (due to discreteness) could be represented by a term that is proportional to the square root of the species abundance (sqrt multiplicative noise) (Walczak et al., 2012). In this case, the discretization is:

$$x_i(t + \delta t) = x_i(t) + g_i x_i(t) \delta t + \omega_{ij} x_i(t) x_j(t) \delta t + \sigma \delta W \sqrt{x_i(t)}. \tag{D.10}$$

We also considered a constant noise term (additive noise), which represents noise due to stochastic immigration and emigration of species. The corresponding discretization is:

$$x_i(t + \delta t) = x_i(t) + g_i x_i(t) \delta t + \omega_{ij} x_i(t) x_j(t) \delta t + \sigma \delta W. \tag{D.11}$$

In Chapter 8, it is shown that the source of the noise, linear multiplicative, square root multiplicative, or additive, has no influence on the relation between the mean abundance, self-interaction, and noise color.

## D.2 Analysis of experimental data

We studied time series of different microbial communities. References for all of the time series can be found in Table D.1.

### Rank abundance

Rank abundances of all studied time series can be found in Figure D.5. In nature, the abundance distributions are often power law, lognormal, or logarithmic series distributions. The abundance distribution of the microbial communities we study fits best a lognormal distribution (Figure D.6).

### Distribution of the differences between abundances at successive time points

We study the differences between abundances at successive time points of the time series in two ways. First, we look at the mean absolute value of the difference between abundances at successive time points $\langle |\ x(t+\delta t) - x(t)\ | \rangle$ and second we fit the distribution of the ratios of abundances at successive time points $x(t+\delta t)/x(t)$ with a lognormal. More details about the latter can be found in Section D.1. The slope of the mean absolute difference $\langle |\ x(t+\delta t) - x(t)\ | \rangle$ as a function of the mean abundance in the log-log scale ranges between 0.84 and 0.99 (Figure D.7). The order of magnitude of the width of the distribution of the ratios of abundances at successive time points $x(t + \delta t)/x(t)$ is one and there is no correlation between the mean abundance and the width of this distribution (Figure D.8). The goodness of the fit is characterized by the p-value of the Kolmogorov-Smirnov test. A higher p-value corresponds to a better fit of the distribution to a lognormal function (see Section D.1). The p-value is represented by the color of the points.

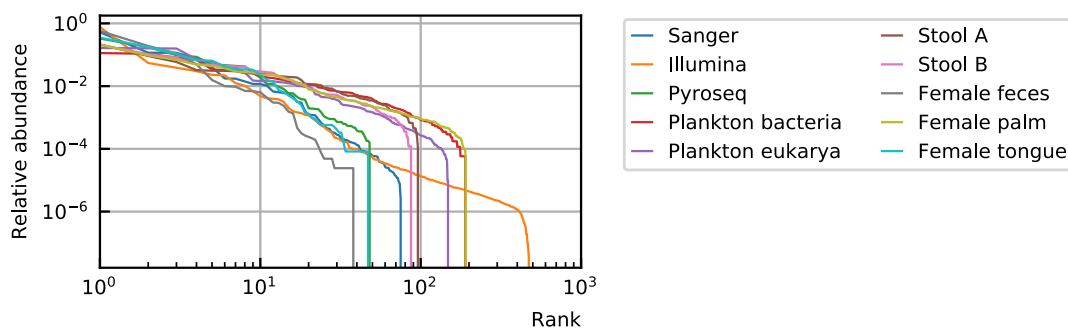

**Figure D.5:** The rank abundance curve of the heavy-tailed abundance distribution of experimental measurements of microbial communities, from human microbiomes to plankton.

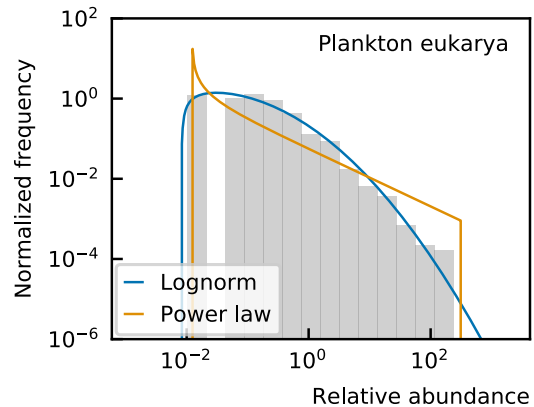

**Figure D.6:** The abundance distribution of experimental data is heavy-tailed. It fits the lognormal distribution better than the power law.

## Neutrality test

Both neutrality tests, the Kullback-Leibler divergence and the p-value of the neutral covariance test, show that most of the experimental time series are in the niche regime (Figure D.9).

## Noise color

The noise color is predominantly in the pink to white region. It is also independent of the mean abundance (Figure D.10).

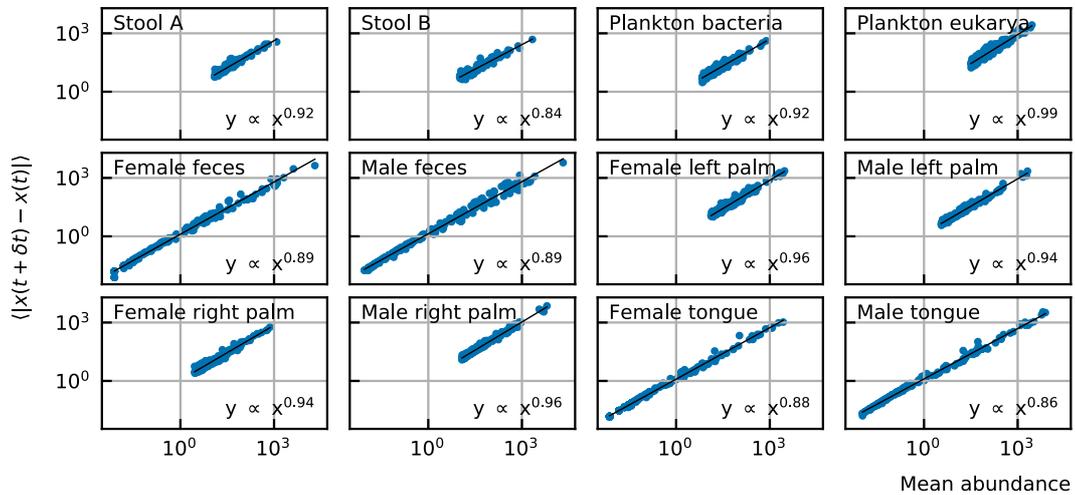

**Figure D.7:** For experimental time series, the slope of the mean absolute difference between abundances at successive time points as a function of the mean abundance in the log-log scale ranges between 0.84 and 0.99.

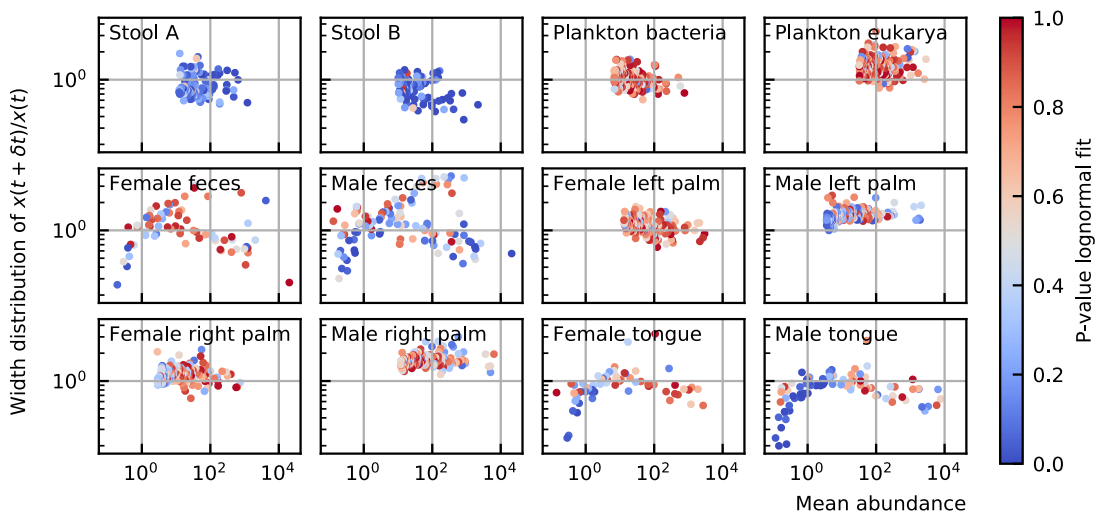

**Figure D.8:** For experimental time series, the width of the distribution of the ratios of abundances at successive time points $x(t + \delta t)/x(t)$ is in the order of 1, which means that the fluctuations are large.

**Table D.1:** References for all time series and compositions of microbial communities.

| Label | Label in Figure 1 of the main article | Source |
| --- | --- | --- |
| Stool A | Microbiome stool | Subject A gut of David et al., 2014 |
| Stool B | | Subject B gut of David et al., 2014 |
| Plankton bacteria | | Bacteria relative abundance of Martin-Platero et al., 2018 |
| Plankton eukarya | Plankton eukarya | Eukaryota relative abundance of Martin-Platero et al., 2018 |
| Female feces | | Feces microbiome of subject F4 at the genus level (L6) Caporaso et al., 2011 |
| Male feces | | Feces microbiome of subject M3 at the genus level (L6) Caporaso et al., 2011 |
| Female left palm | Microbiome palm | Left palm microbiome of subject F4 at the genus level (L6) Caporaso et al., 2011 |
| Male left palm | | Left palm microbiome of subject M3 at the genus level (L6) Caporaso et al., 2011 |
| Female right palm | | Right palm microbiome of subject F4 at the genus level (L6) Caporaso et al., 2011 |
| Male right palm | | Right palm microbiome of subject M3 at the genus level (L6) Caporaso et al., 2011 |
| Female tongue | Microbiome tongue | Tongue microbiome of subject F4 at the genus level ((L6) Caporaso et al., 2011 |
| Male tongue | | Tongue microbiome of subject M3 at the genus level (L6) Caporaso et al., 2011 |
| Sanger | | Feces composition of DA-AD-1 of Arumugam et al., 2011 |
| Illumina | | Feces composition of MH0001 of Arumugam et al., 2011 (original data Qin et al., 2010) |
| Pyroseq | | Feces composition of TS1_V2_turnbaugh of Arumugam et al., 2011 (original data Turnbaugh et al., 2009) |

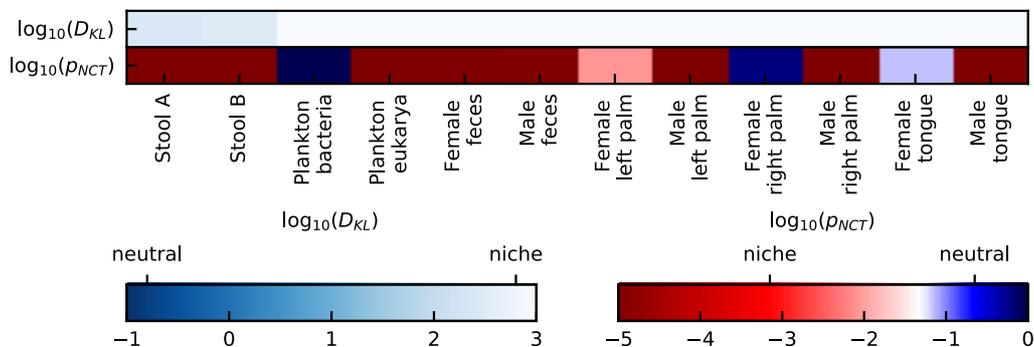

**Figure D.9:** The values of the Kullback-Leibler divergence (KL) and the p-value of the neutral covariance test (NCT) are represented for the different microbial communities. The value can be read from the colorbars at the bottom. The neutral regime is given by a dark blue color. The niche regime is represented by a light colors for the KL and red for the NCT. For the NCT scale, a p-value of 0.05 is represented by white. Both neutrality tests show that most of the experimental time series are in the niche regime.

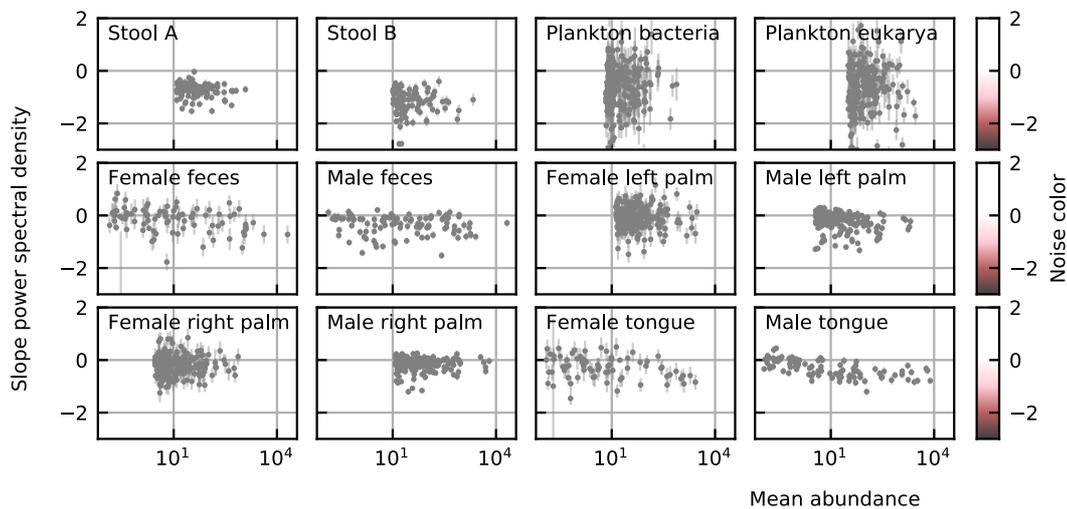

**Figure D.10:** For experimental time series, the noise color is predominantly in the pink to white region. It is also independent of the mean abundance.

## D.3  Supporting results

**The noise color depends on the product of the mean abundance and the self-interactions.**

To study the noise color, we represented it as a function of different variables. We conclude that the noise color depends on the product of the mean abundance and the self-interaction (Figure D.11). Given the sampling rate, this curve can be fitted by a sigmoid with the *x*-axis in the log scale (Figure D.12) and the fitted function can be used to calculate the self-interaction of a species given its noise color and mean abundance.

**All noise characteristics can be obtained with the logistic model.**

Given the abundance and the slope of the power spectral density (noise color), we can determine the self-interaction and growth rate of a species via the fitted sigmoid curve (Figure D.13). The fitted sigmoid curve gives the value of the product of the mean abundance and self-interaction given the slope of the power spectral density (Figure D.13B). Using this value and the relation between the self-interaction and the product of the self-interaction and mean abundance (colored lines of Figure D.13F), the self-interaction can be calculated (Figure D.13E). For noninteracting species, the growth rate equals the product of the mean abundance and self-interaction (black line in Figure D.13H). Therefore, the growth rate is retrieved immediately (Figure D.13G).

Once all the parameters are determined, we can perform simulations with large linear multiplicative noise that have the same characteristics as experimental time series. These results are presented in Figure 4 of the main paper.

**The noise characteristics remain for normalized time series.**

It is experimentally often challenging to determine the absolute abundances of species. Therefore, only fractional abundances are measured. The logistic equation models the absolute abundance of a species. In Figure D.14, we show that normalizing the species abundances for every time point does not influence the characteristics.

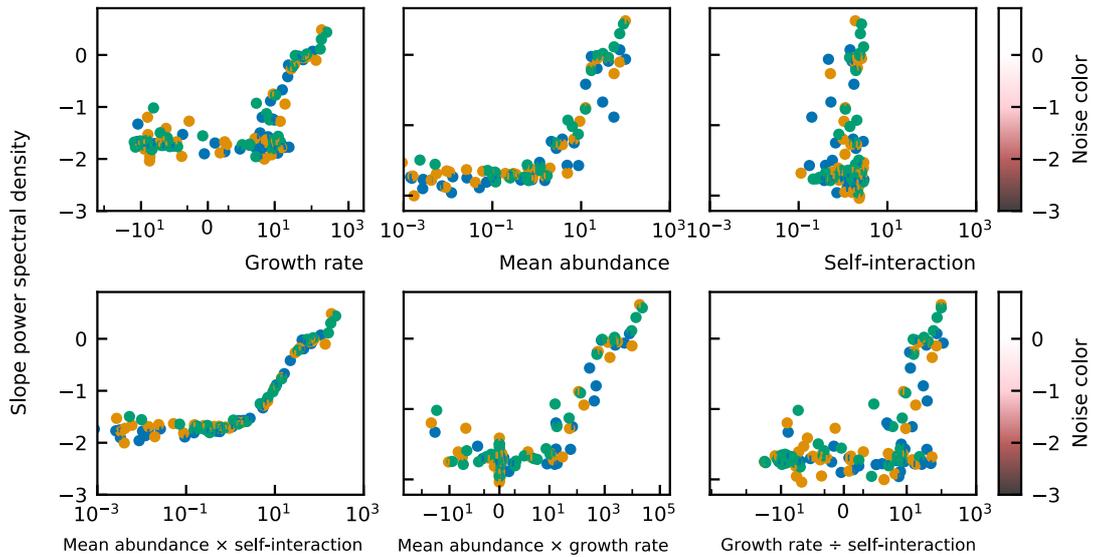

**Figure D.11:** The noise color depends on the product of the mean abundance and the self-interaction. The correspondence with the growth rate or mean abundance are less pronounced.

## The noise characteristics can also be obtained with generalized Lotka-Volterra equations.

In the presence of inter-species interactions, the properties of the experimental time series can also be retrieved. In Figure D.15, an example is given with interaction strength 0.02 ($\omega_{ij} \propto \mathcal{N}(0, 0.02)$) and connectance 0.1 which means that 90% of the interactions are set to 0. The rank abundance and self-interactions are derived from the stool A data and the noise is linear with strength $\sigma_{\text{lin}} = 2.5$. The growth rate is calculated given the steady-state abundances and interaction matrix. Interactions are however not necessary to obtain the characteristics, as seen in Chapter 8. The same results still hold when normalizing the time series (Figure D.16).

## The slope of the differences between abundances at successive time points $x(t + \delta t)/x(t)$ can be fine-tuned by a balanced combination of extrinsic and intrinsic noise.

We can combine extrinsic (linear multiplicative) and intrinsic (square root multiplicative) noise such that the slope of the absolute differences between abundances at successive time points as a function of the mean abundance in the log-log space is smaller than one as observed in the experimental time series. For linear multiplicative noise, the slope has a value around 1, for sqrt multiplicative noise the value is around 0.66 (Figure D.17). A combination where the strengths of both noise sources are equal ($\sigma_{\text{lin}} = \sigma_{\text{lin}} = 0.5$) results in an intermediate slope (Figure D.17).

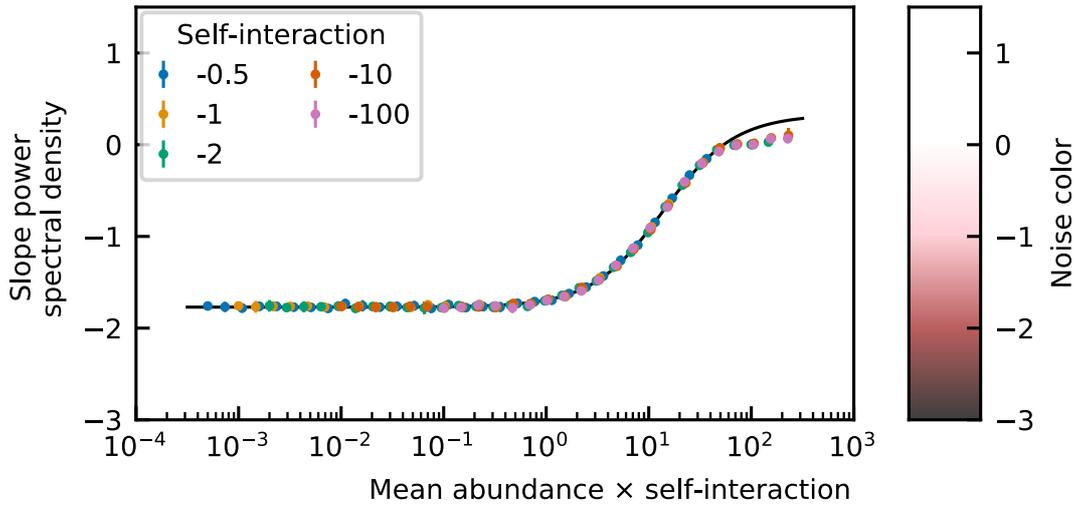

**Figure D.12:** The curve of the noise color as a function of the logarithm of the mean abundance and self-interaction can be fitted with a sigmoid function.

## The width of the distribution of ratios of abundances at successive time points increases with increasing strength of the noise.

The distribution of the ratios of abundances at successive time points $x(t+\delta t)/x(t)$ depends on the rank abundance, self-interaction, and interaction strengths. In a first implementation, we simulate 50 interacting species with linear noise. The interaction matrix has elements drawn from a normal distribution with mean 0 and the standard deviation given by the interaction strength. All self-interactions are imposed to be -1. For an increasing noise strength $\sigma_{\text{lin}}$, the width of the distribution of ratios of abundances at successive time points $x(t+\delta t)/x(t)$ increases (Figure D.18). In a second implementation, we use 50 species with the rank abundance of the experimental human stool data (Stool A) and self-interactions inferred from the noise color as explained in the main paper. The interaction matrix for the interacting case is much more sparse here, the connectance is 0.1 which means that only 10% of the interactions are non-zero. The non-zero interactions are distributed like a normal distribution with mean 0 and a standard deviation of 0.03. As expected, increasing the strength of the noise increases the width of the distribution, and adding interaction also slightly increases the width (Figure D.19).

### Neutrality of generalized Lotka-Volterra models

The higher the interaction strength, the more niche the time series become (Figure D.20).

### Rounding time series to integer numbers makes the noise color lighter.

For low abundances, there is a finite precision effect. When we calculate the noise color for the time series with a precision up to the integer, we see that the noise color becomes lighter for smaller abundances (Figure D.21). This is consistent with the fact that the noise color is correlated with the autocorrelation and the amount of structure in the time series, i.e. the predictability of the time series. In the time series that have smaller precision, there is less information and the time series is less predictable, therefore the noise color is lighter.

### The self-organized instability model can be reproduced by the stochastic generalized Lotka-Volterra model.

In the main text, we discussed that next to stochastic generalized Lotka-Volterra models there exists another algorithm to model interacting species stochastically, the individual-based model. We mention in particular the self-organized instability (SOI) model introduced by Solé et al., 2002. The question arises whether the SOI model is equivalent to the generalized Lotka-Volterra models, or if it can produce more or different characteristics. An elaborate study of the SOI has been performed in Faust et al., 2018. Most results are in agreement with our findings. However; in a supplemental figure of Faust et al., 2018 (S3), it is shown that the percentage of taxa with pink noise increases and the percentage of taxa with brown noise decreases when the stochasticity increases, where the stochasticity is defined as the ratio between the mean of extinctions and immigrations and the mean of the absolute interaction strengths (excluding diagonal values). This cannot be easily concluded from our previous results and we therefore studied this aspect in more detail.

We first evaluated the contributions of the different parameters. We therefore run simulations using the code provided by the authors of Faust et al., 2018, only changing one of the parameters at a time. What we notice is that the shift from brown to pink noise is mostly caused by the extinction rate and not the immigration rate or interaction strength (Figure D.22, parameters are in Table D.2 and Table D.3).

Table D.2: Parameter values of varied parameters.

| Column in figure 1 | 1 | 2 | 3 |
|---|---|---|---|
| Varying parameter | Immigration rate | Extinction rate | Interaction strength |
| Immigration (m) | $\mathcal{U}(0, \text{immigration})$ | $\mathcal{U}(0, 0.02)$ | $\mathcal{U}(0, 0.1)$ |
| Extinction (e) | $\mathcal{U}(0, 0.01)$ | $\mathcal{U}(0, \text{extinction})$ | $\mathcal{U}(0, 0.1)$ |
| Interaction strength (A) | $\mathcal{U}(-0.2, 0.2)$ | $\mathcal{U}(-0.2, 0.2)$ | $\mathcal{U}(-\text{interaction}, \text{interaction})$ |

Table D.3: Parameter values of fixed parameters.

| Parameter | Value |
|---|---|
| Number of individuals / lattice sites (I) | 4000 |
| Number of species (S) | 100 |
| Positive edge percentage (PEP) | 20 |
| Diagonal elements | -0.5 |
| Connectance | 0.02 |

In Figure D.22, we also see that with increasing extinction the mean abundance is decreasing (top panel). At the same time, the noise color is becoming lighter (shifting to pink). To understand the connection between the parameters, we made a figure with both the noise color and mean abundance on the axes for two time series (Figure D.23A). We see that the noise color becomes more pink for smaller mean abundances. How do these results relate to the results for the generalized Lotka-Volterra equations?

There are multiple differences between the SOI and gLV models. First of all, the SOI model follows an individual-based approach. It is based on the Gillespie algorithm with the K-leap method. At every time step, the propensities for all events are calculated and K processes are chosen accordingly. The elapsed time $\tau$ is calculated by the gamma function. Since it is an individual-based model, the abundances will only take integer numbers. GLV models are continuous and the abundances of the time series are real positive numbers. We showed in the previous section that rounding time series to integer numbers makes the noise color lighter Section D.3.

A second difference between the SOI and our gLV model is that the SOI model is based on immigration and that no species can grow in the absence of other species.

In the gLV model, we do not consider immigration and the growth rates of species are often positive. Furthermore, the interaction matrix is interpreted differently. Given the interaction matrix used for the SOI model ($\omega$), the interaction matrix of the gLV model ($\omega'$) needs to obey the following rules for the time series to be comparable:

1. $\omega'_{ii} = 0$ all self-interactions are zero,
2. if $\omega_{ij} > \omega_{ji} > 0$ : $\omega'_{ij} = \omega_{ij} + \omega_{ji}$ and $\omega'_{ji} = 0$ mutualistic interactions are transformed into commensalistic interactions and,
3. if $\omega_{ij} < 0$ and $\omega_{ij} < \omega_{ji}$ : $\omega'_{ij} = \omega_{ji} - \omega_{ij}$ and $\omega'_{ji} = 0$ competitive or parasitic interactions are transformed into amensalistic interactions.

The most important difference is that the self-interactions are zero. For gLV models, the self-interaction needs to be negative such that the abundance remains bounded and cannot go to infinity. In the SOI model, this constraint is imposed by the maximal number of individuals. The propensity for growth decreases linearly with the number of individuals. When the maximal number of individuals is attained all growth propensities are zero.

If we include all the aforementioned elements:

1. add immigration,
2. consider only negative growth rates (extinction),
3. small self-interaction,
4. round the results to integer values after performing the complete time series,

then we obtain similar results as the SOI model (Figure D.23B and Figure D.24).

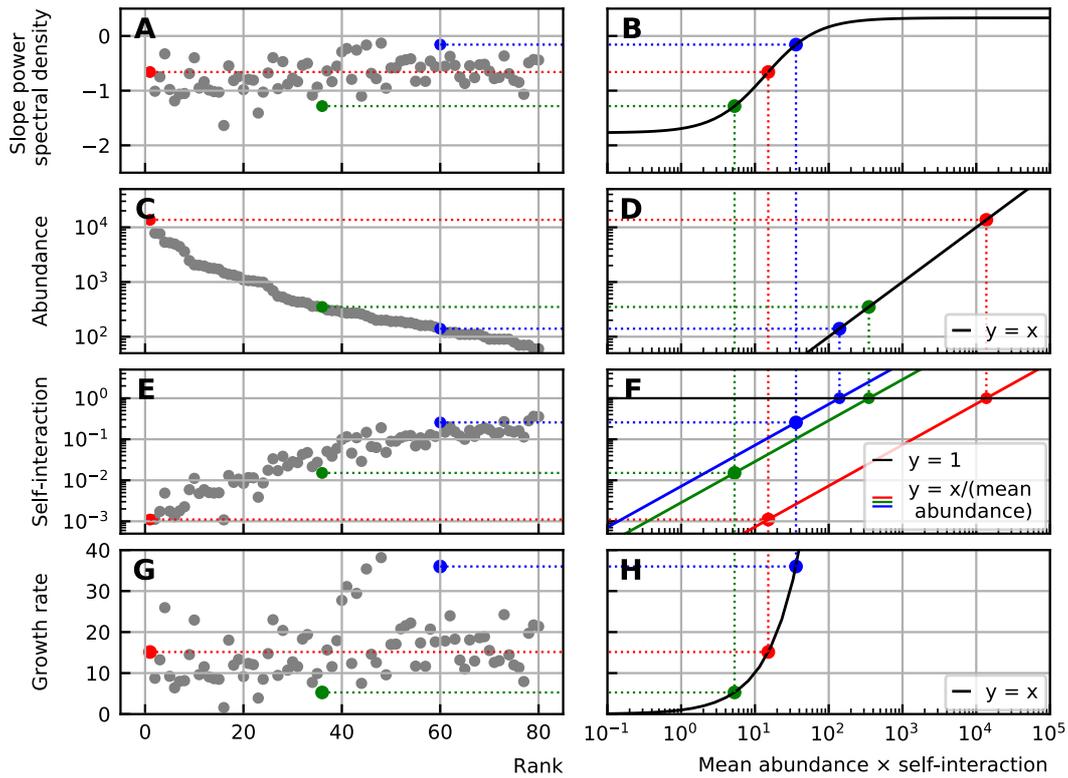

**Figure D.13:** This scheme represents how the self-interaction and growth rate can be retrieved from the noise color and abundance data. For a given sampling step $\delta t$, the relation between the noise color, abundance and self-interaction can be fitted by a sigmoid curve (B). Given the abundance (D), the relation between the mean abundance and self-interaction can be drawn (colored lines in F). The value of the noise color can then be translated to a value of the self-interaction (A-B-F-E). For noninteracting species, the growth rate equals the product of the mean abundance and self-interaction. The growth rate is thus easily determined (A-B-H-G).

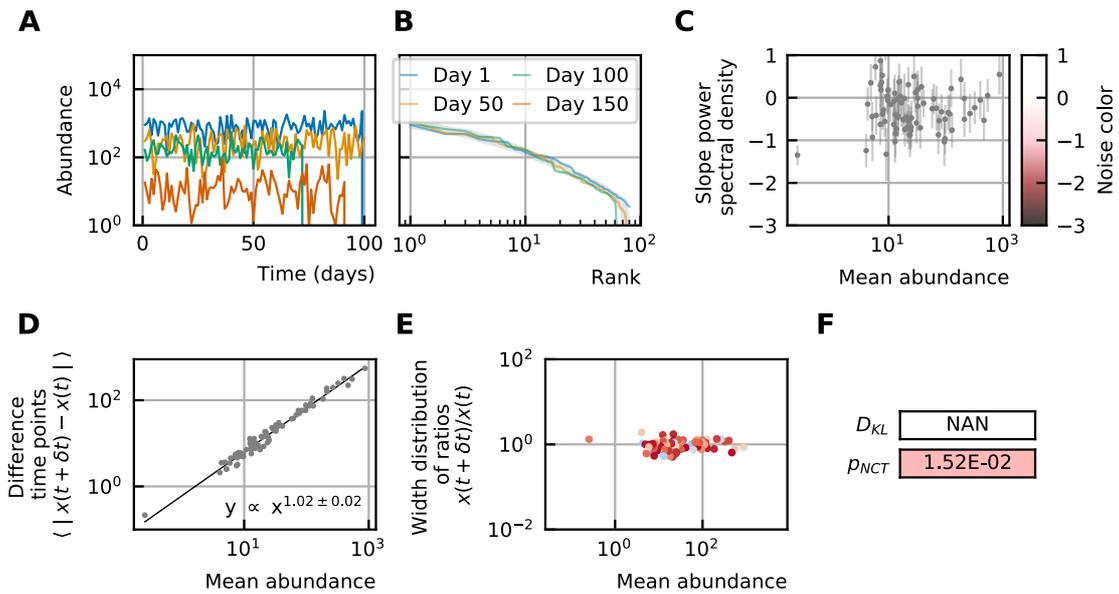

**Figure D.14:** All properties of the noise remain after normalization: (A) Time series. (B) The rank abundance of a heavy-tailed abundance distribution remains stable over time. (C) Noise color in the white-pink region with no dependence on the mean abundance. (D) The slope of the mean absolute difference between abundances at successive time points is around 1. (E) The width of the distribution of the ratios of abundances at successive time points is in the order of 1 and independent of the mean abundance. (F) Results of the neutrality test in the niche regime.

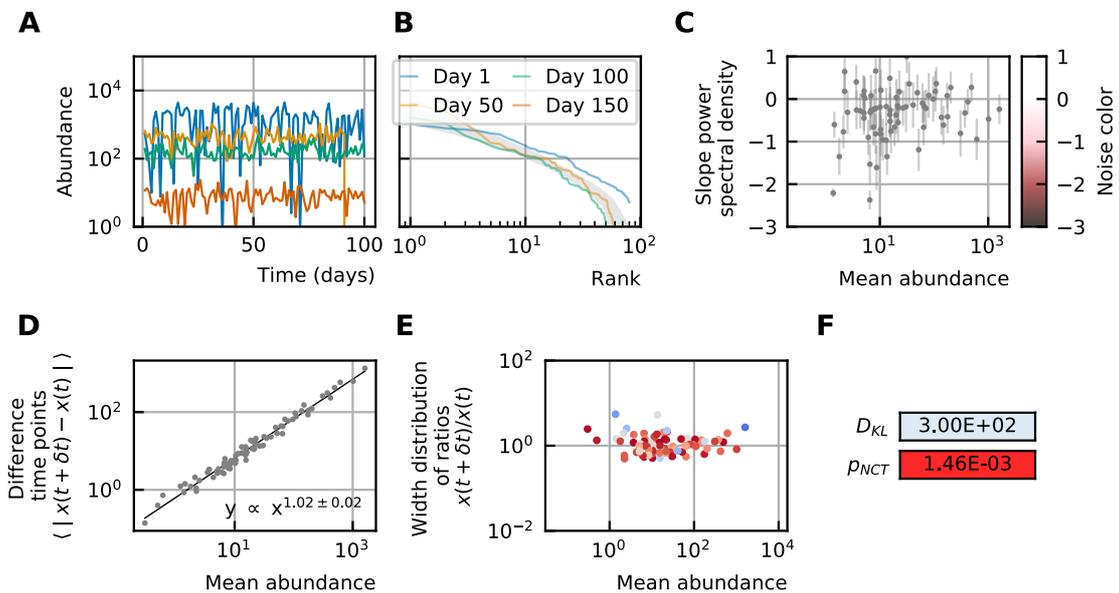

**Figure D.15:** All properties of the noise can still be obtained in the presence of interactions: (A) Time series. (B) The rank abundance of a heavy-tailed abundance distribution remains stable over time. (C) Noise color in the white-pink region with no dependence on the mean abundance. (D) The slope of the mean absolute difference between abundances at successive time points is around 1. (E) The width of the distribution of the ratios of abundances at successive time points is in the order of 1 and independent of the mean abundance. (F) Results of the neutrality test in the niche regime.

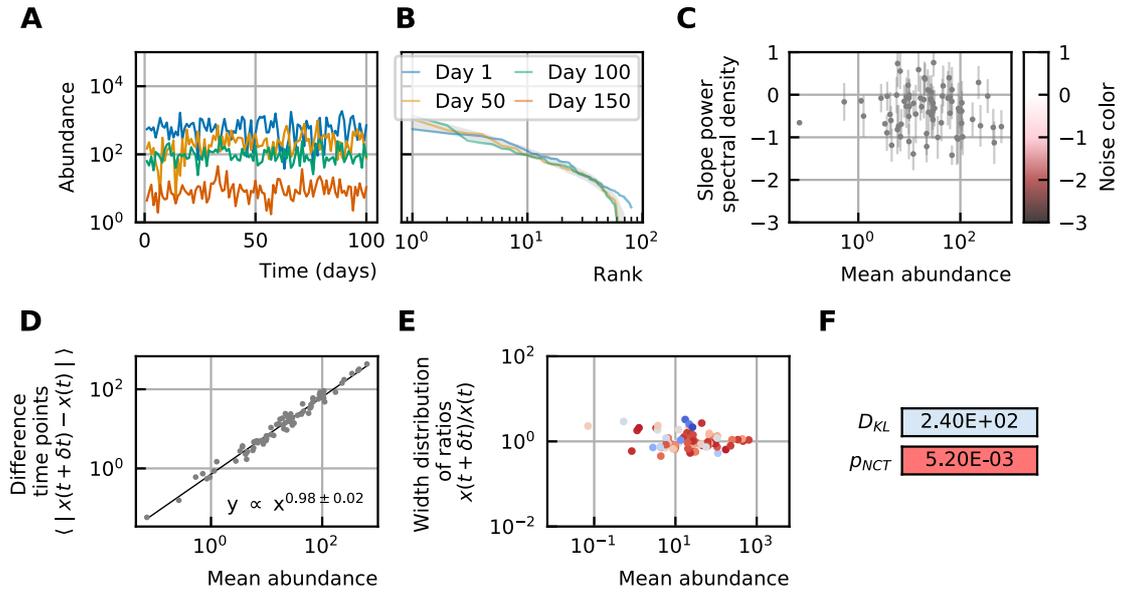

**Figure D.16:** All properties of the noise can still be obtained in the presence of interactions and when the time series are normalized: (A) Time series. (B) The rank abundance of a heavy-tailed abundance distribution remains stable over time. (C) Noise color in the white-pink region with no dependence on the mean abundance. (D) The slope of the mean absolute difference between abundances at successive time points is around 1. (E) The width of the distribution of the ratios of abundances at successive time points is in the order of 1 and independent of the mean abundance. (F) Results of the neutrality test in the niche regime.

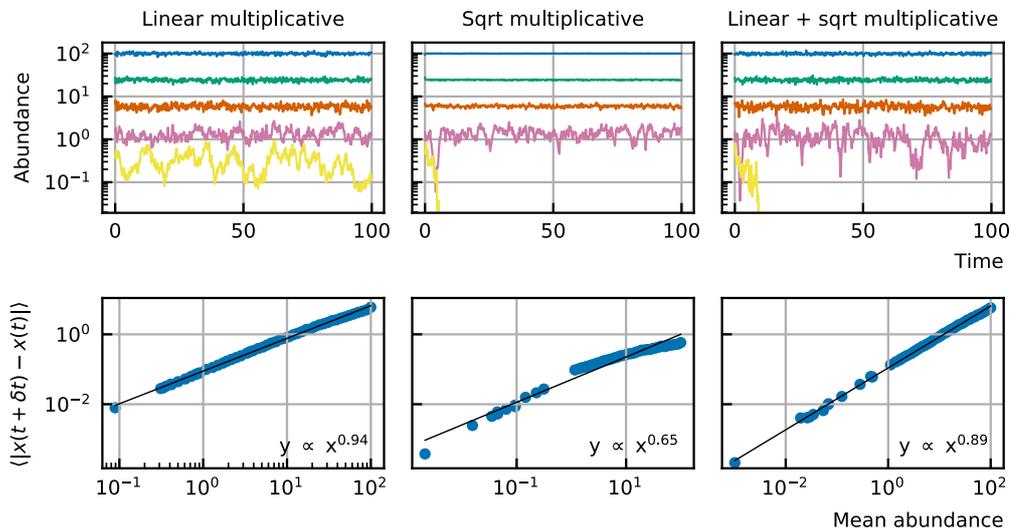

**Figure D.17:** The slope of the mean absolute difference between abundances at successive time points with respect to the mean abundance in log-log scale is around 1 (left). For square root noise, this slope is around 0.66 (middle). A mix of linear and square root noise results in intermediate slopes (right). Here, a combination of linear and square root multiplicative noise with equal strengths ($\sigma_{\text{lin}} = \sigma_{\text{sqrt}} = 0.5$) was used to obtain a slope of 0.89.

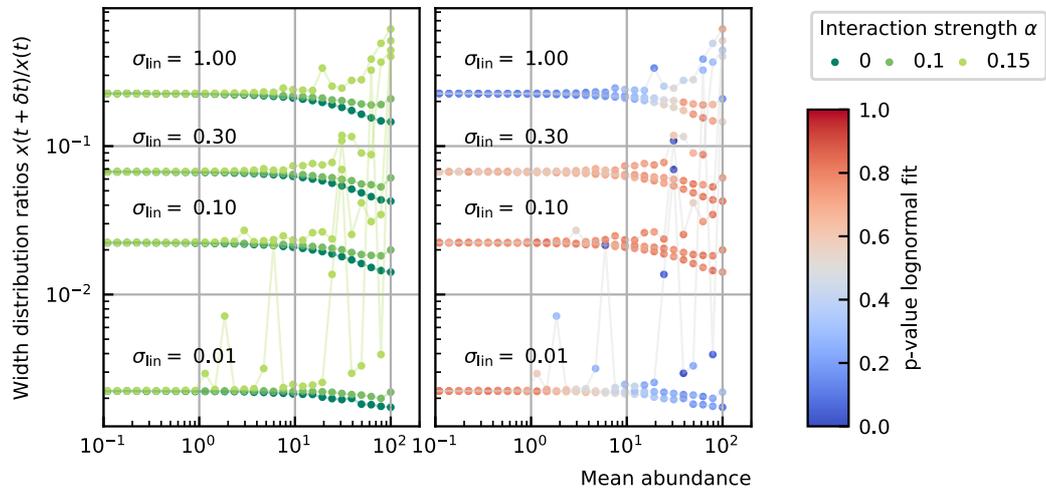

**Figure D.18:** The width of the distribution of the ratios of abundances at successive time points $x(t+\delta t)/x(t)$ for time series of 50 interacting species with equal abundances increases with increasing strength of the linear noise. For high abundances, the width also increases with the interaction strength.

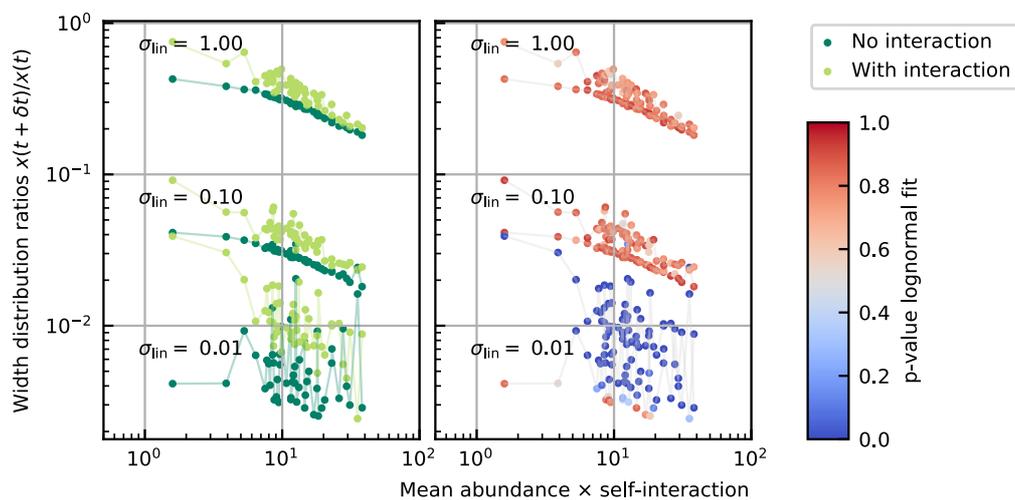

**Figure D.19:** The width of the distribution of the ratios of successive time points for time series for 50 interacting species with a rank abundance and inferred self-interaction of the stool A data. The width increases with increasing strength of the linear noise.

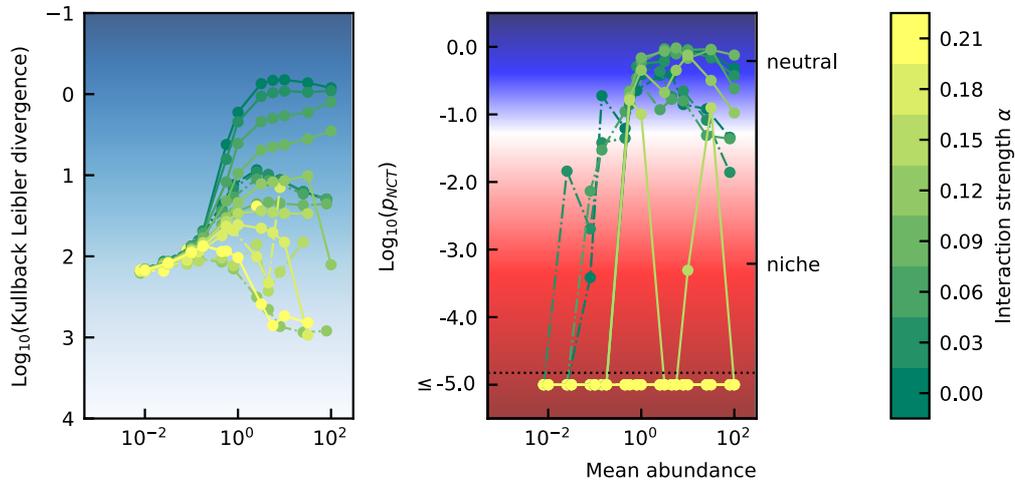

**Figure D.20:** The higher the interaction strength the more niche the time series become.

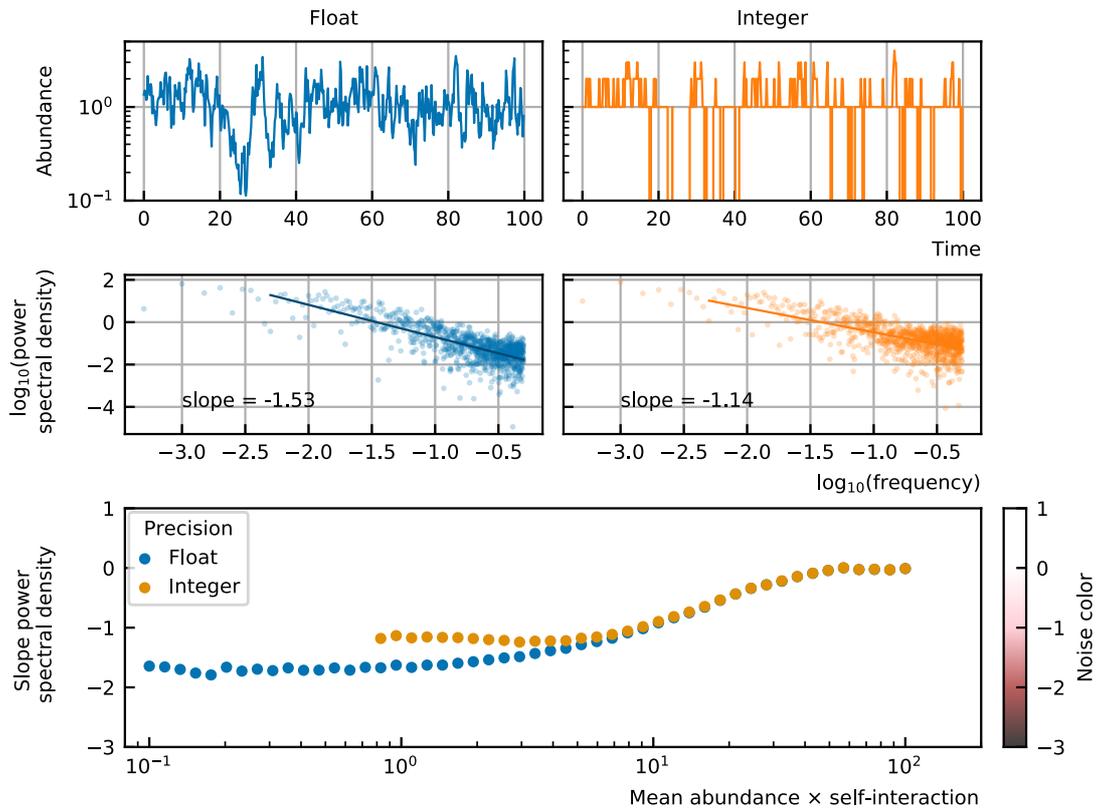

**Figure D.21:** Finite precision makes the noise color lighter for species with small abundances.

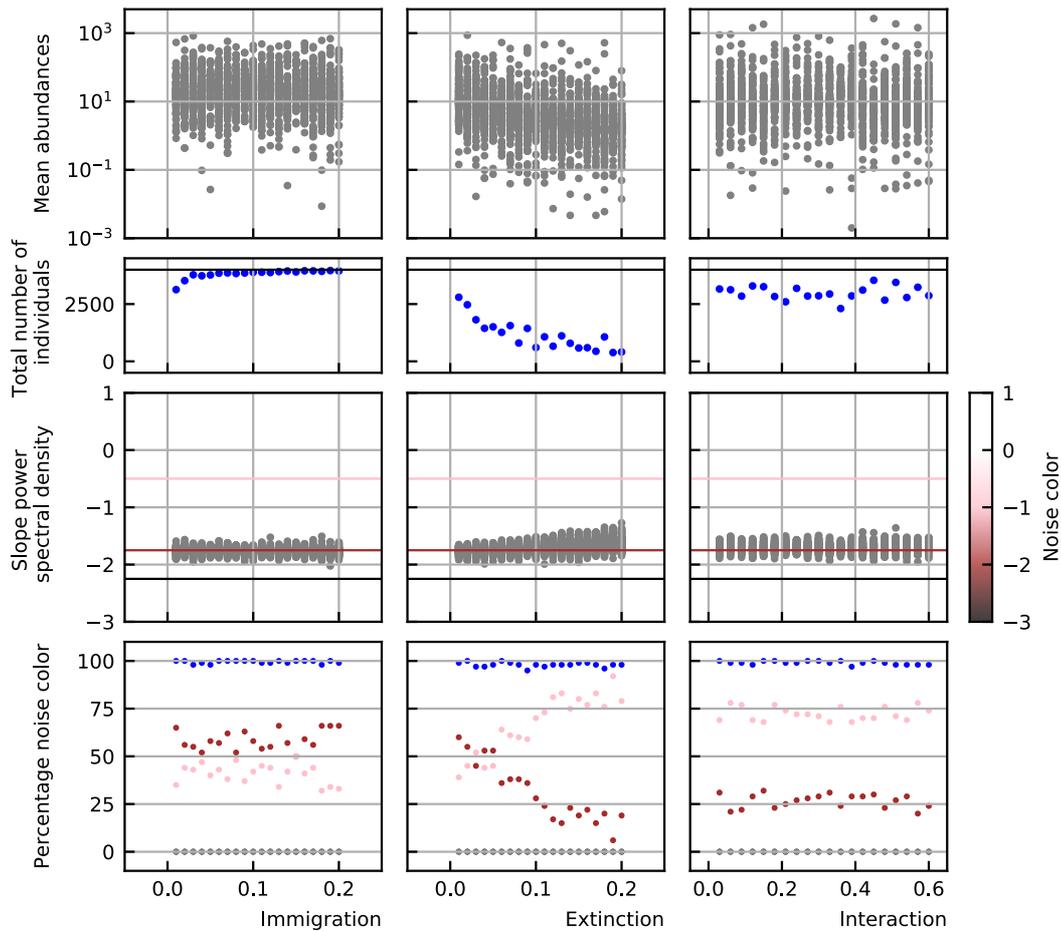

**Figure D.22:** Time series simulations were performed for multiple values of the immigration, extinction and interaction rate. Shown are the mean abundance values of these simulations (top panels), the total number of individuals (second row), the distribution of the noise colors (third row) and the percentages of all noise colors as delimited by Faust et al., 2018 (bottom row). For the SOI model, an increase in the extinction rate causes a shift from brown to pink noise. At the same time, the mean abundances and total number of individuals decrease.

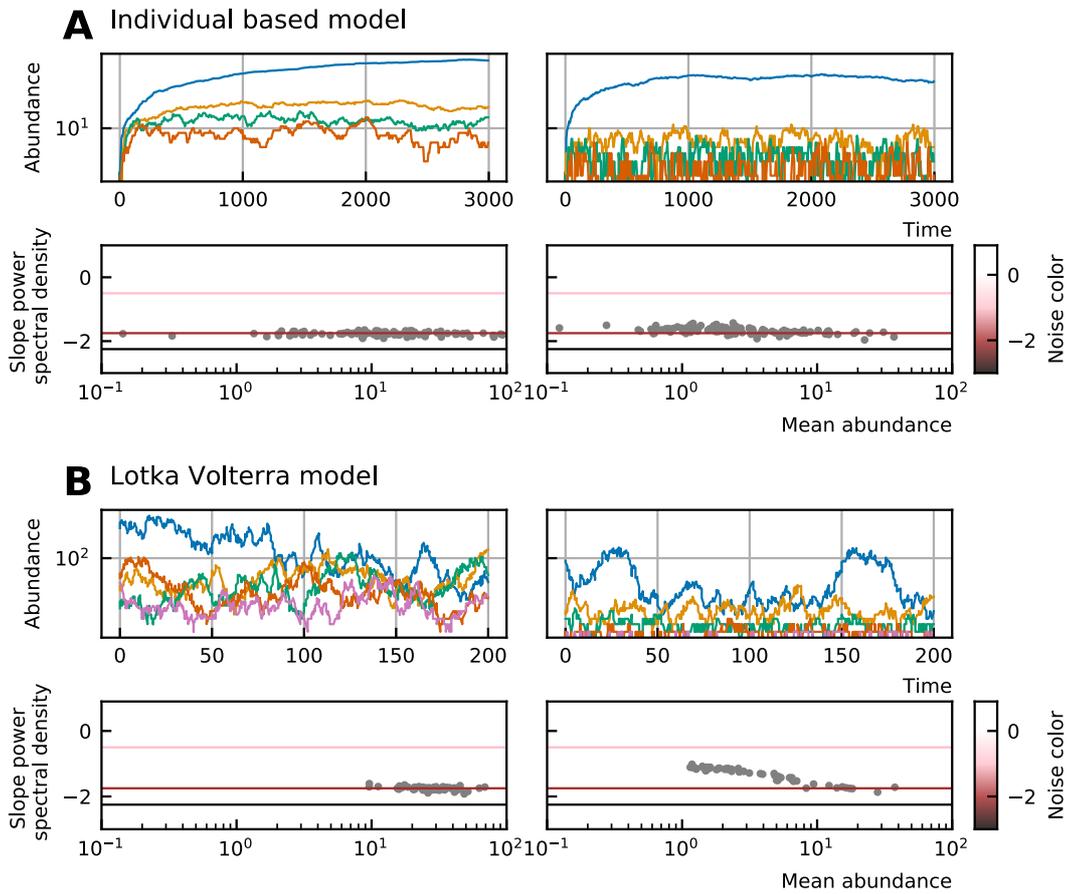

**Figure D.23:** (A) Time series generated by the SOI model for different extinction rates, $e = 0.01$ for the left column and $e = 0.16$ for the right column (additional parameters can be found in the second column of Table D.2 and Table D.3). The lower figures show the noise color as a function of the mean abundance for the corresponding time series. For higher extinction (right side) the mean abundances are lower and the noise color becomes lighter with decreasing mean abundance. (B) Time series generated with the gLV model and rounded to integer precision after the simulation for different extinction rates, $e = 0.001$ on the left and $e = 0.8$ on the right. The noise color becomes lighter for low abundances.

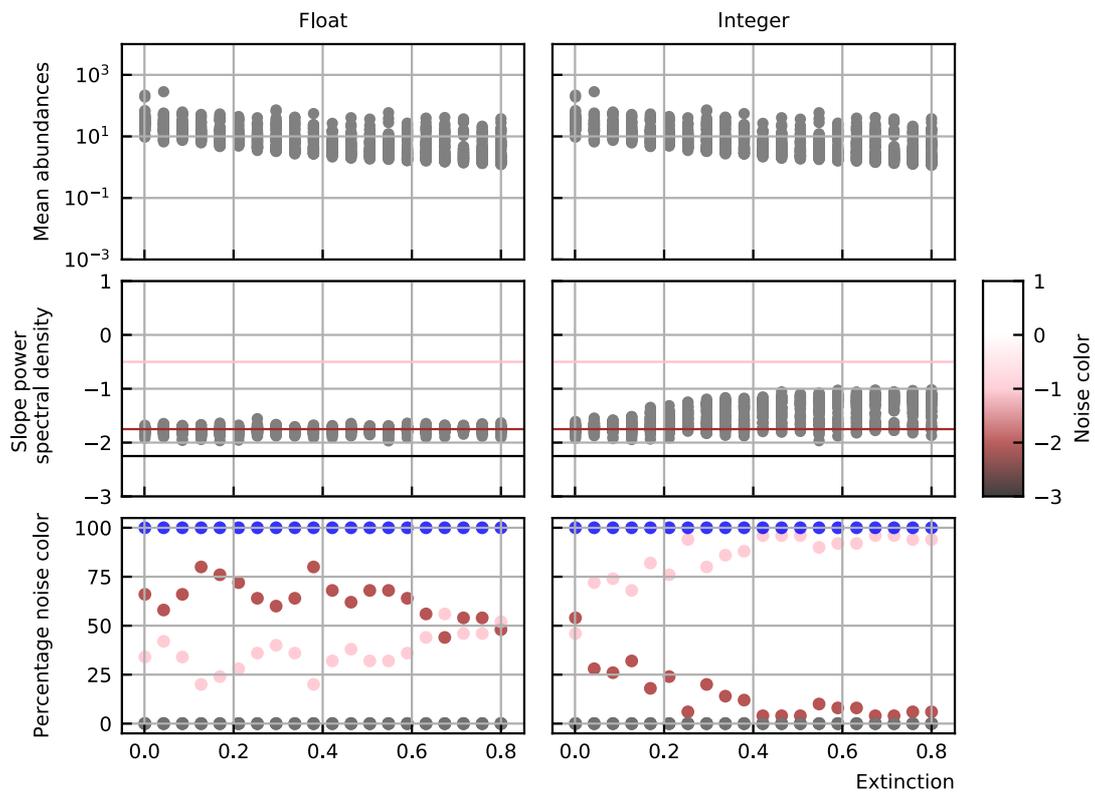

**Figure D.24:** When imposing different constraints—addition of immigration, negative growth rates, small self-interaction, small precision—on the gLV model, the noise color shifts for increasing extinction as in the SOI model.

# Supplementary material of "Heavy-tailed abundance distributions with stochastic generalized Lotka-Volterra models"

E

## E.1 Individual-based models

The concepts speak for themselves, individual-based model models (IBM) consider individuals whereas population-level models (PLM) consider populations. Individual-based models are a useful tool when one wants to take into account the differences between individuals even when they belong to the same species. They have the set-up of an automaton, *i.e.* all individuals obey a some simple rules. Although this set of rules determines only the fate of individuals, patterns can emerge on larger scales.

**Model of Solé**

We, here, shortly present the IBM of Solé et al., 2002 (called model B in the reference) which leads to heavy-tailed distributions and the species-diversity relation. This model considers a lattice filled with a fixed number of lattice sites. Every site can either be empty or harbor one individual of a species. There is a propagule supply defined with all species present, such that individuals from any species can immigrate. With $\Sigma = \{1, 2, \ldots, S\}$ the set of all species and 0 to denote the empty site as a pseudo-species, three possible transitions are defined:

1. *Immigration*: a individual of species $A \in \Sigma$ can immigrate and replace one empty site 0 with probability $\mu_A$

$$0 \xrightarrow{\mu_A} A$$

2. *Death*: one individual of species $A \in \Sigma$ can die or emigrate, thereby emptying a lattice site. This occurs with probability $e_A$

$$A \xrightarrow{e_A} 0$$

3. *Interaction*: An interaction matrix $\Omega_{ij}$ determines the effect of species $i$ on species $j$. For two random individuals of species $A \in \Sigma$ and $B \in \Sigma$, interaction occurs when $\Omega_{AB} > \Omega_{BA}$ with probability $\Omega_{AB} - \Omega_{BA}$. The interaction consists of an individual of $A$ replacing one of $B$

$$A + B \xrightarrow{\Omega_{AB} - \Omega_{BA}} 2A.$$

Notice that there is no growth process, no self-interaction and that the implementation of the interactions is such that there are effectively only predator-prey interactions. The equivalent interaction matrix in deterministic models would be an anti-symmetric matrix.

Because the details of the exact implementation of the model are not clear from the manuscript of Solé et al., 2002 and because growth processes, self-interaction, and interactions other than predation are absent in this model, we use a derivative of this model proposed by Heyvaert, 2017

## Model of Heyvaert

The algorithm we here describe was developed by Heyvaert, 2017 based on the model of Solé et al., 2002. We consider a system with $S$ different species and a maximal number of individuals $N$. This is equivalent to a system with a lattice with $N$ sites that can each be occupied by at most one individual. We will use this image of the lattice with species and empty sites to explain the algorithm as it is easy to understand, although the spatial aspect of the lattice is unimportant for our model and no spatial dependences are taken into account. The state of the system can be described by $\vec{x} = (x_0, x_1, x_2, ..., x_S)$ where $x_i$ represents the number of individuals of species $i$ and $x_0$ the number of empty lattice sites, equivalently we can describe the system by a lattice $X$. By the construction of the model the following constraint is satisfied at all time steps:

$$\sum_{i=0}^{S} x_i = N. \tag{E.1}$$

At every time step, the state vector of the system is adapted by a given number of events similar to a Monte Carlo simulation. The algorithm is repeated for a fixed number of time steps. The length of every time step is equal. The simulation rules for every time step are explained in the next paragraph.

At every time step, $N$ sites of the lattice $X(t)$ are visited at random:

▶ If the chosen site is empty, there is a possibility for *immigration*. One species of the species is pool is chosen and the probabilities for all species are proportional to their immigration probabilities $\mu_i$. The immigration event occurs with a probability $\mu_a$. In the case of an immigration event by species $a$, the state of the system at the subsequent time point is changed:

$$x_0 \mapsto x_0 - 1$$
$$x_a \mapsto x_a + 1$$

▶ If the chosen site is occupied by an individual of species $a$, a second site is chosen from the lattice at time $t$, $X(t)$.

- *Interaction:* If the second site is an individual of species $b$, this individual has a probability of interacting with the first individual of species $a$. The probability of interaction is $|\omega_{ab}|$. For $\omega_{ab} = 0$, there is no interaction. For $\omega_{ab} \neq 0$, the sign of the interaction coefficient $\omega_{ab}$ denotes whether the abundance of species $a$ will grow or decrease:

$$\left.\begin{array}{r} x_0 \mapsto x_0 \pm 1 \\ x_a \mapsto x_a \mp 1 \end{array}\right\} \text{ if } \omega_{ab} \lessgtr 0$$

- *Growth:* If the second site is empty, the abundance of species $a$ can grow with a probability denoted by the growth rate $r_a$:

$$x_0 \mapsto x_0 - 1$$
$$x_a \mapsto x_a + 1$$

- *Extinction:* If the individual of first site did not die through interaction, it can still die with probability $e_a$:

$$x_0 \mapsto x_0 + 1$$
$$x_a \mapsto x_a - 1$$

To calculate the state of the system at time point $t + 1$, we start from the state at time $t$ and subsequently visit $N$ times a random site of the lattice at time point $t$. Because the process is random, one site can be visited twice and another is not considered at a given time step. The events associated with this lattice site change the state at time $t + 1$. Because none of the species numbers or the number of empty sites can become negative, the events can only occur if they do not violate this assumption. The implementation of this algorithm can be found at https://github.com/lanadescheemaeker/sglv_timeseries.

## E.2 Fit distributions

### Power law distribution

As discussed in Appendix 2, there are multiple descriptions of a power law distribution defined on different domains. In Chapter 9, we fit all three distribution—power law distribution, pareto distribution, and truncated power law distribution—and choose the distribution with the best fit for the Kolmogorov-Smirnov test (see Appendix B.3). The fit of the three distributions is done by the scipy.stats.powerlaw and scipy.stats.pareto of the scipy package, and the powerlaw package in Python, respectively. Performing a linear fit through the logarithm of the values using a least square estimator would be statistically uncorrect. The most robust method to fit a power law is the maximal likelihood estimator (Goldstein et al., 2004). This is the method used by the python functions we cited previously.

### Lognormal distribution

We used the scipy.stats.lognorm function of the scipy package in Python to fit lognormal distributions. Before fitting the abundance distribution, we scaled all values such that the median is 1. We then imposed $\mu = 0$ and $x_0 = 1$ in Equation 9.1.

### Normal distribution

We used the scipy.stats.norm function of the scipy package in Python to fit normal distributions.

### Gamma distribution

The Gamma PDF is described by

$$f_\Gamma(x) = \frac{x^{\beta-1}}{\Gamma(\beta)} \left(\frac{\beta}{\bar{x}}\right)^\beta \exp\left(\frac{-\beta x}{\bar{x}}\right) \tag{E.2}$$

where $\bar{x}$ represents the average of $x$ and $\beta = (\bar{x}/\sigma_x)^2$ is the inverse coefficient of variation squared. To fit the Gamma distributions, we used the scipy.stats.gamma function of the scipy package in Python.

## E.3 Derivations of inverses and ratio distributions

In nature, we find variables that are the inverse or ratios of other variables. Given the probability density function (PDF) of the random variables, the PDF of the inverse or ratio can be calculated. The former is also called the *reciprocal distribution*. In this section we focus on the inverse of the uniform distribution, the ratio of uniform distributions and the ratio of lognormal distributions.

### Inverse of uniform distribution

If $z = 1/x$, a random variable, we can determine the PDF of $z$, $f_Z$, given the distribution of $x$, $f_X$. For the derivation, we rely on the cumulative density functions (CDFs) $F_X$ and $F_Z$:

$$F_Z(z) = \Pr(Z \leq z) = \Pr(X \geq z^{-1}) = 1 - \Pr(X < z^{-1}) = 1 - F_X(z^{-1}). \quad \text{(E.3)}$$

Because the probability density function $f$ is the derivative of the cumulative distribution $F$, we have

$$f_Z(z) = z^{-2} f_X(z^{-1}). \quad \text{(E.4)}$$

Therefore, if $x$ is a random variable uniformly distributed between $a$ and $b$ ($X \propto \mathcal{U}(a,b)$), or

$$f_X(x) = \begin{cases} \frac{1}{b-a} & a \leq x \leq b, \\ 0 & \text{otherwise,} \end{cases} \quad \text{(E.5)}$$

we obtain

$$f_Z(z) = \begin{cases} \frac{1}{z^2(b-a)} & 1/b \leq x \leq 1/a, \\ 0 & \text{otherwise.} \end{cases} \quad \text{(E.6)}$$

### Ratio of uniform distributions

Suppose we have a variable $Z = Y/X$ which is the ratio of variables $X$ and $Y$, then the probability distribution of $Z$ can be determined from the joint probability of $X$ and $Y$, $f_{X,Y}$. We will, in our derivation, assume that the distributions of $X$ and $Y$ are independent such that $f_{X,Y} = f_X \cdot f_Y$. To determine the distribution of the ratio $Z$, we use the following formula (Curtiss, 1941):

$$f_Z(z) = \int_{-\infty}^{+\infty} |x| f_X(x) f_Y(xz) dx. \tag{E.7}$$

With $X$ and $Y$ uniformly distributed between 0 and a and b respectively— $X \propto \mathcal{U}(0,a), Y \propto \mathcal{U}(0,b)$, we have

$$f_X(x) = \begin{cases} \frac{1}{a} & 0 \leq x \leq a, \\ 0 & \text{otherwise}, \end{cases} \tag{E.8}$$

and

$$f_Y(y) = \begin{cases} \frac{1}{b} & 0 \leq y \leq b, \\ 0 & \text{otherwise}, \end{cases} \tag{E.9}$$

and the distribution of $Z$ becomes

$$f_Z(z) = \int_0^a |x| f_X(x) f_Y(xz) dx \tag{E.10}$$

$$= \begin{cases} \int_0^a \frac{|x|}{ab} dx = \frac{a}{2b} & \text{for } 0 \leq z < \frac{b}{a} \\ \int_0^{b/z} \frac{|x|}{ab} dx = \frac{b}{2az^2} & \text{for } z \geq \frac{b}{a}. \end{cases} \tag{E.11}$$

### Ratio and inverse of lognormal distributions

Suppose $X$ and $Y$ follow a lognormal distributed, such that $\ln(X)$ and $\ln(Y)$ describe normal distributions. The ratio distribution of these lognormal distributions $Y/X$ can also be written as $\exp(\ln(Y/X)) = \exp(\ln(Y) - \ln(X))$. If we define the mean of $\ln(X)$ as $\mu_X$ and similarly for $\ln(Y)$, and the covariance matrix between both distributions is

$$\text{cov} = \begin{bmatrix} \sigma_X^2 & \sigma_{XY} \\ \sigma_{XY} & \sigma_{X'}^2 \end{bmatrix} \tag{E.12}$$

then $Z = \ln(Y) - \ln(X)$ is normally distributed with mean $\mu_Y - \mu_X$ and variance $\sigma_X^2 + \sigma_Y^2 - 2\sigma_{XY}$. Therefore, $Y/X = \exp(Z)$ follows a lognormal distribution.

Following the same logic, one obtains that the inverse distribution of a lognormal distribution is a lognormal distribution:

$$\ln(X) \propto \mathcal{N}(\mu, \sigma^2) \Leftrightarrow \ln(1/X) \propto \mathcal{N}(-\mu, \sigma^2). \tag{E.13}$$

## E.4 Supporting results

### The JS distance between cross-sectional data ranges from 0.4 to 0.5.

We compare the gut microbial composition of different individuals using the Sanger, Illumina, and Pyroseq datasets (see Table D.1 for the references). The median JS distance between compositions of distinct individuals is between 0.4 and 0.5 (Figure E.1).

### The width of the lognormal distribution depends on the maximal number of individuals in IBMs.

The width of the lognormal distribution increases for smaller maximum numbers of individuals and immigration (Figure E.2). As the width of the lognormal increases, the number of species decreases. Simultaneously, the average coefficient of variation and the average JS distance increase.

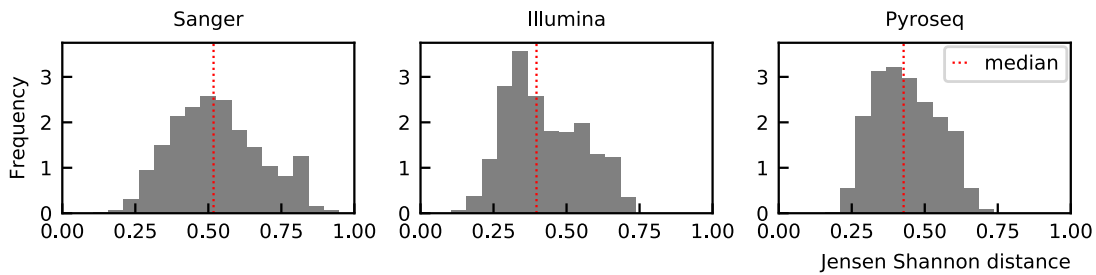

**Figure E.1:** Jensen-Shannon distances between different individuals of datasets Sanger, Illumina and Pyroseq. The median JS distance is 0.4 to 0.5.

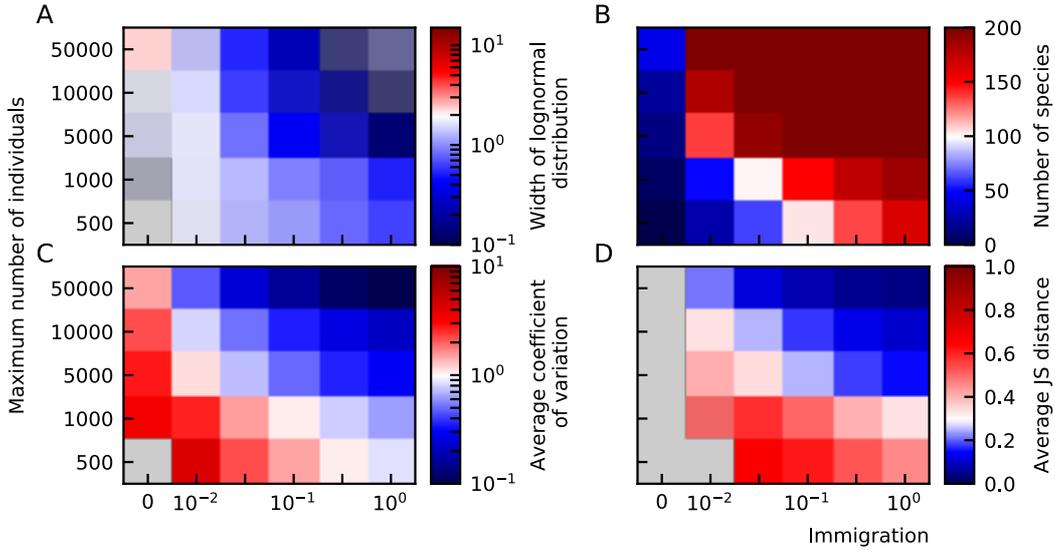

**Figure E.2:** Individual-based models result in heavy-tailed abundance distributions. (A) The width of the lognormal distribution increases for smaller maximum numbers of individuals and immigration. As the width of the lognormal increases, the number of species decreases (B). Simultaneously, the average coefficient of variation (C) and the average JS distance (D) increase. The fixed parameters of these plots are the interaction strength 0.5 and the connectance 0.5.

## In the absence of interactions and immigration, maximal capacity rescales the steady state.

For gLV equations (Equation 9.6) without immigration, we know that the non-trivial steady state is given by $x_i^\dagger = -\sum_j (\omega^{-1})_{ij} \cdot g_j$. This state is possibly non-feasible if the steady state of one of the species is negative. Because we know that the logistic equations with a maximal capacity reduce to gLV equations, we can calculate the steady state for logistic equations with a maximal capacity. The interaction matrix is

$$\omega_{ij} = -\frac{g_i^+}{N_{\max}} - \beta_{ii}\delta_{ij} \tag{E.14}$$

where $\delta_{ij}$ is the Kronecker delta. The Sherman-Morrison formula allows us to invert this interaction matrix and we obtain the expression for the steady state (Sherman & Morrison, 1950)

$$x_j^* = \beta_j^{-1} g_j (1 + \sum_i \beta_i^{-1} g_i / N_{\max})^{-1}. \tag{E.15}$$

Because the steady state without maximal capacity is $g_i/\beta_i$, we see that the maximal capacity rescales the steady state by a factor $(1 + \sum_i x_i^\dagger / N_{\max})^{-1}$. Therefore the evenness of the species abundances or the steepness of the rank abundance is not changed when a maximal capacity is added in the absence of immigration and interactions.

## Intra-competition must be stronger than inter-competition for stability

Consider a system of gLV equations

$$\dot{x}_i = (g_i + \sum_j \omega_{ij} x_j) x_i \tag{E.16}$$

The Jacobian of this equation is

$$J_{ij} = x_i^* \omega_{ij} \tag{E.17}$$

where $x_i^*$ represents the steady-state value of $x_i$. For all species equivalent the Jacobian takes the shape

$$J = x^* \begin{bmatrix} d & o & \cdots & o \\ o & d & \cdots & o \\ \vdots & \vdots & \ddots & \vdots \\ o & o & \cdots & d \end{bmatrix} \tag{E.18}$$

where the diagonal elements $d$ denote the self-interaction and the off-diagonal elements $o$ the interaction between species. The eigenvalues of this matrix are $x^*(d - o)$ and $x^*((N - 1)o + d)$. Stability thus requires that $d < o$ for otherwise the first eigenvalue is positive and $d < o < 0$ to keep the second eigenvalue negative. When the inter-species competition is larger than the self-interaction, i.e. $d > o$ (since both values are negative), the system is unstable and a small perturbation of the steady state will result in one of the species taking over (winner-take-all). Because the Jacobian of the logistic model with maximal capacity is given by Equation E.14, and we have $-\frac{g_i^+}{N_{max}} - \beta_{ii} < -\frac{g_i^+}{N_{max}} < 0$, the logistic model with maximal capacity is stable.

**Smaller maximal capacity leads to slower dynamics.**

The steady state of Equation 9.9 is

$$x_i^* = \frac{g_i}{\beta_i} + \frac{g_i^+}{\beta_i} \frac{\sum_j g_j \beta_j^{-1}}{N_{\max} - \sum_j g_j^+ \beta_j^{-1}}. \tag{E.19}$$

When $g = g^+$, this simplifies to $x_i = y_i/(1 + \sum_j y_j/N_{\max})$ where $y$ is the steady-state value of $x$ in the absence of a maximal capacity $y_i = g_i \beta_i^{-1}$. The speed of the dynamics to steady state can be estimated by evaluating $\frac{d\dot{x}_i}{dx_i}$ in $x = x^*$. The derivative is

$$\frac{d\dot{x}_i}{dx_i} = g_i - \frac{g_i^+}{N_{\max}}\left(\sum_j x_j + x_i\right) - 2\beta_i x_i \tag{E.20}$$

Evaluated this equation in the steady state, we estimate the speed to the steady state by

$$\left.\frac{d\dot{x}_i}{dx_i}\right|_{x^*} = \left(-g_i + g_i^+ \frac{\Sigma^- + \Sigma^+}{N_{\max} + \Sigma^+}\right) \cdot \left(1 + \frac{g_i^+}{\beta_i N_{\max}}\right) \tag{E.21}$$

with $\Sigma^\pm = \sum_j g_j^\pm \beta_j^{-1}$. For large $N_{\max}$, the speed reduces to $-g_i$ which is the speed for the normal logistic equation at steady state. If we assume that all species are equivalent $\beta_i = \beta$ and $g_i = g_i^+ = g$ for all $i$ (or equivalently $g_i^- = 0$), we can estimate the speed for small $N_{\max}$ by $-g/N_{\text{spec}}$ with $N_{\text{spec}}$ the number of species (Figure E.3).

## Interaction strength and connectance both make abundance distribution wider.

The interaction matrix is an important factor for the shape of the abundance distribution. Both the interaction strength and connectance strengthen each other (Figure E.5). When the maximal capacity increases, the width of the lognormal distribution increases. In the absence of a maximal capacity, the solution for large interaction and connectance is unstable and many species go extinct.

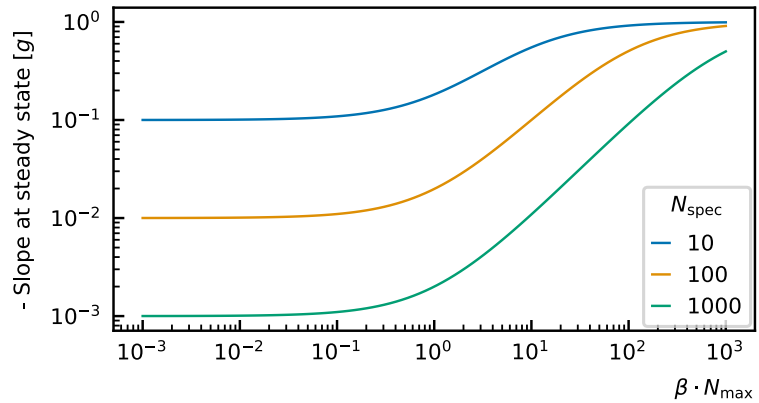

**Figure E.3:** The speed with which a system goes to steady state is related to the slope of the derivative at the steady state. The smaller the maximal capacity the smaller this slope, the lower the speed. The speed is shown for systems where all species are equal ($\beta_i = \beta$, $g_i = g_i^+ = g$ for all $i$).

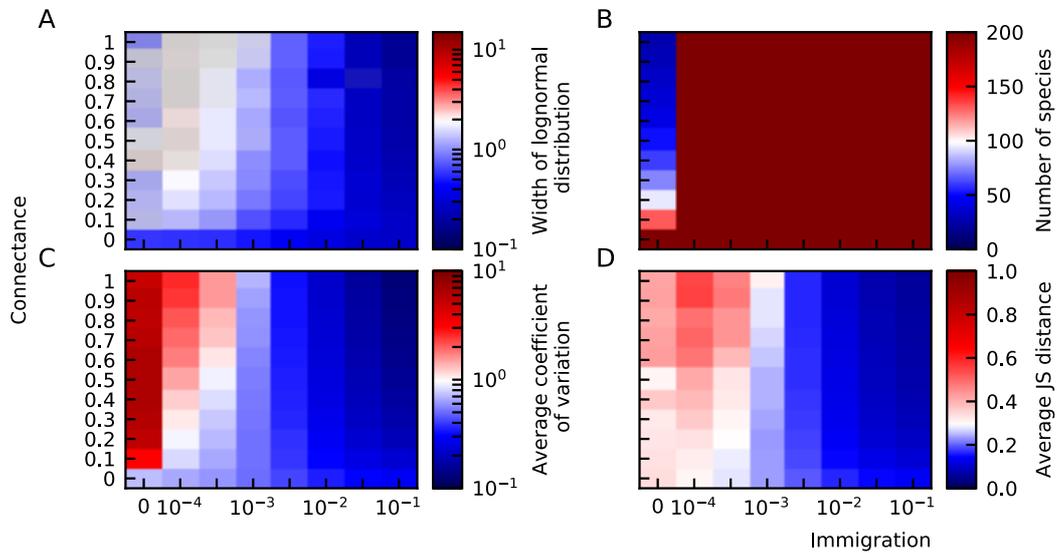

**Figure E.4:** Generalized Lotka-Volterra models with a maximal capacity result in heavy-tailed abundance distributions. (A) The width of the lognormal distribution becomes larger for decreasing immigration and increasing connectance. (B) Diversity is maintained in the presence of immigration. The average coefficient of variation (C) and the average Jensen Shannon distance (D) grow for decreasing immigration. The number of species is $N_{\text{spec}} = 200$. The fixed parameters of these plots are the interaction strength $\omega = 0.5$, the noise strength $\eta = 0.5$ and the maximal capacity 100.

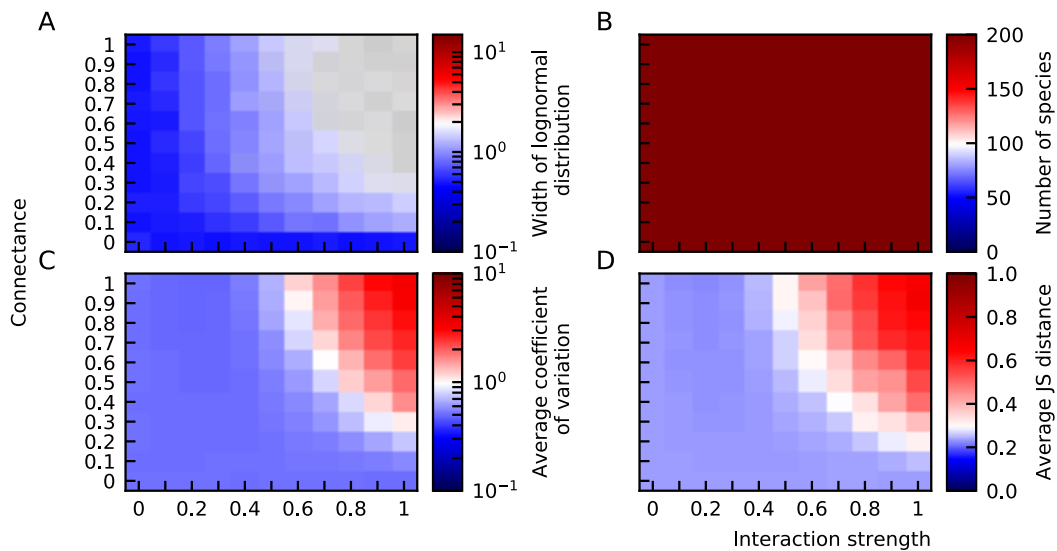

**Figure E.5:** The width of the abundance distribution (A), the average coefficient of variation (C) and average JS distance (D) increase for larger interaction strength and connectance. The diversity is maximal (D). The number of species is $N_{spec} = 200$. The fixed parameters of these plots are strength of the noise $\eta = 0.5$, the immigration rate $\lambda = 0.1$ and the maximal capacity $N_{max} = 100$.

## Experimental data does not fit the ratio of uniform distributions.

For a logistic model with uniformly distributed growth rates and self-interactions, the abundances follow a ratio of uniform distributions (RUD). Its description is given by Equation 9.12 where the parameter $\phi$ is the median of the distribution. Without free parameters, we can assess how likely it is that the abundances follow this distribution. We compare the RUD to a lognormal distribution, they are denoted by the full and dashed lines in Figure E.7. Notice that the latter distribution needs to be fitted. We asses the likelihood for both distributions by the Kolmogorov-Smirnov test (see Appendix B.3) and conclude that we can reject the hypothesis that the abundances follow a RUD for many of the experimental distributions (blue lines) and that the lognormal distribution is more likely for all experimental data but the microbiome of the female palm.

One important element is that not all reads of the sample could be classified and that a considerable percentage of the data is missing. In the Sanger, Illumina and Pyroseq data, it amounts up to 50%. When we assume that these reads come from species that are undersampled and have a lower abundance than the smallest abundance measured, we need to add a lot of small abundance species to the histogram shifting the median value to lower values and, therefore the knee of the RUD to the left. In that case, the hypothesis that the abundance distribution follows a RUD can be rejected for all experimental data considered.

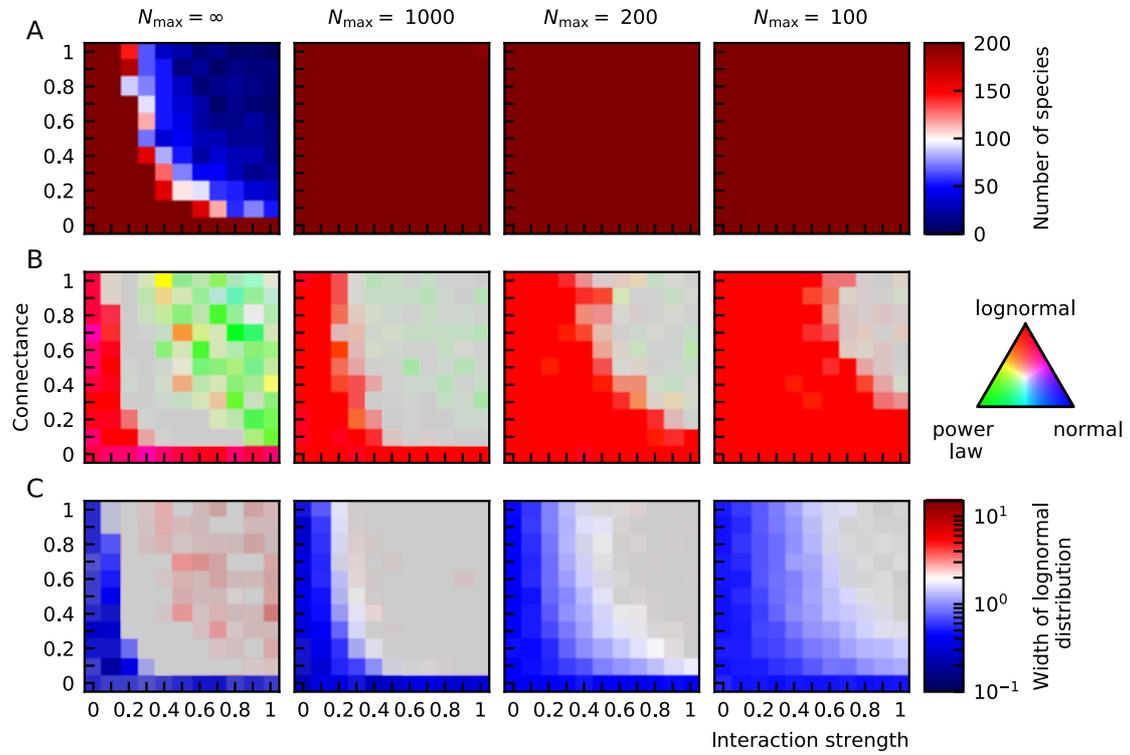

**Figure E.6:** The width of the lognormal distribution as a function of the interaction strength and connectance for different values of the maximal capacity $N_{\max}$. The grey areas denote unstable solutions. Smaller maximal capacities allow for stable solutions with more and larger interactions. More and larger interactions lead to wider abundance distributions.

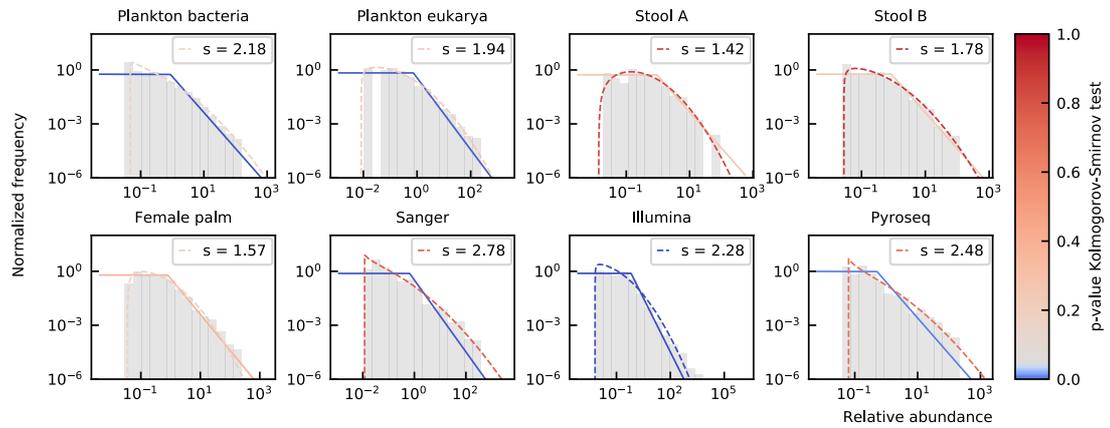

**Figure E.7:** Experimental data compared to their RUD curve using $\phi$ as the median value of the measured abundances.

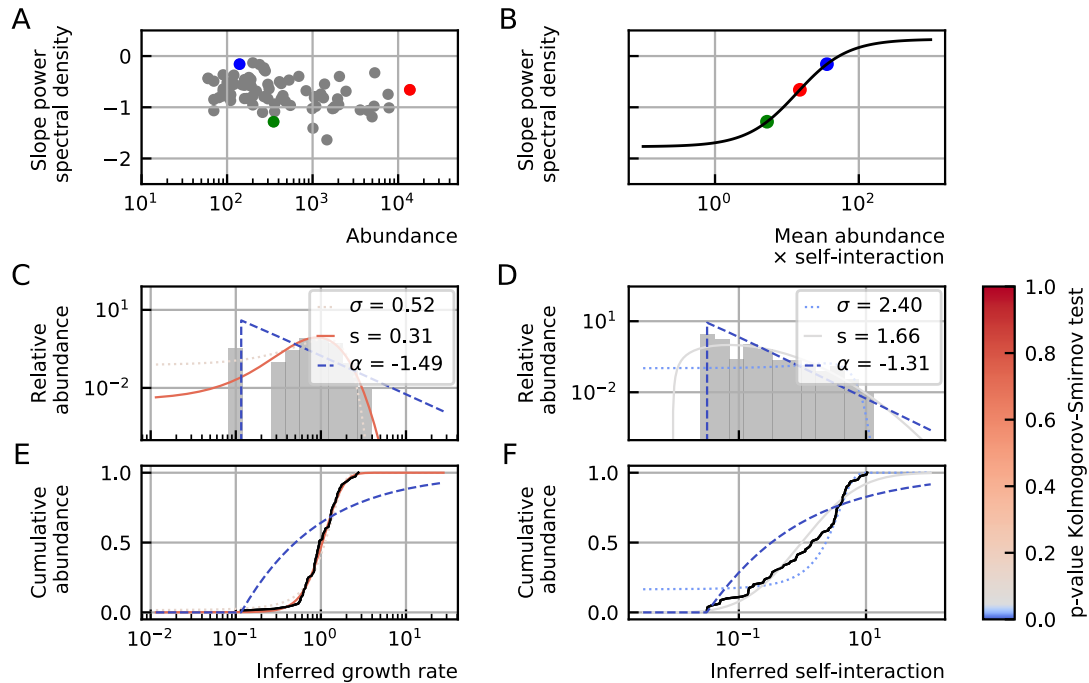

**Figure E.8:** (A) The noise color as a function of the abundance for the David Stool A data. (B) The simulated relation between the noise color, mean abundance and self-interaction. (C) The self-interaction inferred from the noise color and abundance. (D) The distribution of the inferred self-interactions fitted by a normal (dotted line), lognormal (full line) and power law (dashed line). The color of the lines denotes the p-value of the Kolmogorov-Smirnov test. The lognormal curve has the best fit. (E) The cumulative density function of the self-interactions (black line) with the fitted lines corresponding to the ones in (E)

## Self-interactions derived from noise color are compatible with lognormal distribution.

In Chapter 8, we have shown that, assuming the experimental data can be modeled by logistic equations, the self-interaction rate can be determined given the abundance and noise color, which is defined as the slope of the power spectral density (Figure E.8B). We determined the self-interactions and growth rates assuming a certain sampling rate ($\delta t = 0.05$), and concluded that these results are not incompatible with a lognormal distribution for both variables. This result is not highly surprising, because the abundance distribution of the experimental data, which is lognormal, was used to infer the parameters. Nevertheless, it shows that the results of the noise color are compatible with lognormal distributions of the parameters.

# List of abbreviations

**2DS**  two dimer system.
**3DS**  three dimer system.

**CDF**  cumulative distribution function.

**DNA**  deoxyribonucleic acid.

**gLV**  generalized Lotka-Volterra.

**IBM**  individual-based modeling.

**JS**  Jensen-Shannon.

**KL**  Kullback-Leibler.
**KS**  Kolmogorov-Smirnov.

**MDS**  monomer dimer system.
**mRNA**  messenger ribonucleic acid.

**ODE**  ordinary differential equation.
**OTU**  operational taxonomic unit.

**PDF**  probability density function.
**PLM**  population-level modeling.

**RAD**  relative abundance distribution (not to be confused with rank abundance distribution).
**RUD**  ratio of uniform distributions.

**Ss-LrpB**  Leucine responsive protein B of the *Sulfolobus solfataricus*.

# Alphabetical Index

# Figures credits